\documentclass [12pt,twoside,openright,a4paper]{reportnew}
\usepackage [dvips]{graphicx}
\usepackage {amssymb}
\usepackage {verbatim}
\usepackage {amsmath}
\usepackage {mathrsfs}

\usepackage{Ind_in_Ind} %Inserisce l'indice in se stesso

\usepackage{Frontispiece_cita} %Fa il frontespizio ed altre cosine

\newcommand{\be}{\begin{eqnarray}}
\newcommand{\ee}{\end{eqnarray}}

\newcommand{\expecl}[1]
{\mbox{$\left\langle\,\strut\displaystyle{#1}\,\right\rangle$}}

\renewcommand{\d}{\mbox{${\rm d}$}}

\begin{document}
\title{Semiclassical Approximations to Cosmological Perturbations}
\author{Mattia Luzzi}

\supervisor{Prof. Giovanni Venturi}

\corelator{Dott. Roberto Casadio}

\cocorelator{Dott. Fabio Finelli}

\cocorelatorBIS{Dott. Alexander Kamenshchik\\
\vspace{0.5cm}
Sigla del Settore Scientifico Disciplinare: FIS/02\\
\vspace{0.3cm}
Visto del Coordinatore di Dottorato, Prof. Fabio Ortolani}

\sess{XIX}

\yrs{2005-2006}

\cita{``Would you tell me, please, which way I ought go from here?''\\
``That depends a good deal on where you want to get to'', said the Cat.\\
``I don't much care where -'' said Alice.\\
``Then it doesn't matter which way you go,'' said the Cat.\\
``- so long as I get somewhere,'' Alice added as an explanation.\\
``Oh, you're sure to do that,'' said the Cat, ``if you only walk long enough.''}

\autorcita{from ``Alice's Adventures in Wonderland'' by Lewis Carroll}

\ackno{I would like to thank my supervisor Prof. Giovanni Venturi
for his support and encouragement. I would also like to thank the whole INFN Bologna group BO11.
 I warmly thank Dr. Roberto Casadio for being always so kind and
ready--to--help me in any respect. For the many interactions I had and specially for the
fact that working with them was lots of fun, I want to express my acknowledgement to
Dr. Fabio Finelli and Dr. Alexander Kamenshchik.
Our fruitful collaboration forms the basis of much of the work presented here.}

\beforepreface

\nonumchapter{Introduction}
One of the main ideas of modern cosmology is that there was an epoch
in the history of the Universe when vacuum energy dominated other forms
of energy density such as radiation or matter.
During this vacuum-dominated (or potential-dominated) era, the scale factor grew
exponentially (or almost exponentially) in time and, for this reason, it is known as
{\em inflation\/}.
During inflation, a small and smooth spatial region of size less than the Hubble radius at
that time could have grown so large to easily encompass the comoving volume
of the entire Universe presently observable.
If the early Universe underwent a period of so rapid expansion,
one can then understand why the observed Universe is homogeneous
and isotropic to such a high accuracy.
All  these virtues of inflation were already noted when it was first proposed in a seminal paper
by Guth in 1981~\cite{guth}.
Other, and more dramatic consequences of the inflationary paradigm were discovered
soon after.
Starting with a Universe which is absolutely homogeneous and isotropic at the classical
level, the inflationary expansion will ``freeze in'' the vacuum fluctuations of the ``inflaton''
(the field responsible for the rapid expansion) which then becomes a classical quantity.
On each comoving scale, this happens soon after the fluctuation exits the Hubble horizon.
A primordial energy density, which survives after the end of inflation, is associated with these vacuum
fluctuations and may be responsible for both the anisotropies in the Cosmic Microwave Background
(CMB) and for the large-scale structures of the Universe, like galaxies, clusters of galaxies, and dark matter.
Inflation also generates primordial gravitational waves, which may contribute
to the low multipoles of the CMB anisotropy.
The amazing prediction of inflation is therefore that all of the
structures we see in the present Universe are the result of quantum-mechanical
fluctuations during the inflationary epoch.
\markboth{Introduction}{}
\nonumchapter{Units and Conventions}
The so--called natural units, namely $\hbar = c = 1$, will be employed, so that
$$[{\rm Energy}]=[{\rm Mass}] =[{\rm Length}]^{-1}=[{\rm Time}]^{-1}~.$$
$m_{\rm Pl} =1/ \sqrt{G}$ is the Planck mass, whereas 
$M_{\rm Pl} =1/ \sqrt{8\,\pi\,G}$ stands for the reduced Planck mass,
and $\ell_{\rm Pl}=\sqrt{8\,\pi}\,M^{-1}_{\rm Pl}$
is the Planck length.
The present Hubble expansion rate is customarily parameterised as
$H_0 = 100\,h\,{\rm km}\,{\rm sec}^{-1}\,{\rm Mpc}^{-1}$ where $h$
parameterises the uncertainty in the present value of the Hubble parameter.
\par
We shall also adopt the Einstein sum rule for repeated indices and the following conventions:
The metric signature is ($+, -, -, -$),
Greek letters denote space-time indices which run from 0 to 3
and Latin letters denote spatial indices which run from 1 to 3.
\par
\vspace{1cm}
\noindent Some abbreviations used in this thesis are the following:

\vspace{0.5cm}
\noindent CMB = Cosmic Microwave Background

\noindent EW = electroweak

\noindent BBN = Big Bang Nucleosintesis

\noindent GR = General Relativity

\noindent COBE = COsmic Background Explorer satellite

\noindent DMR = Differential Microwave Radiometer

\noindent WMAP = Wilkinson Microwave Anisotropy Probe

\noindent ACBAR = Arcminute Cosmology Bolometer Array Receiver

\noindent VSA = Very Small Array

\noindent CBI = Cosmic Background Imager

\noindent $\Lambda$ CDM = $\Lambda$Cold Dark Matter

\noindent FRW= Friedmann-Robertson-Walker

\noindent WKB = Wentzel-Kramers-Brillouin

\noindent MCE = Method of Comparison Equation

\noindent PS = Power Spectrum

\noindent HFF = Horizon Flow Functions

\noindent SR = Slow Roll

\noindent HSR = Hubble Slow Roll

\noindent PSR = Hubble Slow Roll

\noindent GFM = Green Function Method

\markboth{Units and Conventions}{}

\chapter{What's the matter?}
\label{inflation}
What is the origin of the inhomogeneities we observe  in the sky?
What is their typical wavelength?
How come that we are able to observe these inhomogeneities?
The term inhomogeneity is rather generic and indicates fluctuations both
in the background geometry and in the energy content of the Universe.
These inhomogeneities can be represented by  plane waves characterized by a
comoving wave-number $k$.
Since the evolution of the Universe is characterized by a scale factor $a(t)$ which is
a function of the cosmic time $t$, the physical wave-number is given by $k_{\rm ph} = k/a$.
The comoving wave-number is a constant and does not feel the expansion of the Universe,
while the physical momentum is different at different epochs.
Conversely the value of a given physical frequency is fully specified only by stating the time
at which the physical frequency is ``measured''.
\par
For example, if a given fluctuation has a momentum comparable with the present value of the Hubble
parameter, $k_{{\rm ph,}0}\simeq H_{0}\simeq5.6\times10^{-61}\, M_{\rm Pl}$,
we will have that $k_{{\rm ph,}0}\simeq2.3\times10^{-18}$ Hz.
Fluctuations with momentum smaller that $H_{0}$ have a wave-length
larger than the present value of the Hubble radius and are therefore
impossible to detect directly since the distance between two of their maxima (or minima)
is larger than our observable Universe.
\par
At the decoupling epoch, when the radiation became free to propagate,
the temperature of the Universe was of the order of a fraction of ${\rm eV}$
and the Hubble rate was $H_{\rm dec}\simeq6.7\times10^{-56}\,M_{\rm Pl}$.
The corresponding decoupling frequency was then red-shifted to the present value
$k_{\rm ph, dec}\simeq3\times10^{-16}\,{\rm Hz}$. 
Fluctuations of this typical frequency appear as 
temperature inhomogeneities of the CMB induced
by the primordial fluctuations in the background geometry and in the matter density.
The latter fluctuations have already collapsed due to gravitational
instability and formed galaxies and clusters of galaxies.
At the epoch of the formation of light nuclei, the Universe was hotter, with a temperature
of a fraction of ${\rm MeV}$ and the corresponding
value of the Hubble expansion rate was about $8\times10^{-44}\, M_{\rm Pl}$.
When the electroweak (EW) phase transition took place,
the temperature of the Universe was of the order of $100\,{\rm GeV}$ leading,
approximately, to $H_{\rm EW}\sim10^{-32}\,M_{\rm Pl}$.
Comoving wave-numbers of the order of the Hubble rate at the EW phase transition
or at the epoch of the Big Bang Nucleosintesis (BBN)
correspond, today, to physical frequencies of the order of
$k_{\rm ph, EW}\sim2\times10^{-5}\,{\rm Hz}$ or $k_{\rm ph, BBN}\simeq9\times10^{-11}\,{\rm Hz}$.
Finally, in the context of the inflationary paradigm the value of the
Hubble rate during inflation was $H_{\rm inf}\sim 10^{-6}\,M_{\rm Pl}$.
Under the assumption that the Universe was dominated by radiation right after inflation,
the corresponding physical frequency today is of the order of $10^{8}\,$Hz.
\par
The detectors developed to measure anisotropies in the CMB are sensitive
to frequency scales only slightly higher than $10^{-16}\,$Hz.
Therefore, they cannot be used to ``observe'' fluctuations originated at earlier epochs and
whose physical frequencies presently lie in the range $10^{-3}\,{\rm Hz}<k_{\rm ph}<1\,{\rm MHz}$.
For frequencies in this interval, one can only hope to detect the stochastic
background of gravitational radiation which is therefore the only means we have to obtain
direct information about the EW phase transition, the BBN and inflation itself.
\par
The cosmological inhomogeneities are usually characterized
by their correlation functions.
The simplest correlation function encoding informations on the nature
of the fluctuations is the two-point function which is computed by taking averages
of the fluctuation amplitude at two spatially separated points but at the same time.
Its Fourier transform is usually called the {\em power spectrum}.
On considering the physical scales discussed above,
one is led to conclude that the power spectrum of cosmological inhomogeneities
is defined over a huge interval of frequencies.
For instance, the tensor fluctuations of the geometry, which only couple to the curvature
and not to the matter sources, have a spectrum
ranging, in some specific models, from $10^{-18}\,$Hz up to
about $1\,$GHz.
\par
From the history of the Universe sketched so far, some questions may be naturally raised.
For our purposes, we just quote two of them:
\begin{itemize}
\item
Given that the precise thermodynamical history of the Universe above $10\,$MeV is unknown,
how it is possible to have a reliable framework describing the evolution of the cosmological
fluctuations?
\item
Does it make sense to apply General Relativity (GR) to curvatures which
can reach up to almost the Planck scale?
\end{itemize}
As recalled above, we do not have any direct evidence for the thermodynamical state
of the Universe for temperatures above the MeV, although some indirect predictions
can be made.
For example, the four lighter isotopes (${\rm D}$, $^{3}{\rm He}$, $^{4}{\rm He}$ and $^{7} {\rm Li}$)
should have been mainly produced at the BBN below a typical temperature of $0.8\,$MeV
when neutrinos decouple from the plasma and the neutron abundance
evolves via free neutron decay.
The abundances calculated in the
homogeneous and isotropic BBN model agree fairly well with the
astronomical observations.
This is the last indirect test of the
thermodynamical state of the Universe.
For temperatures larger than $10\,$MeV, collider physics and the success of the
standard model of particle physics offer a precious handle up to the epoch of the
EW phase transition occurring for temperatures
of the order of $100\,$GeV.
There is then hope that our knowledge
of particle physics could help us to fill the gap between the epoch
of nucleosynthesis and that of the EW phase transition.
\par
The evolution of inhomogeneities
over cosmological distances may well be unaffected by our ignorance of the
specific history of the expansion rate.
The key concept is that of conservation laws and,
in the framework of metric theories of gravity (and of GR
in particular), one can show that there exist conservation laws for the evolution
of the fluctuations which become exact in the limit
of vanishing comoving frequency $ k\to0$.
This conclusion holds under the assumption that GR
is a valid description throughout the whole evolution of the Universe
and similar properties also hold, with some modifications, in scalar-tensor theories 
of gravity (of the Brans-Dicke type) like those suggested by the low-energy
limit of superstring theory.
The conservation laws of cosmological perturbations apply
strictly in the case of the tensor modes of the geometry whereas,
for scalar fluctuations which couple directly to the (scalar) matter sources,
the precise form of the conservation laws depends upon the matter content
of the early Universe.
Experiments aimed at the direct detection of stochastic backgrounds
of relic gravitons which regard the high-frequency behaviour of cosmological fluctuations
should therefore yield valuable clues on the early evolution of the Hubble rate.
\section{A short (incomplete) introduction to a long history}
The history of the studies about the CMB anisotropies and cosmological
fluctuations is closely related with the history of the standard
cosmological model (see, for instance, the textbooks by Weinberg \cite{weinberg}
and Kolb and Turner \cite{KTbook}).
\par
During the development of the standard model of  cosmology, one of the recurrent
issues has been the explanation of the structures we observe in the present Universe,
which goes under the name of the structure formation.
The reviews of Kodama and Sasaki~\cite{KS} and of Mukhanov, Feldman and
Brandenberger \cite{mukh_thory} contain some historical surveys that are a valuable
introduction to this subject.
\par
A seminal contribution to the understanding of cosmological perturbations is the paper of
Sachs and Wolfe~\cite{sachs} on the possible generation of anisotropies in the CMB,
where the basic general relativistic effects were estimated assuming that the CMB itself
was of cosmological origin.
The authors argued that galaxy formation would imply $\Delta T /T\sim10^{-2}$.
Subsequent calculations~\cite{silk1a,silk1b,PeeblesYu,doro,wilson} brought
the estimate down to $10^{-4}$, closer to what is experimentally observed.
A second foundamental paper is the contribution of Sakharov~\cite{sakharov}
where the concept of acoustic oscillations was introduced.
This paper contained two ideas whose legacy, somehow unexpectedly,
survived until the present.
The first idea is summarized in the abstract
\vspace{0.2cm}
\par
{\em A hypothesis of the creation of astronomical bodies as a result
of gravitational instability of the expanding Universe is investigated.
It is assumed that the initial inhomogeneities arise as a result of {\rm quantum fluctuations}...}
\vspace{0.2cm}
\par\noindent
The model indeed assumed that the initial state of the baryon-lepton fluid was cold.
So the premise was wrong, but, still, it is amusing to notice that a variety of  models today
assume quantum fluctuations as initial conditions for the fluctuations of the geometry
at a totally different curvature (or energy).
\par
To summarise, Refs.~\cite{sachs} and~\cite{sakharov} highlight three important problems
regarding the fluctuations of the gravitational inhomogeneities:
\begin{itemize}
\item their initial conditions;
\item their evolution;
\item their imprint on the CMB temperature fluctuations.
\end{itemize}
\section{CMB experiments}
Although we shall just study a particular aspect of the theory of cosmological perturbations,
it is appropriate to give a swift account of our recent experimental knowledge of CMB anisotropies.
A new twist in the study of CMB anisotropies came from the observations of the COsmic Background Explorer satellite (COBE)
and, more precisely, from the Differential Microwave Radiometer (DMR).
The DMR was able to probe the angular power spectrum $C_{\ell }$~\footnote{We just
recall here that $\ell(\ell +1)\,C_{\ell}/(2\,\pi)$ measures the
degree of inhomogeneity in the temperature distribution per logarithmic interval of $\ell$.
Consequently, a given multipole $\ell$ can be related to a given spatial structure in the
microwave sky:
small $\ell$ will correspond to smaller and high $\ell$ to larger wave-numbers 
around a present physical frequency of the order of $10^{-16} $ Hz.} up to $\ell\simeq25$
(see, for instance, \cite{cobe,cob2} and references therein).
DMR was a differential instrument measuring temperature differences in the microwave sky.
The angular separation $\vartheta$ of the two horns of the antenna is related to the
largest multipole according to the approximate relation $\vartheta\simeq\pi/\ell$.
\par
COBE explored angular separations $\vartheta \gtrsim 7^{0}$ and, after the COBE mission,
various experiments have probed smaller angular separation, i.e., larger multipoles.
A finally convincing evidence of the existence
and location of the first peak in $C_{\ell}$ came from Boomerang~\cite{boom1,boom2},
Maxima~\cite{maxima} and Dasi~\cite{dasi}.
These last three experiments explored multipoles up to $\ell=1000$, determining the
first acoustic oscillation for $\ell \simeq 220$.
Another important experiment was Archeops~\cite{archeops} providing
interesting data for the region characterizing the first rise of the angular spectrum.
\par
\begin{figure}[t!]
\includegraphics[height=10cm,angle=90]{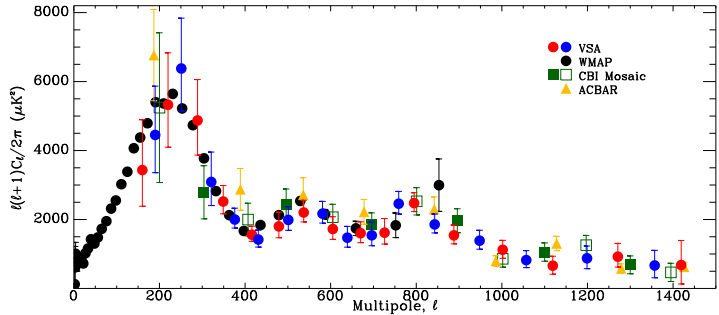}
\caption{Some of the most recent  CMB anisotropy data from~\cite{DICK}:
WMAP (filled circles); VSA (shaded circles)~\cite{DICK}; CBI (squares)~\cite{mason,pearson}
and ACBAR (triangles)~\cite{kuo}.}
\label{F13}
\end{figure}
The angular spectrum, as measured by different recent experiments is reported in
Fig.~\ref{F13}.
The data from the Wilkinson Microwave Anisotropy Probe (WMAP) are displayed as
filled circles in Fig. \ref{F13} and provide, among other important evidences,
a precise determination of the position of the first peak
(i.e., $\ell = 220.1 \pm 0.8$~\cite{WMAP02}) and
a clear evidence of the second peak.
Further, once all the four-years data are collected, the spectrum will be obtained with
good resolution up to $\ell \sim 1000$, i.e., around the third peak. 
In Fig.~\ref{F13}, we also display the results of  the
Arcminute Cosmology Bolometer Array Receiver (ACBAR) as well as the data points of the
Very Small Array (VSA) and of the Cosmic Background Imager (CBI).
\par
In recent years, thanks to the combined observations of CMB anisotropies \cite{WMAP01},
Large scale structure~\cite{LSS01,CL01}, supernovae of type Ia, BBN~\cite{BBN01},
some kind of paradigm for the evolution of the late time (or even present) Universe has
emerged.
This is normally called $\Lambda$Cold Dark Matter ($\Lambda$ CDM) model or, sometimes, ``concordance model".
The term $\Lambda$CDM refers to the fact that, in this model, the dominant
(present) component of the energy density of the Universe is given by a cosmological
constant $\Lambda$ and a fluid of cold dark matter particles interacting only gravitationally
with the other (known) particle species such as baryons, leptons, photons.
According to this paradigm, our understanding of the Universe can be summarized in two sets
of cosmological parameters, one referring to the homogeneous background and
one to the inhomogeneities.
As far as the homogeneous background is concerned, the most important parameter
is the indetermination on the (present) Hubble expansion rate which is customarily
parameterised as $H_{0} = h\times100{\rm km}/[{\rm sec}\times{\rm Mpc}]$ ($h \simeq 0.72$
in the concordance model).
There are various other parameters, such as the dark and CDM energy densities in critical units,
$\Omega_{\rm cdm}\simeq0.224$ and $\Omega_{\Lambda}\simeq0.73$, respectively;
the radiation and baryon energy densities in critical units, $\Omega_{\rm r}\simeq 8\times10^{-5}$
and $\Omega_{\rm b}\simeq 0.046$; the number of neutrino species and their masses;
the possible contribution of the spatial curvature ($\Omega_{\kappa}=0$ in the concordance model)
the optical depth of the plasma at recombination (of the order of $0.16$ in the concordance model).
 \par
The second set of parameters refers to the inhomogeneities.
In the $\Lambda$CDM model the initial conditions for the inhomogeneities
belong to a specific class of scalar fluctuations of the geometry which are called adiabatic
and are characterized by the Fourier transform of their two-point function.
This is the power spectrum which is usually parametrized in terms of an amplitude and
a spectral slope.
Besides these two parameters, one usually considers the ratio between the amplitudes of
the scalar and tensor modes of the geometry.
\par
The purpose of this thesis is to report the development of a method for the resolution of the differential equation that drives the modes of scalar and tensorial perturbations of the cosmological inhomogeneities.
This work is organized as follows: in Chapter~\ref{cosm_pert} we review
the Friedmann-Robertson-Walker (FRW) space-time and remark some aspects about gauge choices.
Chapter~\ref{eq_motion} deals with the process of generation of cosmological perturbations
by quantum fluctuations of a scalar field and a second-order differential equation for the quantities
of interest is obtained.
Such a kind of equations will be treated with semiclassical approximations in Chapter~\ref{WKBa}.
In Chapter~\ref{WKBb} we give new general results for the power spectra of cosmological perturbations, while in Chapter~\ref{WKBresutls} we use those results to find, in detail, explicit expressions for the power spectra in several inflationary models.
The thesis ends with our Conclusions and several Appendicies.

\chapter{FRW Universes and their inhomogeneities}
\label{cosm_pert}
\section{FRW Universes}
A Friedmann-Robertson-Walker (FRW) metric can be written in terms of the cosmic time $t$ as
\be
^{(0)}g_{\mu\,\nu}\,\d x^{\mu}\,\d x^{\nu}
=\d t^2-a^2(t)\,\left[\frac{\d r^2}{1-\kappa\,r^2}
+r^2\,\left(\d\theta^2+\sin^2{\theta}\d\varphi^2\right)\right]
\ ,
\label{FRW_t}
\ee
where $\kappa$ can be $1$, $0$ and $-1$ for spherical, eucledian or hyperbolic
spatial sections, respectively; $a$ is the scale factor, while $r, \theta, \varphi$ are the
spherical coordinates.
We can also write the metric in terms of the conformal time $\eta$,
defined by $\d \eta\equiv\d t/a(t)$, as
\be
^{(0)}g_{\mu\,\nu}\,\d x^{\mu}\,\d x^{\nu}
=a^2(\eta)\,
\left\{\d\eta^2-\left[\frac{\d r^2}{1-\kappa\,r^2}
+r^2\,\left(\d\theta^2+\sin^2{\theta}\d\varphi^2\right)\right]\right\}
\ .
\label{FRW_eta}
\ee
The evolution of the geometry is driven by the matter sources
according to the Einstein equations
\be
G_{\mu\,\nu}
\equiv R_{\mu\,\nu}-\frac{1}{2}\,\delta_{\mu\,\nu}\,R
=8\,\pi\,G\,T_{\mu\,\nu}
\ ,
\label{ein}
\ee
where $G_{\mu\nu}$ is the Einstein tensor, $R_{\mu\nu}$ and $R$ are the Ricci tensor
and Ricci scalar, respectively; $T_{\mu\nu}$ is the energy-momentum tensor of the sources.
Under the hypothesis of homogeneity and isotropy, we can always write the energy-momentum
tensor in the form $T_{\mu\nu}={\rm diag}\left(\rho,p,p,p\right)$ where $\rho$ is
the energy density of the system and $p$ its pressure and are functions of the time.
Using Eq.~(\ref{ein}) we obtain the Friedmann equations which, in conformal time, read
\be
{\cal H}^2=\frac{8\,\pi\,G\,\rho\,a^2}{3}-\kappa
\ ,\quad
{\cal H}^\prime=-\frac{4\,\pi\,G}{3}\,\left(\rho+3\,p\right)\,a^2
\label{FrEQ_eta}
\ee
along with the continuity equation
\be
\rho^\prime+3{\cal H}\,(\rho+p)=0
\ .
\label{contEQ_eta}
\ee
In the above formulae ${\cal H}\equiv a'/a$ and a prime denotes a
derivation with respect to the conformal time.
In cosmic time, the same equations become
\be
H^2=\frac{8\,\pi\,G\,\rho}{3}-\frac{\kappa}{a^2}
\ ,\quad
\dot{H}=-\,4\,\pi\,G\,\left(\rho+p\right)+\frac{\kappa}{a^2}
\label{FrEQ_t}
\ee
and the continuity equation is given by
\be
\dot{\rho}+3\,H\,(\rho+p)=0
\ ,
\label{contEQ_t}
\ee
where $H\equiv\dot{a}/a$ is the Hubble expansion rate and dot denotes a
derivation with respect to $t$.
The relation between the Hubble factors in the two time parametrizations is ${\cal H}=a\,H$.
\section{Fluctuations of the geometry in flat FRW Universes}
Let us now consider a conformally flat (i.e., $\kappa =0$) background metric of
FRW type,
\be
g_{\mu\nu}(\eta)=
a^2(\eta)\,\eta_{\mu\nu}
\ ,
\ee
where $\eta_{\mu\nu}$ is the flat Minkowski metric.
Its first-order fluctuations can be written as
\be
\delta\,g_{\mu\nu}(\eta,\vec{x})=
\delta_{\rm s}\,g_{\mu\nu}(\eta,\vec{x})
+\delta_{\rm v}\,g_{\mu\nu}(\eta,\vec{x})
+\delta_{\rm t}\,g_{\mu\nu}(\eta,\vec{x})
\ ,
\ee
where the subscripts define, respectively, the scalar, vector and  tensor
perturbations. The perturbed metric $\delta\,g_{\mu\nu}$ has ten independent
components whose explicit form can be parametrized as
\be
\delta\,g_{00}&=&
2\,a^2\,\phi
\nonumber\\
\delta\,g_{ij}&=&
2\,a^2\,\left(\psi\,\delta_{ij}-\partial_{i}\,\partial_{j}\,E\right)
-a^2\,h_{ij}-a^2\,\left(\partial_{i}\,F_{j}+\partial_{j}\,F_{i}\right)
\nonumber\\
\delta\,g_{0i}&=&
-a^2\,\partial_{i}\,B+a^2\,S_{i}
\label{g_00_g_ij_g_0i}
\ ,
\ee
with the conditions
\be
\partial_{i}\,S^{i}=\partial_{i}\,F^{i}=0
\ ,\quad
h_{i}^{i}=\partial_{i}\,h^{i}_{j}=0
\ .
\label{div_cond}
\ee
The decomposition given in Eqs.~(\ref{g_00_g_ij_g_0i})
is that normally employed in  the Bardeen formalism~\cite{bardeen}
(see Ref.~\cite{mukh_thory}).
\par
The scalar fluctuations are parametrized by the scalar functions
$\phi$, $\psi$, $B$ and $E$. The vector fluctuations are described by
two divergenceless (see Eqs.~(\ref{div_cond})) vectors in three (spatial)
dimensions $F_{i}$ and $S_{i}$ (i.e., by four independent degrees of freedom).
Finally the tensor modes are described by $h_{ij}$, leading to two independent
components (see Eqs.~(\ref{div_cond})).
\par
Under general infinitesimal coordinate transformations
\be
\eta\to\tilde{\eta}=\eta+\epsilon_{0}
\ ,\quad
{x}^{i}\to\tilde{x}^{i}=x^{i}+\epsilon^{i}
\label{inf_trasf}
\ee
the fluctuations of the geometry defined in Eqs.~(\ref{g_00_g_ij_g_0i}) transform as
\be
\delta\,{g}_{\mu\nu}\to\delta\,\tilde{g}_{\mu\nu}=
\delta\,g_{\mu\nu}-\nabla_{\mu}\,\epsilon_{\nu}-\nabla_{\nu}\,\epsilon_{\mu}
\ ,
\label{inf_trasf_og_g}
\ee
where $\nabla_{\mu}$ is the covariant derivative with respect to the
background goemetry and $\epsilon_{\mu}=a^2(\eta)\left(\epsilon_{0},-\epsilon_{i}\right)$.
These infinitesimal coordinate transformations form a group and the functions
$\epsilon_{0}$ and $\epsilon_{i}$ are the gauge parameters.
The gauge-fixing procedure amounts to fixing the four gauge parameters.
Further, the spatial components $\epsilon_{i}$ can be separated into divergenceless and
divergencefull parts
\be
\epsilon_{i}=\partial_{i}\,\epsilon+\zeta_{i}
\ ,
\label{esp_i_scomp}
\ee
where $\partial_{i}\,\zeta^{i}=0$. The gauge transformations
involving $\epsilon_{0}$ and $ \epsilon$ preserve the scalar nature of
the fluctuations while those parametrized by $\zeta_{i}$
preserve the vector nature of the fluctuation.
\par
The background covariant derivatives in Eq.~(\ref{inf_trasf_og_g}) are
given by $\nabla_{\mu}\,\epsilon_{\nu}=\partial_{\mu}\,\epsilon_{\nu}-
^{(0)}\Gamma_{\mu\nu}^{\sigma}\,\epsilon_{\sigma}$ and, in terms
of the unperturbed connections (see Eqs.~(\ref{Chris_background})),
the fluctuations for the scalar modes can be written
in the tilded coordinate system as
\be
&&
\phi\to\tilde{\phi}=\phi-{\cal H}\,\epsilon_0-\epsilon_{0}'
\label{phi_transf}\\
&&
\psi\to\tilde{\psi}=\psi+{\cal H}\,\epsilon_{0}
\label{psi_transf}\\
&&
B\to\tilde{B}=B+\epsilon_{0}-\epsilon'
\label{B_transf}\\
&&
E\to\tilde{E}=E-\epsilon
\ .
\label{E_transf}
\ee
Under a coordinate transformation
preserving the vector nature of the fluctuation,
$x^{i}\to\tilde{x}^{i}=x^{i}+\zeta^{i}$ (with $\partial_{i}\,\zeta^{i}=0$),
the rotational modes of the geometry transform as
\be
&&
S_{i}\to\tilde{S}_{i}=S_{i}+\zeta_{i}'
\label{S_transf}\\
&&
F_{i}\to\tilde{F}_{i}=F_{i}-\zeta_{i}
\ .
\label{F_transf}
\ee
Finally, the tensor fluctuations in the parametrization of Eqs.~(\ref{g_00_g_ij_g_0i})
are automatically gauge invariant.
\par
The perturbed components of the energy-momentum tensor can be written as
\be
\delta\,T_{0}^{0}=\delta\,\rho
\ ,\quad
\delta\,T_{i}^{j}=-\delta\,p\,\delta_{i}^{j}
\ ,\quad
\delta\,T_{0}^{i}=\left(p+\rho\right)\,\partial^{i}\,v
\ ,
\label{T_pert}
\ee
where $v$ is the velocity field and $\delta u_{i}=\partial_{i}\,v$.
Under the infinitesimal coordinate transformations~(\ref{inf_trasf}), the fluctuations
given in Eq.~(\ref{T_pert}) transform as
\be
&&
\delta\,\rho\to\delta\,\tilde{\rho}=
\delta\,\rho-\rho'\,\epsilon_{0}
\label{rho_pert}\\
&&
\delta\,p\to\delta\,\tilde{p}=
\delta\,p-w\,\rho'\,\epsilon_{0}
\label{p_pert}\\
&&
v\to\tilde{v}=v+\epsilon'
\ .
\label{vel_L}
\ee
Using the covariant conservation equation~(\ref{contEQ_eta}) for the background
fluid density, the gauge transformation for the density perturbation
$\delta\equiv\delta\,\rho/\rho$ follows from Eq.~(\ref{rho_pert}) and reads
\be
\tilde{\delta}=
\delta-3\,{\cal H}\,\left(1+\frac{P}{\rho}\right)\,\epsilon_{0}
\ .
\label{delta_contr}
\ee
We now have two possible ways of proceeding:
\begin{itemize}
\item{} fix, completely or partially, the coordinate system (i.e., the gauge);
\item{} use a gauge-invariant approach.
\end{itemize}
We shall briefly review both.
\section{Gauge fixing and gauge-invariant variables}
As we recalled above, the tensor modes are already gauge-invariant,
whereas the vector and scalar modes change for
infinitesimal coordinate transformations.
Since the gauge fixing of the vector fluctuations is analogous, but simpler,
to the one of the scalar modes, vectors will be discussed before scalars.
\subsection{Vector modes}
We have two different gauge choices for the vector modes.
The first is to set $\tilde{S}_{i}=0$ so that, from Eq.~(\ref{S_transf}), we have
\be
\zeta_{i}(\eta,\vec{x})=
-\int^{\eta}\,S_{i}(\eta',\vec{x})\,\d\eta'+C_{i}(\vec{x})
\ .
\ee
Since the gauge function is determined up to arbitrary (space-dependent)
constants $C_{i}$, the coordinate system is not completely fixed by this
choice.
The second choice is $\tilde{F}_{i}=0$ and, from Eq.~(\ref{F_transf}), 
$\zeta_{i}=F_{i}$.
The gauge freedom is therefore completely fixed in this manner.
\par
Instead of fixing a gauge, we could work directly with gauge-invariant quantities.
We note that, from Eqs.~(\ref{S_transf}) and (\ref{F_transf}), the combination
$F_{i}'+S_{i}$ is invariant under infinitesimal coordinate transformations preserving
the vector nature of the fluctuations.
This is the vector counterpart of the Bardeen potentials~\cite{bardeen}.
\subsection{Scalar modes}
More options are available in the literature for scalar modes and we briefly review
the more common ones (see Ref.~\cite{mukh_thory} for more details).
\subsubsection{Conformally Newtonian (or Longitudinal) gauge}
In this gauge we eliminate all the off-diagonal entries $\tilde E$ and $\tilde B$ of the
perturbed metric in Eq.~(\ref{inf_trasf_og_g}), i.e. $\tilde E_{\rm L}=\tilde B_{\rm L}=0$.
From Eqs.~(\ref{B_transf}) and (\ref{E_transf}) the gauge
functions are then determined as $\epsilon=E$ and $\epsilon_{0}=E'-B$ and
the only non-vanishing entries of the perturbed metric are $\tilde \phi$ and $\tilde \psi$.
\subsubsection{Off-diagonal (or Uniform Curvature) gauge}
This gauge is characterized by the conditions $\tilde E_{\rm od}=0$ and $\tilde\psi_{\rm od}=0$.
From Eqs.~(\ref{psi_transf}) and (\ref{E_transf}) we find $\epsilon_0=-\psi/{\cal H}$
and $\epsilon=E$. The two non-vanishing metric fluctuations are then $\tilde B$ and $\tilde\phi$.
Since $\tilde\psi_{\rm od}=0$, the fluctuation of the spatial curvature also vanishes.
\subsubsection{Synchronous gauge}
This coordinate system is defined by $\tilde\phi_{\rm S}=\tilde B_{\rm S}=0$, which
however does not fix the gauge completely.
In fact, Eqs.~(\ref{phi_transf}) and (\ref{B_transf}) imply that the gauge functions are
determined by
\be
(\epsilon_{0}\,a)'=a\,\phi
\ ,\quad
\epsilon'=(B+\epsilon_{0})
\ ,
\label{sync_gauge}
\ee
leaving, after integration over $\eta$, two undetermined space-dependent functions.
This gauge freedom is closely related with unphysical gauge modes.
\subsubsection{Comoving Orthogonal gauge}
In this gauge the quantity $\tilde v_{\rm C}+\tilde B_{\rm C}$ is set to zero and the expression
of the curvature fluctuations coincides with a relevant gauge-invariant expression
whose evolution obeys a conservation law.
%(see section 5).
%
%
\subsubsection{Uniform Density gauge}
With this choice the total matter density is unperturbed, $\delta\,\tilde\rho_{\rm D}=0$.
From Eq.~(\ref{rho_pert}) the gauge function is then fixed as
$\epsilon_0=(\delta\,\rho)/\rho'$. The curvature fluctuations on constant density
hypersurfaces also obey a simple conservation law. If the total energy-momentum
tensor of the sources is represented by a single scalar field one can also define the
{\em Uniform Field Gauge}, i.e. the gauge in which the scalar field fluctuation vanishes
%(see section 5).
%
%
\subsubsection{Gauge-Invariant approach}
The set of gauge-invariant fluctuations can be written as~\cite{bardeen,mukh_thory}
\be
&&
\Phi=\phi+(B-E')'+{\cal H}\,(B-E')
\label{phi_Bard}\\
&&
\Psi=\psi-{\cal H}\,(B-E')
\ ,
\label{psi_Bard}
\ee
for the metric perturbations and
\be
&&
\delta\,\rho_{\rm g}=\delta\,\rho+\rho'\,(B-E')
\label{delta_rho_Bard}\\
&&
\delta\,p_{\rm g}=\delta\,p+p'\,(B-E')
\label{delta_p_Bard}\\
&&
V_{\rm g}^{i}=v^{i}+\partial^{i}\,E'
\ ,
\label{v_Bard}
\ee
for the fluid inhomogeneities.
Using Eqs.~(\ref{phi_transf})--(\ref{E_transf}) and~(\ref{rho_pert})--(\ref{vel_L}),
it can be verified that the quantities defined in
the above equations are gauge-invariant.
Eq.~(\ref{v_Bard}) can also
be written as $ V_{\rm g}=v+E'$ since $V_{\rm g}^{i}=\partial^{i}\,V_{\rm g}$
and $v^{i}=\partial^{i}\,v$.
Sometimes it is also convenient to introduce the divergence of
the peculiar velocity $\theta=\partial_{i}\,v^{i}$, whose associated gauge-invariant
quantity is $\Theta=\partial_{i}\,V_{\rm g}^{i}$.

\chapter{Evolution of metric fluctuations}
\label{eq_motion}
We have seen in the previous Chapter that the tensor modes of the geometry are gauge-invariant.
The vector modes are not
invariant under gauge transformation, however, their description is rather simple
if the Universe expands and in the absence of vector sources. The scalar modes are
the most difficult ones: they are not gauge-invariant and couple directly to the
sources of the background geometry. The analysis will be given for the case of a
spatially flat background geometry.
\section{Evolution of the tensor modes}
\label{tensor_modes}
For the tensor modes the perturbed Einstein equations imply
that~\footnote{We recall that $\delta_{\rm t}$, $\delta_{\rm v}$
and $\delta_{\rm s}$ denote the first-order fluctuation with respect
to tensor, vector and scalar modes respectively.} $\delta_{\rm t} R_{i}^{j}=0$.
Hence, according to  Eq.~(\ref{tensorial_Ricci_cov_contra}), we have in Fourier space
\be
{h_{i}^{i}}'' +2\,{\cal H}\,{h_{i}^{j}}'+k^2\,h_{i}^{j}=0
\ ,
\label{tensor_Ricci_eq}
\ee
which is satisfied by the two tensor polarizations.
In fact, since $\partial_{i} h^{i}_{j} = h_{k}^{k} =0$, the direction of propagation can be
chosen along the third axis and, in this case, the two physical polarizations of the graviton
will be
\be
h_{1}^{1}=-h_{2}^{2}\equiv h_{\oplus}
\  ,\quad
h_{1}^{2}=h_{2}^{1}\equiv h_{\otimes}
\label{tens_pol}
\ee
where $h_{\oplus}$ and $h_{\otimes}$ obey the same evolution Eq.~(\ref{tensor_Ricci_eq}) and will be denoted by $h$.
Eq.~(\ref{tensor_Ricci_eq}) can  be written in two slightly different, but mathematically equivalent forms
\be
&&
\left(a^2\,h_{k}'\right)'=-k^2\,h_{k}
\label{tensor_eq1}
\\
&&
\mu_{k}''+\left[k^2-\frac{a''}{a}\right]\,\mu_{k}=0
\ ,
\label{tensor_eq2}
\ee
where $\mu_{k} = a\,h_{k}$.
For $k^2\gg |a''/a|$ the solutions of Eq.~(\ref{tensor_eq2}) are oscillatory, and
the amplitudes of the tensor modes decrease as $1/a$, if the background expands.
For $k^2\ll |a''/a|$ the solution is non-oscillatory and is given by
\be
\mu_{k}(\eta)\simeq A_{k}\,a(\eta)+B_{k}\,a(\eta)\,\int^{\eta}\,\frac{d\eta'}{a^2(\eta')}
\ ,
\ee
where $A_k$ and $B_k$ are numerical coefficients.
\subsection{Hamiltonians for the tensor problem}
The variable $\mu_{k}= a\,h_{k}$ introduced above plays a special role in the theory of primordial
fluctuations, since it is the canonical variable that diagonalises the action for the tensor fluctuations.
Such a quadratic action for the tensor modes
of the geometry can be obtained by perturbing the Einstein-Hilbert action to second-order
in the amplitude of the tensor fluctuations of the metric.
\par
To this end, it is convenient to
start  from a form of the gravitational action which excludes, from the beginning,
the known total derivative. In the case of tensor modes this calculation is rather
straightforward and the result can be written as
\be
S_{\rm gw}=
\frac{1}{64\,\pi\,G}\,\int\,\d^4x\,a^2\,\eta^{\alpha\beta}\,\partial_{\alpha}h_{i}^{j}\,
\partial_{\beta}h_{j}^{i}
\ ,
\label{tensor_action1}
\ee
where $\eta_{\alpha\beta}$ is the Minkowski metric.
After a redefinition of the tensor amplitude through the
reduced Planck mass~\footnote{In our notation, $M_{\rm Pl}^{-1}= \sqrt{8\,\pi\,G}$.},
\be
h=\frac{h_{\oplus}\,M_{\rm Pl}}{\sqrt{2}}
\quad
{\rm or}
\quad
h=\frac{h_{\otimes}\,M_{\rm Pl}}{\sqrt{2}}
\ ,
\ee
the action~(\ref{tensor_action1}) for a single tensor polarization becomes
\be
S_{\rm gw}^{(1)}=\frac{1}{2}\,\int\,\d^{4}x\,a^2\, \eta^{\alpha\beta}\,\partial_{\alpha}h\,\partial_{\beta}h
\ ,
\label{tensor_action2}
\ee
and its canonical momentum is simply given by $\Pi= a^2\,h'$.
The classical Hamiltonian associated with
Eq.~(\ref{tensor_action2}) is
\be
H^{(1)}_{\rm gw}(\eta)
=\frac{1}{2}\,\int\,\d^3x\,\left[\frac{\Pi^2}{a^2}
+a^2\,\left(\partial_{i}h\right)^2\right]
\ ,
\label{tensor_hamilt1}
\ee
which is time-dependent.
It is therefore possible to perform a
time-dependent canonical transformation, leading to a different Hamiltonian that will be classically
equivalent to~(\ref{tensor_hamilt1}).
Defining the rescaled field $\mu=a\,h$, the action~(\ref{tensor_action2}) becomes
\be
S^{(2)}_{\rm gw}=
\frac12\,\int\,\d^4x\,\left[{\mu '}^2-2\,{\cal H}\,\mu\,\mu'
+{\cal H}^2\,\mu^2-\left(\partial_{i}\mu\right)^2\right]
\ ,
\label{tensor_action3}
\ee
which yields the Hamiltonian
\be
H^{(2)}_{\rm gw}(\eta)=
\frac12\,\int\,\d^{3}x\,\left[\pi^2+2\,{\cal H}\,\mu\,\pi
+\left(\partial_{i}\mu\right)^2\right]
\ ,
\label{tensor_hamilt2}
\ee
for $\mu$ and of its canonically conjugate momentum $\pi =\mu'-{\cal H}\,\mu$.
A further canonical transformation can be performed starting from~(\ref{tensor_hamilt2})
which is defined by the generating functional 
\be
{\cal G}_{2\to 3}( \mu,\tilde{\pi},\eta)=
\int\,\d^{3}x\,\left( \mu\,\tilde{\pi}-\frac{{\cal H}}{2}\,\mu^2\right)
\ ,
\label{gen_funct}
\ee
where $\tilde{\pi}=\mu'$ is  the new momentum conjugated to $\mu$.
The new Hamiltonian can now be obtained by taking the partial time derivative
of~(\ref{gen_funct}) and reads
\be
H^{(3)}_{\rm gw}(\eta)=
\frac12\,\int\,\d^{3}x\,\left[\tilde{\pi}^2+
\left(\partial_{i}\,\mu\right)^2-\left({\cal H}^2+{\cal H}'\right)\,\mu^2\right]
\ .
\label{tensor_hamilt3}
\ee
\par
The possibility of defining different Hamiltonians is connected with the possibility
of dropping total (non covariant) derivatives from the action.
To illustrate this point, let us consider the action given in Eq.~(\ref{tensor_action2}) for
a single polarization which can be rewritten as
\be
S_{\rm gw}=\frac12\,\int\,\d^{3}x\,\d\eta\,a^2\,\left[{h'}^2-\left(\partial_{i}h\right)^2\right]
\ .
\label{tensor_action4}
\ee
Recalling that, from the natural definition of the canonical field $\mu=a\,h$,
we also have $a\,h'=\mu'-{\cal H}\,\mu$,
Eq.~(\ref{tensor_action4}) becomes
\be
S_{\rm gw}=
\frac12\,\int\,\d^{3}x\,\d\eta\,\left[{\mu'}^2+{\cal H}^2\,\mu^2-2\,{\cal H}\,\mu\,\mu'
-\left(\partial_{i}\mu\right)^2\right]
\ .
\label{tensor_action5}
\ee
In Eq.~(\ref{tensor_action5}), the canonically conjugate momentum is obtained by functionally
differentiating the associated Lagrangian density ${\cal L}\left(\eta,\vec{x}\right)$, where
\be
S_{\rm gw}=\int\,\d\eta\,L(\eta)
\ ,\quad
L(\eta)=\int\,\d^{3}x\,{\cal L}\left(\eta,\vec{x}\right)
\ ,
\label{lagrangian}
\ee
with respect to $\mu'$ and the result is $\pi=\mu'-{\cal H}\,\mu$.
Hence, the Hamiltonian will be
\be
H_{\rm gw}(\eta)=\int\,\d^3x\,\left[\pi\,\mu'-{\cal L}_{\rm gw}\,\left(\eta,\vec{x}\right)\right]
\ .
\label{hamiltonian}
\ee
Consequently, from Eqs.~(\ref{tensor_action5}) and recalling the expression of $\tilde{\pi}$,
one exactly obtains the result given in Eq. (\ref{tensor_hamilt2}).
We now note that one can write
\be
{\cal H}\,\mu\,\mu'=
-\frac{{\cal H}'}{2}\,\mu^2+\frac{\d}{\d\eta}\,\left[\frac{{\cal H}\,\mu^2}{2}\right]
\ ,
\label{rewrite_mu_mud}
\ee
which can be replaced in Eq.~(\ref{tensor_action5}) and leads to
\be
\tilde{S}_{\rm gw}=
\frac12\,\int\,\d^{4}x\,\left[{\mu'}^2+\left({\cal H}^2+{\cal H}'\right)\,\mu^2
-\left(\partial_{i}\mu\right)^2\right]
\ .
\label{tensor_action6}
\ee
The latter will be classically equivalent to the action~(\ref{tensor_action5}) for $\mu$ and
its conjugate momentum $\tilde{\pi}=\mu'$.
Hence, the Hamiltonian for $\mu$ and $\tilde\pi$ can be easily obtained
from Eq.~(\ref{hamiltonian}) with $\pi$ replaced by $\tilde{\pi}$ and
will exactly be the one given in Eq.~(\ref{tensor_hamilt3}).
\par
The same conclusion also applies to the scalar fluctuations.
In the Lagrangian approach,
this aspect reflects in the possibility of defining classical actions that differ
by the addition of a total time derivative.
While at the classical level the equivalence among the different actions is complete,
at the quantum level this equivalence is somehow broken since the minimization of an action
may lead to computable differences with respect to another.
\section{Evolution of the vector modes}
\label{vector_modes}
This paragraph is inserted for completeness, but in the following we only focus
on the vector and scalar perturbations.
\par
The evolution of the vector modes of the geometry can be obtained by perturbing the
Einstein equations and the covariant conservation equation with respect to the vector
modes of the geometry
\be
&&
\delta_{\rm v}\,{\cal G}_{\mu}^{\nu}
=8\,\pi\,G\,\delta_{\rm v}\,T_{\mu}^{\nu}
\label{vector_eq1}
\\
&&
\nabla_{\mu}\,\delta_{\rm v}\,T^{\mu\nu}=0
\ .
\label{vector_eq2}
\ee
Using  Eqs.~(\ref{vector_Chris}) and (\ref{vector_Ricci}) of Appendix~\ref{appendice_metric_fluc},
the explicit form of the above equations can be obtained as
\be
&&
\nabla^2\,S_{i}=16\,\pi\,G\,\left(p+\rho\right)\,a^2\,{\cal V}_{i}
\label{vector_eq3}
\\
&&
S_{i}'+2\,{\cal H}\,S_{i}=0
\label{vector_eq4}
\\
&&
{\cal V}_{i}'+\left(1-3\,w\right)\,{\cal H}\,{\cal V}_{i}=0
\ ,
\label{vector_eq5}
\ee
where ${\cal V}_{i}$ is the divergenceless part of the velocity field and, as
discussed in Chapter~\ref{cosm_pert}, the gauge has been completely fixed by requiring
$\tilde{F}_{i}= 0$, implying, according to Eqs.~(\ref{S_transf}) and (\ref{F_transf}),
that  $\zeta_{i}=F_{i}$. 
Note also that, according to our conventions for the velocity field (see Eq.~(\ref{v_def})),
$\delta_{\rm v}\,T_{0}^{i}=\left(p+\rho\right)\,{\cal V}^{i}$.
Eqs.~(\ref{vector_eq3})--(\ref{vector_eq5}) follow, respectively,
from the $0-i$ and $i-j$ components of the Einstein equations and from the spatial component
of the covariant conservation equation.
\par
Eqs.~(\ref{vector_eq3}) and (\ref{vector_eq4})
are not independent.
In fact, it is easy to check, using Eq.~(\ref{vector_eq3}), that $S_{i}$
can be eliminated from Eq.~(\ref{vector_eq4}) and Eq.~(\ref{vector_eq5}) is then obtained.
By Fourier transforming Eqs.~(\ref{vector_eq3}) and (\ref{vector_eq4}), we find the solution
\be
&&
S_{i}(\eta)=\frac{C_{i}(k)}{a^2}
\label{vector_eq6}
\\
&&
{\cal V}_{i}\left(\eta\right)=\frac{k^2\,C_{i}(k)}{16\,\pi\,G\, a^4\,\left(p+\rho\right)}
\ ,
\label{vector_eq7}
\ee
where $C_{i}(k) $ is an integration constant.
Two distinct situations are possible.
If the Universe is expanding, $S_{i}$ is always decreasing.
However, ${\cal V}_{i}$ may also increase.
Consider, for example, the case of a single barotropic fluid $p=w\,\rho$.
Since, according to Eqs.~ (\ref{FrEQ_eta}) and (\ref{contEQ_eta}) we have
$\rho\simeq a^{-3\,\left(w+1\right)}$, then
\be
{\cal V}_{i}\propto a^{3\,w-1}
\ .
\ee
If $w =1$, ${\cal V}_{i}$  increases like $\eta$ (since $ a(\eta)\sim\sqrt{\eta}$),
while $S_{i}(\eta)\simeq\eta^{-1}$ decays for large conformal times and is negligible
for our purposes of describing the CMB.
This observation was indeed raised in the past, in connection with the idea that the early
Universe could be dominated by vorticity.
If the Universe contracts in the future, the evolution of
the vector modes will be reversed and $S_{i}$ may increase again.
\par
Finally, the situation may change even more radically if the theory is not of
Einstein-Hilbert type or if it is higher-dimensional.
Note that, if the energy-momentum
tensor of the fluid sources possesses a non-vanishing torque,
rotational perturbations can be copiously produced.
\section{Evolution of the scalar modes}
In analogy with the case of the tensor modes, large-scale scalar
fluctuations follow a set of  conservation laws that hold approximately for
$k\,\eta\ll 1$ and become exact only in the limit $k\,\eta\to 0$.
\par
The evolution equations for the fluctuations of the geometry and of the sources
can be written without fixing a specific coordinate system. Then, by using the definitions
of the Bardeen potentials the wanted gauge-invariant form of the equations can be obtained.
Inserting the definitions~(\ref{phi_Bard})--(\ref{v_Bard})
into Eqs.~(\ref{00_without_gauge}) and (\ref{0i_without_gauge}), the gauge-invariant
Hamiltonian and momentum constraint become
\be
&&
\nabla^2\Psi-3\,{\cal H}\,\left(\Psi'+{\cal H}\,\Phi\right)
=4\,\pi\,a^2\,G\,\delta\,\rho_{\rm g}
\label{Ham_gi_EQ}
\\
&&
\nabla^2\,\left(\Psi'+{\cal H}\,\Phi\right)
=-4\,\pi\,G\,a^2\,\left(p+\rho\right)\,\Theta
\ ,
\label{mom_gi_EQ}
\ee
where $\Theta=\partial_{i}V^{i}_{\rm g}$ is the divergence of the gauge-invariant
(scalar) peculiar velocity field. Inserting the expressions of the gauge-invariant
fluctuations given in Eqs.~(\ref{phi_Bard})--(\ref{v_Bard}) into
Eqs.~(\ref{ij_tracefull_and_traceless}) we can derive
\be
&&
\!\!\!\!\!\!\!\!\!\!\!\!\!\!\!\!\!\!\!\!\!\!\!\!
\Psi''+{\cal H}\,\left(\Phi'+2\,\Psi'\right)+\left({\cal H}^2+2\,{\cal H}'\right)\,\Phi
+\frac13\,\nabla^2\left(\Phi-\Psi\right)
=4\,\pi\,G\,a^2\,\delta\,p_{\rm g}
\label{tracefull_gi_EQ}
\\
&&
\!\!\!\!\!\!\!\!\!\!\!\!\!\!\!\!\!\!\!\!\!\!\!\!
\nabla^2\left(\Phi-\Psi\right)
=12\,\pi\,G\,a^2\,\left(p+\rho\right)\,\sigma
\ .
\label{traceless_gi_EQ}
\ee
Finally, the gauge-invariant form of the perturbed covariant
conservation equations become
\be
&&
\!\!\!\!\!\!\!\!\!\!\!\!\!\!\!\!\!\!\!\!\!\!\!\!
\delta\,\rho_{\rm g}'-3\,\left(p+\rho\right)\Psi'+\left(p+\rho\right)\,\Theta
+3\,{\cal H}\,\left(\delta\,\rho_{\rm g}+\delta\,p_{\rm g}\right)=0
\label{drgi}
\\
&&
\!\!\!\!\!\!\!\!\!\!\!\!\!\!\!\!\!\!\!\!\!\!\!\!
\left(p+\rho\right)\,\Theta'+\left[\left(p'+\rho'\right)
+4\,{\cal H}\,\left(p+\rho\right)\right]\,\Theta
+\nabla^2\delta\,p_{\rm g}+\left(p+\rho\right)\,\nabla^2\Phi=0
\ ,
\label{THgi}
\ee
as follows from inserting Eqs.~(\ref{phi_Bard})--(\ref{v_Bard}) into Eqs.~(\ref{cons_eq_pert_0_expl})
and (\ref{cons_eq_pert_i_expl}) without imposing any gauge condition.
Eqs.~(\ref{Ham_gi_EQ}) and (\ref{mom_gi_EQ}) and Eqs.~(\ref{tracefull_gi_EQ}) and
(\ref{traceless_gi_EQ}) have the same form of the evolution equations
in the longitudinal gauge.
We also define the quantity 
\be
{\cal R}\equiv-\left(\Psi-{\cal H}\,V_{\rm g}\right)
=-\Psi-\frac{{\cal H}\,\left(\Psi'+{\cal H}\,\Phi\right)}{{\cal H}^2-{\cal H}'}
\ ,
\label{R_def}
\ee
which is gauge-invariant and will be used extensively later.

\subsection{Physical interpretation of curvature perturbations}
In the comoving  gauge, the three-velocity of the fluid vanishes, i.e. $v_{\rm C}=0$.
Since hypersurfaces of constant (conformal) time should be orthogonal to the
four-velocity, we will also have that $B_{\rm C}=0$. In this gauge the curvature
perturbation can be computed directly from the expressions of the perturbed
Christoffel connections bearing in mind that we want to compute the fluctuations
in the spatial curvature,
\be
\delta\,R^{(3)}_{\rm C}=
\delta_{\rm s}\,\gamma^{ij}\,\overline{R}_{ij}^{(3)}
+\overline{\gamma}^{ij}\,\delta_{\rm s}\,R^{(3)}_{ij}
\equiv\frac{4}{a^2}\,\nabla^2\psi_{\rm C}
\ ,
\label{curv_pert_Chr}
\ee
where the subscript ${\rm C}$ is to remind that the expression holds
on comoving hypersurfaces.
The curvature fluctuations in the comoving gauge
can be connected to the fluctuations in a different gauge characterized by a different
value of the time coordinate, i.e. $\eta_{\rm C}\to\eta=\eta_{\rm C}+\epsilon_0$.
Under this shift
\be
&&
\psi_{\rm C}\to\psi=\psi_{\rm C}+{\cal H}\,\epsilon_{0}
\label{comov_repar1}
\\
&&
\left(v_{\rm C}+B_{C}\right)\to v+B=\left(v_{\rm C}+B_{\rm C}\right)+\epsilon_0
\ .
\label{comov_repar2}
\ee
Since in the comoving orthogonal gauge $v_{\rm C}+B_{\rm C}=0$,
Eqs.~(\ref{comov_repar1}) and (\ref{comov_repar2}) imply, in the new coordinate system,
that
\be
\psi_{\rm C}
=\psi-{\cal H}\,\left(v+B\right)
=\left(\Psi-{\cal H}\,V_{\rm g}\right)
\ ,
\label{psiC}
\ee
where the second equality follows from the definitions of gauge-invariant
fluctuations given in Eqs~(\ref{phi_Bard})--(\ref{v_Bard}).
From Eq.~(\ref{psiC}), one can conclude that ${\cal R}$ in Eq.~(\ref{R_def}) 
corresponds to
the curvature fluctuations of the spatial curvature on comoving orthogonal
hypersurfaces.
\par
Other quantities, defined in specific gauges, turn out to have a gauge-invariant
interpretation. Take, for instance, the curvature fluctuations on constant density
hypersurfaces $\psi_{\rm D}$. Under infinitesimal gauge transformations
we have
\be
&&
\psi_{\rm D}\to\psi=\psi_{\rm D}+{\cal H}\,\epsilon_{0}
\nonumber\\
&&
\delta\,\rho_{\rm D}\to\delta\,\rho=\delta\,\rho_{\rm D}-\rho'\,\epsilon_{0}
\ ,
\label{const_dens_hyper}
\ee
where, on constant density hypersurfaces, $\delta\,\rho_{\rm D}=0$ by definition of the
uniform density gauge.
Hence, from Eq.~(\ref{const_dens_hyper}), we obtain
\be
\psi_{\rm D}=\psi+{\cal H}\,\frac{\delta\,\rho}{\rho'}
=\Psi+{\cal H}\,\frac{\delta\,\rho_{\rm g}}{\rho'}
\ ,
\label{psiD}
\ee
where the second equality follows from Eqs.~(\ref{phi_Bard})--(\ref{v_Bard}).
Hence,  the (gauge-invariant) curvature fluctuations on constant density hypersurfaces
can be defined as
\be
\zeta\equiv-\left(\Psi+{\cal H}\,\frac{\delta\,\rho_{\rm g}}{\rho'}\right)
\ ,
\label{zeta_def}
\ee
which coincides with the curvature fluctuations in the uniform density gauge.
\par
The values of $\zeta$ and ${\cal R}$ are equal up to terms proportional
to the Laplacian of $\Psi$.
This can be shown by using the definitions of ${\cal R}$
and $\zeta$, whose difference gives
\be
\zeta-{\cal R}=-{\cal H}\,\left(V_{\rm g}+\frac{\delta\,\rho_{\rm g}}{\rho'}\right)
\ .
\label{diff_Rzeta}
\ee
From the Hamiltonian and momentum constraints and from the conservation of the energy
density of the background
\be
&&
V_{\rm g}=-\frac{1}{4\,\pi\,G\,a^2\,\left(p+\rho\right)}\,\left({\cal H}\,\Phi+\Psi'\right)
\nonumber\\
&&
\frac{\delta\,\rho_{\rm g}}{\rho'}= \frac{1}{4\,\pi\,G\,a^2\,\left(p+\rho\right)}\,
\left[\nabla^2\Psi-3\,{\cal H}\,\left({\cal H}\,\Phi+\Psi'\right)\right]
\ ,
\ee
we in fact obtain
\be
\zeta-{\cal R}=\frac{\nabla^2\Psi}{12\,\pi\,G\,a^2\,\left(p+\rho\right)}
\ .
\label{diff_Rzeta2}
\ee
The density fluctuation on comoving orthogonal hypersurfaces can be also defined
as \cite{bardeen}
\be
\epsilon_{\rm m}
\equiv\frac{\delta\,\rho_{\rm C}}{\rho}
=\frac{\delta\,\rho+\rho'\,\left(v+B\right)}{\rho}
=\frac{\delta\,\rho_{\rm g}-3\,{\cal H}\,\left(p+\rho\right)\,V_{\rm g}}{\rho}
\ ,
\ee
where the second equality follows from the first one by using the definitions of
gauge-invariant fluctuations.
Again, using the Hamiltonian and momentum constraints, one obtains
\be
\epsilon_{\rm m}=\frac{1}{4\,\pi\,G\,a^2\,\rho}\,\nabla^2\Psi
\ ,
\label{epsil_m}
\ee
which means that $\left(\zeta-{\cal R}\right)\propto\epsilon_{\rm m}$.
\subsection{Curvature perturbations induced by scalar fields}
Some of the general considerations developed earlier in this section
will now be specialized to the case of scalar field  sources characterized by a
potential $V(\varphi)$.
For a single scalar field, the background Einstein
equations In a spatially flat FRW geometry read
\be
&&
{\cal H}^2=\frac{8\,\pi\,G}{3}\,\left(\frac{{\varphi'}^2}{2}+V\,a^2\right)
\label{Fried_eq_scal1}\\
&&
\left({\cal H}^2-{\cal H}'\right)=4\,\pi\,G\,\varphi'^2
\label{Fried_eq_scal2}\\
&&
\varphi''+2\,{\cal H}\,\varphi'+\frac{\partial V}{\partial \varphi}\,a^2=0
\ .
\label{Fried_eq_scal3}
\ee
The fluctuation of $\varphi$ defined in Appendix~\ref{appendice_metric_fluc} under a coordinate
transformation~(\ref{inf_trasf}) changes as
\be
\chi\to\tilde{\chi}=\chi-\varphi'\,\epsilon_{0}
\ .
\label{chi_trasf}
\ee
Consequently, the associated gauge-invariant scalar field fluctuation is given by
\be
X=\chi+\varphi'\,\left(B-E'\right)
\ .
\label{chi_gi}
\ee
The two Bardeen  potentials and the gauge-invariant scalar field fluctuation $X$
define the coupled system of scalar fluctuations of the geometry,
\be
&&\!\!\!\!\!\!\!\!\!\!\!\!\!\!\!\!\!\!\!\!\!\!
\nabla^2\Psi-3\,{\cal H}\,\left({\cal H}\,\Phi+\Psi'\right)
=4\,\pi\,G\,\left[-\varphi'^2\,\Phi+\varphi'\,X'
+\frac{\partial V}{\partial \varphi}\,a^2\,X\right]
\label{00_scalar}\\
&&\!\!\!\!\!\!\!\!\!\!\!\!\!\!\!\!\!\!\!\!\!\!
{\cal H}\,\Phi+\Psi'=4\,\pi\,G\,\varphi'\,X
\label{0i_scalar}\\
&&\!\!\!\!\!\!\!\!\!\!\!\!\!\!\!\!\!\!\!\!\!\!
\Psi''+{\cal H}\,\left(\Phi'+2\,\Psi'\right)
+\left({\cal H}^2+2\,{\cal H}'\right)\,\Phi
=4\,\pi\,G\,\left[-\varphi'^2\,\Phi+\varphi'\,X'
-\frac{\partial V}{\partial \varphi}\,a^2\,X\right]
\ ,
\label{ij_scalar}
\ee
where Eqs.~(\ref{00_scalar}), (\ref{0i_scalar}) and (\ref{ij_scalar}) are, respectively,
the perturbed $0-0$, $0-i$ and $i-j$ components of the Einstein equations.
Although these equations are sufficient to determine
the evolution of the system, it is also appropriate to recall the gauge-invariant
form of the perturbed Klein-Gordon equation
\be
X''+2\,{\cal H}\,X'-\nabla^2X+\frac{\partial^2 V}{\partial \varphi^2}\,a^2\,X
+2\,\Phi\,\frac{\partial V}{\partial \varphi}\,a^2
-\varphi'\,\left(\Phi'+3\,\Psi'\right)=0
\ .
\label{Klein_Gordon_scalar}
\ee
The above equation can be obtained from the perturbed Klein-Gordon equation without
gauge fixing, reported in Eq.~(\ref{Klein-Gordon_pert_expl}) of
Appendix~\ref{appendice_metric_fluc}, by inserting
the gauge-invariant fluctuations of the metric given in
Eqs.~(\ref{phi_Bard})--(\ref{psi_Bard}) into
Eq.~(\ref{Klein-Gordon_pert_expl}) together with Eq.~(\ref{chi_gi}).
\par
If the perturbed energy-momentum tensor has vansihing anisotropic stress, the $(i\neq j)$
component of the perturbed Einstein equations leads to $\Phi=\Psi$.
The gauge-invariant curvature fluctuations on comoving orthogonal  hypersurfaces are,
for a scalar field source,
\be
{\cal R}=-\Psi-{\cal H}\,\frac{X}{\varphi'}
=-\Psi-\frac{{\cal H}\,\left({\cal H}\,\Phi+\Psi'\right)}{{\cal H}^2-{\cal H}'}
\ .
\label{R_chi}
\ee
Eq.~(\ref{R_chi}) and a linear combination of Eqs.~(\ref{00_scalar})--(\ref{ij_scalar})
lead to
\be
{\cal R}'=-\frac{{\cal H}}{4\,\pi\,G\,{\varphi'}^2}\,\nabla^2\Psi
\ ,
\label{R_der}
\ee
which implies that ${\cal R}$ is constant for the modes $k$ such that $k\,\eta \ll 1$~\cite{bardeen}.
The power spectrum of the scalar modes amplified during the inflationary phase is
customarily expressed in terms of ${\cal R}$, which is conserved on super-horizon
scales.
\par
Curvature perturbations on comoving spatial hypersurfaces can also be simply related
to curvature perturbations on the constant density hypersurfaces,
\be
\zeta
=-\Psi-{\cal H}\,\frac{\delta\,\rho_{\varphi}}{\rho_{\varphi}'}
=-\Psi+\frac{a^2\,\delta\,\rho_{\varphi}}{3\,{\varphi'}^2}
\ .
\label{zeta_def2}
\ee
Taking Eqs.~(\ref{R_def}) and (\ref{zeta_def2}), and using Eq. (\ref{00_scalar}) we have
\be
\zeta-{\cal R}
={\cal H}\,\frac{X}{{\varphi}'}+\frac{a^2\,\delta\,\rho_{\varphi}}{3\,{\varphi'}^2}
=\frac{2\,m_{\rm Pl}^2}{3}\,\frac{\nabla^2\Psi}{{\varphi '}^2}
\ ,
\label{zeta-R_scal}
\ee
which confirms that ${\cal R}$ and $\zeta$ differ by the Laplacian
of a Bardeen potential.
Taking the time derivative of Eq.~(\ref{R_der}) and using Eq.~(\ref{R_chi}) and
Eqs.~(\ref{00_scalar})--(\ref{ij_scalar}), we obtain the second-order equation
\be
{\cal R}''+2\,\frac{z'}{z}\,{\cal R}'-\nabla^2{\cal R}=0
\ ,
\label{scal_eq_R}
\ee
where the function $z(\eta)$ is
\be
z(\eta)=\frac{a\,\varphi'}{\cal H}
\ .
\label{z_def}
\ee
Eq.~(\ref{scal_eq_R}) is formally similar to Eq.~(\ref{tensor_Ricci_eq}),
i.e. the evolution equation of a single tensor polarization.
We can eliminate the first time derivative appearing in Eq.~(\ref{scal_eq_R}) by introducing
the gauge-invariant variable
\be
q=-z\,{\cal R}
\ ,
\label{from_R_to_q}
\ee
which leads to
\be
q''-\frac{z''}{z}\,q-\nabla^2q=0
\ .
\label{scal_eq_q}
\ee
After a Fourier transformation, we have
\be
q_{k}''+\left[k^2-\frac{z''}{z}\right]\,q_{k}=0
\ ,
\label{scal_eq2}
\ee
i.e., the scalar analog of Eq.~(\ref{tensor_eq2}).
\subsection{Horizon flow functions}
\label{Hff}
The driving fields of the scalar and tensor modes are usually referred to as
{\em pump fields}.
They appear explicitly as the term $a''/a$ in Eq.~(\ref{tensor_eq2}) and 
$z''/z$ in Eq.~(\ref{scal_eq2}).
For general potentials leading to a
quasi-de Sitter expansion of the geometry, they are conventionally described
in terms of the hierarchy of horizon flow functions (HFF; see Appendix~\ref{appendice_SR-HFF}
for a comparison with different conventions), defined according to
\be
\epsilon_0&\equiv&\frac{H(N_{i})}{H(N)}
\nonumber
\\
\epsilon_{j+1}&\equiv&
\frac{\d\ln |\epsilon_j|}{\d N}
\ ,
\quad\quad
j\geq 0
%\ ,
\label{hor_flo_fun}
\ee
where $N$ is the number of e-folds, $N\equiv\ln (a/a_i)$ [where $a_i=a(\eta_i)$]
after the arbitrary initial time $\eta_i$.
A general result is that inflation takes place for $\epsilon_1<1$.
\par
To find the expressions of the pump fields in terms of the HFF it is useful to
write the background equations as a first-order set of non-linear differential
equations
\be
&&
\left(\frac{\partial H}{\partial \varphi}\right)^2
-\frac{3}{2\,m_{\rm Pl}^2}\,H^2\left(\varphi\right)
=-\frac{V(\varphi)}{2\,m_{\rm Pl}^4}
\label{eq1_for_pot}\\
&&
\dot{\varphi}=-2\,m_{\rm Pl}^2\,\frac{\partial H}{\partial \varphi}
\ .
\label{eq2_for_pot}
\ee
Bearing in mind Eq.~(\ref{hor_flo_fun}), the explicit relation determining the form
of the pump field for tensor modes follows from
\be
\frac{a''}{a}
={\cal H}^2+{\cal H}'
=2\,a^2\,H^2\,+a^2\,\dot{H}
=a^2\,H^2\,\left(2-\epsilon_1\right),
\label{tens_pot}
\ee
where in the last equality, we use $\epsilon_1=-\dot{H}/H^2$.
In the same manner, for the scalar case we obtain
\be
\frac{z''}{z}
=a^2\,H^2\,\left(2-\epsilon_1+\frac32\,\epsilon_2
-\frac12\,\epsilon_1\,\epsilon_2+\frac14\,\epsilon_2^2
+\frac12\,\epsilon_2\,\epsilon_3\right)
\ .
\label{scal_pot}
\ee
\par
We conclude this section with a useful expression that relates $\eta$, $a$, $H$ and the HFF.
Starting from the definition of the conformal time $\d \eta\equiv\d t/a(t)$, we can integrate
by parts and find that
\be
\eta=-\frac{1+\epsilon_1\,\epsilon_2+{\cal O}\left(\epsilon_j^3\right)}
{a\,H\,\left(1-\epsilon_1\right)}
\label{eta}
\ .
\ee
\subsection{Hamiltonians for the scalar problem}
Different Hamiltonians for the evolution of the scalar fluctuations can be defined.
\par
We begin by expressing the action of the scalar fluctuations of the geometry
in terms of the curvature fluctuations
\be
S^{(1)}_{\rm scal}=\frac{1}{2}\,\int\d^4 x\,z^2\,\left[{{\cal R}'}^2
-\left(\partial_{i}{\cal R}\right)^2\right]
\ .
\label{action_for_R}
\ee
Using the canonical momentum $\pi_{\cal R}\equiv z^2\,{\cal R}'$,
the Hamiltonian related to the above action becomes
\be
H^{(1)}_{\rm scal}(\eta)=
\frac{1}{2}\,\int\d^{3}x\,\left[\frac{\pi_{{\cal R}}^2}{z^2}
+z^2\,\left(\partial_{i}{\cal R}\right)^2\right]
\ ,
\label{ham_scalar1}
\ee
and the Hamilton equations
\be
\pi_{\cal R}'=z^2\,\nabla^2{\cal R}
\ ,\quad
{\cal R}'=\frac{\pi_{{\cal R}}}{z^2}
\ .
\label{ham_equation}
\ee
Eqs.~(\ref{ham_equation}) can then be combined in the single second order
equation~(\ref{scal_eq_R}).
\par
The canonically conjugate momentum $\pi_{{\cal R}}$ is related to the
density perturbation on comoving hypersurfaces.
In particular, for a single scalar field source, one finds~\cite{bardeen}
\be
\epsilon_{\rm m}
=\frac{\delta\,\rho_{\varphi}+3\,{\cal H}\,\left(\rho_{\varphi}
+p_{\varphi}\right)\,V}{\rho_{\varphi}}
=\frac{a^2\,\delta\,\rho_{\varphi}+3\,{\cal H}\,\varphi'\,X}{a^2\,\rho_{\varphi}}
\ ,
\label{contrast_sing_scal}
\ee
where the second equality can be obtained using
$\rho_{\varphi}+p_{\varphi}={\varphi'}^2/a^2$ and the fact that the effective ``velocity''
in the case of a scalar field is $ V = X/\varphi'$.
Inserting now Eq.~(\ref{0i_scalar})
into Eq.~(\ref{00_scalar}), Eq.~(\ref{contrast_sing_scal}) can be expressed as
\be
\epsilon_{\rm m}
=\frac{2\,m_{\rm Pl}^2\,\nabla^2\Psi}{a^2\,\rho_{\varphi}}
=\frac23\,\frac{\nabla^2\Psi}{{\cal H}^2}
\ ,
\label{contrast_sing_scal_BIS}
\ee
where the last equality follows from the first equation in~(\ref{FrEQ_eta}).
From Eq.~(\ref{contrast_sing_scal_BIS}), it also follows that
\be
\pi_{{\cal R}}=-6\,a^2\,{\cal H}\,\epsilon_{\rm m}
\ ,
\label{pi_R}
\ee
in which we used Eq.~(\ref{R_der}) to express ${\cal R}'$.
Hence, in this description, the canonical field is the curvature
fluctuation on comoving spatial hypersurfaces and the canonical momentum is
the density fluctuation on the same hypersurfaces.
\par
To bring the second-order action in the simple form of Eq.~(\ref{action_for_R}),
various (non-covariant) total derivatives have been dropped. Hence, there
is always the freedom of redefining the canonical fields through time-dependent
functions of the background geometry. In particular, the action~(\ref{action_for_R})
can be rewritten in terms of the variable $q$ defined in Eq.~(\ref{from_R_to_q}).
Then~\cite{mukh1,mukh2}
\be
S^{(2)}_{\rm scal}
=\frac12\,\int\d^4 x\,\left[{q'}^2-2\,\frac{z'}{z}\,q\,q'
-\left(\partial_{i}q\right)^2+\left(\frac{z'}{z}\right)^2\,q^2\right]
\ ,
\label{action_for_R_bis}
\ee
whose related Hamiltonian and canonical momentum are, respectively
\be
H_{\rm scal}^{(2)}(\eta)
=\frac12\,\int\d^{3} x\,\left[\pi_{q}^2+2\,\pi_{q}\,q
+\left(\partial_{i}q\right)^2\right]
\ ,\quad
\pi=q'-\frac{z'}{z}\,q
\ .
\label{ham_equation2}
\ee
In Eq.~(\ref{action_for_R_bis}) another total derivative term can be dropped,
leading to the action
\be
S^{(3)}_{\rm scal}
=\frac12\,\int\d^4 x\,\left[{q'}^2-\left(\partial_{i}q\right)^2
+\frac{z''}{z}\,q^2\right]
\ ,
\label{action_for_R_tris}
\ee
and Hamiltonian
\be
H^{(3)}_{\rm scal}(\eta)
=\frac12\,\int\d^{3} x\,\left[\tilde{\pi}_{q}^2+\left(\partial_{i}q\right)^2
-\frac{z''}{z}\,q^2\right]
\ .
\label{ham_equation3}
\ee
where $\tilde{\pi}=q'$. As in the case of the Hamiltonians for the tensor modes,
Eqs.~(\ref{ham_scalar1}), (\ref{ham_equation2})  and (\ref{ham_equation3}) are all
related by canonical transformations.
\section{Quantum amplification of metric fluctuations}
\label{quantization}
In the framework of quantum theory, the evolution of the metric fluctuations
can be described either in the Schr\"odinger formalism  or the Heisenberg
picture.
\par
In the Schr\"odinger description,  the quantum evolution can be pictured as the
spreading of a quantum mechanical wave-functional.
The initial wave-functional is constructed as the direct product of states
minimising the uncertainty relations for the harmonic oscillator modes forming
the quantum field.
This initial vacuum state has zero momentum and
each Fourier mode of the field will evolve into a (two mode)
squeezed quantum state.
The uncertainty relations are still minimised on such states,
but in such a way that one of the two canonically
conjugate operators will have a variance much larger than the quantum limit,
while the other canonically conjugate operator will have a variance
much smaller than the quantum uncertainty limit.
As a background pumping electromagnetic field (a laser)
is able to produce, under some circumstances, squeezed states of photons,
the classical gravitational field (for instance the curvature)
is able to produce squeezed states of gravitons (for the tensor modes)
or of phonons (in the case of the scalar modes).
\par
In fully equivalent terms, the evolution of the fluctuations can be
described in the Heisenberg representation.
This is the description which will be adopted here.
The expectation values of the Hamiltonians of the scalar and tensor modes
of the geometry will then be minimized for  $\eta\to-\infty$, which is
physically identified with the beginning of inflation.
\subsection{Large-scale power spectra of tensor fluctuations}
Let us consider the Hamiltonian given in Eq.~(\ref{tensor_hamilt3}) and drop
the tildes in the momenta,
\be
H(\eta)=\frac12\,\int\d^3x\,\left[\pi^2-\frac{a''}{a}\,\mu^2
+(\partial_{i}\,\mu)^2\right]
\ ,
\label{ham_4quant}
\ee
where $\mu=a\,h_{\oplus}\,M_{\rm Pl}/\sqrt{2}$ (an identical expression
holds for the other polarization).
After imposing the  commutation relations for the canonically conjugate quantum
fields (in natural units with $\hbar=1$)
\be
\left[\hat{\mu}(\vec{x},\eta),\hat{\pi}(\vec{y},\eta)\right]=i\,\delta^{(3)}(\vec{x}-\vec{y}),
\label{commut}
\ee
the operator corresponding to the Hamiltonian~(\ref{ham_4quant}) becomes
\be
\hat{H}(\eta)=\frac12\,\int\d^3x\,\left[\hat{\pi}^2-\frac{a''}{a}\,\hat{\mu}^2
+(\partial_{i}\,\hat{\mu})^2\right]
\ .
\label{ham_4quant2}
\ee
In Fourier space the quantum fields can be written as
\be
\hat{\mu}(\vec{x},\eta)&=&
\frac{1}{2\,(2\,\pi)^{3/2}}\,\int\d^3k\,\left[\hat{\mu}_{\vec{k}}\,e^{-i\,\vec{k}\cdot\vec{x}}
+\hat{\mu}_{\vec{k}}^{\dagger}\,e^{i\,\vec{k}\cdot\vec{x}}\right]
\nonumber\\
\hat{\pi}(\vec{x},\eta)&=&
\frac{1}{2\,(2\,\pi)^{3/2}}\,\int\d^3k\,\left[\hat{\pi}_{\vec{k}}\,e^{-i\,\vec{k}\cdot\vec{x}}
+\hat{\pi}_{\vec{k}}^{\dagger}\,e^{i\,\vec{k}\cdot\vec{x}}\right]
\ ,
\label{Four_expan}
\ee
Inserting now Eqs.~(\ref{Four_expan}) into Eq.~(\ref{commut}) and demanding the validity
of the latter implies the following canonical commutation relations for the Fourier
components of the quantum operators
\be
&&
\left[\hat{\mu}_{\vec{k}}(\eta),\hat{\pi}_{\vec{p}}^{\dagger}(\eta)\right]
=\left[\hat{\mu}_{\vec{k}}^{\dagger}(\eta),\hat{\pi}_{\vec{p}}(\eta)\right]
=i\,\delta^{(3)}(\vec{k}-\vec{p})
\nonumber
\\
\label{f_commut}
\\
&&
\left[\hat{\mu}_{\vec{k}}(\eta),\hat{\pi}_{\vec{p}}(\eta)\right]
=
\left[\hat{\mu}_{\vec{k}}^{\dagger}(\eta),\hat{\pi}_{\vec{p}}^{\dagger}(\eta)\right]
=i\,\delta^{(3)}(\vec{k}+\vec{p})
\ .
\nonumber
\ee
Inserting now Eq.~(\ref{Four_expan}) into Eq.~(\ref{ham_4quant2}) we get the Fourier
space representation of the quantum Hamiltonian~\footnote{Note that, in order
to derive the following equation, the relations $\hat{\mu}_{-\vec{k}}^{\dagger}
\equiv\hat{\mu}_{\vec{k}}$ and $\hat{\pi}_{-\vec{k}}^{\dagger}\equiv\hat{\pi}_{\vec{k}}$
should be used.}
\be
\hat{H}(\eta)=\frac14\,\int\d^3k\,
\left[\left(\hat{\pi}_{\vec{k}}\,\hat{\pi}^{\dagger}_{\vec{k}}
+\hat{\pi}_{\vec{k}}^{\dagger}\,\hat{\pi}_{\vec{k}}\right)
+\left(k^2-\frac{a''}{a}\right)\,\left(\hat{\mu}_{\vec{k}}\,\hat{\mu}^{\dagger}_{\vec{k}}
+\hat{\mu}_{\vec{k}}^{\dagger}\,\hat{\mu}_{\vec{k}}\right)\right]
\ .
\label{ham_4quant3}
\ee
In the Heisenberg representation the field operators obey
\be
i\,\hat{\mu}'=\left[\hat{\mu},\hat{H}\right]
\quad
{\rm and}
\quad
i\,\hat{\pi}'=\left[\hat{\pi},\hat{H}\right]
\ .
\ee
Using now the mode expansion~(\ref{Four_expan}) and the
Hamiltonian~(\ref{ham_4quant3}),
the evolution for the Fourier components of the operators is
determined by
\be
\hat{\mu}_{\vec{k}}'&=&\hat{\pi}_{\vec{k}}
\\
\hat{\pi}_{\vec{k}}'&=&-\left(k^2-\frac{a''}{a}\right)\,\hat{\mu}_{\vec{k}}
\ ,
\ee
whose general solution is
\be
\hat{\mu}_{\vec{k}}(\eta)&=&\hat{a}_{\vec{k}}(\eta_0)\,f_{k}(\eta)
+\hat{a}_{-\vec{k}}^{\dagger}(\eta_0)\,f^{\ast}_{k}(\eta)
\label{sol_mu}\\
\hat{\pi}_{k}(\eta)&=&\hat{a}_{\vec{k}}(\eta_0)\,g_{k}(\eta)
+\hat{a}_{-\vec{k}}^{\dagger}(\eta_0)\,g^{\ast}_{k}(\eta)
\ ,
\label{sol_pi}
\ee
where the mode function $f_{k}$ obeys
\be
f_{k}'' + \left[ k^2 - \frac{a''}{a}\right] f_{k} =0,
\label{f_eqT}
\ee
and $g_{k} = f_{k}'$.
In Eqs.~(\ref{sol_mu}) and (\ref{sol_pi}) the operators
$\hat{a}_{\vec{k}}(\eta_{0})$ annihilate the vacuum, 
\be
\hat{a}_{\vec{k}}(\eta_{0})| 0_{-\infty} \rangle =0
\quad
{\rm for}\
\eta_{0}\to-\infty
\ .
\ee
As anticipated, this is the state which minimises the Hamiltonian~(\ref{ham_4quant2})
close to the onset of inflation.
We remark that the time $\eta_{0}$ is the same for all the modes $k$.
The second remark is that the vacuum has zero total momentum during the entire
evolution.
This means that particles can only be produced in pairs
from the vacuum triggered by the pumping action of the gravitational field.
Inserting Eqs.~(\ref{sol_mu}) and (\ref{sol_pi}) into Eq.~(\ref{Four_expan}),
the following Fourier expansions can be simply obtained
\be
\hat{\mu}(\vec{x},\eta)&=&
\frac{1}{(2\,\pi)^{3/2}}\,\int\d^3k\,\left[\hat{a}_{\vec{k}}(\eta_0)\,f_{k}(\eta)\,
e^{-i\,\vec{k}\cdot\vec{x}}
+\hat{a}_{\vec{k}}^{\dagger}(\eta_0)\,f^{\ast}_{k}(\eta)\,e^{i\,\vec{k}\cdot\vec{x}}\right]
\\
\hat{\pi}(\vec{x},\eta)&=&
\frac{1}{(2\,\pi)^{3/2}}\,\int\d^3k\,\left[\hat{a}_{\vec{k}}(\eta_0)\,g_{k}(\eta)\,
e^{-i\,\vec{k}\cdot\vec{x}}
+\hat{a}_{\vec{k}}^{\dagger}(\eta_0)\,g^{\ast}_{k}(\eta)\,e^{i\,\vec{k}\cdot\vec{x}}\right]
\ ,
\ee
since, upon integration by parts, one has
$\hat{a}_{-\vec{k}}\,e^{i\,\vec{k}\cdot\vec{x}}
=\hat{a}_{\vec{k}}e^{-i\,\vec{k}\cdot\vec{x}}$  and
$\hat{a}_{-\vec{k}}^{\dagger}\,e^{-i\,\vec{k}\cdot\vec{x}}
=\hat{a}_{\vec{k}}^{\dagger}e^{i\,\vec{k}\cdot\vec{x}}$.
The two point function for the tensor fluctuations can then be obtained as
\be
\expecl{0_{-\infty}\left|\hat{h}(\vec{x},\eta)\,\hat{h}(\vec{y},\eta)\right|0_{-\infty}}
=\frac{2\,l_{\rm Pl}^2}{(2\,\pi)^3\,a^2(\eta)}\,\int\d^{3}k\,|f_{k}(\eta)|^2\,
e^{-i\,\vec{k}\cdot\vec{r}}
\ee
where $\vec{r}=(\vec{x}-\vec{y})$. After angular integration, the previous expression
becomes
\be
\expecl{0_{-\infty}\left|\hat{h}(\vec{x},\eta)\,\hat{h}(\vec{y},\eta)\right|0_{-\infty}}
=\int\d\ln{k}\,{\cal P}_h\,\frac{\sin{k\,r}}{k\,r}
\ee
where
\be
{\cal P}_h=\frac{l_{\rm Pl}^2}{a^2(\eta)}\,\frac{k^3}{\pi^2}\left|f_{k}(\eta)\right|^2,
\ee
is the tensor power spectrum.
In the study of inflation, the tensor spectral index is usually defined as
\be
n_{\rm T}=\frac{\d\ln{{\cal P}_h}}{\d\ln{k}}
\ ,
\ee
and its $\alpha$-running is
\be
\alpha_{\rm T}=\frac{\d^2\ln{{\cal P}_h}}{(\d\ln{k})^2}
\ .
\ee
\subsection{Large-scale power spectra of scalar fluctuations}
The calculation of the scalar power spectrum follows exactly the same
steps discussed in the case of the tensor modes.
\par
In full analogy with the calculation of the tensor power spectrum, the Hamiltonian
given in Eq.~(\ref{ham_equation3}) can be chosen.
Promoting the canonical mode $q$ and its conjugate momentum to quantum operators
which satisfy
\be
\left[\hat{q}(\eta,\vec{x}),\hat{\pi}_{q}(\eta,\vec{y})\right]
=i\,\delta^{(3)}(\vec{x}-\vec{y})
\ ,
\ee
one can also define a vacuum state $| 0_{-\infty} \rangle$ for the scalar modes
at $\eta\to-\infty$.
Recalling the relation between the canonical modes and the gauge-invariant
curvature fluctuations, obtained in Eq.~(\ref{from_R_to_q}), we can then write the scalar
two-point function as
\be
\expecl{0_{-\infty}\left|\hat{{\cal R}}(\eta,\vec{x})\,\hat{{\cal R}}(\eta,\vec{y})\right|0_{-\infty}}
=\frac{1}{z^2}\,\int\frac{\d^{3}k}{2\,\pi^3}\left|f_{{\rm s}\,\,k}(\eta)\right|^2\,
e^{-i\,\vec{k}\cdot\vec{r}}
\ ,
\ee
where $f_{{\rm s}\,k}^{q}(\eta)$ are the mode functions pertaining to the scalar problem and
obeying the equation
\be
f_{{\rm s}\,k}''+\left[k^2-\frac{z''}{z}\right]\,f_{{\rm s}\,k}=0
\ .
\label{f_eqS}
\ee
The scalar power spectrum, the scalar spectral index and its running will be defined,
with a convenient redefinition of the tensor quantities, in the Chapter~\ref{WKBb}.

\chapter{WKB-type approximations}
\label{WKBa}
In this chapter we give a review of some Wentzel-Kramers-Brillouin (WKB)-type approaches
used to solve second-order differential equations.
We start in the first section by reviewing the standard WKB approximation and then proceed,
in the second section, by presenting the problem that arises around the turning point(s)
of the frequency and how it can be avoided in other WKB-type approximations.
In particular, we give a first improved version of the WKB analysis in the third section and
a generalized WKB method, also called Method of Comparison Equations, in the last section.
\section{Standard WKB approximation}
\label{stWKBa}
In this Section we begin by recalling the standard WKB method, as is
introduced in almost all
Quantum Mechanics textbooks (See, for example~\cite{bender}).
\par
The WKB approximation is defined by first introducing a ``small''
parameter $\delta>0$ in order to perform the formal adiabatic
(or semiclassical) expansion of the mode
functions~\cite{birrel}. 
Let us briefly recall that such
an expansion is called adiabatic when $\sqrt{\delta}$ is the
inverse of a (typically very large) time over which the system
evolves slowly, and semiclassical when $\sqrt{\delta}\sim\hbar$.
Further, the limit $\delta\to 1$ must be taken at the end of the computation.
Consequently, second-order differential equations such as
Eqs.~(\ref{tensor_eq2}) and (\ref{scal_eq2})
[or equivalently Eqs.~(\ref{f_eqT}) and (\ref{f_eqS})] are formally
replaced by
\be
\left[\delta\,\frac{\d^2}{\d\eta^2}+\Omega^2(k,\eta)\right]\,
\mu=0
\ .
\label{eq_wkb_expa}
\ee
We then denote the leading order term in such an expansion
of $\mu$ by
\be
\mu_{{\rm WKB}}(k,\eta )=
\frac{{\rm e}^{\pm\frac{i}{\sqrt{\delta}}\,\int^{\eta }\Omega(k,\tau)
{\rm d}\tau}}{\sqrt{2\,\Omega(k,\eta)}}
\ ,
\label{wkb_sol}
\ee
which satisfies the following differential equation
\be
\left[\delta\,\frac{\d^2}{\d \eta^2}
+\Omega^2(k,\eta)-\delta\,Q_\Omega(k,\eta)\right]\,
\mu_{{\rm WKB}}=0
\ ,
\label{eq_wkb}
\ee
where
\be
Q_\Omega(k,\eta)\equiv
\frac{3}{4}\,\frac{(\Omega')^2}{\Omega^2}
-\frac{\Omega''}{2\,\Omega}
\ .
\label{Q}
\ee
We now observe that $\mu_{{\rm WKB}}$ is the exact solution of
Eq.~(\ref{eq_wkb_expa}) in the ``adiabatic limit'' $\delta\to 0$,
and is expected to be a good approximation to the exact $\mu$
(in the limit $\delta\to 1$) if
\be
\Delta\equiv
\left\vert\frac{Q_\Omega}{\Omega^2}\right\vert
\ll 1
\ .
\label{WKB_cond}
\ee
Sometimes, to respect the above condition, we need to perform a suitable change of variables.
This will indeed be the case for the cosmological perturbation and
in Chapter~\ref{WKBb} we shall give
the correct transformation to improve all WKB-type approximations.
After that change of variables we
shall use the following form for the differential equation
\be
\left[\frac{{\d}^2}{{\d} x^2}+\omega^2(x)\right]
\,\chi=0
\ ,
\label{new_eqa}
\ee
which is itroduced here to keep track of that transformation.
Finally, we recall that Eq.~(\ref{new_eqa}), but also Eq.~(\ref{eq_wkb_expa}),
in general suffer of some problems near the turning point of the frequency $\omega(x)$
which we now discuss.
\section{The ``Turning Point'' problem}
\label{TPproblem}
\begin{figure}[!ht]
\includegraphics[width=0.5\textwidth]{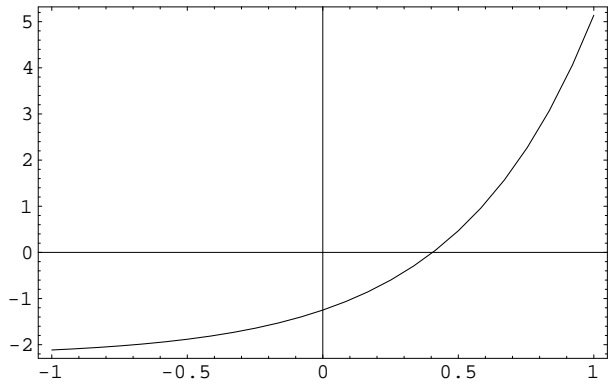}
\caption{Sketch of a typical frequency with one turning point (as is the case
for cosmological perturbations).
The independent variable is irrelevant for the discussion developed here.}
\label{freq_TP}
\end{figure}

We describe here a kind of problem that can arise around the turning point of the frequency
for the standard WKB approximation and how we can avoid it in other WKB-type
approximations.
First of all, a ``Turning Point'' is defined as that point where the frequency of a
second-order differential equation like~(\ref{eq_wkb_expa}) or~(\ref{new_eqa})
vanishes (see Fig.~\ref{freq_TP}).
A solution of the type given in Eq.~(\ref{wkb_sol}) (or the analogue for Eq.~(\ref{new_eqa}))
diverges in this point (as well as the error given by the term in Eq.~(\ref{Q}))
and it is well known that one therefore needs a different branch
of solution there.
\par
\begin{figure}[!hb]
\includegraphics[width=0.5\textwidth]{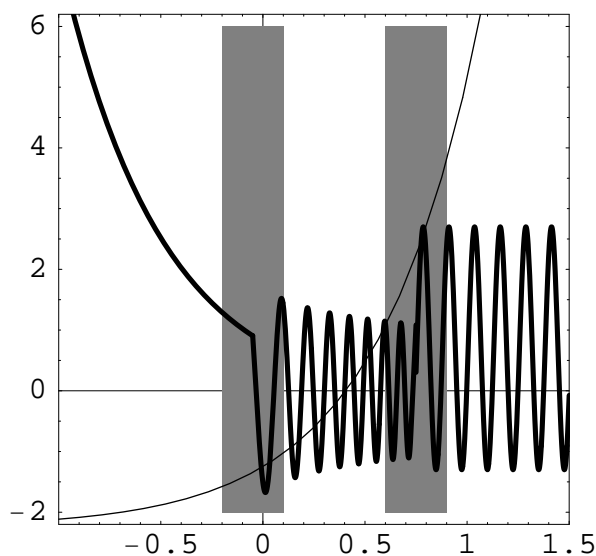}
\caption{Standard WKB approximation.}
\label{TPandWKB}
\end{figure}

Schematically, for the standard WKB approximation we have the situation
shown in Fig.~\ref{TPandWKB}, where we see the solution for an ordinary second-order
differential equation with only one turning point.
The standard procedure to address this problem is to divide
the domain of the independent variable into three regions:
on the left and on the right of the turning point, where the standard
WKB solution is valid, and the ``turning point region'', where we use a non-WKB solution.
Usually, around the turning point, the true frequency is then replaced by a linear approximation
and the approximate solution is then given by a linear combination of Airy functions.
\par
The next step is to connect the three branches to obtain a ``global'' solution on
the whole range of the independent variable.
We usually match the WKB solutions (i.e.,~the oscillatory and exponential branches) to that around
the turning point using two asymptotic expansions of the latter branch.
The overlap regions, where this asymptotic match is performed, are not {\em a priori} defined
but should be determined so as to minimise the error, which is usually a very difficult task,
although it does not normally affect the accuracy of the solution to leading order in $\delta$
(in Fig.~\ref{TPandWKB} we use two grey bands to indicate their possible location).
This subtlety has already been studied by the physicists and mathematicians who first used
the WKB approximation and it is known that the method can give some problems, for example,
if one needs a more accurate solution at next-to-leading order in the expansion in $\delta$.
In fact, if we used next-to-leading order expressions of the standard WKB expansion, we could find
that the overlap regions, where the match is to be performed, become too narrow
(or disappear at all!) and an accurate match of the different branches is no more
possible~\cite{jeffreys,langer34}.
This could even lead to next-to-leading order results less accurate than the leading ones~\cite{hunt}.
\par
\begin{figure}[!ht]
\includegraphics[width=0.5\textwidth]{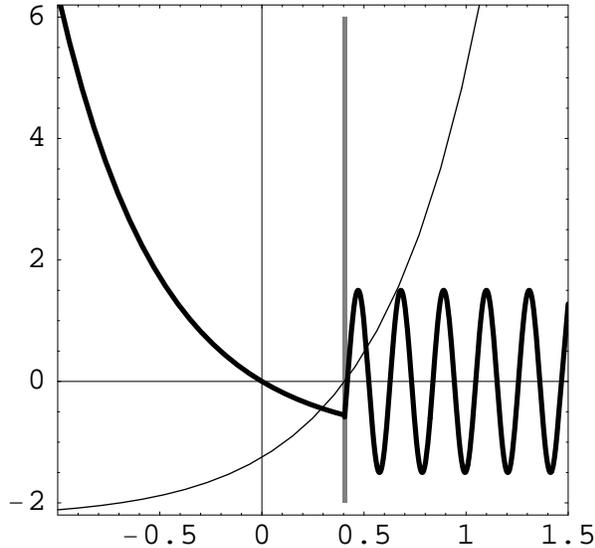}
\caption{Improved WKB approximation.}
\label{TPandImprovedWKB}
\end{figure}
The improved WKB approach, that we shall present in detail in the next section,
will give us an approximation which is also valid at the turnig point.
The solution will be divided into only two branches and the procedure of matching will be clearer
and more accurate.
Actually, the match will be performed exactly at the turning point, which is well located, and,
without ambiguous matching regions, the results at next-to-leading order are always improved (see Fig.~\ref{TPandImprovedWKB}).
\par
\begin{figure}[!ht]
\includegraphics[width=0.5\textwidth]{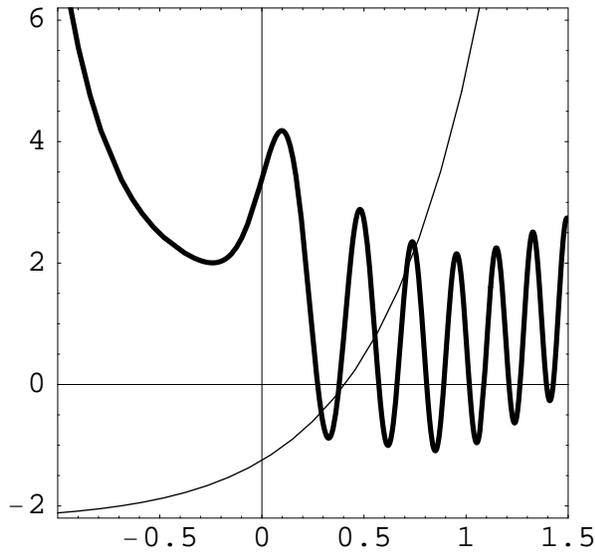}
\caption{MCE approximation.}
\label{TPandMCE}
\end{figure}
Finally, the Method of Comparison Equations, presented in the last part of this chapter,
is a generalised WKB method that will not need any matching between branches of solutions.
The solution will be valid on the whole range of the independent variable and could be used
in every problems where we need a global solution of our differential equation (see Fig.~\ref{TPandMCE}).
\par
So far we haven only considered one turning point, since this is the case for the equations
governing cosmological perturbations we shall solve for later.
We however note that, if the frequency has two, or more, turning points the solution of this problem
could be enormously complicated (a case with two turning points will be described in
Appendix~\ref{MCE_BH}).
\section{Improved WKB approximation}
\label{imprWKBa}
We shall now describe a method to improve the WKB approximation
for equations of the form in Eq.~(\ref{new_eqa})
to all orders in the adiabatic parameter $\delta$ by following the scheme
outlined in Ref.~\cite{langer} (see also Ref.~\cite{schiff}).
The basic idea is to use approximate expressions which are valid
for all values of the coordinate $x$ and then expand around them.
Two different such expansions we shall consider, namely
one based on the Green's function technique and the other
on the usual adiabatic expansion~\cite{langer}.
\par
Let us first illustrate the main idea of Ref.~\cite{langer}
for the particular case $\omega^{2}=C\,\left(x-x_0\right)^{n}$,
with $C$ a positive costant.
All the solutions to Eq.~(\ref{new_eqa}) for $x>x_0$ can be written
as linear combinations of the two functions
\be
u_{\pm}(x)=\sqrt{\frac{\xi(x)}{\omega(x)}}\,
J_{\pm m}\left[\xi(x)\right]
\ ,
\label{particular_sol}
\ee
where
\be
\xi(x)\equiv\int_{x_0}^{x}\omega(y)\,\d y
\ ,
\quad
\quad
m=\frac{1}{n+2}
\ ,
\label{xi_and_m}
\ee
and $J_\nu$ are Bessel functions~\cite{abram}.
Moreover, for a general frequency $\omega$,
the expressions in Eq.~(\ref{particular_sol})
satisfy
\be
\left[\frac{{\d}^2}{{\d} x^2}+\omega^2(x)-\sigma(x)\right]
\,u_{\pm}=0
\ ,
\label{eq_with_eta}
\ee
where the quantity (primes will denote
derivatives with respect to the argument of the given
function from here on)
\be
\sigma(x)
%&=&
=
\frac{3}{4}\,\frac{(\omega')^2}{\omega^2}
-\frac{\omega''}{2\,\omega}
+\left(m^{2}-\frac14\right)\,\frac{\omega^{2}}{\xi^{2}}
%\nonumber\\&=&
=
Q_\omega+\left(m^{2}-\frac14\right)\,\frac{\omega^{2}}{\xi^{2}}
\ ,
\label{sigma}
\ee
contains the term $Q$ defined in Eq.~(\ref{Q}),
whose divergent behavior at the turning point $x=x_0$ we
identified as the possible cause of failure of the standard WKB
approach.
For a general (finite) frequency, which can be expanded in powers
of $x-x_0$~\footnote{In the following we shall consider the
general case in Eq.~(\ref{omega_expand}), although one might restrict
to $n=1$, since the frequencies for cosmological perturbations
usually exhibit a linear turning point.},
\be
\omega^{2}(x)=C\,\left(x-x_0\right)^{n}
\left[1+\sum_{q\ge 1}\,c_q\,(x-x_0)^q\right]
\ ,
\label{omega_expand}
\ee
one finds that the extra term in Eq.~(\ref{sigma}) precisely
``removes'' the divergence in $Q$ at the turning point.
In fact, the residue
\be
\lim_{x\to x_0}\sigma(x)=\frac{3\,(n+5)\,c_1^{2}}{2\,(n+4)\,(n+6)}
-\frac{3\,c_2}{n+6}
\ ,
\label{limit_eta_at_TP}
\ee
is finite~\cite{langer,schiff}.
The finiteness of $\sigma$ at the turning point is crucial in
order to extend the WKB method to higher orders.
For slowly varying frequencies, with $|\sigma(x)|$ small
everywhere, the functions~(\ref{particular_sol}) are thus
expected to be good approximations to the solutions of
Eq.~(\ref{new_eqa}) for the whole range of $x$, including
the turning point.
\par
We can now introduce both the adiabatic expansion of the previous
Section and a new formal expansion of the mode functions by
replacing Eq.~(\ref{new_eqa}) by
\be
\left[\delta\,\frac{{\d}^2}{{\d} x^2}+
\omega^2(x)-\delta\,\sigma(x)\right]\,\chi
=-\delta\,\varepsilon\,\sigma(x)\,\chi
\ ,
\label{new_eq_rewrite}
\ee
where $\delta$ and $\varepsilon$ are ``small'' positive
parameters.
For clarity, we shall refer to expressions proportional to $\delta^n$ as
the $n$-th adiabatic order and to those proportional to
$\varepsilon^n$ as the $n$-th perturbative order.
\par
It is convenient to consider the solutions to
Eq.~(\ref{new_eq_rewrite}) to the left and right of the
turning point $x_0$ separately.
We shall call:
\begin{subequations}
\begin{description}
\item[region~I)]
where $\omega^{2}>0$ (on the right of $x_0$), with
\be
\omega_{\rm I}(x)\equiv\sqrt{\omega^{2}(x)}
\ ,
\quad
\xi_{\rm I}(x)\equiv\int_{x_0}^{x}\omega_{\rm I}(y)\,\d y
\ ,
\label{xi_I}
\ee
and the solutions to the corresponding homogeneous equation
(\ref{eq_with_eta}) are given in Eq.~(\ref{particular_sol});
\item[region~II)]
where $\omega^{2}<0$ (on the left of
$x_0$), with
\be
\omega_{\rm II}(x)\equiv\sqrt{-\omega^{2}(x)}
\ ,
\quad
\xi_{\rm II}(x)\equiv\int_{x}^{x_0}\omega_{\rm II}(y)\,\d y
\ ,
\label{xi_II}
\ee
and the solutions to the corresponding homogeneous equation
(\ref{eq_with_eta}) are obtained from those in
Eq.~(\ref{particular_sol}) by replacing the $J_\nu$ with the $I_\nu$~\cite{abram}.
\end{description}
\end{subequations}
Corresponding expressions are obtained for all relevant
quantities, for example $\sigma_{\rm I}$ and $\sigma_{\rm II}$,
and we shall omit the indices I and II whenever it is not
ambiguous.
\par
Although it is possible to expand in both parameters,
so as to obtain terms of order $\delta^p\,\varepsilon^q$
(with $p$ and $q$ positive integers),
in the following we shall just consider each expansion separately,
that is we shall set $\delta=1$ in the perturbative expansion
(see~Section~\ref{pert}) and $\varepsilon=1$ in the adiabatic
expansion (see~Section~\ref{adiaba}).
\subsection{Perturbative expansion}
\label{pert}
For $\delta=1$, the limit $\varepsilon\to 0$ is exactly solvable
and yields the solutions (\ref{particular_sol}), whereas the case
of interest (\ref{new_eqa}) is recovered in the limit
$\varepsilon\to 1$, which must always be taken at the end of
the computation.
The solutions to Eq.~(\ref{new_eq_rewrite}) can be further
cast in integral form as
\be
\chi(x)=u(x)-\varepsilon\,\int G(x,y)\,\sigma(y)\,\chi(y)\,\d y
\ ,
\label{complete_sol}
\ee
where $u(x)=A_+\,u_{+}(x)+A_-\,u_{-}(x)$ is a linear combination
of solutions (\ref{particular_sol}) to the corresponding homogeneous
equation~(\ref{eq_with_eta}), and $G(x,y)$ is the Green's function
determined by
\be
\left[\frac{{\d}^2}{{\d} x^2}+\omega^2(x)-\sigma(x)\right]
\,G(x,y)
=\varepsilon\,\delta(x-y)
\ .
\label{green_eq}
\ee
The solutions of the homogeneous equation~(\ref{eq_with_eta})
in the two regions are then given by
\be
&&
\!\!\!\!\!\!\!\!\!\!u_{\rm I}(x)=
\displaystyle\sqrt{\frac{\xi_{\rm I}(x)}{\omega_{\rm I}(x)}}\,
\left\{
A_{+}\,J_{+m}\left[\xi_{\rm I}(x)\right]
+A_{-}\,J_{-m}\left[\xi_{\rm I}(x)\right]
\right\}
%\nonumber\\&\!\!\!\equiv\!\!\!&
\!\equiv\!
A_+\,u_{{\rm I}+}(x)+A_-\,u_{{\rm I}-}(x)
\nonumber
\\
\label{sol_in}
\\
&&
\!\!\!\!\!\!\!\!\!\!u_{\rm II}(x)=
\displaystyle\sqrt{\frac{\xi_{\rm II}(x)}{\omega_{\rm II}(x)}}\,
\left\{
B_{+}\,I_{+m}\left[\xi_{\rm II}(x)\right]
+B_{-}\,I_{-m}\left[\xi_{\rm II}(x)\right]
\right\}
%\nonumber\\&\!\!\!\equiv\!\!\!&
\!\equiv\!
B_+\,u_{{\rm II}+}(x)+B_-\,u_{{\rm II}-}(x)
\nonumber
\ee
and the Green's functions can be written as
\be
&&
\!\!\!\!\!\!\!\!\!\!\!\!\!\!G_{\rm I}(x,y)=
-\displaystyle\frac{\pi}{\sqrt{3}}\,
\theta(x-y)\,u_{{\rm I}-}(x)\,u_{{\rm I}+}(y)
%\nonumber\\&&
-\displaystyle\frac{\pi}{\sqrt{3}}\,
\theta(y-x)\,u_{{\rm I}+}(x)\,u_{{\rm I}-}(y)
\nonumber
\\
\label{green_funct}
\\
&&
\!\!\!\!\!\!\!\!\!\!\!\!\!\!G_{\rm II}(x,y)=
\displaystyle\frac{i\,\pi}{\sqrt{3}}\,
\theta(x-y)\,u_{{\rm II}-}(x)\,u_{{\rm II}+}(y)
%\nonumber\\&&
+\,\displaystyle\frac{i\,\pi}{\sqrt{3}}\,
\theta(y-x)\,u_{{\rm II}+}(x)\,u_{{\rm II}-}(y)
\ .
\nonumber
\ee
\par
With the ansatz
\be
\!\!\!\!\!\!\!\!\!\!
\chi_{\rm I}(x)&\!\!=\!\!&
A_{+}\,a_{+}(x)\,u_{{\rm I}+}(x)
+A_{-}\,a_{-}(x)\,u_{{\rm I}-}(x)
\nonumber
\\
\label{ansatz}
\\
\!\!\!\!\!\!\!\!\!\!
\chi_{\rm II}(x)&\!\!=\!\!&
B_{+}\,b_{+}(x)\,u_{{\rm II}+}(x)
+B_{-}\,b_{-}(x)\,u_{{\rm II}-}(x)
\ ,
\nonumber
\ee
from Eqs.~(\ref{sol_in}) and~(\ref{green_funct}),
the problem is now reduced to the determination
of the $x$-dependent coefficients
\be
&&
\!\!\!\!\!\!\!\!\!\!
a_{+}(x)\equiv\left[1+\varepsilon\,J^+_{+-}(x,x_i)\right]
+\varepsilon\,\frac{A_{-}}{A_+}\,J^-_{--}(x,x_i)
\nonumber
\\
\nonumber
\\
&&
\!\!\!\!\!\!\!\!\!\!
a_{-}(x)\equiv
\left[1+\varepsilon\,J^-_{+-}(x_0,x)\right]
+\varepsilon\,\frac{A_{+}}{A_-}\,J^+_{++}(x_0,x)
\nonumber
\\
\label{Abar_Bbar}
\\
&&
\!\!\!\!\!\!\!\!\!\!
b_{+}(x)\equiv
\left[1-i\,\varepsilon\,I^+_{+-}(x,x_0)\right]
-i\,\varepsilon\,\frac{B_{-}}{B_+}\,I^-_{--}(x,x_0)
\nonumber
\\
\nonumber
\\
&&
\!\!\!\!\!\!\!\!\!\!
b_{-}(x)\equiv
\left[1-i\,\varepsilon\,I^-_{+-}(x_f,x)\right]
-i\,\varepsilon\,\frac{B_{+}}{B_-}\,I^+_{++}(x_f,x)
\ ,
\nonumber
\ee
in which, for the sake of brevity (and clarity),
we have introduced a shorthand notation for the following
integrals
\be
%&&
\!\!\!\!\!\!\!\!\!\!
J^{w}_{s\bar s}(x_1,x_2)&\equiv&
\displaystyle\frac{\pi}{\sqrt{3}}\,
\int_{x_1}^{x_2}
\sigma_{\rm I}(y)\,a_w(y)\,
u_{{\rm I}s}(y)\,u_{{\rm I}\bar s}(y)\,\d y
\nonumber
\\
%&&\!\!\!\!\!\!\!\!\!\!
&=&
\displaystyle\frac{\pi}{\sqrt{3}}\,\int_{x_1}^{x_2}
\!\!\!\!
\sigma_{\rm I}(y)
\,\frac{\xi_{\rm I}(y)}{\omega_{\rm I}(y)}\,a_{w}(y)\,
\,J_{sm}\left[\xi_{\rm I}(y)\right]\,J_{\bar sm}\left[\xi_{\rm I}(y)\right]
\d y
\nonumber
\\
\phantom{A}
\\
%&&
\!\!\!\!\!\!\!\!\!\!
I^{w}_{s\bar s}(x_1,x_2)&\equiv&
\displaystyle\frac{\pi}{\sqrt{3}}\,
\int_{x_1}^{x_2}
\sigma_{\rm II}(y)\,b_w(y)\,
u_{{\rm II}s}(y)\,u_{{\rm II}\bar s}(y)
\,\d y
\nonumber
\\
%&&\!\!\!\!\!\!\!\!\!\!\!\!
&=&
\displaystyle\frac{\pi}{\sqrt{3}}\,\int_{x_1}^{x_2}
\!\!\!\!
\sigma_{\rm II}(y)
\,\frac{\xi_{\rm II}(y)}{\omega_{\rm II}(y)}\,b_{w}(y)\,
\,I_{sm}\left[\xi_{\rm II}(y)\right]\,
I_{\bar sm}\left[\xi_{\rm II}(y)\right]
\d y
\, ,
\nonumber
\ee
where the symbols $w$, $s$ and $\bar s$ are $+$ or $-$.
\par
Before tackling this problem, we shall work out relations
which allow us to determine the constant coefficients $A_\pm$
and $B_\pm$ uniquely.
\subsection{Adiabatic expansion}
\label{adiaba}
Let us now apply the usual adiabatic expansion with the
assumption that the leading order be given by the
functions~(\ref{particular_sol})
[rather than the more common expression~(\ref{wkb_sol})].
This leads one to replace
\be
\xi(x)\to
\frac{\xi(x)}{\sqrt{\delta}}
\equiv\frac{1}{\sqrt{\delta}}\,\int^x
\omega(y)\,\d y
\ ,
\ee
and consider the forms~\cite{langer}.
\begin{subequations}
\be
U_\pm=F_\pm(x;\delta)\,u_\pm(x)+G_\pm(x;\delta)\,u'_\pm(x)
\ ,
\ee
where the $x$-dependent coefficients $F$ and $G$ must now
be determined.
The particular case of a linear turning
point~\cite{langer49}
(precisely the one which mostly concerns us here),
can then be fully analyzed as follows.
\subsubsection{Recursive (adiabatic) relations}
The $x$-dependent coefficients $F$ and $G$ can be
expanded in powers of $\delta$ as
\be
&&
F(x;\delta)=\sum_{j\ge 0} \delta^j\,\phi_{(j)}(x)
\nonumber
\\
\label{adiab_exp}
\\
&&
G(x;\delta)=\sum_{j\ge 0} \delta^j\,\gamma_{(j)}(x)
\nonumber
\ .
\ee
Upon substituting into Eq.~(\ref{new_eq_rewrite}) with
$\varepsilon=1$, one therefore obtains that the coefficients
$\phi_{(j)}$ and $\gamma_{(j)}$ are given recursively
by the formulae
\be
\phi_{(j)}(x)&\!\!=\!\!&-\frac{1}{2}
\int^{x}
\left[\gamma''_{(j)}(y)+\sigma(y)\,\gamma_{(j)}(y)\right]
\d y
\nonumber
\\
\label{delta_j}
\\
\gamma_{(j)}(x)&\!\!=\!\!&\frac{1}{2\,\omega(x)}
\int^x
\!\!
\left\{
\sigma(y)\,\left[2\,\gamma'_{(j-1)}(y)+\phi_{(j-1)}(y)\right]
%\,\frac{\d y}{\omega(y)}
\right.
\nonumber
\\
&&
\phantom{\frac{1}{2\,\omega(x)}\,\int\!\!\!}
\left.
+\sigma'(y)\,\gamma_{(j-1)}(y)+\phi''_{(j-1)}(y)\right\}
\frac{\d y}{\omega(y)}
\ ,
\nonumber
\ee
\end{subequations}
where it is understood that the integration must be performed
from $x_0$ to $x$ in region~I and from $x$ to $x_0$ in region~II.
Let us also remark that, for $\omega^2\sim (x-x_0)$ near the
turning point, the above expressions are finite~\cite{langer49},
whereas for more general cases one expects divergences as
with the more standard WKB approach.\\
\\
{\em Leading order}\\
\\
The lowest order solutions (\ref{particular_sol}) are correctly
recovered on setting
\be
\phi_{(0)}=1
\quad
{\rm and}
\quad
\gamma_{(0)}=0
\ ,
\label{delta0}
\ee
in the limit $\delta=1$.
The relevant leading order perturbations are therefore obtained
on imposing the initial conditions~(\ref{init_cond_on_chi})
and matching conditions between region~I and region~II at the
turning point, which therefore yield for $\chi_{\rm I}$ and
$\chi_{\rm II}$ the same linear combinations $u_{\rm I}$
of Eq.~(\ref{sol_in_I_leading}) and $u_{\rm II}$ of
Eq.~(\ref{sol_in_II_leading}) previously obtained.\\
\\
{\em Higher orders}\\
\\
The condition~(\ref{delta0}) makes the formal expressions of
the coefficients $\phi_{(1)}$ and $\gamma_{(1)}$ particularly
simple,
\be
\phi_{(1)}(x)&\!\!=\!\!&-\frac{1}{2}\,\int^{x}
\left[\gamma''_{(1)}(y)+\sigma(y)\,\gamma_{(1)}(y)\right]\,
\d y
\nonumber
\\
\label{delta1}
\\
\gamma_{(1)}(x)&\!\!=\!\!&\frac{1}{2\,\omega(x)}\,\int^x
\frac{\sigma(y)}{\omega(y)}\,
\d y
\ ,
\nonumber
\ee
and the first order solutions are then given by linear combinations
of the two functions
\be
U_\pm(x)=\left[1+\delta\,\phi_{(1)}(x)\right]\,u_\pm(x)
+\delta\,\gamma_{(1)}(x)\,u'_\pm(x)
\ .
\ee
\section{Method of Comparison Equation}
\label{sMCEa}
In this Section we briefly review the Method of Comparison Equation
(MCE).
The name is due to Dingle and it was independently introduced
in Refs.~\cite{MG,dingle} and 
applied to wave mechanics by Berry and Mount in Ref.~\cite{berry}.
The standard WKB approximation and its improvement by Langer~\cite{langer}
are just particular cases of this method.
In Section~\ref{appKLV} we shall also present the connection of the MCE with
the Ermakov-Pinney equation.
\par
We start from the well-known second-order differential equation~(\ref{new_eqa})
and suppose that we know an exact solution to a
similar second-order differential equation,
\be
\left[
\frac{{\d}^2}{{\d}\sigma^2}+\Theta^2(\sigma)
\right]
\,U(\sigma)=0
\ ,
\label{aux_EQ}
\ee
where $\Theta$ is the ``comparison function''.
One can then represent an exact solution of Eq.~(\ref{new_eqa})
in the form
\be
\chi(x)=\left(\frac{{\d}\sigma}{{\d}x}\right)^{-1/2}\,U(\sigma)
\ ,
\label{exact_SOL}
\ee
provided the variables $x$ and $\sigma$ are related by
\be
\omega^2(x)\!=\!\left(\frac{{\d}\sigma}{{\d}x}\right)^{2}\Theta^2(\sigma)
-\left(\frac{{\d}\sigma}{{\d}x}\right)^{1/2}
\frac{{\d}^2}{{\d}x^2}\left(\frac{{\d}\sigma}{{\d}x}\right)^{-1/2}
\ .
\label{new_EQ}
\ee
Eq.~(\ref{new_EQ}) can be solved by using some iterative
scheme, in general cases as~\cite{hecht} and Appendix~\ref{app}
or Refs.~\cite{mori,pechukas}  for specific problems.
If we choose the comparison function sufficiently similar to
$\omega$, the second term in the right hand side of
Eq.~(\ref{new_EQ}) will be negligible with respect to the first
one, so that
\be
\omega^2(x)\simeq\left(\frac{{\d}\sigma}{{\d}x}\right)^{2}
\Theta^2(\sigma)
\ .
\label{new_EQ_appr}
\ee
On selecting a pair of values $x_0$ and $\sigma_0$ such that
$\sigma_0=\sigma(x_0)$, the function $\sigma(x)$ can be implicitly
expressed as
\be
-\xi(x)\equiv
\int_{x_0}^x\sqrt{\pm\,\omega^2(y)}\,{\d}\,y
\simeq
\int_{\sigma_0}^{\sigma}\sqrt{\pm\,\Theta^2(\rho)}\,{\d}\,\rho
\ ,
\label{new_EQ_int}
\ee
where the signs are chosen conveniently~\footnote{Here
$\xi(x)$ corresponds to the quantity $\xi_{\rm II}(x)$ used in
Section~\ref{imprWKBa} when the signs are both chosen as minus.}.
The result in Eq.~(\ref{exact_SOL}) leads to a uniform approximation
for $\chi(x)$, valid in the whole range of the variable $x$, including
``turning points''.
The similarity between $\Theta$ and $\omega$ is clearly very
important in order to implement this method.
\subsection{MCE vs Ermakov\&Pinney}
\label{appKLV}
%
%\subsubsection{The general case}
%\label{intro}
In Ref.~\cite{KLV}, we studied the relation between the MCE~\cite{MG,dingle}
and the Ermakov - Pinney equation~\cite{Ermakov,Pinney}.
Let us consider again the second-order differential equation~(\ref{new_eqa})
but with the small parameter $\varepsilon$ defined by~\footnote{With respect to
Eq.~(\ref{new_eqa}), we have absorbed a minus sign in $\varepsilon$,
as this change of sign is irrelevant for the present topic and for its application in
Appendix~\ref{app}.}
\begin{eqnarray}
\frac{d^2\chi(x)}{dx^2} = \frac{1}{\varepsilon}\omega^2(x)\chi(x)
\ .
\label{equation}
\end{eqnarray}
Let us suppose that one knows the solution of another 
second-order differential equation
\begin{eqnarray}
\frac{d^2U(\sigma)}{d\sigma^2} = \frac{1}{\varepsilon}
\Theta^2(\sigma)U(\sigma)
\ .
\label{equation1}
\end{eqnarray}
In this case one can represent an {\it exact} solution of
Eq.~(\ref{equation}) in the form
\begin{eqnarray}
\chi(x) = \left(\frac{d\sigma}{dx}\right)^{-1/2}U(\sigma)
\ ,
\label{form}
\end{eqnarray}
where the relation between the variables $x$ and $\sigma$ is given 
by the equation
\begin{eqnarray}
\omega^2(x) = \left(\frac{d\sigma}{dx}\right)^2\Theta^2(\sigma) + 
\varepsilon \left(\frac{d\sigma}{dx}\right)^{1/2}\frac{d^2}{dx^2}
\left(\frac{d\sigma}{dx}\right)^{-1/2}
\ .
\label{compare}
\end{eqnarray}
On solving Eq.~(\ref{compare}) for $\sigma(x)$ and substituting
for it into Eq.~(\ref{form}) one finds the solution to Eq.~(\ref{equation}).
Eq.~(\ref{compare}) can be solved by using some iterative scheme.
The application of such a scheme is equivalent to the application of the 
MCE or, in other words, to the construction of the uniform asymptotic
expansion  for the solution of Eq.~(\ref{equation}).
The method of construction of the solution to Eq.~(\ref{equation})
by means of the solutions of Eqs.~(\ref{equation1}) and (\ref{compare})
is called the method of comparison equations and the function 
$\Theta^2(\sigma)$ is called the comparison function~\cite{dingle}.
\par
The iterative scheme of solution of Eq.~(\ref{compare}) depends essentially
on the form of the comparison function $\Theta(\sigma)$.
A reasonable approach consists in the elimination of
$\sigma$ from Eq.~(\ref{compare}) and its reduction to a form,
where the only unknown function is $\left({d\sigma}/{dx}\right)$.
The equation obtained for the case of the simple comparison function 
$\Theta^2(\sigma) = 1$ coincides with the Ermakov-Pinney equation, this  will
be explicitly shown in the next section. For more complicated forms
of the comparison function $\Theta^2(\sigma)$ useful for the description 
of various physical problems the corresponding equation will acquire a more
intricate form and we shall call it a generalised Ermakov-Pinney equation.
The role of an unknown function in this equation is played by the function
\begin{eqnarray}
y(x) \equiv    
\left(\frac{d\sigma}{dx}\right)^{-1/2}
\ ,
\label{y-define}
\end{eqnarray}
while the equation itself has the form 
\begin{eqnarray}
K\left(y,\frac{dy}{dx}\right) = \varepsilon L\left(y,\frac{dy}{dx},
\frac{d^2y}{dx^2},\frac{d^3y}{dx^3}\right)
\ ,
\label{gen-eq}
\end{eqnarray}
where the explicit forms for the functions $K$ and $L$ will be defined 
in the next sections. We shall represent the function $y(x)$ as a series
\begin{eqnarray}
y = \sum_{n=0}^{\infty} y_n \varepsilon^n
\ ,
\label{asymp}
\end{eqnarray}  
and the general recurrence relation connecting 
the different coefficient functions
$y_n$ can be represented as 
\begin{eqnarray}
\left.\left(\frac{d^nK}{d\varepsilon^n} - \frac{d^{n-1}L}{d\varepsilon^{n-1}}
\right)\right|_{\varepsilon = 0} = 0
\ .
\label{gen-rec}
\end{eqnarray}
To implement this formula one may use the combinatorial relation~\cite{Grad}
\begin{eqnarray}
\frac{d^n(uv)}{dx^n} = (u+v)^{(n)}
\ ,
\label{Grad}
\end{eqnarray}
where on the right-hand side one has a binomial expression wherein
the powers are replaced by the derivatives of the same order.
When the function $y(x)$  is constructed up to some level of approximation
one can, in principle,  find $\sigma(x)$ and substitute for it into 
Eq.~(\ref{form}). 
\par
We shall consider the implementation of this scheme 
for four choices of the comparison function: $\Theta^2(\sigma) = 1,
\Theta^2(\sigma) = \sigma, \Theta^2(\sigma) = \exp(\sigma) -1$ and
$\Theta^2(\sigma) = \sigma^2 - a^2$. At the end we shall
discuss the relation between the different WKB-type methods
and equations of the Ermakov-Pinney type. In Appendix~\ref{app} we shall give a pure numerical
example of application  of the method of comparison equations.
\subsubsection{No turning points}
\label{NO_TP}
In the case for which the function $\omega^2(x)$ does not have zeros (turning
points) it is convenient to choose a comparison function 
$\Theta^2(\sigma) = 1$.
In this case the function $U(\sigma)$ is simply an exponent, while
Eq.~(\ref{compare}) can be rewritten as
\begin{eqnarray}
\varepsilon\frac{d^2y(x)}{dx^2} = \omega^2(x)y(x) - \frac{1}{y^3(x)}
\ ,
\label{Pinney}
\end{eqnarray}
where the function $y(x)$ is defined by Eq.~(\ref{y-define}).
The above equation is nothing more then the well known Pinney or
Ermakov-Pinney Eq.~\cite{Ermakov,Pinney}, which can be solved perturbatively
with respect to the parameter $\varepsilon$~\cite{Lewis}.
We shall give here some general formulae for such a solution.
It is convenient to rewrite Eq.~(\ref{Pinney}) as 
\begin{eqnarray}
\omega^2 y^4 - 1 = \varepsilon y^3\ddot{y}
\ ,
\label{Pinney1}
\end{eqnarray}
where a ``dot'' denotes the derivative with respect to $x$. 
We shall search for the solution of Eq.~(\ref{Pinney1}) in the
form~(\ref{asymp}). The zero-order solution of Eq.~(\ref{Pinney1}) is
\begin{eqnarray}
y_0 = \omega^{-1/2}
\ ,
\label{zero-order}
\end{eqnarray}
and the general recurrence relation following from Eq.~(\ref{Pinney1}) 
for $n \geq 1$ is
\begin{eqnarray}
y_n &=&
\frac{1}{4y_0^3}\sum_{k_1 = 0}^{n-1}
\cdots
\sum_{k_4 = 0}^{n-1}
\left\{\frac{1}{\omega^2}\,
\delta\left[\left(n-1\right)-\left(k_1+k_2+k_3+k_4\right)\right]
\ddot{y}_{k_1}y_{k_2}y_{k_3}y_{k_4}\right.\nonumber
\\
&&
\quad\quad\quad\quad\quad\quad\quad\,\,\,\,\,
-\left.\delta\left[n-\left(k_1+k_2+k_3+k_4\right)\right]
y_{k_1}y_{k_2}y_{k_3}y_{k_4}\right\}
\ ,
\label{recurrent}
\end{eqnarray}
where $\delta()$ is the Kronecker delta symbol. 
It is now easy to obtain from the general expression~(\ref{recurrent})
expressions for particular values of $y_n$
\begin{subequations}
\begin{eqnarray}
y_1 &=& \frac{\ddot{y}_0}{4\omega^2}
\label{y1}
\\
y_2 &=& -\frac{3y_1^2}{2y_0} + \frac{\ddot{y}_1}{4\omega^2}
+\frac{3\ddot{y_0}y_1}{4\omega^2 y_0}
\label{y2}
\\
y_3 &=& -\frac{y_1^3}{y_0^2} - \frac{3y_2 y_1}{y_0}
 +\frac{\ddot{y}_2}{4\omega^2}
+\frac{3\ddot{y}_0 y_2}{4\omega^2 y_0} +
\frac{3\ddot{y}_1 y_1}{4\omega^2 y_0} +\frac{3\ddot{y}_0 y_1^2}
{4\omega^2 y_0^2}
\ .
\label{y3}
\end{eqnarray}
\end{subequations}
Let us end  by noting that the standard WKB approximation 
corresponds to the case discussed here and is, of course, only
valid away from turning points. 
\subsubsection{One turning point: the Langer solution}
\label{langer_TP}
In the case of the presence of a linear zero in the 
function $\omega^2(x)$ one can use 
the Langer solution~\cite{langer}. In terms of the method of comparison
equations it means that one chooses the comparison function 
$\Theta^2(\sigma) = \sigma$. In this case Eq.~(\ref{compare}) becomes
\begin{eqnarray}
\omega^2(x) = \left(\frac{d\sigma}{dx}\right)^2\sigma + 
\varepsilon \left(\frac{d\sigma}{dx}\right)^{1/2}\frac{d^2}{dx^2}
\left(\frac{d\sigma}{dx}\right)^{-1/2}
\ .
\label{compare1}
\end{eqnarray}
Dividing this equation by $\left({d\sigma}/{dx}\right)^2$ 
and differentiating the result with respect to $x$ we get an equation
which depends only on the derivative $\dot{\sigma}$ and not on the function
$\sigma$. Such an equation can be rewritten in the form 
\begin{eqnarray}
2\dot{\omega}\omega y^6 + 4\omega^2\dot{y}y^5 = 
1 + \varepsilon\left(3\ddot{y}\dot{y}y^4
+y^{(3)}y^5\right)
\ ,
\label{Langer}
\end{eqnarray}
where the function $y(x)$, as  in the preceding section,  
is defined by Eq.~(\ref{y-define}).
Again we shall search for the solution of Eq.~(\ref{Langer}) in the
form~(\ref{asymp}). The equation for $y_0$ is
\begin{eqnarray}
2\dot{\omega}\omega y_0^6 + 4\omega^2\dot{y_0}y_0^5 =
\left(\omega^2 y_0^4\right)^{.}y_0^2 = 1
\ .
\label{Langer0}
\end{eqnarray}
This equation may be rewritten as
\begin{eqnarray}
\sqrt{\omega^2 y_0^4}\,\frac{d}{dx}\left(\omega^2 y_0^4\right) = \omega
\ ,
\end{eqnarray}
and integrated 
\begin{eqnarray}
\omega^2 y_0^4 = \left(\frac32\int \omega  dx\right)^{2/3}
\ .
\label{langer01}
\end{eqnarray}
Finally
\begin{eqnarray}
y_0 = \frac{1}{\omega^{1/2}} \left(\frac32\int \omega dx\right)^{\frac16}
\ .
\label{Langer02}
\end{eqnarray}
Comparing the coefficients of $\varepsilon^n$ with $n \geq 1$ in
Eq.~(\ref{Langer}) we have
\begin{eqnarray}
4\omega^2 y_0^5 \dot{y}_n + 
\left(12\dot{\omega}\omega y_0^5 + 20\omega^2\dot{y}_0y_0^4\right)y_n =
4\omega^{-1}\left(\omega^{3}y_0^5y_n\right)^{.} \equiv F_n
\ ,
\label{Langern}
\end{eqnarray}
where $F_n$ contains the terms which depend on $y_k,0\leq  k \leq n-1$,
and can be written as
\begin{eqnarray}
F_n &\!=\!& \sum_{k_1=0}^{n-1}
\!\!\cdots\!\!
\sum_{k_6=0}^{n-1}
\left\{
-\delta\left[n\!-\!\left(k_1\!+\!k_2\!+\!k_3\!+\!k_4\!+\!k_5\!
+\!k_6\right)\right]
\left(4\omega^2\dot{y}_{k_1}+2\dot{\omega}\omega y_{k_1}\right)
y_{k_2}\!\cdots\!y_{k_6}
\right.
\nonumber \\
&&+\left.\delta[\left(n\!-\!1\right)\!-\!(k_1\!+\!k_2\!+\!k_3\!+\!k_4\!
+\!k_5\!+\!k_6)]
\left(3\ddot{y}_{k_1}\dot{y}_{k_2}
+y^{(3)}_{k_1}y_{k_2}\right)
y_{k_3}\!\cdots\!y_{k_6}\right\}
\ .
\label{Fn}
\end{eqnarray}
The general expression for $y_n$ is
\begin{eqnarray}
y_n =\frac{1}{4\omega^{3}y_0^5}\int dx \omega F_n
\ .
\label{Langern1}
\end{eqnarray}
Let us also give  the explicit expressions for the first terms in the
expansion~(\ref{asymp})
\begin{subequations}
\begin{eqnarray}
y_1 &=&\frac{1}{4\omega^{3}y_0^5}\int dx \omega 
\left(3\ddot{y}_0\dot{y}_0y_0^4 + y^{(3)}_0 y_0^5\right)
\label{Langer1}
\\
y_2 &=&\frac{1}{4\omega^{3}y_0^5}\int dx \omega 
\left(-30\dot{\omega}\omega y_1^2y_0^4 - 20\omega^2\dot{y}_1y_1y_0^4
-40\omega^2\dot{y}_0y_1^2y_0^3 \right.\label{Langer2} \\
&&\,\,\quad\quad\quad\quad
+\left.3\ddot{y}_1\dot{y}_0y_0^4 + 3\ddot{y}_0\dot{y}_1y_0^4 + 
y^{(3)}_1y_0^5 + 5y^{(3)}_0y_1y_0^4+12\ddot{y}_0\dot{y}_0y_1y_0^3\right)
\nonumber
\\
y_3 &=&\frac{1}{4\omega^{3}y_0^5}\int dx \omega 
\left(-60\dot{\omega}\omega y_2y_1y_0^4 - 40\dot{\omega}\omega y_1^3y_0^3-
20\omega^2\dot{y}_2y_1y_0^4-80\omega^2\dot{y}_0y_2y_1y_0^3 
\right.\nonumber \\
&&\quad\quad
\left.-40\omega^2\dot{y}_0y_1^3y_0^2
-40\omega^2\dot{y}_1y_1^2y_0^3
-20\omega^2\dot{y}_1y_2y_0^4
+3\ddot{y}_2\dot{y}_0y_0^4
+3\ddot{y}_0\dot{y}_2y_0^4
\right.\nonumber \\
&&\quad\quad
+12\ddot{y}_0\dot{y}_0y_2y_0^3 
+3\ddot{y}_1\dot{y}_1y_0^4+\left.12\ddot{y}_1\dot{y}_0y_1y_0^3
+12\ddot{y}_0\dot{y}_1y_1y_0^3
+18\ddot{y}_0\dot{y}_0y_1^2y_0^2
\right.\nonumber \\
&&\quad\quad
\left.+y^{(3)}_2y_0^5
+5y^{(3)}_0y_2y_0^4
+5y^{(3)}_1y_1y_0^4
+10y^{(3)}_0y_1^2y_0^3\right)
\ .
\label{Langer3}
\end{eqnarray}
\end{subequations}
Before closing this section we may  compare  our results 
with some formulae, obtained by Dingle~\cite{dingle}.
In the vicinity of the turning point, which we choose as $x=0$,
the function $\omega^2(x)$ can be represented as 
\begin{eqnarray}
\omega^2(x) = \gamma_1 x + \frac{\gamma_2}{2}x^2 + \frac{\gamma_3}{3!}x^3
+ \frac{\gamma_4}{4!}x^4+ \cdots
\label{gamma-turn}
\end{eqnarray}
while the function $\sigma(x)$ looks like  
\begin{eqnarray}
\sigma(x) = \sigma_0 + \sigma_1 x + \frac{\sigma_2}{2}x^2
+\frac{\sigma_3}{3!}x^3
+ \frac{\sigma_4}{4!}x^4 + \cdots
\label{sigma-turn}
\end{eqnarray}
In reference~\cite{dingle} the first five coefficients $\sigma_0,
\sigma_1,\sigma_2, \sigma_3$ and $\sigma_4$, as functions of the coefficients
$\gamma_1,\gamma_2,\gamma_3$ and $\gamma_4$, were obtained in the
lowest approximation by using a system of recurrence relations. We can
reproduce
all these formulae on only using the  general formula~(\ref{Langer02}).
Indeed, on writing
\begin{eqnarray}
\frac{d\sigma}{dx} = \frac{1}{y^2} = \frac{\omega(x)}
{\left[\frac32\int \omega(x)\right]^{1/3}}
\label{Langer-turn}
\end{eqnarray}
using expansions~(\ref{gamma-turn}) and (\ref{sigma-turn}) and
comparing the coefficients of different powers of $x$ we get the formulae
\begin{eqnarray}
\sigma_0 &=& 0 \nonumber \\
\sigma_1 &=& \gamma_1^{1/3} \nonumber \\
\sigma_2 &=& \frac15\gamma_1^{-2/3}\gamma_2 \nonumber \\
\sigma_3 &=& \frac17\gamma_1^{-2/3}\gamma_3 - \frac{12}{175}\gamma_1^{-5/3}
\gamma_2^2\nonumber \\
\sigma_4 &=& \frac19\gamma_1^{-2/3}\gamma_4 -
\frac{44}{315}\gamma_1^{-5/3}\gamma_2\gamma_3 + \frac{148}{2625}
\gamma_1^{-8/3}\gamma_2^3
\ ,
\label{sigma-turn1}
\end{eqnarray}
which coincide with those obtained in~\cite{dingle}. The value of $\sigma_0$
comes directly from Eq.~(\ref{compare1}) on ignoring the $\epsilon$-term and
using the relations~(\ref{gamma-turn}) and (\ref{sigma-turn}).
\par
We can also find the $\varepsilon$-dependent corrections to the function
$\sigma(x)$ in the neighbourhood of the point $x = 0$ by using the
results~(\ref{sigma-turn1}) and the recurrence relations~(\ref{Langer1}) etc.
We shall only give here the first correction, proportional 
to $\varepsilon$, to the coefficients $\sigma_0$ and $\sigma_1$.
The value of $\sigma_0$ comes directly from Eq.~(\ref{compare1}) using the
relations~(\ref{gamma-turn}), (\ref{sigma-turn}) and (\ref{sigma-turn1})
\begin{eqnarray}
\sigma_0 = \varepsilon\left(\frac{\gamma_3\gamma_1^{-5/3}}{14}
-\frac{9\gamma_2^2\gamma_1^{-8/3}}{140}\right)
\ .
\label{sigma0-next}
\end{eqnarray}
In order to obtain $\sigma_1$ we introduce into the definition of the
function $y(x)$, Eq.~(\ref{y-define}), the expansion~(\ref{sigma-turn}) and
\begin{eqnarray} 
y_0\!=\!\sigma_1^{-1/2}\left[1\!-\!\frac{\sigma_2}{2\sigma_1}x
\!+\!\left(\frac{3\sigma_2^2}{8\sigma_1^2}\!
-\!\frac{\sigma_3}{4\sigma_1}\right)x^2
%\right. \nonumber \\&&\left
\!+\!\left(\frac{3\sigma_2\sigma_3}{8\sigma_1^2}\!
-\!\frac{\sigma_4}{12\sigma_1}
\!-\!\frac{5\sigma_2^3}{16\sigma_1^3}\right)x^3\right]\!+\cdots
\label{y0-neig}
\end{eqnarray}
Now, on using the recurrence formula~(\ref{Langer1}) we find
the leading (constant) term in the expression for $y_1$
\begin{eqnarray}
y_1 = \frac{1}{\gamma_1}\left(-\frac{1}{2}\sigma_2^3\sigma_1^{-7/2}
+\frac{1}{2}\sigma_3\sigma_2\sigma_1^{-5/2}
-\frac{\sigma_4\sigma_1^{-3/2}}{12}\right)
+\cdots
\label{y1-neig}
\end{eqnarray}
From the formula
\begin{eqnarray}
\frac{d\sigma}{dx} = \frac{1}{y^2} = \frac{1}{(y_0 + \varepsilon y_1)^2}
\label{sigma-y}
\end{eqnarray}
it is easy to find the first correction to the first coefficient
$\sigma_1$ in the expansion~(\ref{sigma-turn})
\begin{eqnarray}
\sigma_1 = \gamma_1^{1/3} - 2\varepsilon \frac{y_1}{y_0^3}
\ .
\label{sigma-cor}
\end{eqnarray}
Substituting into Eq.~(\ref{sigma-cor}) the leading term of
$y_1$ from Eq.~(\ref{y1-neig}), of $y_0$ from~(\ref{y0-neig}) and
using the explicit expressions for the coefficients $\sigma_i$ from
Eq.~(\ref{sigma-turn1}) we finally obtain
\begin{eqnarray}
\sigma_1 =
\gamma_1^{1/3}
%\nonumber\\&&
+ \varepsilon\left(\frac{7}{225}\gamma_2^3
\gamma_1^{-11/3} -\frac{7}{135}\gamma_3\gamma_2\gamma_1^{-8/3}
+\frac{1}{54}\gamma_4\gamma_1^{-5/3}\right)
\ .
\label{sigma-cor1}
\end{eqnarray}
The results~(\ref{sigma0-next}) and (\ref{sigma-cor1}) was found by
Dingle~\cite{dingle}, but in his article there were two sign errors in the
second line of equation (69).
\subsubsection{One turning point: more complicated comparison function}
\label{cosmpert_TP}
Let us consider the differential Eq.~(\ref{equation}),
when the function $\omega^2(x)$ has a linear turning point, but
instead of Langer's comparison function $\Theta^2(\sigma) = \sigma$
consider a more complicated comparison function, also having linear turning
point, namely
\begin{eqnarray}
\Theta^2(\sigma) = e^{\sigma}-1
\ .
\label{gammanew}
\end{eqnarray}
Now, Eq.~(\ref{compare}) will have the form
\begin{eqnarray}
\omega^2(x) = \left(\frac{d\sigma}{dx}\right)^2\left(e^{\sigma}-1\right)+
\varepsilon \left(\frac{d\sigma}{dx}\right)^{1/2}\frac{d^2}{dx^2}
\left(\frac{d\sigma}{dx}\right)^{-1/2}
\ .
\label{compare2}
\end{eqnarray}
On isolating the function $e^{\sigma}$, taking its logarithm and
differentiating it with respect to $x$ we finally get an equation which
involves only the derivative $\dot{\sigma}$. This equation can be written as
\begin{eqnarray}
\omega^2 y^4 -2\dot{\omega}\omega y^6 -4\omega^2\dot{y}y^5 + 1
= \varepsilon\left(\ddot{y}y^3 - 3\ddot{y}\dot{y}y^4 - y^{(3)}y^5\right)
\ .
\label{nonLanger}
\end{eqnarray}
To find the function $y_0$ one should solve the equation
\begin{eqnarray}
\omega^2 y_0^4 -2\dot{\omega}\omega y_0^6 -4\omega^2\dot{y}_0y_0^5 + 1 = 0
\ .
\label{nonLanger0}
\end{eqnarray} 
On introducing the new variable 
\begin{eqnarray}
z \equiv \omega^2 y_0^4
\label{z-define}
\end{eqnarray}
one can rewrite Eq.~(\ref{nonLanger0}) as 
\begin{eqnarray}
\dot{z} = \frac{(z+1)\omega}{z^{1/2}}
\ .
\label{z-eq}
\end{eqnarray}
It is also convenient  to introduce the variable
\begin{eqnarray}
v \equiv \sqrt{z}
\ .
\label{v-define}
\end{eqnarray}
Eq.~(\ref{z-eq}) now becomes
\begin{eqnarray}
\frac{2v^2\dot{v}}{1+v^2} = \omega
\label{v-eq}
\end{eqnarray}
which can be integrated, obtaining 
\begin{eqnarray}
2v -2\arctan v = \int \omega dx
\ .
\label{v-sol}
\end{eqnarray}
Finally, for $y_0$ we have the implicit representation
\begin{eqnarray}
y_0^2 - \frac{1}{\omega} \arctan \omega y_0^2 =
\frac{1}{2\omega}\int \omega dx
\ .
\label{nonLanger01}
\end{eqnarray}
We may now write the equation defining the recurrence relation
for $y_n, n\geq 1$
\begin{eqnarray}
-4\omega^2 y_0^5 \dot{y}_n\!+\!\left(4\omega^2 y_0^3 - 20\omega^2
\dot{y}_0y_0^4-12\dot{\omega}\omega y_0^5\right)y_n
= -4\frac{\left(\omega^3\,y_0^5\,G\,y_n\right)^{.}}{\omega\,G} \equiv  F_n
\ ,
\label{nonLangern}
\end{eqnarray}
where
\begin{eqnarray}
F_n &\!=\!& 
\sum_{k_1=0}^{n-1}
\!\!\cdots\!\!
\sum_{k_4=0}^{n-1}
\left\{\delta\left[\left(n\!-\!1\right)\!-\!\left(k_1\!+\!k_2\!+\!k_3\!
+\!k_4\right)\right]\,
\ddot{y}_{k_1}y_{k_2}y_{k_3}y_{k_4}\right.
\label{nonLangerF}\\
&&\quad\quad\quad\quad\quad
\left.-\omega^2\,\delta\left[n\!-\!\left(k_1\!+\!k_2\!+\!k_3\!
+\!k_4\right)\right]y_{k_1}y_{k_2}y_{k_3}y_{k_4}\right\}
\nonumber \\
&\!+\!&
\sum_{k_1=0}^{n-1}
\!\!\cdots\!\!
\sum_{k_6=0}^{n-1}
\left\{
\delta\left[n\!-\!\left(k_1\!+\!k_2\!+\!k_3\!+\!k_4\!+\!k_5\!
+\!k_6\right)\right]\!
\left(2\,\dot{\omega}\,\omega\,y_{k_1}+4\,\omega^2\,
\dot{y}_{k_1}\right)\!y_{k_2}\!\cdots\!y_{k_6}\right.
\nonumber \\
&&\,
\left.-\delta\left[\left(n\!-\!1\right)\!-\!\left(k_1\!+\!k_2\!+\!k_3\!
+\!k_4\!+\!k_5\!+\!k_6\right)\right]\!
\left(3\,\ddot{y}_{k_1}\dot{y}_{k_2}+y^{(3)}_{k_1}y_{k_2}\right)\!
y_{k_3}\!\cdots\! y_{k_6}\right\}
\ ,\nonumber
\end{eqnarray}
and
\begin{eqnarray}
G = \exp\left(-\int^{x}\frac{dx'}{y_0^2(x')}\right)
\ .
\label{G-define}
\end{eqnarray}
The solution for $y_n$ is
\begin{eqnarray}
y_n = -\frac{1}{4\omega^{3}y_0^5G}\int dx\,\omega G F_n
\ ,
\label{nonLangernsol}
\end{eqnarray}
Here we give the explicit expressions for the first two functions $F_1$
and $F_2$
\begin{subequations}
\begin{eqnarray}
F_1 &=& \ddot{y}_0y_0^3 - 3\ddot{y}_0\dot{y}_0y_0^4 - y^{(3)}_0
y_0^5
\label{nonLangerF1}\\
F_2 &=& -6\omega^2 y_1^2 y_0^2 + 30\dot{\omega}\omega y_1^2y_0^4
+20\omega^2\dot{y}_1y_1y_0^4
+40\omega^2\dot{y}_0y_1^2y_0^3 
+\ddot{y}_1y_0^3
\label{nonLangerF2} \\
&+&3\ddot{y}_0y_1y_0^2
-3\ddot{y}_1\dot{y}_0y_0^4
-3\ddot{y}_0\dot{y}_1y_0^4
-12\ddot{y}_0\dot{y}_0y_1y_0^3
-y^{(3)}_1y_0^5
-5y^{(3)}_0y_1y_0^4
\ .
\nonumber
\end{eqnarray}
\end{subequations}
\subsubsection{Two turning points}
\label{2_TP}
Let us now consider the differential Eq.~(\ref{equation}), when the
function $\omega^2(x)$ has two turning points. This situation, describing
the tunneling through a potential barrier was considered
in~\cite{MG,berry}. In this case the comparison function can be chosen
to be
\begin{eqnarray}
\Theta^2(\sigma) = \sigma^2 - a^2
\ .
\label{comp-two}
\end{eqnarray}
The solutions  $U(\sigma)$ of Eq.~(\ref{equation1}) are the well-known
parabolic functions~\cite{berry}. Substituting  $\Theta^2(\sigma)$
of Eq.~(\ref{comp-two}) into Eq.~(\ref{compare}) one can isolate
$\sigma$ and after the subsequent differentiation one can write down the
equation for the function $y(x)$
\begin{eqnarray}
2\sqrt{\omega^2 y^4 + a^2 - \varepsilon \ddot{y}y^3} =
2\dot{\omega}\omega y^6 +4\omega^2\dot{y}y^5 -
\varepsilon\left(3\ddot{y}\dot{y}y^4
+y^{(3)}y^5\right)
\ .
\label{comp-two1}
\end{eqnarray}
The lowest-approximation $y_0$ can be found from the equation
\begin{eqnarray}
2\sqrt{\omega^2 y_0^4 + a^2} = 
2\dot{\omega}\omega y_0^6 +4\omega^2\dot{y}_0y_0^5
\ .
\label{comp-two2}
\end{eqnarray}
This equation can be integrated and the solution $y_0$ can be written down
in the implicit form
\begin{eqnarray}
\omega y_0^3 -\frac{a^2}{\omega}{\rm arcsinh}
\frac{\omega y_0}{a} = \frac{2}{\omega}\int\omega dx
\ .
\label{implicit}
\end{eqnarray}
To get the recurrence relations for the functions $y_n, n \geq 1$ it is
convenient to take the square of Eq.~(\ref{comp-two1})
\begin{eqnarray}
&&4\dot{\omega}^2\omega^2 y^{12}+16\omega^4\dot{y}^2y^{10}
+16\dot{\omega}\omega^3\dot{y}y^{11}-4\omega^2 y^4-4a^2
=
\nonumber \\
&&\quad\quad\quad\quad\quad-\varepsilon(4\ddot{y}y^3-12\dot{\omega}\omega
\ddot{y}\dot{y}y^{10}
-24\omega^2 \ddot{y}\dot{y}^2y^9 - 4\dot{\omega}\omega y^{(3)}y^{11}
-8\omega^2y^{(3)}\dot{y}y^{10})
\nonumber \\
&&\quad\quad\quad\quad\quad-\varepsilon^2(9\ddot{y}^2\dot{y}^2y^8
+{y^{(3)}}^2y^{10}
+6y^{(3)}\ddot{y}\dot{y}y^9)
\ .
\label{comp-two3}
\end{eqnarray}
On comparing the terms containing the $n^{\rm th}$ power of the small
parameter $\varepsilon$ one obtains
\begin{eqnarray}
f \dot{y}_n + g y_n 
= \frac{f}{G}\,\left(G\,y_n \right)^{.}
\equiv  F_n
\ ,
\label{rec-two}
\end{eqnarray}
where
\begin{subequations}
\begin{eqnarray}
f&=& 32\omega^4\dot{y}_0y_0^{10} + 16\dot{\omega}\omega^3 y_0^{11}
\\
\label{f-def}
g&=& 48\dot{\omega}^2\omega^2 y_0^{11} + 160\omega^4\dot{y}_0^2y_0^9
+176\dot{\omega}\omega^3\dot{y_0}y_0^{10}-16\omega^2 y_0^3
\\
\label{g-def}
F_n&=&
4\sum_{k_1=0}^{n-1}\!\cdots\!\sum_{k_4=0}^{n-1}
\left\{\omega^2\delta\left[n-\left(k_1+k_2+k_3+k_4\right)\right]
y_{k_1}y_{k_2}y_{k_3}y_{k_4}\right.\nonumber\\
&&\left.\quad\quad\quad\quad\quad\quad
-\delta\left[(n-1)-\left(k_1+k_2+k_3+k_4\right)\right]\ddot
{y}_{k_1}y_{k_2}y_{k_3}y_{k_4}\right\}\nonumber\\
&-&4\sum_{k_1=0}^{n-1}\!\cdots\!\sum_{k_{12}=0}^{n-1}
\left\{\delta\left[n-\left(k_1+\cdots+k_{12}\right)\right]\right.\nonumber\\
&&\left.
\times \left(\dot{\omega}^2\omega^2 y_{k_1}y_{k_2}
+4\omega^4\dot{y}_{k_1}\dot{y}_{k_2}
+4\dot{\omega}\omega^3\dot{y}_{k_1}y_{k_2}\right)
y_{k_3}\!\cdots\! y_{k_{12}}\right.\nonumber \\
&&\left.\quad\quad\quad\quad\quad\quad
-\delta\left[(n-1)-\left(k_1+\cdots+k_{12}\right)\right]\right.\nonumber\\
&&\left.
\times\,\left(3\dot{\omega}\omega\ddot{y}_{k_1}\dot{y}_{k_2}y_{k_3}+
6\omega^2\ddot{y}_{k_1}\dot{y}_{k_2}\dot{y}_{k_3}
+\dot{\omega}\omega y^{(3)}y_{k_2}y_{k_3}+
2\omega^2y^{(3)}_{k_1}\dot{y}_{k_2}y_{k_3}\right)y_{k_4}\!\cdots\! y_{k_{12}}
\right\}\nonumber \\
&-&\sum_{k_1=0}^{n-2}\!\cdots\!\sum_{k_{12}=0}^{n-2}
\delta\left[(n-2)-\left(k_1+\cdots+k_{12}\right)\right]\nonumber\\
&&
\times\,\left(9\ddot{y}_{k_1}\ddot{y}_{k_2}\dot{y}_{k_3}\dot{y}_{k_4}
+y^{(3)}_{k_1}y^{(3)}_{k_2}
y_{k_3}y_{k_4} + 6y^{(3)}_{k_1}\ddot{y}_{k_2}\dot{y}_{k_3}
y_{k_4}\right) y_{k_5}\!\cdots\! y_{k_{12}}
\ .
\label{F-two}
\\
G&=&\exp\left(\int^x dx' \frac{g}{f}\right)
\ .
\label{G-two}
\end{eqnarray}
\end{subequations}
The integration of Eq.~(\ref{rec-two}) gives
\begin{eqnarray}
y_{n} = \frac{1}{G}\int dx \frac{G F_n}{f}
\ .
\label{yntwo}
\end{eqnarray}
The explicit expressions for the coefficients $F_n$ are rather
cumbersome and we shall only write down  the coefficient $F_1$
\begin{eqnarray}
F_1=12\dot{\omega}\omega\ddot{y}_0\dot{y}_0y_0^{10}
+24\omega^2\ddot{y}_0\dot{y}_0^2y_0^9+4\dot{\omega}\omega y^{(3)}_0y_0^{11}
+8\omega^2y^{(3)}_0\dot{y}_0y_0^{10}-4\ddot{y}_0y_0^3
\ .
\label{F1-two}
\end{eqnarray}
\subsubsection{Historical and physical remarks}
\label{WKB_ErmPinn}
We have seen that, for some non-trivial applications of the MCE,
instead of the standard form of the Ermakov-Pinney equation,
one is lead to its generalizations which are rather 
different from the known generalizations of Ermakov systems. 
Let us note that although the Ermakov equation was already reproduced 
in the MCE context (first of all in the important paper by
Milne~\cite{Milne}), its generalizations arising in the process of
application of the MCE~\cite{MG,dingle} were
not yet studied, at least to the best of our knowledge.
We shall therefore try to give
a short review of the history of the Ermakov-Pinney equation
and its applications to quantum mechanics and to similar WKB-type problems
and discuss the physical significance of its non-trivial generalization
(for a more detailed account of the history of the Eramkov-Pinney equation
see e.g.~\cite{Espinosa}).
\par
The relation between the second-order linear differential equation 
\begin{eqnarray}
\frac{d^2 u}{dx^2} + \omega^2(x)\chi(x) = 0
\label{Ermakov}
\end{eqnarray}
and the non-linear differential equation 
\begin{eqnarray}
\frac{d^2 \rho}{dx^2} + \omega^2(x)\rho(x) = \frac{\alpha}{\rho^3(x)}
\ ,
\label{Ermakov1}
\end{eqnarray} 
where $\alpha$ is  some constant
was noticed by Ermakov~\cite{Ermakov}, who  showed 
that any two  solutions $\rho$ and $u$  of the above equations are 
connected by the formula
\begin{eqnarray}
C_1 \int \frac{dx}{u^2} + C_2 = \sqrt{C_1\frac{\rho^2}{u^2} - \alpha}
\ .
\label{Ermakov2}
\end{eqnarray}
The couple of Eqs.~(\ref{Ermakov}) and (\ref{Ermakov1}) constitute
the so called Ermakov system. An important corollary was
derived~\cite{Ermakov} from the formula~(\ref{Ermakov2}).
Namely, on having a particular solution $\rho(x)$ of Eq.~(\ref{Ermakov1}) 
one can construct the general solution of Eq.~(\ref{Ermakov})
which is given by
\begin{eqnarray}
\chi(x) = c_1 \rho(x) \exp\left[\sqrt{-\alpha}\int \frac{dx}{\rho^2(x)}\right]+
c_2 \rho(x) \exp\left[-\sqrt{-\alpha}\int \frac{dx}{\rho^2(x)}\right]
\ .
\label{Ermakov3}
\end{eqnarray}
It is easy to see that the function $\rho(x)$ plays of the role of an
``amplitude'' of the function $\chi(x)$, while the integral
$\int \left[{dx}/{\rho^2(x)}\right]$ represents some kind of ``phase''.
Thus, it is not surprising that the Ermakov equation was re-discovered by
Milne~\cite{Milne} in a quantum-mechanical context. On introducing the
amplitude obeying Eq.~(\ref{Ermakov1}) and the phase, Milne  constructed
a formalism for the solution of the Schr\"odinger equation which was
equivalent to the MCE. Milne's MCE technique was
extensively used for the solution of quantum mechanical problems
(see e.g.~\cite{Fano}). In the paper by Pinney~\cite{Pinney} the general form
of the  solution of the Ermakov equation was presented,
while the most general expression for this solution was written down by
Lewis~\cite{Lewis}.
\par
A very simple physical example rendering trasparent the derivation 
of the Ermakov-Pinney equation and its solutions was given in the paper 
by Eliezer and Gray~\cite{Gray}. The motion of a two-dimensional 
oscillator with  time-dependent frequency was considered. In this case 
the second-order linear differential Eq.~(\ref{Ermakov})
describes the projection of the 
 motion of 
this two-dimensional oscillator on one of its Cartesian coordinates,
while the Ermakov-Pinney Eq.~(\ref{Ermakov1}) describes the evolution 
of the radial coordinate $\rho$. The parameter $\alpha$ is nothing more than
the square of the conserved angular momentum of the two-dimensional
oscillator. Thus, the notion of the amplitude and phase acquire in this
example a simple geometrical and physical meaning. Similarly, the
supersymmetric generalization of the Ermakov-Pinney equation was considered
in the paper by Ioffe and Korsch~\cite{ioffe}.
\par
An additional generalization of the notion of an Ermakov system of equations
was suggested in the paper by Ray and Reid~\cite{Ray}. Thy consider
the system of two
equations 
\begin{eqnarray}
\frac{d^2 u}{dx^2} + \omega^2(x)\chi(x) = \frac{1}{\rho x^2}g
\left(\frac{\rho}{x}\right)
\label{Ray}
\end{eqnarray}
and  
\begin{eqnarray}
\frac{d^2 \rho}{dx^2} + \omega^2(x)\rho(x) = \frac{1}{x\rho^2(x)}
f\left(\frac{x}{\rho}\right)
\ ,
\label{Ray1}
\end{eqnarray} 
where $g$ and $f$ are arbitrary functions of their arguments.
The standard Ermakov system~(\ref{Ermakov}), (\ref{Ermakov1})
corresponds to the choice of functions $g(\xi) = 0$ and 
$f(\xi) = \alpha\xi$. One can show that the generalized Ermakov
system~(\ref{Ray}), (\ref{Ray1}) has an invariant
(zero total time derivative)
\begin{eqnarray}
I_{f,g} = \frac12\left[\phi\left(\frac{u}{\rho}\right) 
+\theta\left(\frac{\rho}{u}\right)
+\left(u\frac{d\rho}{dx}-\rho\frac{du}{dx}\right)^2\right]
\ ,
\label{invar}
\end{eqnarray}
where 
\begin{subequations}
\begin{eqnarray}
\phi\left(\frac{u}{\rho}\right) &=& 2\int^{\frac{u}{\rho}}f(x)dx
\label{invar1}
\\
\theta\left(\frac{\rho}{u}\right) &=& 2\int^{\frac{\rho}{u}}g(x)dx
\ .
\label{invar2}
\end{eqnarray}
\end{subequations}
The invariant~(\ref{invar}) establishes the connection between
the solutions of Eqs.~(\ref{Ray}) and (\ref{Ray1}) and sometimes
allows one to  find the solution  of one of these equations provided
the solution of the other one is known (just as in the case of the standard
Ermakov system). However, for the case of an arbitrary couple of functions
$f$ and $g$ a simple physical or geometrical interpretation of the
Ermakov system analogous to that given in~\cite{Gray} is not known.
\par
Here we have established the connection between
the generalized WKB method (MCE~\cite{MG,dingle})
and the Ermakov - Pinney equation and on the other hand we have obtained a
generalization of the Ermakov - Pinney equation, different to those
studied in the literature. 
One has two second-order differential Eqs.~(\ref{equation}) and
(\ref{equation1}) and the solution of one of them~(\ref{equation1}) is
well known and described. Using the previous convenient analogy~\cite{Gray},
one can say that in this case  
one has two time-dependent oscillators, evolving with two different
time-parameters. One can try to find the solution of Eq.~(\ref{equation})
 by representing it as a known  solution of Eq.~(\ref{equation1})
multiplied by a correction factor. 
This correction factor plays the role of the prefactor
in the standard WKB approach while the known solution of
Eq.~(\ref{equation1}) represents some kind of  generalized phase term.
Further, the prefactor is expressed in terms of the derivative
between two variables $x$ and $\sigma$ (or in terms of the oscillator
analogy, between two times). 
On writing down the equation definining this factor, which we have denoted
 by $y(x)$ (see Eq.~(\ref{y-define})) we arrive to Eq.~(\ref{compare})
which could be called a generalized Ermakov-Pinney equation.
For the case when the function $\omega^2(x)$ does not have turning points
the comparison function $\Theta^2(\sigma)$ can be chosen constant
(the second oscillator has a time-independent frequency) and the
the equation defining the prefactor becomes the standard
Ermakov-Pinney Eq.~(\ref{Pinney}). In terms of the two-dimensional
oscillator analogy~\cite{Gray}
this means that we exclude from the equation for the radial
coordinate the dependence on the angle coordinate $\sigma$ by using its
cyclicity, i.e. the conservation of the angular moment.
\par
In cases for which the function $\omega^2(x)$ has
turning points, as in Secs.~\ref{langer_TP}, \ref{cosmpert_TP} and
\ref{2_TP}, the comparison functions are non-constant and instead of the
standard Ermakov-Pinney equation we have its non-trivial generalizations.
The common  feature of these generalizations consists in the fact that the
corresponding equations for the variable $y$ also depend explicitly on the
parameter $\sigma$ which does not disappear automatically. To get a
differential equation for $y$, one should isolate the parameter $\sigma$
and subsequently differentiate ti with respect to $x$. As a result one gets a
differential equation of higher-order for the function $y(x)$. Remarkably,
for the perturbative solution of these equations one can again construct
a reasonable iterative procedure.
\par
Again, it is interesting to look at the generalized Ermakov-Pinney equation
as an equation describing a two-dimensional physical system in the spirit
of the reference~\cite{Gray}. In our case one has
\begin{eqnarray}
\frac{d^2y(x)}{dx^2} + \omega^2(x) y(x) = \frac{\Theta^2(\sigma)}{y^3(x)}
\ ,
\label{Gray}
\end{eqnarray}
where $y$ plays the role of a radial coordinate, $\sigma$
resembles a phase or 
an angle (e.g. the position of a hand of a clock)
and $x$ is a time parameter. It is important to notice
that according to the definition of the variable $y$~(\ref{y-define})
there is a relation between the radial and angle coordinates 
\begin{eqnarray}
\frac{d\sigma}{dx} = \frac{1}{y^2}
\ ,
\label{angular}
\end{eqnarray}
which 
on interpreting $\sigma$ as a phase corresponds to constant (unit) angular
momentum.
In our case however
the right-hand side of the radial Eq.~(\ref{Gray}) 
contains an explicit dependence on the ``angle'' $\sigma$. Thus,
the couple of Eqs.~(\ref{Gray}) and (\ref{angular})
cannot be represented as a system of equations of motion corresponding to
some Lagrangian or Hamiltonian as in Ref.~\cite{Gray}.
Perhaps, this system could be described
in terms of non-Hamiltonian dissipative dynamics, but this question
requires a further study.
\chapter{Analysis of Cosmological Perturbations}
\label{WKBb}
In this chapter, we shall apply the different WKB approaches reviewed previously
to the equations that drive the scalar and tensorial cosmological perturbations.
In the first Section, we recall the main expressions and review 
one fo the first applications of the WKB approximation to
cosmolgical perturbations~\cite{martin_schwarz2}.
In the second Section, we present the improved version of the WKB analysis~\cite{WKB1}
and, in the last part, we employ the MCE~\cite{MCE}
\section{Standard WKB approximation}
\label{stWKBb}
In this Section we begin by recalling that the straightforward
application of the standard WKB method leads to a poor
approximation and that the results can be improved to first adiabatic
order by means of the transformation~(\ref{transf}), as was shown in
Ref.~\cite{martin_schwarz2}.
However, the extension to higher orders is more subtle, as we shall
discuss in some details.
\par
We have seen that the general equation for cosmological perturbations is
\be
\left[\delta\,\frac{\d^2}{\d\eta^2}+\Omega_j^2(k,\eta)\right]\,
\mu_j=0
\ ,
\label{eq_wkb_expb}
\ee
where
\be
\Omega_j^2(k,\eta)=k^2-\frac{z''_j}{z_j}
\ ,
\label{freq_eta}
\ee
with $j=S$ for scalar perturbations and $j=T$ for tensorial ones, so $z_S=z$ and $z_T=a$.
We can now solve Eq.~(\ref{eq_wkb_expb})
together with the condition that the modes are initially plane waves for
wavelengths much shorter than the Hubble radius,
\be
\lim_{\frac{k}{a\,H}\rightarrow +\infty} \mu(k,\eta)
=d\,\frac{{\rm e}^{-i\,k\,\eta}}{\sqrt{2\,k}}
\ ,
\label{init_cond_on_mu}
\ee
where $d=d_{\rm S}=1$ and $d=d_{\rm T}=\sqrt{8\,\pi}/m_{\rm Pl}$
for scalar and tensor perturbations ($m_{\rm Pl}$ is
the Planck mass).
So the dimensionless power spectra of scalar and tensor
fluctuations are given by the usual definitions
\be
\mathcal{P}_{\zeta}=
\displaystyle\frac{k^{3}}{2\,\pi^{2}}\,
\left|\frac{\mu_{\rm S}}{z_{\rm S}}\right|^{2}
\ ,
\ \
\mathcal{P}_{h}=
\displaystyle\frac{4\,k^{3}}{\pi^{2}}\,
\left|\frac{\mu_{\rm T}}{z_{\rm T}}\right|^{2}
\ .
\label{spectra_def}
\ee
The spectral indices and their runnings give the standard parametrization for the power spectra.
They are usually defined at an arbitrary pivot scale $k_*$~\footnote{Sometimes the spectral
indices and their runnings are defined without assuming $k=k_*$.
We shall always make it clear where we use this assumption.} as
\be
n_{\rm S}-1=
\left.\displaystyle\frac{\d\ln \mathcal{P}_{\zeta}}
{\d\ln k}\right|_{k=k_{*}}
\ ,
\ \
n_{\rm T}=
\left.\displaystyle\frac{\d\ln \mathcal{P}_{h}}
{\d\ln k}\right|_{k=k_{*}}
\ ,
\label{n_def}
\ee
and
\be
\alpha_{\rm S}=\left.
\frac{\d^{2}\ln\mathcal{P}_{\zeta}}
{(\d\ln k)^{2}}\right|_{k=k_{*}}
\ ,
\ \
\alpha_{\rm T}=\left.
\frac{\d^{2}\ln \mathcal{P}_{h}}
{(\d\ln k)^{2}}\right|_{k=k_{*}}
\ .
\label{alpha_def}
\ee
We also define the tensor-to-scalar ratio
\be
R=\frac{\mathcal{P}_{h}}{\mathcal{P}_{\zeta}}
\ .
\label{Rat_def}
\ee
In the following we shall drop the subscript $j$, $S$ and $T$ if they are not necessary.
On using the potentials in Eqs.~(\ref{tens_pot}) and (\ref{scal_pot}) in terms of
$a$ and $H$, one finds, for the quantities defined in Eq.~(\ref{WKB_cond})
\be
\Delta_{\rm S,T}=
\left\{ \begin{array}{ll}
\displaystyle\left(\frac{a\,H}{k}\right)^{4}
\mathcal{O}(\epsilon_{n})\sim 0
&
\textrm{sub-horizon scales}
\\
&
\\
\displaystyle\frac{1}{8}+\mathcal{O}(\epsilon_{n})\
&
\textrm{super-horizon scales.}
\end{array}
\right.
\label{cond_prechange}
\\
\nonumber
\ee
The result for super-horizon scales suggests that a change of
variable and function are necessary in order to obtain better
accuracy.
\par
A convenient change of dependent variable and funtion is the
Langer transformation~\cite{langer}
\be
x\equiv\ln\left(\frac{k}{H\,a}\right)
\ ,\quad
\chi\equiv(1-\epsilon_1)^{1/2}\,{\rm e}^{-x/2}\,\mu
\ ,
\label{transf}
\ee
introduced in Ref.~\cite{wang_mukh} is used~\footnote{Note
that we define $x$ with the opposite sign with respect to that in
Ref.~\cite{martin_schwarz2}.
This of course implies that our $x_i\to+\infty$ and $x_f\to-\infty$,
whereas $x_i\to-\infty$ and $x_f\to+\infty$ in
Ref.~\cite{martin_schwarz2}.}
in order to improve the accuracy.
Such a transformation bring Eqs.~(\ref{tensor_eq2}) and (\ref{scal_eq2})
into the form
\be
\left[\frac{{\d}^2}{{\d} x^2}+\omega^2(x)\right]
\,\chi=0
\ ,
\label{new_eqb}
\ee
where the (new) frequency $\omega(x)$ in general vanishes at
the classical ``turning point'' $x=x_0$. In fact, on employing
the transformation (\ref{transf}), one finds (for $\delta=1$)
the new equations of motion (\ref{new_eqb}), which explicitly read
\be
&&\!\!\!\!\!\!\!\!\!\!\!\!
\frac{\d^2\chi_{\rm S}}{\d x^2}
+\!\!\left[\frac{{\rm e}^{2\,x}}{(1-\epsilon_1)^2}
-\left(\frac{3-\epsilon_1}{2-2\,\epsilon_1}\right)^2
-\frac{(3-2\,\epsilon_1)\,\epsilon_2}{2\,(1-\epsilon_1)^2}
-\frac{(1-2\,\epsilon_1)\,\epsilon_2\,\epsilon_3}{2\,(1-\epsilon_1)^3}
-\frac{(1-4\,\epsilon_1)\,\epsilon_2^2}{4\,(1-\epsilon_1)^4} \right]\,
\chi_{\rm S}=0
\nonumber
\\
\label{eqST_x}
\\
&&\!\!\!\!\!\!\!\!\!\!\!\!
\frac{\d^2 \chi_{\rm T}}{\d x^2}
+\!\!\left[\frac{{\rm e}^{2\,x}}{(1-\epsilon_1)^2}
+\left(\frac{3-\epsilon_1}{2-2\epsilon_1}\right)^2
+\frac{\epsilon_1\,\epsilon_2}{2\,(1-\epsilon_1)^2}
+\frac{\epsilon_1\,\epsilon_2\,\epsilon_3}{2\,(1-\epsilon_1)^3}
+\frac{(2+\epsilon_1)\,\epsilon_1\,\epsilon_2^2}{4\,(1-\epsilon_1)^4}
\right]\,
\chi_{\rm T}=0
\nonumber
\ .
\ee
From $\omega_{\rm S}^2(x)$ and $\omega_{\rm T}^2(x)$, respectively
given by the expressions in the square brackets of
Eqs.~(\ref{eqST_x}), we now obtain
\be
\left\{\begin{array}{ll}
\!\!\!\Delta_{\rm S,T}
\!=\!e^{-2\,x}\mathcal{O}(\epsilon_{n})\sim 0
&
%\quad
\textrm{sub-horizon scales}
\\
\\
\!\!\!\Delta_{\rm S}
\!=\!\displaystyle\frac{4}{27}\,
\left(\epsilon_{1}\,\epsilon_{2}^2
+\epsilon_{1}\,\epsilon_{2}\,\epsilon_{3}+
\frac{\epsilon_{2}\,\epsilon_{3}^2}{2}
+\frac{\epsilon_{2}\,\epsilon_{3}\,\epsilon_{4}}{2}\right)
+\mathcal{O}(\epsilon_{n}^{4})
&
%\quad
\textrm{super-horizon scales}
\\
\\
\!\!\!\Delta_{\rm T}
\!=\!\displaystyle\frac{4}{27}\,
\left(\epsilon_{1}\,\epsilon_{2}^2
+\epsilon_{1}\,\epsilon_{2}\,\epsilon_{3} \right)
+\mathcal{O}(\epsilon_{n}^{4})
&
%\quad
\textrm{super-horizon scales}
\ ,
\end{array}
\right.
\label{rapp_Q/omega2}
\ee
and it is thus obvious that $\Delta$ becomes very small for any
inflationary potential satisfying the slow-roll
conditions~\cite{wang_mukh} in both limits of interest
(but, remarkably, not at the turning point).
As a consequence, the WKB approximation is now also valid on
super-horizon scales.
\par
One might naively expect that the transformation~(\ref{transf})
will improve the WKB approach to all orders (in the formal
expansion in $\delta$), thus yielding increasingly
better results.
However, since $Q$ diverges at the turning point $x=x_0$,
whether the function $\mu_{\rm WKB}$ can still be regarded
as a solution to Eq.~(\ref{eq_wkb_expb}) for $x\sim x_0$ in
the adiabatic limit $\delta\to 0$ becomes a subtle issue.
It is precisely for this reason that one usually considers
particular solutions around the turning point (for example,
combinations of Airy functions for a linear turning point)
which must then be matched with the asymptotic
form~(\ref{wkb_sol}) on both sides of the turning point.
The problem with this standard procedure is that there is no
general rule to determine the actual matching points
(which cannot be $x_0$ where $Q$ diverges) in such a way that
the overall error be small.
As a consequence, the standard adiabatic expansion for
$\delta\ll 1$ may not lead to a reliable approximation
around the turning point, which would then result in large
errors to higher orders and a questionable further
extension of all the expressions to the interesting case
$\delta=1$.
\par
It was in fact shown in Ref.~\cite{hunt} that a simple
extension to higher orders of the above procedure~\cite{bender}
actually seems to reduce the accuracy with respect to the
leading order results of~Ref.~\cite{martin_schwarz2}.
We shall see in the next Section that one can overcome this
problem by introducing mode functions which remain good
approximations also around the turning point.
\section{Improved WKB approximation}
\label{imprWKBb}
We consider here the formalism given in Section~\ref{imprWKBa} and
apply it to cosmological perturbations.
The basic idea is to use approximate expressions which are valid for all
values of the coordinate $x$ and then expand around them.
It is also easy to show that, for the cosmological case,
$|\sigma/\omega^2|\simeq\Delta$ as given in Eq.~(\ref{rapp_Q/omega2})
for $x\to\pm\infty$, so that the new approximate solutions remain very accurate
for large $|x|$.
\subsection{Perturbative expansion}
We start by considering the initial conditions and the matching procedure
for the perturbative expansion given in Section~\ref{imprWKBa}.
\subsubsection{Initial and matching conditions}
We first use the initial conditions~(\ref{init_cond_on_mu})
in order to fix the constant coefficients $A_\pm$.
\par
For the function $\chi$ such conditions become
\be
\lim_{x\rightarrow x_i}\chi(x)
%&=&
=
d\,\sqrt{\frac{1-\epsilon_{1}(x_i)}{2k}}\,
{\rm e}^{-i\,k\,\eta_i-x_i/2}
%\nonumber\\ &=&
=
d\,\sqrt{\frac{1-\epsilon_{1}(x_i)}{2k}}\,
{\rm e}^{i\,\xi_{\rm I}(x_i)-x_i/2}
\ ,
\label{init_cond_on_chi}
\ee
where, from the definitions of $\eta$ and $x$,
$k\,\eta_i\simeq -i\,\xi_{\rm I}(x_i)$.
We then recall the asymptotic form of the Bessel functions $J_\nu$
for $x\to+\infty$~\cite{abram,schiff}~\footnote{Since the initial
conditions must be imposed at $x$ in deep region~I,
we can consider $x_i\sim+\infty$.},
\be
J_{\pm m}\left[\xi_{\rm I}(x)\right]
\sim
\sqrt{\frac{2}{\pi\,\xi_{\rm I}(x)}}
\,\cos\left(\xi_{\rm I}(x)\mp\frac{\pi}{2}\,m-\frac{\pi}{4}\right)
\ ,
\label{J_asymp}
\ee
from which we obtain the asymptotic expression of the mode
function in region~I,
\be
&&\!\!\!\!\!\!\!\!\chi_{\rm I}(x_i)\!\sim\!
\sqrt{\frac{1-\epsilon_{1}(x_i)}{2\,\pi}}\,{\rm e}^{-{x_i}/{2}}\,\times
\label{init_sol in_I}
\\
&&\!\!\!\!\!\!\!\!\left\{{\rm e}^{+i\,\xi_{\rm I}(x_i)}
\,\left[{\rm e}^{-i\,\left(\frac{1}{4}+\frac{m}{2}\right)\,\pi}\,A_{+}
+{\rm e}^{-i\,\left(\frac{1}{4}-\frac{m}{2}\right)\,\pi}
\left(A_{-}+\varepsilon\,A_{-}\,J_{+-}^{-}(x_0,x_i)
+\varepsilon\,A_{+}\,J_{++}^{+}(x_0,x_i)\right)
\right]\right.
\nonumber
\\
&&\!\!\!\!\!\!\!\!\left.+{\rm e}^{-i\,\xi_{\rm I}(x_i)}\,
\left[{\rm e}^{+i\,\left(\frac{1}{4}+\frac{m}{2}\right)\,\pi}\,A_+
+{\rm e}^{+i\,\left(\frac{1}{4}-\frac{m}{2}\right)\,\pi}\,
\left(A_{-}+\varepsilon\,A_{-}\,J_{+-}^{-}(x_0,x_i)
+\varepsilon\,A_{+}\,J_{++}^{+}(x_0,x_i)\right)
\right]
\right\}
\nonumber
\ .
\ee
The initial conditions~(\ref{init_cond_on_chi}) therefore yield
\be
&&
\!\!\!\!\!\!
A_{+}
=d\,\sqrt{\frac{\pi}{k}}\,
\frac{{\rm e}^{+i\,\left(\frac{1}{4}+\frac{m}{2}\right)\,\pi}}
{1-{\rm e}^{2\,i\,m\,\pi}}
\ ,\
A_{-}
=-A_{+}\,\displaystyle\frac{{\rm e}^{+i\,m\,\pi}
+\varepsilon\,J_{++}^{+}(x_0,x_i)}
{1+\varepsilon\,J_{+-}^{-}(x_0,x_i)}
\ .
\label{A+-}
\ee
Note that $A_+$ does not depend on $\varepsilon$, so it holds to all orders.
\par
We next impose continuity at the turning point in order to determine
the coefficients $B_\pm$.
For this we shall need the (asymptotic) expressions of Bessel functions
for $x\to x_0$~\cite{abram,schiff}~\footnote{The Bessel functions are
actually regular around $x=x_0$, therefore the following expressions
are just the leading terms in the Taylor expansion of $J_\nu$ and $I_\nu$.}.
In particular,
\be
\begin{array}{ll}
J_{\pm m}\left[\xi_{\rm I}(x)\right]
\simeq
\displaystyle
\left[\frac{\xi_{\rm I}(x)}{2}\right]^{\pm m}\,
\frac{1}{\Gamma\left(1\pm m\right)}
&
{\rm for}\
x\to x_0^+
\\
\\
I_{\pm m}\left[\xi_{\rm II}(x)\right]
\simeq
\displaystyle
\left[\frac{\xi_{\rm II}(x)}{2}\right]^{\pm m}\,
\frac{1}{\Gamma\left(1\pm m\right)}
&
{\rm for}\
x\to x_0^-
\ ,
\end{array}
\label{JI_taylor}
\ee
%\end{subequations}
where $\Gamma$ is Euler's gamma function.
Near the turning point, we just keep the leading term in the
expansion~(\ref{omega_expand}), so that
\be
&&
\omega(x)\simeq\displaystyle
\sqrt{C}\,\left|x-x_0\right|^{\frac{1-2\,m}{2\,m}}
\ ,\
\xi(x)\simeq
2\,m\,{\sqrt{C}}\,\left|x-x_0\right|^{\frac{1}{2\,m}}
\ .
\label{DX_TP}
\ee
We thus obtain the following forms of the mode functions near
the turning point,
\begin{subequations}
\be
\!\!\!\!
\chi_{\rm I}(x\simeq x_0)\!\!&\simeq&\!\!
A_{+}\,a_+(x_0)\,
\frac{\sqrt{2\,m}\,\left(m\,\sqrt{C}\right)^{m}}
{\Gamma\left(1+m\right)}\,
\left(x-x_0\right)
\nonumber
\\
&&
\!\!\!\!
+A_{-}\,a_-(x_0)\,
\frac{\sqrt{2\,m}\,\left(m\,\sqrt{C}\right)^{-m}}
{\Gamma\left(1-m\right)}
\label{sol_IX_at_TP}
\\
\phantom{A}
\nonumber
\\
\!\!\!\!\!\!\!\!
\chi_{\rm II}(x\simeq x_0)\!\!&\simeq&\!\!
B_{+}\,b_+(x_0)\,
\frac{\sqrt{2\,m}\,\left(m\,\sqrt{C}\right)^{m}}
{\Gamma\left(1+m\right)}\,
\left(x_0-x\right)
\nonumber
\\
&&
\!\!\!\!
+B_{-}\,b_-(x_0)\,
\frac{\sqrt{2\,m}\,\left(m\,\sqrt{C}\right)^{-m}}
{\Gamma\left(1-m\right)}
\ .
\label{sol_IIX_at_TP}
\ee
\end{subequations}
Continuity across $x_0$ then implies that
\be
B_{+}\,b_+(x_0)=
-A_{+}\,a_+(x_0)
\ ,\ \
B_{-}\,b_-(x_0)=
A_{-}\,a_-(x_0)
\label{connection_at_TP}
\ .
%\end{array}
\ee
From the definitions~(\ref{Abar_Bbar}) at $x=x_0$,
Eqs.~(\ref{A+-}) and the relations~(\ref{connection_at_TP})
we therefore obtain
\be
\frac{B_{+}}{A_+}&\!\!=\!\!&
-1
+\varepsilon\,
\frac{{\rm e}^{+i\,m\,\pi}
+\varepsilon\,J_{++}^{+}(x_0,x_i)}
{1+\varepsilon\,J_{+-}^{-}(x_0,x_i)}\,
J_{--}^{-}(x_0,x_i)
%\nonumber\\&&
-\varepsilon\,J_{+-}^{+}(x_0,x_i)
\nonumber
\\
\label{B+-}
\\
\frac{B_{-}}{A_+}&\!\!=\!\!&
-\frac{{\rm e}^{+i\,m\,\pi}
+\varepsilon\,J_{++}^{+}(x_0,x_i)}
{1+\varepsilon\,J_{+-}^{-}(x_0,x_i)}
%\nonumber\\&&\ \ \times
\frac{1-i\,\varepsilon^2\,J_{--}^{-}(x_0,x_i)\,
I_{++}^{+}(x_f,x_0)}
{1-i\,\varepsilon\,I_{+-}^{-}(x_f,x_0)}
\nonumber
\\
&&
-i\,\varepsilon\,
\displaystyle\frac{1+\varepsilon\,J_{+-}^{+}(x_0,x_i)}
{1-i\,\varepsilon\,I_{+-}^{-}(x_f,x_0)}\,I_{++}^{+}(x_f,x_0)
\ .
\nonumber
\ee
\par
It is apparent from the above derivation that, since the matching
between approximate solutions in the two regions is always
performed at $x=x_0$, no ambiguity hinders the evaluation of
the coefficients $B_\pm$ and the accuracy of the approximate
solution is therefore expected to increase with increasing order.
\subsubsection{Recursive (perturbative) relations}
We can finally obtain expressions for the coefficients
${a}_{\pm}(x)$ and ${b}_{\pm}(x)$ by solving the integral
relations (\ref{Abar_Bbar}).
We remark that the r.h.s.'s of such equations contain
$a_\pm(x)$ and $b_\pm(x)$, which can therefore be determined
recursively by expanding them (as well as any function of them)
in $\varepsilon$, that is
\be
a_\pm(x)=1+\displaystyle\sum_{q\geq1} \varepsilon^q\,a_{\pm}^{(q)}\ ,\ \
b_\pm(x)=1+\displaystyle\sum_{q\geq1} \varepsilon^q\,b_{\pm}^{(q)}
\ .
\ee
It will also be useful to define the following integrals
%\begin{subequations}
\be
&&
\!\!\!\!\!\!\!\!\!\!
J^{w(q)}_{s\bar s}(x_1,x_2)\equiv
\displaystyle\frac{\pi}{\sqrt{3}}\,
\int_{x_1}^{x_2}\!\!\!
\sigma_{\rm I}(y)\,a_w^{(q)}(y)\,
u_{Is}(y)\,u_{I\bar s}(y)\,\d y
\nonumber
\\
\\
&&
\!\!\!\!\!\!\!\!\!\!
I^{w(q)}_{s\bar s}(x_1,x_2)\equiv
\displaystyle\frac{\pi}{\sqrt{3}}\,
\int_{x_1}^{x_2}\!\!\!\!
\sigma_{\rm II}(y)\,b_w^{(q)}(y)\,
u_{IIs}(y)\,u_{II\bar s}(y)\,\d y
\,,
\nonumber
\ee
where again $w$, $s$ and $\bar s=\pm$, and $q$ is a
non-negative integer.\\
\\
{\em Leading order}\\
\\
We begin by considering the leading~($\varepsilon^0$) order
\be
a_\pm^{(0)}(x)=b_\pm^{(0)}(x)=1
\ .
\label{0-th}
\ee
The corresponding solutions are simply given by
\begin{subequations}
\be
&&\!\!\!\!\!\!\!\!\!\!\!\!\!\!\!\!\!\!\!\!\!\!\!\!\!\!
\chi_{\rm I}(x)\simeq
u_{\rm I}(x)
=A_+\,\sqrt{\frac{\xi_{\rm I}(x)}{\omega_{\rm I}(x)}}\,
\left\{
J_{+m}\left[\xi_{\rm I}(x)\right]
-\,{\rm e}^{+i\,m\,\pi}\,J_{-m}\left[\xi_{\rm I}(x)\right]
\right\}
\label{sol_in_I_leading}
\\
\nonumber
\\
&&\!\!\!\!\!\!\!\!\!\!\!\!\!\!\!\!\!\!\!\!\!\!\!\!\!\!
\chi_{\rm II}(x)\simeq
u_{\rm II}(x)
=-A_+\,\sqrt{\frac{\xi_{\rm II}(x)}{\omega_{\rm II}(x)}}\,
\left\{
\,I_{+m}\left[\xi_{\rm II}(x)\right]
+{\rm e}^{+i\,m\,\pi}\,I_{-m}\left[\xi_{\rm II}(x)\right]
\right\}
\label{sol_in_II_leading}
\ ,
\ee
\end{subequations}
where we used the zero~order forms of
Eqs.~(\ref{A+-}) and~(\ref{B+-}),
\be
A_{-}=B_{-}=-A_{+}\,{\rm e}^{+i\,m\,\pi}
\ ,\ \
B_{+}=-A_{+}
\ .
\label{relation_AB_leading}
\ee
\par
We know that these expressions are also good solutions
of Eq.~(\ref{new_eqb}) near the turning point and,
deep in region~II, we can use the asymptotic form of
$I_\nu$ for $x\sim x_f\to -\infty$~\cite{schiff,abram},
\be
I_{\pm m}\left[\xi_{\rm II}(x)\right]
&\!\!\sim\!\!&
\frac{1}{\sqrt{2\,\pi\,\xi_{\rm II}(x)}}\,
\left[{\rm e}^{\xi_{\rm II}(x)}
+{\rm e}^{-\xi_{\rm II}(x)-i\,\left(\frac12\pm m\right)\,\pi}\right]
\ .
\label{I_asymp}
\ee
On neglecting the non-leading mode ${\rm e}^{-\xi_{\rm II}(x)}$,
we then obtain
\be
\chi_{\rm II}(x_f)\sim
-A_{+}
\,\frac{\left(1+{\rm e}^{+i\,m\,\pi}\right)}
{\sqrt{2\,\pi\,\omega_{\rm II}(x_f)}}
\,{\rm e}^{\xi_{\rm II}(x_f)}
\ .
\label{sol_deep_in II}
\ee
If we now use Eqs.~(\ref{spectra_def}),~(\ref{A+-})
and~(\ref{sol_deep_in II}), and the values of $d$,
we recover the results of Ref.~\cite{martin_schwarz2}
for all quantities of interest,
i.e.~power spectra, spectral indices and
$\alpha$-runnings
\begin{subequations}
\be
&&
\!\!\!\!\!\!\!\!\!\!\!\!\!\!\!\!\!
{\cal P}_{\zeta}
\simeq
\frac{H^2}{\pi\,\epsilon_1\,m_{\rm Pl}^2}
\left(\frac{k}{a\,H}\right)^3
\frac{{\rm e}^{2\,\xi_{{\rm II,S}}(k,\eta)}}
{\left[1-\epsilon_1(\eta)\right]\,\omega_{\rm II,S}(k,\eta)}
\equiv{\cal P}_{\zeta}^{(0)}
\label{spectra_S_M&S}
\\
\nonumber
\\
&&
\!\!\!\!\!\!\!\!\!\!\!\!\!\!\!\!\!
n_{\rm S}-1\simeq
3+2\,\left.\frac{\d\,\xi_{{\rm II,S}}}{\d\,\ln k}
\right|_{k=k_*}
\label{indices_S_M&S}
\\
\nonumber
\\
&&
\!\!\!\!\!\!\!\!\!\!\!\!\!\!\!\!\!
\alpha_{\rm S}\simeq
2\,\left.
\frac{\d^2\,\xi_{{\rm II,S}}}{\left(\d\,\ln k\right)^2}
\right|_{k=k_*}
\ ,
\label{runnings_S_M&S}
\ee
and
\be
&&
\!\!\!\!\!\!\!\!\!\!\!\!\!\!\!\!\!
{\cal P}_{h}\simeq
\frac{16\,H^2}{\pi\,m_{\rm Pl}^2}
\,\left(\frac{k}{a\,H}\right)^3\,
\frac{{\rm e}^{2\,\xi_{{\rm II,T}}(k,\eta)}}
{\left[1-\epsilon_1(\eta)\right]\,\omega_{\rm II,T}(k,\eta)}
\equiv{\cal P}_{h}^{(0)}
\label{spectra_T_M&S}
\\
\nonumber
\\
&&
\!\!\!\!\!\!\!\!\!\!\!\!\!\!\!\!\!
n_{\rm T}\simeq
3+2\,\left.\frac{\d\,\xi_{{\rm II,T}}}{\d\,\ln k}
\right|_{k=k_*}
\label{indices_T_M&S}
\\
\nonumber
\\
&&
\!\!\!\!\!\!\!\!\!\!\!\!\!\!\!\!\!
\alpha_{\rm T}\simeq
2\,\left.\frac{\d^2\,\xi_{{\rm II,T}}}{\left(\d\,\ln k\right)^2}
\right|_{k=k_*}
\ ,
\label{runnings_T_M&S}
\ee
\end{subequations}

where we have transformed back to the original
variables $k$ and $\eta$, and all quantities are
evaluated in the super-horizon limit (i.e.~for
$k\ll a\,H$).\\
\\
{\em Next-to-leading order}\\
\\
We now insert the leading order expressions~(\ref{0-th})
into the integral relations~(\ref{Abar_Bbar}) to compute the
next-to-leading order expressions
%\begin{subequations}
%
\be
&&
{a}_{+}^{(1)}(x)=
J_{+-}^{(0)}(x,x_i)
-{\rm e}^{+i\,m\,\pi}\,J^{(0)}_{--}(x,x_i)
\nonumber %\label{a+_tilde_1}
\\
\nonumber
\\
&&
{a}_{-}^{(1)}(x)=
J^{(0)}_{+-}(x_0,x)
-{\rm e}^{-i\,m\,\pi}\,J^{(0)}_{++}(x_0,x)
\nonumber %\label{a-_tilde_1}
\\
\label{ab+-1}
\\
&&
{b}_{+}^{(1)}(x)=
-i\,I^{(0)}_{+-}(x,x_0)
-i\,{\rm e}^{+i\,m\,\pi}\,I^{(0)}_{--}(x,x_0)
\nonumber %\label{b+_tilde_1}
\\
\nonumber
\\
&&
{b}_{-}^{(1)}(x)=
-i\,I^{(0)}_{+-}(x_f,x)
-i\,{\rm e}^{-i\,m\,\pi}\,I^{(0)}_{++}(x_f,x)
\ ,
\nonumber %\label{b-_tilde_1}
\ee
%\end{subequations}
%
where we used the simplified notation
$J^{+(0)}_{s\bar s}=J^{-(0)}_{s\bar s}\equiv J^{(0)}_{s\bar s}$
and $I^{+(0)}_{s\bar s}=I^{-(0)}_{s\bar s}\equiv I^{(0)}_{s\bar s}$,
and also expanded the coefficients $A_-$ and $B_\pm$ to zero
order in $\varepsilon$.
\par
From Eq.~(\ref{I_asymp}), again neglecting the non-leading mode
${\rm e}^{-\xi_{\rm II}(x)}$, we obtain the mode function
at $x=x_f$ to first order in $\varepsilon$,
\be
&&\!\!\!\!\!\!\!\!\!\!
\chi_{\rm II}(x_f)\!\sim\!
-\frac{A_+\,{\rm e}^{\xi_{\rm II}(x_f)}}
{\sqrt{2\,\pi\,\omega_{\rm II}(x_f)}}\,
\left\{1+{\rm e}^{+i\,m\,\pi}
%\phantom{\frac{A}{B}}
\right.
%\nonumber\\&&\ \ \
+i\,\varepsilon\,
\left[I_{++}^{(0)}(x_f,x_0)-I_{+-}^{(0)}(x_f,x_0)\right]
\nonumber
\\
&&\!\!\!\!\!\!\!\!
-i\,\varepsilon\,{\rm e}^{+i\,m\,\pi}\,
\left[I_{--}^{(0)}(x_f,x_0)-I_{+-}^{(0)}(x_f,x_0)\right]
%\nonumber\\&&\ \ \
-\varepsilon\,{\rm e}^{+i\,m\,\pi}\,
\left[J_{+-}^{(0)}(x_0,x_i)+J_{--}^{(0)}(x_0,x_i)\right]
\nonumber
\\
&&\!\!\!\!\!\!\!\!
\left.
+\varepsilon\left[J_{+-}^{(0)}(x_0,x_i)+J_{++}^{(0)}(x_0,x_i)\right]
\right\}
\ ,
\label{sol_deep_in II_next_leading}
\ee
which, using Eqs.~(\ref{spectra_def}), (\ref{A+-}),
(\ref{sol_deep_in II_next_leading}) and the values of $d$,
yields all quantities of interest to first order in
$\varepsilon$.
For the spectra we obtain
\be
{\cal P}_{\zeta}\simeq
{\cal P}_{\zeta}^{(0)}\,
\left[1+g_{(1) \, {\rm S}}^{\rm GREEN} (x_f)\right]
\ ,\ \
{\cal P}_{h}\simeq
{\cal P}_{h}^{(0)}\,
\left[1+g_{(1) \, {\rm T}}^{\rm GREEN} (x_f)\right]
\ ,
\label{spectra_next-to-leading}
\ee
where the relative corrections to the leading order
expressions~(\ref{spectra_S_M&S}) and~(\ref{spectra_T_M&S})
are now given by
\be
g_{(1) \, {\rm S,T}}^{\rm GREEN} (x)&=&
\left\{
J_{++}^{(0)}(x_0,x_i)-J_{--}^{(0)}(x_0,x_i)
\right.
\nonumber
\\
&&
\left.
+\frac{1}{\sqrt{3}}
\left[I_{++}^{(0)}(x,x_0)
-2\,I_{+-}^{(0)}(x,x_0)
+I_{--}^{(0)}(x,x_0)\right]
\right\}_{\rm S,T}
\ ,
\label{g(1)}
\ee
in which we set $\varepsilon=1$ and $n=1$ as required, and
write S and T to recall the use of the corresponding
frequencies.
The expressions for the spectral indices and their runnings
to this order can finally be derived from the
definitions~(\ref{n_def}) and~(\ref{alpha_def}).\\
\\
{\em Higher orders}\\
\\
The above procedure can be extended to all orders.
The coefficients up to $a_\pm^{(q)}$ and $b_\pm^{(q)}$ must
be used to compute the integrals $J^{w(q)}_{s\bar s}$ and
$I^{w(q)}_{s\bar s}$ which determine $a_\pm^{(q+1)}$ and
$b_\pm^{(q+1)}$.
However, general expressions soon become very involved,
and we shall test the effectiveness of our method by
applying it to several cases of interest.
\subsection{Adiabatic expansion}
\label{adiabb}
It is important to observe that the initial conditions~(\ref{init_cond_on_chi})
and matching conditions at the turning point to first order in the adiabatic
parameter $\delta$ yield different results with respect
to the perturbative expansion in $\varepsilon$.
First of all, the coefficient $A_+$, which was left unaffected
by the expansion in $\varepsilon$ [see~Eq.~(\ref{A+-})],
acquires a correction.
The relations between $A_-$, $B_\pm$ and $A_+$ instead
remain those given in Eq.~(\ref{relation_AB_leading})
to all orders, whereas in the perturbative
expansion they were modified [see Eqs.~(\ref{A+-})
and~(\ref{B+-})].
To summarize, to first order in $\delta$,
one obtains
\be
A_+&\!\!\simeq\!\!&
d\,\sqrt{\frac{\pi}{k}}\,
\frac{{\rm e}^{i\,\left(\frac{1}{4}+\frac{m}{2}\right)\,\pi}}
{1-{\rm e}^{2\,i\,m\,\pi}}\,
\left\{1-\delta\,\left[
\phi_{{\rm I}(1)}(x_i)
+\gamma_{{\rm I}(1)}(x_i)\,\left(
i\,\omega_{\rm I}(x_i)-\frac{1}{2}\right)
\right]\right\}
\nonumber
\\
A_-&\!\!=\!\!&-A_+\,{\rm e}^{+i\,m\,\pi}
\ ,\ \
B_\pm=\mp A_\pm
\ ,
\label{ABadia1}
\ee
and the perturbation modes deep in Region~II
(at $x\ll x_0$) are finally given by
%
%\begin{subequations}
\be
\chi_{\rm II}(x)
\simeq
u_{\rm II}(x)\,
\left\{1+
\delta\,\left[
\phi_{{\rm II}(1)}(x)
-\gamma_{{\rm II}(1)}(x)\,\left(
\omega_{\rm II}(x)
+\frac{\omega_{\rm II}'(x)}{2\,\omega_{\rm II}(x)}\right)
\right.\right.
\nonumber
\\
\left.\left.
-\phi_{{\rm I}(1)}(x_i)
+\gamma_{{\rm I}(1)}(x_i)\,\left(
\frac{1}{2}-i\,\omega_{\rm I}(x_i)\right)
\right]
\right\}
\ .
\label{chi_1}
\ee
We note that, in order to obtain the above result,
one must first take the asymptotic expansion of the
functions $u_\pm$ and then take the derivative,
since the reverse order would lead to larger errors.
The corrected power spectra are then given by
\begin{subequations}
\be
{\cal P}_\zeta\simeq
{\cal P}^{(0)}_\zeta\,\left[1+g_{(1){\rm S}}^{\rm AD}(x_f)
\right]
\ ,\ \
{\cal P}_h\simeq
{\cal P}^{(0)}_h\,\left[1+g_{(1){\rm T}}^{\rm AD} (x_f)\right]
\ ,
\label{spectra_correct}
\ee
in which we set $\delta=1$ at the end~\footnote{Since our method has some similarity with that
of Ref.~\cite{HHHJM} (see also Ref.~\cite{olver}),
it is worth pointing out that the first order corrections
shown here differ from those of Ref.~\cite{HHHJM} (at least)
in that they contain contributions from both Regions~I and~II.}.
The spectral indices are also given by
\be
&&
n_{\rm S}-1=
3+2\,
\frac{\d\,\xi_{\rm II,S}}{\d\,\ln k}
+
\frac{\d\,g_{(1){\rm S}}^{\rm AD}}{\d\,\ln k}
\nonumber
\\
\label{index_correct}
\\
&&
n_{\rm T}=
3+2\,
\frac{\d\,\xi_{\rm II,T}}{\d\,\ln k}
+
\frac{\d\,g_{(1){\rm T}}^{\rm AD}}{\d\,\ln k}
\ ,
\nonumber
\ee
and their runnings by
\be
&&
\alpha_{\rm S}=
2\,
\frac{\d^2\,\xi_{\rm II,S}}{\left(\d\,\ln k\right)^2}
+
\frac{\d^2\,g_{(1){\rm S}}^{\rm AD}}{\left(\d\,\ln k\right)^2}
\nonumber
\\
\label{runnings_correct}
\\
&&
\alpha_{\rm T}=
2\,
\frac{\d^2\,\xi_{\rm II,T}}{\left(\d\,\ln k\right)^2}
+
\frac{\d^2\,g_{(1){\rm T}}^{\rm AD}}{\left(\d\,\ln k\right)^2}
\ .
\nonumber
\ee
Finally, the tensor-to-scalar ratio takes the form
\be
R=16\,\epsilon_1\,
\frac{{\rm e}^{2\,\xi_{\rm II,T}}\,\left(1+g_{(1){\rm T}}^{\rm AD}\right)\,
\omega_{\rm II,S}}
{{\rm e}^{2\,\xi_{\rm II,S}}\,\left(1+g_{(1){\rm S}}^{\rm AD}\right)\,
\omega_{\rm II,T}}
\ ,
\label{R_correct}
\ee
\end{subequations}
where all quantities are evaluated in the super-horizon limit
($x\ll x_0$) and we used the results
\be
g_{(1){\rm S,T}}^{\rm AD}(x)
=
%&\!\!\!=\!\!\!&
2\left[
\phi_{{\rm II}(1)}(x)
-\gamma_{{\rm II}(1)}(x)
\left(\omega_{\rm II}(x)
+\frac{\omega_{\rm II}'(x)}{2\,\omega_{\rm II}(x)}\right)
%\right.\nonumber\\&&\left.\phantom{2\,[}
+\frac{\gamma_{{\rm I}(1)}(x_i)}{2}
-\phi_{{\rm I}(1)}(x_i)
\right]_{\rm S,T}
\ ,
\label{g_1}
\ee
where the indices S and T recall the use of the
corresponding frequencies.
\par
Given the complicated expressions for $\omega$, it is
usually impossible to carry out the double integration that
yields $\phi_{(1)}$ analytically, let alone higher order terms.
One must therefore rely on numerical computations.
In the context of power-law inflation we can compare results obtained
in next-to-leading order from the two expansions with the exact result.
\section{Method of Comparison Equation}
\label{sMCEb}
In order to apply the MCE to cosmological perturbations, we note that
the frequency $\omega^2(x)$ is
\be
\omega^2(x)=\frac{{\rm e}^{2\,x}}{\left[1-\epsilon_1(x)\right]^2}-\nu^2(x)
\ ,
\label{our_freq}
\ee
with $\nu^2(x)$ given, respectively for scalar and tensor
perturbations, by
\begin{subequations}
\be
\!\!\!\!\!\!\!\!\!\!\!\!\!\!\!\!\!\!\!\!\!\!\!\!\!
\nu_{\rm S}^2(x)&=&
\frac{1}{4}\,\left(\frac{3-\epsilon_1}{1-\epsilon_1}\right)^2
+\frac{(3-2\,\epsilon_1)\,\epsilon_2}{2\,(1-\epsilon_1)^2}
%\nonumber\\&&
+\frac{(1-2\,\epsilon_1)\,\epsilon_2\,\epsilon_3}{2\,(1-\epsilon_1)^3}
+\frac{(1-4\,\epsilon_1)\,\epsilon_2^2}{4\,(1-\epsilon_1)^4}
\label{nu2_S}
\\
\!\!\!\!\!\!\!\!\!\!\!\!\!\!\!\!\!\!\!\!\!\!\!\!\!
\nu_{\rm T}^2(x)&=&
\frac{1}{4}\,\left(\frac{3-\epsilon_1}{1-\epsilon_1}\right)^2
-\frac{\epsilon_1\,\epsilon_2}{2\,(1-\epsilon_1)^2}
-\frac{\epsilon_1\,\epsilon_2\,\epsilon_3}{2\,(1-\epsilon_1)^3}
-\frac{(2+\epsilon_1)\,\epsilon_1\,\epsilon_2^2}{4\,(1-\epsilon_1)^4}
\ ,
\label{nu2_T}
\ee
\end{subequations}
where we omit the dependence on $x$ in the $\epsilon_i$ for the sake of
brevity. From Eqs.~(\ref{our_freq}), (\ref{nu2_S}) and (\ref{nu2_T}) we can recognise
our second-order differential basic equations~(\ref{eqST_x}).
The point $x=x_0$ where the frequency vanishes, $\omega(x_0)=0$
(i.e.~the classical ``turning point''), is given by the expression
\be
x_0=\ln\left[\bar{\nu}\,\left(1-\bar{\epsilon}_1\right)\right]
\ ,
\ee
where we have defined $\bar{\nu}\equiv\nu(x_0)$ and
$\bar{\epsilon}_1\equiv\epsilon_1(x_0)$.
We now choose the comparison function
\be
\Theta^2(\sigma)=
\frac{{\rm e}^{2\,\sigma}}{(1-\bar{\epsilon}_1)^2}-\bar{\nu}^2
\ ,
\label{aux_FREQ}
\ee
and note that then $\sigma_0=x_0$.
Solutions to Eq.~(\ref{new_eqa}) can now be expressed, by means of
Eqs.~(\ref{aux_EQ}), (\ref{new_EQ_appr}) and (\ref{aux_FREQ}), as
\be
\chi_{\pm}(x)\simeq\sqrt{\frac{\Theta(\sigma)}{\omega(x)}}\,
J_{\pm\bar{\nu}}\left(\frac{{\rm e}^{\sigma}}{1-\bar{\epsilon}_1}\right)
\ ,
\label{exact_SOL_bis}
\ee
where the $J$'s are Bessel functions~\cite{abram}, and the
initial condition~(\ref{init_cond_on_mu}) can be satisfied by taking
a linear combination of them.
However, in contrast with the improved WKB method presented in
Section~\ref{imprWKBa}, MCE solutions need not be matched at the turning
point, since the functions~(\ref{exact_SOL_bis}) are valid solutions
for the whole range of the variable $x$.
Eq.~(\ref{new_EQ_int}) at the end of inflation, $x=x_{\rm f}$, becomes
\be
\xi(x_{\rm f})
&\simeq&
-\Theta(\sigma_{\rm f})
-\frac{\bar{\nu}}{2}\,\ln\left[
\frac{\bar{\nu}-\Theta(\sigma_{\rm f})}
{\bar{\nu}+\Theta(\sigma_{\rm f})}\right]
\nonumber
\\
&\simeq&
-\bar{\nu}\,
\left[1+\ln\left(\frac{{\rm e}^{\sigma_{\rm f}}}{1-\bar{\epsilon}_1}\right)
-\ln\left(2\,\bar{\nu}\right)\right]
\ ,
\label{new_EQ_integrate}
\ee
where the super-horizon limit
$x_{\rm f}\ll x_0$ ($\sigma_{\rm f}\to-\infty$)
has been taken in the second line.
One then has
\be
\frac{{\rm e}^{\sigma_{\rm f}}}{1-\bar{\epsilon}_1}
\simeq
\frac{2\,\bar{\nu}}{\rm e}\,{\rm exp}
\left[-\frac{\xi(x_{\rm f})}{\bar{\nu}}\right]
\ .
\label{arg_exp}
\ee
Finally, on using Eq.~(\ref{arg_exp}), we obtain the general expressions
for the power spectra to leading MCE order,
\begin{subequations}
\be
{\cal P}_\zeta
&=&
\left[\frac{H^2}{\pi\,\epsilon_1\,m_{\rm Pl}^2}
\left(\frac{k}{a\,H}\right)^3
\frac{{\rm e}^{2\,\xi_{\rm S}}}
{\left(1-\epsilon_1\right)\,\omega_{\rm S}}\right]_{x=x_{\rm f}}
g_{\rm S}(x_0)
\label{spectra_S}
\\
{\cal P}_h
&=&
\left[\frac{16\,H^2}{\pi\,m_{\rm Pl}^2}\,
\left(\frac{k}{a\,H}\right)^3\,
\frac{{\rm e}^{2\,\xi_{\rm T}}}
{\left(1-\epsilon_1\right)\,\omega_{\rm T}}\right]_{x=x_{\rm f}}
g_{\rm T}(x_0)
\label{spectra_T}
\ ,
\ee
where $m_{\rm Pl}$ is the Planck mass and the quantities inside the
square brackets are evaluated in the super-horizon limit, whereas
the functions
\be
g(x_0)
\equiv
\frac{\pi\,e^{2\,\bar{\nu}}\,\bar{\nu}^{1-2\,\bar{\nu}}}
{\left[1-\cos\left(2\,\pi\,\bar{\nu}\right)\right]\,
\left[\Gamma\left(1-\bar{\nu}\right)\right]^2}
\ ,
\label{corr_TP}
\ee
\end{subequations}
describe corrections that just depend on quantities evaluated
at the turning point and represent the main result of the
MCE applied to cosmological perturbations~\footnote{Inside the
square brackets we recognize the general results ${\cal P}_{\zeta}^{(0)}$
and ${\cal P}_{h}^{(0)}$ given by the WKB (leading-order) approximation
in Section~\ref{imprWKBb}.
The ``correction'' $g(x_0)$ accounts for the fact that $\Theta^2$
in Eq.~(\ref{aux_FREQ}) is a better approximation than
Langer's~\cite{WKB1,langer}.}.
The expression in Eq.~(\ref{corr_TP}) is obtained by simply making use of
the approximate solutions~(\ref{exact_SOL_bis}) and their asymptotic
expansion at $x\to\infty$ to impose the initial
conditions~(\ref{init_cond_on_mu}).
In Section~\ref{imprWKBb} one finds a similar
factor but, in that case, using the Bessel functions of order $1/3$ leads
to a large error in the amplitudes.
The MCE instead uses Bessel functions of order $\bar{\nu}$, with $\bar\nu=3/2$
to leading order in the HFF (i.e.~the right index for the de~Sitter model),
which yields a significantly better value for the amplitudes of inflationary
spectra.
\par
The MCE allows one to compute approximate perturbation modes with errors
in the asymptotic regions (i.e.~in the sub- and super-horizon limits)
which are comparable with those of the standard (or improved) WKB
approximation (see Sec.~\ref{stWKBb}).
Since these methods usually give large errors at the turning
point (which produce equally large errors in
the amplitude of the power spectra) it will suffice to estimate the
error at the turning point in order to show that the MCE is indeed
an improvement.
To leading order (that is, on using the approximate
solution~(\ref{exact_SOL_bis})), the MCE gives an error at the
turning point of the second order in the HFF, which means that we
have a small error in the amplitudes of the power spectra.
Unfortunately, this error remains of second order
in the HFF also for next-to-leading order in the MCE.
We shall see this by applying Dingle's analysis~\cite{dingle}
for linear frequencies to our case~(\ref{our_freq}).
We start by rewriting Eq.~(\ref{new_EQ}) as
\be
\left\{\omega^2-\sigma_1^2\,\left[\frac{{\rm e}^{2\,\sigma}}
{(1-\bar{\epsilon}_1)^2}-\bar{\nu}^2\right]
\right\}
+
\left[
\frac34\,\frac{\sigma_2^2}{\sigma_1^2}-\frac12\,\frac{\sigma_3}{\sigma_1}
\right]
=0
\ ,
\label{new_EQ_4_DING}
\ee
where we dropped the $x$ dependence in $\omega$ and $\sigma$
and the order of the derivatives is given by their subscripts
($\sigma_1\equiv \d\sigma/\d x$, $\omega_1^2\equiv \d\omega^2/\d x$,
etc.).
Note that the term in square brackets is the error obtained on using
the solutions~(\ref{exact_SOL_bis}).
We then evaluate Eq.~(\ref{new_EQ_4_DING}) and its subsequent
derivatives at the turning point (i.e.~at $x=x_0$),
\begin{subequations}
\be
\!\!\!\!\!\!\!\!\!\!\!\!\!\!\!\!\!\!\!\!\!\!\!\!\!\!\!\!\!\!\!\!\!\!\!\!\!\!\!\!
\left\{
\omega^2-\sigma_1^2\,\Theta^2\left(\sigma\right)
\right\}
+
\left[
\frac34\,\frac{\sigma_2^2}{\sigma_1^2}-\frac12\,\frac{\sigma_3}{\sigma_1}
\right]
=0
\label{new_EQ_4_DING_TP}
\ee
\be
\!\!\!\!\!\!\!\!\!\!\!\!\!\!\!\!\!\!\!\!\!\!\!\!\!\!\!\!\!\!\!\!\!\!\!\!\!\!\!\!
\left\{
\omega^2_1
-2\,\sigma_2\,\sigma_1\,\Theta^2\left(\sigma\right)
-\frac{2\,{\rm e}^{2\,\sigma}\,\sigma_1^3}{\left(1-\bar{\epsilon}_1\right)^2}
\right\}
+
\left[
2\,\frac{\sigma_3\,\sigma_2}{\sigma_1^2}
-\frac32\,\frac{\sigma_2^3}{\sigma_1^3}
-\frac12\,\frac{\sigma_4}{\sigma_1}
\right]
=0
\label{new_EQ_deriv1}
\ee
\be
\!\!\!\!\!\!\!\!\!\!\!\!\!\!\!\!\!\!\!\!\!\!\!\!\!\!\!\!\!\!\!\!\!\!\!\!\!\!\!\!
&&
\left\{
\omega^2_2
-2\,\left(\sigma_3\,\sigma_1+\sigma_2^2\right)\,\Theta^2\left(\sigma\right)
-\frac{2\,\sigma_1^2\,{\rm e}^{2\,\sigma}}{\left(1-\bar{\epsilon}_1\right)^2}\,
\left(2\,\sigma_1^2+5\,\sigma_2\right)
\right\}
\nonumber\\
\!\!\!\!\!\!\!\!\!\!\!\!\!\!\!\!\!\!\!\!\!\!\!\!\!\!\!\!\!\!\!\!\!\!\!\!\!\!\!\!
&&
+
\left[
\frac92\,\frac{\sigma_2^4}{\sigma_1^4}
-\frac{17}{2}\frac{\sigma_3\,\sigma_2^2}{\sigma_1^3}
+\frac52\,\frac{\sigma_4\,\sigma_2}{\sigma_1^2}
+2\,\frac{\sigma_3^2}{\sigma_1^2}
-\frac12\,\frac{\sigma_5}{\sigma_1}
\right]
=0
\label{new_EQ_deriv2}
\ee
\be
\!\!\!\!\!\!\!\!\!\!\!\!\!\!\!\!\!\!\!\!\!\!\!\!\!\!\!\!\!\!\!\!\!\!\!\!\!\!\!\!
&&
\left\{
\omega^2_3
-2\,\left(\sigma_4\,\sigma_1+3\,\sigma_2\,\sigma_3\right)\,\Theta^2\left(\sigma\right)
-\frac{2\,\sigma_1\,{\rm e}^{2\,\sigma}}{\left(1-\bar{\epsilon}_1\right)^2}\,
\left(4\,\sigma_1^4
+18\,\sigma_2\,\sigma_1^2
+7\,\sigma_3\,\sigma_1
+12\,\sigma_2^2\right)
\right\}
\nonumber
\\
\!\!\!\!\!\!\!\!\!\!\!\!\!\!\!\!\!\!\!\!\!\!\!\!\!
&&
+\left[
\frac{87}{2}\,\frac{\sigma_3\,\sigma_2^3}{\sigma_1^4}
-18\,\frac{\sigma_2^5}{\sigma_1^5}
-\frac{27}{2}\,\frac{\sigma_4\,\sigma_2^2}{\sigma_1^3}
-21\,\frac{\sigma_3^2\,\sigma_2}{\sigma_1^3}
+3\,\frac{\sigma_5\,\sigma_2}{\sigma_1^2}
+\frac{13}{2}\,\frac{\sigma_3\,\sigma_4}{\sigma_1^2}
-\frac12\,\frac{\sigma_6}{\sigma_1}
\right]
=0
\ ,
\label{new_EQ_deriv3}
\ee
\end{subequations}
where $\Theta^2\left(\sigma\right)$ was defined in Eq.~(\ref{aux_FREQ})
and we omit two equations for brevity.
In order to evaluate the error at the turning point
\be
\Delta_{\rm TP}=
\left[\frac34\,\frac{\sigma_2^2}{\sigma_1^2}
-\frac12\,\frac{\sigma_3}{\sigma_1}\right]_{x=x_0}
\ ,
\ee
we ignore the terms in square brackets and equate to zero the
expressions in the curly brackets in
Eqs.~(\ref{new_EQ_4_DING_TP})-(\ref{new_EQ_deriv3}) and so on.
This leads to
\begin{subequations}
\be
\!\!\!\!\!\!\!\!\!\!\!\!\!\!\!\!\!\!\!\!\!\!\!\!\!
\sigma
&\!=\!&
\ln\left[\left(1-\bar{\epsilon}_1\right)\,\bar{\nu}\right]
\label{sigma0_0order}
\\
\!\!\!\!\!\!\!\!\!\!\!\!\!\!\!\!\!\!\!\!\!\!\!\!\!
\sigma_1
&\!=\!&
\left(\frac{\omega^2_1}{2\,\bar{\nu}^{2}}\right)^{1/3}
\label{sigma1_0order}
\\
\!\!\!\!\!\!\!\!\!\!\!\!\!\!\!\!\!\!\!\!\!\!\!\!\!
\sigma_2
&\!=\!&
\frac{1}{5\,\left(2\,\bar{\nu}^2\right)^{1/3}}
\left[
\frac{\omega^2_2}{\left(\omega^2_1\right)^{2/3}}
-\left(2\,\frac{\omega^2_1}{\bar{\nu}}\right)^{2/3}
\right]
\label{sigma2_0order}
\\
\!\!\!\!\!\!\!\!\!\!\!\!\!\!\!\!\!\!\!\!\!\!\!\!\!
\sigma_3
&\!=\!&
-\frac{6 \left(2^{1/3}\,\omega^2_2\right)^2}
{175\,\left(\bar{\nu}^{2/5}\,\omega^2_1\right)^{5/3}}
-\frac{3\cdot 2^{1/3}\,\omega^2_2}
{25\,\left(\bar{\nu}^{4}\,\omega^2_1\right)^{1/3}}
+\frac{16\,\omega^2_1}{175\,\bar{\nu}^2}
+\frac{\omega^2_3}
{7 \left(2^{1/2}\,\bar{\nu}\,\omega^2_1\right)^{2/3}}
\ ,
\label{sigma3_0order}
\ee
\end{subequations}
and similar expressions for $\sigma_4$, $\sigma_5$, and $\sigma_6$
which we again omit for brevity.
On inserting Eqs.~(\ref{our_freq}), (\ref{nu2_S}) and (\ref{nu2_T})
in the above expressions, we find the errors to leading MCE order
\begin{subequations}
\be
\Delta^{(0)}_{\rm TP, S}&=&
-\frac{32}{315}\,\epsilon_1\,\epsilon_2
-\frac{22}{315}\,\epsilon_2\,\epsilon_3
\label{corr_TP_S_0order}
\\
\Delta^{(0)}_{\rm TP, T}&=&
-\frac{32}{315}\,\epsilon_1\,\epsilon_2
\ ,
\label{corr_TP_T_0order}
\ee
\end{subequations}
for scalar and tensor modes respectively.
\par
On iterating this procedure we can further obtain the errors
for the next-to-leading MCE order $\Delta^{(1)}_{\rm TP}$.
We first compute next-to-leading solutions to
Eqs.~(\ref{new_EQ_4_DING_TP})-(\ref{new_EQ_deriv3}) and so on
by inserting the solutions found to leading order for
$\sigma_1,\sigma_2,\ldots,\sigma_6$ into the corrections
(i.e.~the square brackets) and into all terms containing
$\Theta^2(\sigma)$~\cite{dingle}.
This leads to
\begin{subequations}
\be
\Delta^{(1)}_{\rm TP, S}&=&
-\frac{31712}{331695}\,\epsilon_1\,\epsilon_2
-\frac{21598}{331695}\,\epsilon_2\,\epsilon_3
\label{corr_TP_S_1order}
\\
\Delta^{(1)}_{\rm TP, T}&=&
-\frac{31712}{331695}\,\epsilon_1\,\epsilon_2
\ ,
\label{corr_TP_T_1order}
\ee
\end{subequations}
which show that the next-to-leading MCE solutions lead to an error of
second order in the HFF, too.
We suspect that this remains true for higher MCE orders, since there
is no {\em a priori\/} relation between the MCE and the
slow-roll expansions.
Let us however point out that the above expressions were obtained
without performing a slow-roll expansion and therefore do not
require that the $\epsilon_i$ be small.

\chapter{Applications}
\label{WKBresutls}
We shall now apply the formalism developed in the previous chapter
to some models of inflation.
We shall start by considering the so called Power-law inflation as a
benchmark.
We shall then proceed by expanding our general expressions to second order in the HFF and
compare with the results obtained from other approximation methods used in the literature.
\section{Power-law Inflation}
\label{Power-law inflation}
In this model, the scale factor is given in conformal time by
\be
a(\eta )=\ell_0|\eta |^{1+\beta }
\ ,
\label{a_PowLaw}
\ee
with $\beta \le -2$. The case $\beta =-2$ is special, since it corresponds to the de~Sitter space-time with constant Hubble
radius equal to $\ell_0=H^{-1}$.
The frequency is given, both for scalar and tensor modes, by
\be
\omega^2(x)=
\left(1+\beta\right)^2\,{\rm e}^{2\,x}-\left(\beta+\frac12\right)^2
\ ,
\label{omega2_PL}
\ee
and the horizon flow functions read
\be
\epsilon_1=\frac{2+\beta }{1+\beta }
\ ,\quad
\epsilon_n = 0
\ ,
\quad
n>1
\ .
\label{eps_PowLaw}
\ee
\par
\begin{figure}[!ht]
\raisebox{2.5cm}{$\omega^2\ $}
\includegraphics[width=0.5\textwidth]{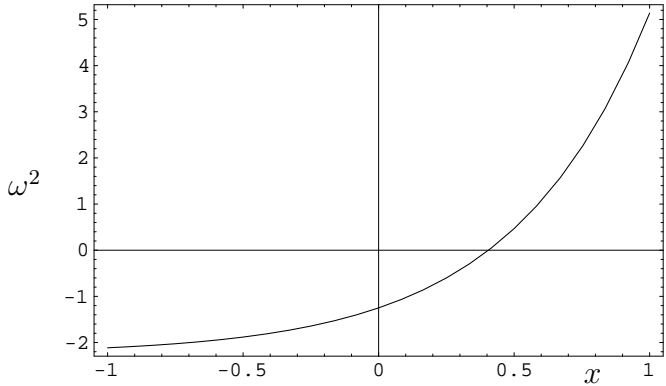}
\hspace{-1.2cm}$x$
\caption{The frequency $\omega^2$ for $\beta=-2$.
The turning point is at $x_*=\ln(3/2)\simeq 0.4$}
\label{omega2_deSitt}
\end{figure}
\begin{figure}[!ht]
\raisebox{2.5cm}{$\xi\ $}
\includegraphics[width=0.5\textwidth]{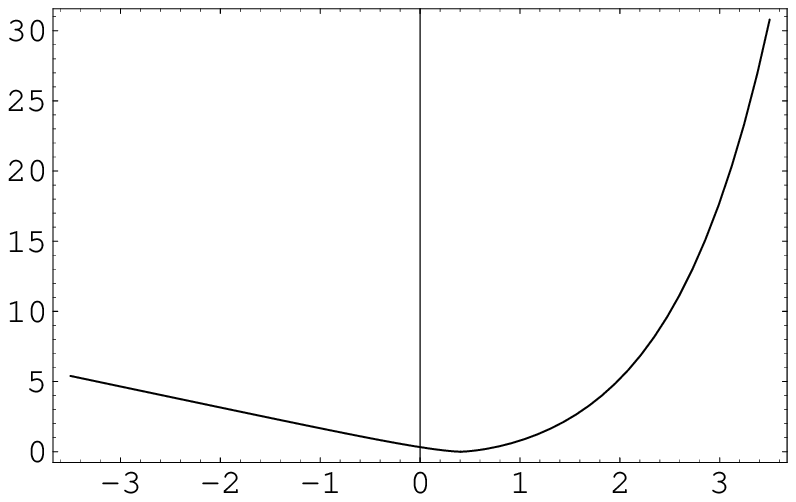}
\hspace{-1.4cm}$x$
\caption{The function $\xi$ for $\beta=-2$.}
\label{xi_deSitt}
\end{figure}
\begin{figure}[!ht]
\raisebox{2.5cm}{$\sigma\ $}
\includegraphics[width=0.5\textwidth]{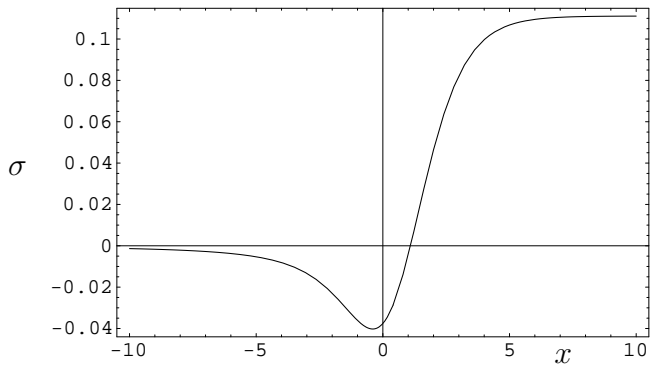}
\hspace{-1.4cm}$x$
\caption{The ``perturbation'' $\sigma$ for $\beta=-2$.}
\label{eta_deSitt}
\end{figure}

Eq.~(\ref{eq_wkb_expb}) with the initial conditions (\ref{init_cond_on_mu})
can be solved analytically and the exact power spectra
at $x_f\to-\infty$ are given by (see,
e.g.~Refs.~\cite{lythstewart,martin_schwarz2,abbott,martin_PL})
\begin{subequations}
\be
&&
\mathcal{P}_{\zeta}=
\frac{1}{\pi\,\epsilon_1\,m_{\rm Pl}^2}\,
\frac{1}{l_0^2}\,f(\beta)\,k^{2\,\beta+4}
\nonumber
\\
\label{P_ST_PawLaw}
\\
&&
\mathcal{P}_h=
\frac{16}{\pi\,m_{\rm Pl}^2}\,
\frac{1}{l_0^2}\,f(\beta)\,k^{2\,\beta+4}
\ ,
\nonumber
\ee
where
\be
f(\beta):=
\frac{1}{\pi}\,\left[
\frac{\Gamma\left(\left|\beta+1/2\right|\right)}
{2^{\beta+1}}\right]^2
\ ,
\label{f_beta}
\ee
\end{subequations}
and $f(\beta=-2)=1$~\footnote{The case $\beta =-2$ is singular
($\epsilon _1=0$ and the expression of the scalar power spectrum
blows up) and should be considered separately.
However, there are no scalar perturbations in De~Sitter space-time
to first order.
%One can in fact show that there are no density perturbations at all
%in the De~Sitter space-time~\cite{martin_PL}.
}.
The spectral indices and their runnings can also be calculated
from Eqs.~(\ref{n_def}) and (\ref{alpha_def}) and one finds
$n_{\rm S}-1=n_{\rm T}=2\,\beta+4$ and
$\alpha_{\rm S}=\alpha_{\rm T}=0$.
Exact scale-invariance is obtained for $\beta =-2$.
For this particular case, we plot the frequency
in Fig.~\ref{omega2_deSitt}, the function $\xi(x)$ in
Fig.~\ref{xi_deSitt} and the perturbation $\sigma(x)$
in Fig.~\ref{eta_deSitt}.
\subsection{Improved WKB leading order results}
\label{Power-law inflation_leading}
\begin{figure}[!ht]
\raisebox{2.5cm}{$\Delta_P\ $}
\includegraphics[width=0.5\textwidth]{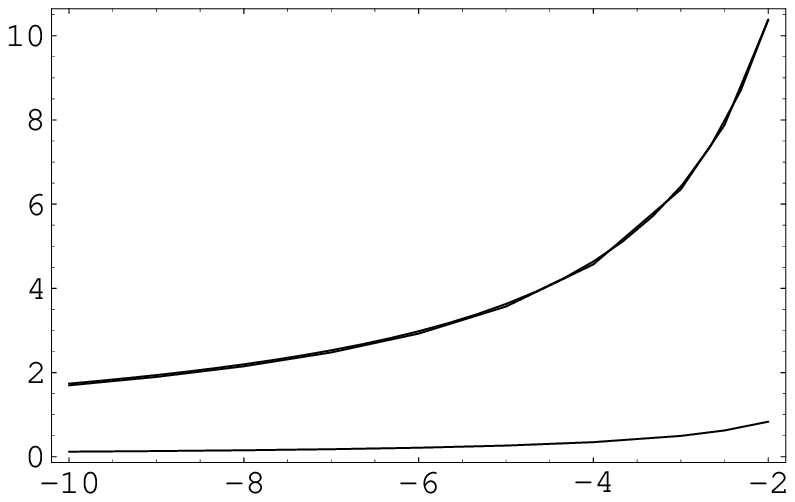}
\hspace{-1.4cm}$\beta$
\caption{The total error on the power spectrum
to leading order and to next-to-leading order for
the perturbative correction (superposed upper lines) and
to next-to-leading order for the adiabatic correction
(lower line).}
\label{errors1}
\end{figure}

The leading order expressions in the perturbative expansion
of Section~\ref{pert}, as well as the adiabatic expansion
of Section~\ref{adiabb}, lead to the same result as in
Ref.~\cite{martin_schwarz2},
\begin{subequations}
\be
&&
\mathcal{P}_{\zeta}\simeq\frac{1}{\pi\,\epsilon_1\,
m_{\rm Pl}^2}\,\frac{1}{l_0^2}\,g^{(0)}(\beta)\,k^{2\,\beta+4}
\nonumber
\\
\label{P_ST_PawLaw_0}
\\
&&
\mathcal{P}_h\simeq\frac{16}{\pi\,m_{\rm Pl}^2}\,
\frac{1}{l_0^2}\,g^{(0)}(\beta)\,k^{2\,\beta+4}
\ ,
\nonumber
\ee
where the function $g^{(0)}(\beta)$ is given by
\be
g^{(0)}(\beta):=
\frac{2\,e^{2\,\beta+1}}{\left|2\,\beta+1\right|^{2\,\beta +2}}
\ .
\label{g_0_beta}
\ee
\end{subequations}
This yields a relative error for the amplitude of
the power spectrum
\be
\Delta_P^{(0)}:=
100\,\left|\frac{f(\beta)-g^{(0)}(\beta)}{f(\beta)}\right|
\%
\ ,
\label{b_error}
\ee
which decreases for increasing $|\beta|$, as can be seen from Fig.~\ref{errors1}, but is rather large (about $10\%$)
for the de~Sitter space-time.
The spectral indices and their runnings are instead predicted
exactly as $n_{\rm S}-1=n_{\rm T}=2\,\beta+4$ and
$\alpha_{\rm S}=\alpha_{\rm T}=0$.
\subsection{Improved WKB next-to-leading order results}
\label{Power-law inflation_next-to-leading}
We shall now compare the corrections to the amplitude
of the power spectra coming from the two different
expansions we described in Chapter~\ref{WKBb}.
\subsubsection{Perturbative expansion}
From the next-to-leading expressions in Section~\ref{pert},
we obtain that the relative correction to the power
spectra is given by the function
$g_{(1)}^{\rm GREEN}:=g_{(1)\,\rm S}^{\rm GREEN}
=g_{(1)\,\rm T}^{\rm GREEN}$ in
Eq.~(\ref{g(1)}) evaluated at $x=x_f\ll x_*$.
This quantity can be determined numerically and one
finds that it approaches an asymptotical finite value
for $x_f\to-\infty$.
Some examples are given in Table~\ref{improveP} for
$x_f=-13$, from which it is clear that the improvement
over the leading order is small.
\begin{table*}[!ht]
\centerline{
\begin{tabular}{|c|c|c|c|c|c|c|c|c|c|}
\hline
$\beta$ 
& $-2$ 
& $-3$ 
& $-4$ 
& $-5$ 
& $-6$
& $-7$
\\
\hline
$g_{(1)}^{\rm GREEN}$ 
& $+9.6\cdot 10^{-5}$ 
& $+7.3\cdot 10^{-4}$
& $+7.1\cdot 10^{-4}$
& $+6.4\cdot 10^{-4}$
& $+5.8\cdot 10^{-4}$
& $+5.2\cdot 10^{-4}$
\\
\hline
$\Delta_P^{(0)}$ 
& $10.4\%$ 
& $6.4\%$
& $4.6\%$
& $3.6\%$
& $3.0\%$
& $2.5\%$
\\
\hline
$\Delta_P^{(1)}$ 
& $10.4\%$ 
& $6.3\%$
& $4.6\%$
& $3.6\%$
& $2.9\%$
& $2.5\%$
\\
\hline
\end{tabular}
}
\caption{Next-to-leading order improvement for the power spectra
($g_{(1)}^{\rm GREEN}$) and total final error
($\Delta_P^{(1)}$) to first order in $\varepsilon$.
The latter is essentially the same as to leading order
($\Delta_P^{(0)}$).}
\label{improveP}
\end{table*}
\par
Once the correction has been included, the total relative
error on the power spectra (denoted by $\Delta_P^{(1)}$)
therefore remains of the same order as the leading order
error $\Delta_P^{(0)}$.
\subsubsection{Adiabatic expansion}
\begin{figure}[!ht]
\includegraphics[width=0.5\textwidth]{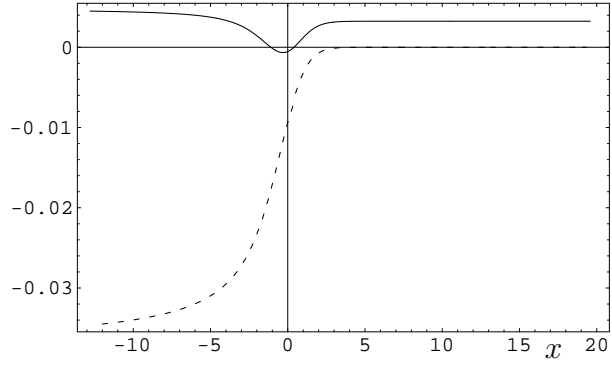}
\hspace{-1cm}$x$
\caption{The functions $\phi_{(1)}(x)$ (solid line) and
$\gamma_{(1)}(x)$ (dotted line) for $\beta=-2$.}
\label{PhiGammaPL}
\end{figure}

We analogously evaluate the coefficients $\phi_{(1)}$
and $\gamma_{(1)}$ numerically.
For the case $\beta=-2$, the result is plotted in
Fig.~\ref{PhiGammaPL}, from which it appears that the
functions $\phi_{(1)}(x)$ and $\gamma_{(1)}(x)$
converge to finite asymptotic values for $x\to\pm\infty$.
Since the same behavior is also found for all $\beta<-2$,
we exhibit the values of such coefficients (with
$x_i=20$ and $x_f=-13$) in Table~\ref{improveA},
together with the corresponding next-to-leading
order relative improvement
[$g_{(1)}^{\rm AD}(x_f)$ as defined in Eq.~(\ref{g_1})]
and the total error on the power spectra ($\Delta_P^{(1)}$).
\par
On comparing with the last row in Table~\ref{improveP},
we can therefore conclude that the adiabatic corrections
are significantly better than those obtained from the
perturbative expansion and, in fact, yield the amplitudes
with extremely high accuracy (see also Fig.~\ref{errors1}).
In fact, this method seems (at least) as accurate as the
next-to-leading order in the uniform approximation employed in
Ref.~\cite{HHHJM}.
\begin{table*}[!ht]
\centerline{
\begin{tabular}{|c|c|c|c|c|c|c|c|c|c|}
\hline
$\beta$ 
& $-2$ 
& $-3$ 
& $-4$ 
& $-5$ 
& $-6$
& $-7$
\\
\hline
$\phi_{{\rm I}(1)}$ 
& $+3.2\cdot 10^{-3}$
& $+1.1\cdot 10^{-3}$
& $+6.0\cdot 10^{-4}$
& $+3.6\cdot 10^{-4}$
& $+2.4\cdot 10^{-4}$
& $+1.7\cdot 10^{-4}$
\\
\hline
$\gamma_{{\rm I}(1)}$ 
& $-2.4\cdot 10^{-19}$
& $-5.9\cdot 10^{-20}$
& $-2.6\cdot 10^{-20}$
& $-1.5\cdot 10^{-20}$
& $-9.4\cdot 10^{-21}$
& $-6.6\cdot 10^{-21}$
\\
\hline
$\phi_{{\rm II}(1)}$ 
& $+4.5\cdot 10^{-3}$ 
& $+1.6\cdot 10^{-3}$ 
& $+8.2\cdot 10^{-4}$ 
& $+5.0\cdot 10^{-4}$ 
& $+3.3\cdot 10^{-4}$ 
& $+2.4\cdot 10^{-4}$ 
\\
\hline
$\gamma_{{\rm II}(1)}$ 
& $-3.5\cdot 10^{-2}$ 
& $-1.2\cdot 10^{-2}$ 
& $-6.4\cdot 10^{-3}$ 
& $-3.8\cdot 10^{-3}$ 
& $-2.6\cdot 10^{-3}$ 
& $-1.8\cdot 10^{-3}$ 
\\
\hline
$g_{(1)}^{\rm AD}$ 
& $+1.1\cdot 10^{-1}$
& $+6.3\cdot 10^{-2}$
& $+4.5\cdot 10^{-2}$
& $+3.5\cdot 10^{-2}$
& $+2.9\cdot 10^{-2}$
& $+2.4\cdot 10^{-2}$
\\
\hline
$\Delta_P^{(1)}$ 
& $0.83\%$ 
& $0.50\%$
& $0.35\%$
& $0.27\%$
& $0.22\%$
& $0.18\%$
\\
\hline
\end{tabular}
}
\caption{First order coefficients $\phi_{(1)}$ and $\gamma_{(1)}$,
correction ($g_{(1)}^{\rm AD}$) and total relative error
($\Delta_P^{(1)}$) for the power spectrum
to first order in $\delta$.}
\label{improveA}
\end{table*}
\subsection{MCE results}
\label{MCE results}
The MCE (developed in Scection~\ref{sMCEb}) yields the exact power spectra,
spectral indices and runnings,
\begin{subequations}
\be
&&
{\cal P}_{\zeta}=
\frac{\ell_{\rm Pl}^2}{\ell_0^2\,\pi\,\epsilon_1}\,f(\beta)\,k^{2\beta+4}
\label{PL_spectra_S}
\\
&&
{\cal P}_h=
\frac{16\,\ell_{\rm Pl}^2}{\ell_0^2\,\pi}\,f(\beta)k^{2\beta+4}
\label{PL_spectra_T}
\ee
where $\ell_{\rm Pl}=m_{\rm Pl}^{-1}$ is the Planck length and
\be
f(\beta)=\frac{\pi}{2^{2\,\beta+1}}\,
\frac{1}{\left[1-\cos\left(2\,\pi\left|\beta+\frac12\right|\right)\right]\,
\Gamma^2\left(\beta+\frac32\right)}
\equiv
\frac{1}{\pi}\,
\left[\frac{\Gamma\left(\left|\beta+\frac12\right|\right)}{2^{\beta+1}}\right]^2
\ ,
\label{f}
\ee
\end{subequations}
with $\Gamma$ the Gamma function.
The spectral indices are $n_{\rm S}-1=n_{\rm T}=2\beta+4$ and
their runnings $\alpha_{\rm S}=\alpha_{\rm T}=0$.
Finally, the tensor-to-scalar ratio becomes
\be
R=16\,\frac{2+\beta}{1+\beta}
\ ,
\label{R_PL}
\ee
which is constant as well.
\section{Slow-Roll Inflation}
\label{Slow-Roll Inflation}
Let us briefly recall the known results obtained
from the slow-roll approximation~\cite{SL,LLMS,martin_sl_roll}.
To first order in the slow-roll parameters~(\ref{hor_flo_fun}),
we have the PS
\begin{subequations}
\be
&&
\mathcal{P}_{\zeta}=\frac{H^2}{\pi\,\epsilon_1\,m_{\rm Pl}^2}
\,\left[1-2\,(C+1)\,\epsilon_1-C\,\epsilon_2
-(2\,\epsilon_1+\epsilon_2)\,\ln\left(\frac{k}{k_*}\right)\right]
\nonumber
\\
\label{P_SlowRoll}
\\
&&
\mathcal{P}_h=\frac{16\,H^2}{\pi\,m_{\rm Pl}^2}
\,\left[1-2\,(C+1)\,\epsilon_1-2\,\epsilon_1\,\ln\left(\frac{k}{k_*}
\right)\right]
\nonumber
\ ,
\ee
where $C\equiv\gamma_{_{\rm E}}+\ln 2-2\approx -\,0.7296$,
$\gamma_{_{\rm E}}$ being the Euler-Mascheroni constant,
and all quantities are evaluated at the Hubble crossing
$\eta_*$, that is the moment of time when
$k_*=(a\,H)(N_*)\equiv(a\,H)_*$.
(The number $k_*$ is usually called ``pivot scale''.)
From Eqs.~(\ref{P_SlowRoll}), we can also obtain the spectral
indices and $\alpha$-runnings,
\be
n_S-1=-2\,\epsilon_1-\epsilon_2 \ ,\quad n_T=-2\,\epsilon_1
\label{n_SlowRoll}
\ee
\be
\alpha_S=\alpha_T=0
\ ,
\label{alpha_SlowRoll}
\ee
on using respectively Eqs.~(\ref{n_def}) and (\ref{alpha_def}).
From Eq.~(\ref{Rat_def}), the tensor-to-scalar ratio becomes
\be
R=16\,\epsilon_1\left[1+C\,\epsilon_2+\epsilon_2
\,\ln\left(\frac{k}{k_*}\right)\right]
\ .
\label{R_SR}
\ee
\end{subequations}
\subsection{Improved WKB leading and second slow-roll order}
\label{1+2}
We now consider the possibility of obtaining consistent second
order results in the parameters $\epsilon_i$'s from the leading
WKB approximation.
\par
We first set $g_{(1)}^{\rm AD}(x)\equiv0$
in Eqs.~(\ref{spectra_correct}), (\ref{index_correct}),
(\ref{runnings_correct}) and (\ref{R_correct}).
We then find it convenient to re-express all relevant
quantities in terms of the conformal time $\eta$.
For this purpose, we employ the relation
\begin{subequations}
\be
-k\,\eta&=&
\frac{k}{a\,H\,(1-\epsilon_1)}\,
\left[1+\epsilon_1\,\epsilon_2+{\cal O}\left(\epsilon_i^3\right)\right]
\label{rel_change_var_2'ord}
\\
&=&
\frac{{\rm e}^{x}}{(1-\epsilon_1)}\,
\left[1+\epsilon_1\,\epsilon_2+{\cal O}\left(\epsilon_i^3\right)\right]
\ .
\label{change_var_2'ord}
\ee
\end{subequations}
We can now expand the frequencies to second order in the HFF,
\be
\omega_{\rm S}^2(\eta)
&=&
k^2\,\eta^2\,\left(1-2\,\epsilon_1\,\epsilon_2\right)
-\left(\frac94+3\,\epsilon_1+4\,\epsilon_1^2 +\frac32\,\epsilon_
2+2\,\epsilon_1\,\epsilon_2
+\frac14\,\epsilon_2^2+\frac12\,\epsilon_2\,\epsilon_3\right)
\nonumber
\\
\label{omega2_2'ord}
\\
\omega _{\rm T}^2(\eta) &=& k^2\,\eta^2\,\left(1-2\,\epsilon_1\,\epsilon_2\right)
-\left(\frac94+3\,\epsilon_1+4\,\epsilon_1^2 -\frac12\,\epsilon_1\,\epsilon_2\right)
\nonumber
\ ,
\ee
and the PS become
\be
{\cal P}_\zeta
&=&\frac{H^2_{\rm f}}{\pi\,\epsilon_{1,{\rm f}}\,m_{\rm Pl}^2}\,
\left(-k\,\eta_{\rm f}\right)^3\,
\frac23\,\left(1-\frac83\,\epsilon_{1,{\rm f}}
-\frac13\,\epsilon_{2,{\rm f}}+\frac{19}{9}\,\epsilon_{1,{\rm f}}^2
-\frac{19}{9}\,\epsilon_{1,{\rm f}}\,\epsilon_{2,{\rm f}}
\right.\nonumber\\&&\left.
\quad\quad\quad\quad\quad\quad\quad\quad\quad\quad\quad\quad\quad\quad
+\frac19\,\epsilon_{2,{\rm f}}^2
-\frac19\,\epsilon_{2,{\rm f}}\,\epsilon_{3,{\rm f}}\right)\,
{\rm e}^{2\xi_{\rm II, S}}
\nonumber
\\
\label{partial_spectra_2'ord}
\\
{\cal P}_h
&=&\frac{16\,H_{\rm f}^2}{\pi\,m_{\rm Pl}^2}\,
\left(-k\,\eta_{\rm f}\right)^3\,
\frac23\,\left(1-\frac83\,\epsilon_{1,{\rm f}}
+\frac{19}{9}\,\epsilon_{1,{\rm f}}^2
-\frac{26}{9}\,\epsilon_{1,{\rm f}}\,\epsilon_{2,{\rm f}}\right)\,
{\rm e}^{2\xi_{\rm II, T}}
\ .
\nonumber
\ee
From now on, the subscript $f$ will denote that the given quantity
is evaluated in the super-horizon limit.
\par
We must now compute the arguments of the exponentials
in the above expressions, which are in general of the form
\be
\xi_{\rm II}(\eta_0,\eta_{\rm f};k)
=\int_{\eta_{\rm f}}^{\eta_0}
\sqrt{A^2(\eta)-k^2\,\eta^2}\,\frac{\d\,\eta}{\eta}
\ ,
\label{int_with_change_2'ord}
\ee
where the function $A^2(\eta)$ contains the HFF,
but does not depend on $k$, and we have used
\be
\d\,x=\frac{\d\,\eta}{\eta}\,
\left[1+\epsilon_1(\eta)\,\epsilon_2(\eta)
+{\cal O}(\epsilon_i^3)\right]
\ .
\label{mesure_2'ord}
\ee
Let us also note that, at the turning point $\eta=\eta_0$,
one has
\be
A(\eta_0)=-k\,\eta_0
\ .
\label{TP}
\ee
It is now clear that, in order to obtain consistent results to
second order in the slow-roll parameters, we must consider
the time-dependence of the $\epsilon_i$'s, and the function
$A^2(\eta)$ may not be approximated by a constant.
This does not allow us to perform the integral unless the
scale factor $a=a(\eta)$ is given explicitly or, as we shall
see, some further approximation is employed.
\par
In Refs.~\cite{WKB_let,WKB_SRprD}, we proposed a procedure which will
now be described in detail.
Let us start with the general exact relation
\be
\!\!\!\!\!\!
\int\limits_{\eta_1}^{\eta_2}
\sqrt{A^2(\eta)-k^2\,\eta^2}\,\frac{\d\,\eta}{\eta}
&=&
\left.
\sqrt{A^2(\eta)-k^2\,\eta^2}
\right|_{\eta_1}^{\eta_2}
\nonumber
\\
&&
+
\left.
\frac{A(\eta)}{2}
\,\ln\left(\frac{A(\eta)-\sqrt{A^2(\eta)-k^2\,\eta^2}}
{A(\eta)+\sqrt{A^2(\eta)-k^2\,\eta^2}}\right)
\right|_{\eta_1}^{\eta_2}
\nonumber
\\
&&
-\int_{\eta_1}^{\eta_2}
\,\ln\left(\frac{A(\eta)-\sqrt{A^2(\eta)-k^2\,\eta^2}}
{A(\eta)+\sqrt{A^2(\eta)-k^2\,\eta^2}}\right)\,
\frac{\left[A^2(\eta)\right]'}{4\,A(\eta)}\,\d\,\eta
\label{xi_gen_exact}
\ ,
\ee
which holds for every function $A^2(\eta)$.
We can derive this relation by initially considering the
integral in Eq.~(\ref{int_with_change_2'ord}) with $A$ constant
(i.e.~independent of $\eta$, which is just power-law inflation).
In this case, the integration can be performed and yields the first
and second term in the r.h.s.~of Eq.~(\ref{xi_gen_exact})
with the function $A$ independent of $\eta$,
whereas the third term vanishes identically.
If we then reinstate the $\eta$-dependence of $A$ in the two
non-vanishing terms and differentiate them with respect to $\eta$,
we obtain the original integrand in Eq.~(\ref{int_with_change_2'ord})
plus another term, originating from $A^2(\eta)$.
We then obtain the result~(\ref{xi_gen_exact}) by (formally) integrating
the latter term and subtracting it from the previous two.
\par
One can also repeat the above procedure for the new integral in the
r.h.s.~of Eq.~(\ref{xi_gen_exact}).
To do this, we define
\be
\left[A^2(\eta)\right]'
\equiv-\frac{1}{\eta}\,B(\eta)
\label{A2'_general}
\ ,
\ee
and, after several simplifications, we can write the general relation
as
\be
\!\!\!\!\!\!\!\!\!
-\int\limits_{\eta_1}^{\eta_2}
\ln\left(\frac{A(\eta)-\sqrt{A^2(\eta)-k^2\,\eta^2}}
{A(\eta)+\sqrt{A^2(\eta)-k^2\,\eta^2}}\right)\,
&&\!\!\!\!\!\!\!\!\!
\frac{\left[A^2(\eta)\right]'}{4\,A(\eta)}\,\d\,\eta
=
\left.
Y(\eta)\,\frac{B(\eta)}{8\,A(\eta)}
\right|_{\eta_1}^{\eta_2}
\nonumber
\\
&&
\!\!\!\!\!\!\!\!\!\!\!\!\!\!\!\!\!\!\!\!\!\!\!\!
-\int_{\eta_1}^{\eta_2}
Y(\eta)\,
\left(\frac{B^2(\eta)}{16\,\eta\,A^3(\eta)}
+\frac{B'(\eta)}{8\,A(\eta)}\right)\,\d\eta
\nonumber
\\
&&
\!\!\!\!\!\!\!\!\!\!\!\!\!\!\!\!\!\!\!\!\!\!\!\!
-\int_{\eta_1}^{\eta_2}
\ln\left(\frac{A(\eta)-\sqrt{A^2(\eta)-k^2\,\eta^2}}
{A(\eta)+\sqrt{A^2(\eta)-k^2\,\eta^2}}\right)\,
\frac{B^2(\eta)}{8\,\eta\,A^3(\eta)}\,\d\eta
\label{last_int}
\ ,
\ee
with
\be
Y(\eta)&\equiv&
\ln\left(\frac{-k\,\eta}{2\,A(\eta)}\right)\,
\ln\left(\frac{A(\eta)-\sqrt{A^2(\eta)-k^2\,\eta^2}}
{A(\eta)+\sqrt{A^2(\eta)-k^2\,\eta^2}}\right)
\nonumber
\\
&&
+{\rm Li}_2\left(\frac{A(\eta)-\sqrt{A^2(\eta)-k^2\,\eta^2}}
{2\,A(\eta)}\right)
\nonumber
\\
&&
-{\rm Li}_2\left(\frac{A(\eta)+\sqrt{A^2(\eta)-k^2\,\eta^2}}
{2\,A(\eta)}\right)
\label{def_Y}
\ ,
\ee
and ${\rm Li}_2(z)$ is the dilogarithm function
(see, e.g.~Ref.~\cite{lewin}),
\be
{\rm Li}_2(z)\equiv\sum^\infty_{k=1}\,\frac{z^k}{k^2}
=-\int_0^z\,\frac{\ln\left(1-z'\right)}{z'}\,\d z'
\ .
\label{Li_def}
\ee
\par
On putting together all the pieces, we finally obtain
\be
\int_{\eta_1}^{\eta_2}
\sqrt{A^2(\eta)-k^2\,\eta^2}\,\frac{\d\,\eta}{\eta}
&\!\!=\!\!&
\left.
\sqrt{A^2(\eta)-k^2\,\eta^2}
\right|_{\eta_1}^{\eta_2}
+
\left.
Y(\eta)\,\frac{B(\eta)}{8\,A(\eta)}
\right|_{\eta_1}^{\eta_2}
\nonumber
\\
&&
+\left.\frac{A(\eta)}{2}
\,\ln\left(\frac{A(\eta)-\sqrt{A^2(\eta)-k^2\,\eta^2}}
{A(\eta)+\sqrt{A^2(\eta)-k^2\,\eta^2}}\right)
\right|_{\eta_1}^{\eta_2}
\nonumber
\\
&&
-\int_{\eta_1}^{\eta_2}
Y(\eta)\,\left(\frac{B^2(\eta)}{16\,\eta\,A^3(\eta)}
+\frac{B'(\eta)}{8\,A(\eta)}\right)\,\d\eta
\label{xi_gen_exact_compl}
\\
&&
-\int_{\eta_1}^{\eta_2}
\ln\left(\frac{A(\eta)-\sqrt{A^2(\eta)-k^2\,\eta^2}}
{A(\eta)+\sqrt{A^2(\eta)-k^2\,\eta^2}}\right)\,
\frac{B^2(\eta)}{8\,\eta\,A^3(\eta)}\,\d\eta
\nonumber
\ .
\ee
\par
For the cases of interest to us [i.e.~for the
frequencies~(\ref{omega2_2'ord})], the functions $A$
for scalar and tensor perturbations to second order
in the HFF are given by
\begin{subequations}
\be
A^2_{\rm S}&=&
\frac94+3\,\epsilon_1+\frac32\,\epsilon_ 2
+4\,\epsilon_1^2+\frac{13}{2}\,\epsilon_1\,\epsilon_2
+\frac14\,\epsilon_2^2+\frac12\,\epsilon_2\,\epsilon_3
\nonumber
\\
\label{A2_2'ord}
\\
A^2_{\rm T}&=&
\frac94+3\,\epsilon_1+4\,\epsilon_1^2
+4\,\epsilon_1\,\epsilon_2
\nonumber
\ ,
\ee
so that
\be
B_{\rm S}=
3\,\epsilon_1\,\epsilon_ 2+\frac32\,\epsilon_ 2\,\epsilon_ 3
\ ,\quad
B_{\rm T}=
3\,\epsilon_1\,\epsilon_ 2
\label{B_2'ord}
\ .
\ee
\end{subequations}
For the results to second order in the HFF, we can
neglect the last two integrals in Eq.~(\ref{xi_gen_exact_compl}),
since $B'(\eta)=\mathcal{O}\left(\epsilon_i^3\right)$
and $B^2(\eta)=\mathcal{O}\left(\epsilon_i^4\right)$.
Thus our approximation consists in neglecting such terms and
not in assuming the HFF be constant (or any specific functional
form for them).
We can finally write $\xi_{\rm II, S}(\eta_0,\eta_{\rm f};k)$
and $\xi_{\rm II, T}(\eta_0,\eta_{\rm f};k)$ as
\begin{subequations}
\be
\xi_{\rm II, S}
&\!\!\!\!\!\simeq\!\!\!\!\!&
-\frac32
+\frac{3\,\ln 3}{2}
+\ln 3\,\epsilon_{1,{\rm f}}
+\frac{\ln 3}{2}\,\epsilon_{2,{\rm f}}
+\left(\ln 3+\frac13\right)\,\epsilon_{1,{\rm f}}^2
+\frac{1}{12}\,\epsilon_{2,{\rm f}}^2
\nonumber
\\
&\!\!\!\!\!\!\!\!\!\!\!\!&
+\left(\frac{11\,\ln 3}{6}-\frac{\ln^2 3}{2}
+\frac{\pi^2}{24}+\frac13\right)\,\epsilon_{1,{\rm f}}\,\epsilon_{2,{\rm f}}
+\left(\frac{\ln 3}{6}-\frac{\ln^2 3}{4}+\frac{\pi^2}{48}\right)\,
\epsilon_{2,{\rm f}}\,\epsilon_{3,{\rm f}}
\nonumber
\\
&\!\!\!\!\!\!\!\!\!\!\!\!&
+\left[
-\frac32-\epsilon_{1,{\rm f}}-\frac12\,\epsilon_{2,{\rm f}}-\epsilon_{1,{\rm f}}^2
+\left(\ln 3-\frac{11}{6}\right)\,\epsilon_{1,{\rm f}}\,\epsilon_{2,{\rm f}}
+\left(\frac{\ln 3}{2}-\frac16\right)\,\epsilon_{2,{\rm f}}\,\epsilon_{3,{\rm f}}
\right]\,
\ln\left(-k\,\eta_{\rm f}\right)
\nonumber
\\
&\!\!\!\!\!\!\!\!\!\!\!\!&
-\frac12\,\left(\epsilon_{1,{\rm f}}\,\epsilon_{2,{\rm f}}
+\frac12\,\epsilon_{2,{\rm f}}\,\epsilon_{3,{\rm f}}\right)
\ln^2\left(-k\,\eta_{\rm f}\right)
\label{xi_2'_order_with_eps_S}
\ee
\be
\xi_{\rm II, T}
&\!\!\!\!\!\simeq\!\!\!\!\!&
-\frac32
+\frac{3\,\ln 3}{2}
+\ln 3\,\epsilon_{1,{\rm f}}
+\left(\ln 3+\frac13\right)\,\epsilon_{1,{\rm f}}^2
+\left(\frac{4\,\ln 3}{3}-\frac{\ln^2 3}{2}+\frac{\pi^2}{24}\right)\,
\epsilon_{1,{\rm f}}\,\epsilon_{2,{\rm f}}
\nonumber
\\
&\!\!\!\!\!\!\!\!\!\!\!\!&
+\left[
-\frac32-\epsilon_{1,{\rm f}}-\epsilon_{1,{\rm f}}^2
+\left(\ln 3-\frac43\right)\,\epsilon_{1,{\rm f}}\,\epsilon_{2,{\rm f}}
\right]\,
\ln\left(-k\,\eta_{\rm f}\right)
\nonumber
\\
&\!\!\!\!\!\!\!\!\!\!\!\!&
-\frac12\,\epsilon_{1,{\rm f}}\,\epsilon_{2,{\rm f}}
\ln^2\left(-k\,\eta_{\rm f}\right)
\ ,
\label{xi_2'_order_with_eps_T}
\ee
\end{subequations}
where the arguments in $\xi_{\rm II}$ have been omitted.
On inserting the above quantities in Eqs.~(\ref{partial_spectra_2'ord}),
we can calculate the PS.
\par
In order to compare the PS from Eqs.~(\ref{partial_spectra_2'ord}),
(\ref{xi_2'_order_with_eps_S}) and (\ref{xi_2'_order_with_eps_T}),
with the slow-roll expressions in Eqs.~(\ref{P_SlowRoll}),
we need a relation between $H(N_{\rm f})$ and
$H(N_*)$, $\epsilon_1(N_{\rm f})$ and $\epsilon_1(N_*)$,
and so on, to second order in the HFF (hereafter, quantities
without a subscript will be evaluated at the Hubble crossing
$\eta_*$ corresponding to the pivot scale $k_*$).
We expand the parameters $\epsilon_{i,{\rm f}}$ to first order
in $\Delta N\equiv N_f-N_*$ in the numerators and to second
order in the denominator of the scalar spectrum,
\be
\frac{\epsilon_i(N_{\rm f})}{\epsilon_i}
\simeq
1+\epsilon_{i+1} \Delta N
+\frac12\,
\left(\epsilon_{i+1}^2+\epsilon_{i+1}\,\epsilon_{i+2}\right)
{\Delta N}^2
\ ,
\label{epsilon_n_expans}
\ee
and, analogously, we find
\be
\frac{H^2(N_{\rm f})}{H^2}
\simeq
1-2\,\epsilon_1\,\Delta N-
\left(\epsilon_1\,\epsilon_2-2\,\epsilon_1^2\right)\,
{\Delta N}^2
\ .
\label{Hquad_expans_2'ord}
\ee
We can eliminate $\eta_{\rm f}$ in the logarithms
by expressing it in terms of $1/(a\,H)_{\rm f}$,
\be
\ln(-k\,\eta_{\rm f})
&\simeq&
\ln\left[\frac{k\,\left(1+\epsilon_{1,{\rm f}}\,
\epsilon_{2,{\rm f}}\right)}
{(a\,H)_{\rm f}\,\left(1-\epsilon_{1,{\rm f}}\right)}
\,\frac{(a\,H)_*}{(a\,H)_*}\right]
\nonumber
\\
&\simeq&
\ln\left(\frac{k}{k_*}\right)
-\Delta N
+\epsilon_{1,{\rm f}}\,\Delta N
+\epsilon_{1,{\rm f}}
\nonumber
\\
&\simeq&
\ln\left(\frac{k}{k_*}\right)
-\Delta N
+\epsilon_1\,\Delta N
+\epsilon_1
\ ,
\label{ln_expans_2'ord}
\ee
where in the first equality we used Eq.~(\ref{rel_change_var_2'ord}),
in the second one the definition of the pivot scale, and,
in the last relation, Eq.~(\ref{epsilon_n_expans}).
If we now use the above expressions we obtain the following PS,
where the superscript $^{(2)}$ stands for second slow-roll order
and the subscript WKB for leading adiabatic order,
\begin{subequations}
\be
\mathcal{P}_{\zeta,\scriptscriptstyle\scriptscriptstyle{\rm WKB}}^{(2)}
&\!\!\!=\!\!\!&
\frac{H^2}{\pi\,\epsilon_1\,m_{\rm Pl}^2}\,
A_{\scriptscriptstyle{\rm WKB}}
\left\{1-2\left(D_{\scriptscriptstyle{\rm WKB}}+1\right)\,\epsilon_1
-D_{\scriptscriptstyle{\rm WKB}}\,\epsilon_2
+\left(\frac12\,D_{\scriptscriptstyle{\rm WKB}}^2+\frac29\right)\,\epsilon_2^2
\right.
\nonumber
\\
&&
+\left(2\,D_{\scriptscriptstyle{\rm WKB}}^2
+2\,D_{\scriptscriptstyle{\rm WKB}}-\frac19\right)\,\epsilon_1^2
+\left(D_{\scriptscriptstyle{\rm WKB}}^2-D_{\scriptscriptstyle{\rm WKB}}
+\frac{\pi^2}{12}-\frac{20}{9}\right)\,\epsilon_1\,\epsilon_2
\nonumber\\&&
+\left(-\frac12\,D_{\scriptscriptstyle{\rm WKB}}^2+\frac{\pi^2}{24}-
\frac{1}{18}\right)\,\epsilon_2\,\epsilon_3
\nonumber
\\
&&\left.
+\left[-2\,\epsilon_1-\epsilon_2
+2\left(2\,D_{\scriptscriptstyle{\rm WKB}}+1\right)\,\epsilon_1^2
+\left(2\,D_{\scriptscriptstyle{\rm WKB}}-1\right)\,\epsilon_1\,\epsilon_2
\right.\right.
\nonumber\\&&\left.\left.
+D_{\scriptscriptstyle{\rm WKB}}\,\epsilon_2^2
-D_{\scriptscriptstyle{\rm WKB}}\,\epsilon_2\,\epsilon_3\right]\,
\ln\left(\frac{k}{k_*}\right)
\right.
\nonumber
\\
&&
\left.
+\frac12\,\left(4\,\epsilon_1^2+2\,\epsilon_1\,\epsilon_2+\epsilon_2^2
-\epsilon_2\,\epsilon_3\right)\,
\ln^2\left(\frac{k}{k_*}\right)\right\}
\nonumber
\\
\label{P_SlowRoll_0_2order}
\\
\mathcal{P}_{h,\scriptscriptstyle{\rm WKB}}^{(2)}
&\!\!\!=\!\!\!&\frac{16\,H^2}{\pi\,m_{\rm Pl}^2}\,
A_{\scriptscriptstyle{\rm WKB}}
\left\{1-2\left(D_{\scriptscriptstyle{\rm WKB}}+1\right)\,\epsilon_1
+\left(2\,D_{\scriptscriptstyle{\rm WKB}}^2
+2\,D_{\scriptscriptstyle{\rm WKB}}-\frac19\right)\,\epsilon_1^2
\nonumber
\right.
\\
&&
+\left(-D_{\scriptscriptstyle{\rm WKB}}^2
-2\,D_{\scriptscriptstyle{\rm WKB}}+\frac{\pi^2}{12}-\frac{19}{9}\right)\,
\epsilon_1\,\epsilon_2
\nonumber\\&&
+\left[-2\,\epsilon_1+2\left(2\,D_{\scriptscriptstyle{\rm WKB}}+1\right)\,
\epsilon_1^2
-2\,\left(D_{\scriptscriptstyle{\rm WKB}}+1\right)\epsilon_1\,
\epsilon_2\right]\,
\ln\left(\frac{k}{k_*}\right)
\nonumber
\\
&&
\left.
+\frac12\,\left(4\,\epsilon_1^2
-2\,\epsilon_1\,\epsilon_2\right)\,\ln^2\left(\frac{k}{k_*}\right)\right\}
\ .
\nonumber
\ee
In the above, we have defined
$D_{\scriptscriptstyle{\rm WKB}}\equiv\frac{1}{3}-\ln 3\approx -0.7653$,
which approximates the coefficient $C$ in Eqs.~(\ref{P_SlowRoll}) with an
error of about $5\%$, and the factor
$A_{\scriptscriptstyle{\rm WKB}}\equiv 18/e^3\approx0.896$ which
gives an error of about $10\%$ on the amplitudes.
An important result is that the PS to second order in the slow-roll parameters do not
depend on $\Delta N$, once the laborious dependence on the Hubble
crossing has been worked out: this fact is in complete agreement with the
constancy of the growing modes of $\zeta$ and $h$ on large scales.
The spectral indices, from Eqs.~(\ref{index_correct}),
and their runnings, from Eqs.~(\ref{runnings_correct}),
are analogously given by
\be
n_{\rm S,\scriptscriptstyle{\rm WKB}}^{(2)}-1&=&
-2\,\epsilon_1-\epsilon_2-2\,\epsilon_1^2
-\left(2\,D_{\scriptscriptstyle{\rm WKB}}+3\right)\,\epsilon_1\,\epsilon_2
-D_{\scriptscriptstyle{\rm WKB}}\,\epsilon_2\,\epsilon_3
\nonumber\\&&
-2\,\epsilon_1\,\epsilon_2\,\ln\left(\frac{k}{k_*}\right)
-\epsilon_2\,\epsilon_3\,\ln\left(\frac{k}{k_*}\right)
\nonumber
\\
\label{n_2'order}
\\
n_{\rm T,\scriptscriptstyle{\rm WKB}}^{(2)}&=&
-2\,\epsilon_1-2\,\epsilon_1^2
-2\,\left(D_{\scriptscriptstyle{\rm WKB}}+1\right)\,\epsilon_1\,\epsilon_2
-2\,\epsilon_1\,\epsilon_2\,\ln\left(\frac{k}{k_*}\right)
\nonumber
\ee
\be
\alpha_{\rm S,\scriptscriptstyle{\rm WKB}}^{(2)}=
-2\,\epsilon_1\,\epsilon_2
-\epsilon_2\,\epsilon_3
\ ,
\quad
\alpha_{\rm T,\scriptscriptstyle{\rm WKB}}^{(2)}=
-2\,\epsilon_1\,\epsilon_2
\label{alpha_2'order}
\ .
\ee
From Eq.~(\ref{R_correct}) the tensor-to-scalar ratio becomes
\be
R^{(2)}_{\scriptscriptstyle{\rm WKB}}&=&
16\,\epsilon_1\left[1+D_{\scriptscriptstyle{\rm WKB}}\,\epsilon_2
+\left(D_{\scriptscriptstyle{\rm WKB}}+\frac19\right)\,\epsilon_1\,\epsilon_2
+\left(\frac12\,D_{\scriptscriptstyle{\rm WKB}}^2-\frac{2}{9}\right)\,\epsilon_2^2
\right.
\nonumber\\&&
+\left(\frac12\,D_{\scriptscriptstyle{\rm WKB}}^2-\frac{\pi^2}{24}
+\frac{1}{18}\right)\,\epsilon_2\,\epsilon_3
\nonumber
\\
&&
+\left(\epsilon_2+\epsilon_1\,\epsilon_2+D_{\scriptscriptstyle{\rm WKB}}\,\epsilon_2^2
+D_{\scriptscriptstyle{\rm WKB}}\,\epsilon_2\,\epsilon_3\right)\,
\ln\left(\frac{k}{k_*}\right)
\nonumber\\&&\left.
+\frac12\,\left(\epsilon_2^2+\epsilon_2\,\epsilon_3\right)\,
\ln^2\left(\frac{k}{k_*}\right)\right]
\ .
\label{R_lead_WKB}
\ee
\end{subequations}
We note that Eqs.~(\ref{P_SlowRoll_0_2order}),
(\ref{n_2'order}) and (\ref{alpha_2'order})
agree with Ref.~\cite{martin_schwarz2}, to first order in the HFF.
\par
We can also compare the above results with others as we
show in Appendix~\ref{app_other_res}.
For example, one could approximate the HFF by their Taylor
expansions around $\eta_0$ as in Ref.~\cite{H_MP,HHHJM,HHHJ},
and give the results in terms of the $\epsilon_i$'s calculated
at that point (see Table~\ref{H_MP_tab_PRD71}
for results only on spectral indices).
\subsection{Improved WKB next-to-leading and first slow-roll order}
\label{2+1}
To first order in the HFF, the function $A$ can be taken as
constant.
In this case, let us then introduce the more convenient
notation
\be
\xi_\pm(x_0,x;k)
&=&\pm\,\int_{x}^{x_0}\,\sqrt{\mp\,\omega^2(y)}\,\d\,y
\nonumber
\\
&\simeq&\pm\,\int_{\eta}^{\eta_0}
\sqrt{\mp\left(k^2\,\tau^2-A^2\right)}\,\frac{\d\,\tau}{\tau}
\nonumber
\\
&=&\pm\,A\,
\left[{\rm atan}_\pm\left(\Theta_\pm\right)-\Theta_\pm\right]
\ ,
\label{xi_Ibis}
\ee
where we have used $\d\,x=\d\,\tau/\tau$, and
\begin{subequations}
\be
A^2_{\rm S}(\epsilon_i)&=&
\frac94+3\epsilon_1+\frac32\,\epsilon_2
\nonumber
\\
\label{B_S}
\\
A^2_{\rm T}(\epsilon_i)&=&
\frac94+3\epsilon_1
\nonumber
\ee
\be
k\,\eta_0=-A
\label{eta*}
\ ,
\ee
\end{subequations}
with $\epsilon_1$ and $\epsilon_2$ constant as well.
It is also useful to define
\begin{subequations}
\be
\Theta_\pm \equiv \sqrt{\pm\left(1-\frac{\eta^2}{\eta_0^2}\right)}
\label{arg_xi_I}
\ee
\be
&&
{\rm atan}_+(x) \equiv {\rm arctanh}(x)
\nonumber
\\
\\
&&
{\rm atan}_-(x) \equiv \arctan(x)
\ ,
\nonumber
\ee
\end{subequations}
where the plus (minus) sign corresponds to region~II (I).
With this notation, the expressions for $\gamma_{(1)}(\eta)$
and $\phi_{(1)}(\eta)$ in Eqs.~(\ref{delta1})
can be explicitly written as
\be
\gamma_{\pm(1)}(\eta)&=&\strut\displaystyle
\frac{1}{A^2}\,\left\{
-\frac{5\mp3\,\Theta_\pm^2}{24\,\Theta_\pm^4}
\pm\frac{5}{72\,\Theta_\pm\,
\left[{\rm atan}_\pm\left(\Theta_\pm\right)-\Theta_\pm\right]}
\right\}
\nonumber
\\
\\
\phi_{\pm(1)}(\eta)&=&\strut\displaystyle
\frac{1}{A^2}\,\left\{
\frac{23}{3150}
\pm\frac{505\mp654\,\Theta_\pm^2+153\,\Theta_\pm^4}{1152\,\Theta_\pm^6}\right.
\nonumber\\
&&
\left.
-\frac{5\,\left[\left(102\mp90\,\Theta_\pm^2\right)\,{\rm atan}_\pm
\left(\Theta_\pm\right)
-\left(102\mp157\,\Theta_\pm^2\right)\,\Theta_\pm\right]}
{10368\,\Theta_\pm^3\,\left[{\rm atan}_\pm\left(\Theta_\pm\right)
-\,\Theta_\pm\right]^2}
\right\}
\ .
\nonumber
\ee
The limits of interest are then given by
\begin{subequations}
\be
&&
\lim_{\eta\to-\infty}\,\gamma_{-(1)}(\eta)=0
\label{limit_gamma_I}
\\
&&
\lim_{\eta\to-\infty}\,\phi_{-(1)}(\eta)
=\frac{23}{3150\,A^2}
\label{limit_phi_I}
\\
&&
\lim_{\eta\to0^-}\,\gamma_{+(1)}(\eta)
=-\,\frac{1}{12\,A^2}
\label{limit_gamma_II}
\\
&&
\lim_{\eta\to0^-}\,\phi_{+(1)}(\eta)
=\frac{181}{16800\,A^2}
\ ,
\label{limit_phi_II}
\ee
where we have used
\be
\lim_{\eta\to-\infty}\,\Theta_-(\eta)=\infty
\label{limit_Theta-}
\\
\nonumber
\\
\lim_{\eta\to0^-}\,\Theta_+(\eta)=1
\ ,
\label{limit_Theta+}
\ee
and
\be
\lim_{\eta\to0^-}\,\left(
\omega_{\rm II}+\frac{\omega_{\rm II}'}{2\,\omega_{\rm II}}
\right)
=A
\ .
\label{limit_fun_in_spectra}
\ee
\end{subequations}
We can now use the values in Eqs.~(\ref{B_S})
to calculate the correction $g_{(1)}^{\rm AD}$ in the
super-horizon limit ($\eta\to 0^-$) for scalar
and tensor perturbations, obtaining
\be
&&
g_{(1){\rm S}}^{\rm AD}
=\frac{37}{324}
-\frac{19}{243}\,
\left(\epsilon_1+\frac12\,\epsilon_2\right)
\nonumber
\\
\label{AD_corrections}
\\
&&
g_{(1){\rm T}}^{\rm AD}
=\frac{37}{324}-\frac{19}{243}\,\epsilon_1
\ .
\nonumber
\ee
Finally, from Eqs.~(\ref{spectra_correct}) and (\ref{AD_corrections}),
we can write the expressions for the scalar and tensor PS,
with the superscript $^{(1)}$ for first slow-roll order and
the subscript WKB$*$ for next-to-leading adiabatic order,
as
\begin{subequations}
\be
&&
\mathcal{P}_{\zeta,\scriptscriptstyle{\rm WKB*}}^{(1)}=
\frac{H^2}{\pi\,\epsilon_1\,m_{\rm Pl}^2}\,
A_{\scriptscriptstyle{\rm WKB*}}\,
\left[1-2\,\left(D_{\scriptscriptstyle{\rm WKB*}}+1\right)\,\epsilon_1
-D_{\scriptscriptstyle{\rm WKB*}}\,\epsilon_2
-\left(2\,\epsilon_1+\epsilon_2\right)\,\ln\left(\frac{k}{k_*}\right)\right]
\nonumber
\\
\label{P_SlowRoll_1}
\\
&&
\mathcal{P}_{h,\scriptscriptstyle{\rm WKB*}}^{(1)}=
\frac{16\,H^2}{\pi\,m_{\rm Pl}^2}\,
A_{\scriptscriptstyle{\rm WKB*}}\,
\left[1-2\,\left(D_{\scriptscriptstyle{\rm WKB*}}+1\right)\,\epsilon_1
-2\,\epsilon_1\,\ln\left(\frac{k}{k_*}\right)\right]
\ ,
\nonumber
\ee
where
$D_{\scriptscriptstyle{\rm WKB*}}\equiv\frac{7}{19}-\ln 3\approx -0.7302$,
which approximates the coefficient $C$ in Eqs.~(\ref{P_SlowRoll}) with
an error of about $0.08\%$, and the factor
$A_{\scriptscriptstyle{\rm WKB*}}\equiv 361/18\,e^3\approx 0.999$
which gives an error of about $0.1\%$ on the amplitudes.
We also obtain the standard slow-roll spectral indices
and $\alpha$-runnings,
\be
\!\!\!\!\!\!\!\!
n_{\rm S,\scriptscriptstyle{\rm WKB*}}^{(1)}-1=
-2\,\epsilon_1-\epsilon_2
\ ,\quad
n_{\rm T,\scriptscriptstyle{\rm WKB*}}^{(1)}=-2\,\epsilon_1
\label{n_NTL_WKB}
\ee
\be
\!\!\!\!\!\!\!\!
\alpha_{\rm S,\scriptscriptstyle{\rm WKB*}}^{(1)}=
\alpha_{\rm T,\scriptscriptstyle{\rm WKB*}}^{(1)}=0
\ ,
\label{alpha_NTL_WKB}
\ee
on respectively using Eqs.~(\ref{index_correct}) and
(\ref{runnings_correct}).
From Eq.~(\ref{R_correct}) the tensor-to-scalar ratio becomes
\be
R^{(1)}_{\scriptscriptstyle{\rm WKB*}}=
16\,\epsilon_1\left[1+D_{\scriptscriptstyle{\rm WKB*}}\,\epsilon_2
+\epsilon_2\,\ln\left(\frac{k}{k_*}\right)\right]
\ .
\label{R_next-to-lead_WKB}
\ee
\end{subequations}
\subsection{Improved WKB next-to-leading and second slow-roll order}
\label{2+2}
In order to give the results to next-to-leading WKB order
and second slow-roll order, we should evaluate the corrections
$g_{(1)}^{\rm AD}$ for scalar and tensor perturbations
to second order in the $\epsilon_i$'s.
This, in turn, would require the computation of the functions
$\phi_{\pm(1)}$ and $\gamma_{\pm(1)}$, in Eqs.~(\ref{delta1}),
whose analytical calculation is extremely difficult.
We shall therefore try to give the expressions for the corrections
by following a heuristic (and faster) method.
\par
Let us start from the first order slow-roll expressions given in
Eqs.~(\ref{AD_corrections}) and add all possible second order terms
in the HFF.
Some of such terms can then be (partially) fixed by making reasonable
requirements, which will be explained below, so that the corrections
to the PS read
\be
1+g_{(1){\rm S}}^{\rm AD}
&\!\!=\!\!&
\frac{361}{324}\,
\left\{1-\frac{4}{57}\,
\left(\epsilon_{1,{\rm f}}+\frac12\,\epsilon_{2,{\rm f}}\right)
-\frac{8}{361}\,\epsilon_{1,{\rm f}}^2
+\left[\frac{4}{57}\,\left({b}_{\rm S}-D_{\scriptscriptstyle{\rm WKB*}}\right)
-\frac{262}{3249}\right]\,\epsilon_{1,{\rm f}}\,\epsilon_{2,{\rm f}}
\right.
\nonumber
\\
&&
+\frac{13}{1083}\,\epsilon_{2,{\rm f}}^2
+\left[\frac{2}{57}\,\left({d}_{\rm S}-D_{\scriptscriptstyle{\rm WKB*}}\right)
-\frac{2}{171}\right]\,\epsilon_{2,{\rm f}}\,\epsilon_{3,{\rm f}}
\nonumber
\\
&&
\left.-\left(\frac{4}{57}\,\epsilon_{1,{\rm f}}\,\epsilon_{2,{\rm f}}
+\frac{2}{57}\,\epsilon_{2,{\rm f}}\,\epsilon_{3,{\rm f}}\right)
\,\ln\left(-k\,\eta_{\rm f}\right)\right\}
\nonumber
\\
\label{AD_corrections_2'order}
\\
1+g_{(1){\rm T}}^{\rm AD}
&\!\!=\!\!&
\frac{361}{324}\,
\left\{1-\frac{4}{57}\,\epsilon_{1,{\rm f}}
-\frac{8}{361}\,\epsilon_{1,{\rm f}}^2
+\left[\frac{4}{57}\,\left({b}_{\rm T}-D_{\scriptscriptstyle{\rm WKB*}}\right)
-\frac{16}{171}\right]\,\epsilon_{1,{\rm f}}\,\epsilon_{2,{\rm f}}\right.
\nonumber
\\
&&
\left.-\frac{4}{57}\,\epsilon_{1,{\rm f}}\,\epsilon_{2,{\rm f}}
\,\ln\left(-k\,\eta_{\rm f}\right)\right\}
\ .
\nonumber
\ee
First of all, we have factorized the number $361/324$ so as
to recover the standard slow-roll amplitudes (\ref{P_SlowRoll})
with a very good accuracy.
As already mentioned, the first order terms are the same as those
in the results (\ref{AD_corrections}) of Section~\ref{2+1},
evaluated however in the super-horizon limit.
The form of the coefficients multiplying the second order monomials
in the HFF have been partially determined by using the expressions given in
Section~\ref{2+1} [Eq.~(\ref{limit_gamma_I})-(\ref{limit_fun_in_spectra})]
with Eqs.~(\ref{A2_2'ord}) replacing Eqs.~(\ref{B_S}).
By this procedure, one expects to obtain results correct up to
derivatives of $A^2$, which can just contain mixed terms to second order.
The numerical coefficients in front of $\epsilon_{1,{\rm f}}^2$ and
$\epsilon_{2,{\rm f}}^2$ are thus uniquely determined,
whereas the coefficients multiplying the mixed terms
$\epsilon_{1,{\rm f}}\,\epsilon_{2,{\rm f}}$ and
$\epsilon_{2,{\rm f}}\,\epsilon_{3,{\rm f}}$ remain partially
ambiguous.
The last number in each square bracket arises directly from
this procedure, whereas the other terms are chosen so as to match
the ``standard'' dependence on $D_{\scriptscriptstyle{\rm WKB*}}$
in the PS, with $b_{\rm S,T}$ and $d_{\rm S}$ left undetermined.
Finally, we have also required that the PS do not depend on
$\Delta N$, consistently with the constancy in time of the growing
modes for $h$ and $\zeta$.
This requirement fixes the coefficients in front of the logarithms in
Eqs.~(\ref{AD_corrections_2'order}) uniquely~\footnote{For the same
reason (i.e.~to avoid the appearance of $\Delta N$), we cannot consider
terms such as
$\mathcal{O}\,(\epsilon^2_i)\,\ln^2\left(-k\,\eta_{\rm f}\right)$
in Eqs.~(\ref{AD_corrections_2'order}).}, and leads to spectral indices
which do not depend on $b_{\rm S,T}$ and $d_{\rm S}$.
\par
Proceeding then as in Section~\ref{1+2}, from the above corrections
we obtain the expressions for the scalar and tensor PS,
\begin{subequations}
\be
\mathcal{P}_{\zeta,\scriptscriptstyle{\rm WKB*}}^{(2)}&\!\!\!=\!\!\!&
\frac{H^2}{\pi\,\epsilon_1\,m_{\rm Pl}^2}\,
A_{\scriptscriptstyle{\rm WKB*}}\,
\left\{1-2\left(D_{\scriptscriptstyle{\rm WKB*}}+1\right)\,\epsilon_1
-D_{\scriptscriptstyle{\rm WKB*}}\,\epsilon_2
+\left(\frac12\,{D^2_{\scriptscriptstyle{\rm WKB*}}}
+\frac{253}{1083}\right)\,\epsilon_2^2
\right.
\nonumber
\\
&&\left.
+\left({D^2_{\scriptscriptstyle{\rm WKB*}}}
-D_{\scriptscriptstyle{\rm WKB*}}+\frac{\pi^2}{12}
+{b}_{\rm S}\,\frac{4}{57}-\frac{2384}{1083}\right)\,
\epsilon_1\,\epsilon_2
\right.
\nonumber
\\
&&\left.
+\left(2\,{D^2_{\scriptscriptstyle{\rm WKB*}}}
+2\,D_{\scriptscriptstyle{\rm WKB*}}-\frac{71}{1083}\right)\,
\epsilon_1^2
+\left(-\frac12\,{D^2_{\scriptscriptstyle{\rm WKB*}}}+\frac{\pi^2}{24}
+{d}_{\rm S}\,\frac{2}{57}-\frac{49}{722}\right)\,
\epsilon_2\,\epsilon_3
\right.
\nonumber
\\
&&\left.
+\left[-2\,\epsilon_1-\epsilon_2+2\left(2\,D_{\scriptscriptstyle{\rm WKB*}}+1\right)\,\epsilon_1^2
+\left(2\,D_{\scriptscriptstyle{\rm WKB*}}-1\right)\,\epsilon_1\,\epsilon_2
\right.\right.
\nonumber\\&&\left.\left.
+D_{\scriptscriptstyle{\rm WKB*}}\,\epsilon_2^2
-D_{\scriptscriptstyle{\rm WKB*}}\,\epsilon_2\,\epsilon_3\right]\,
\ln\left(\frac{k}{k_*}\right)
\right.
\nonumber
\\
&&\left.
+\frac12\,\left(4\,\epsilon_1^2
+2\,\epsilon_1\,\epsilon_2+\epsilon_2^2
-\epsilon_2\,\epsilon_3\right)
\,\ln^2\left(\frac{k}{k_*}\right)\right\}
\nonumber
\\
\label{P_SlowRoll_1_2order}
\\
\mathcal{P}_{h,\scriptscriptstyle{\rm WKB*}}^{(2)}&\!\!\!=\!\!\!&
\frac{16\,H^2}{\pi\,m_{\rm Pl}^2}\,
A_{\scriptscriptstyle{\rm WKB*}}\,
\left\{
1-2\left(D_{\scriptscriptstyle{\rm WKB*}}+1\right)\,\epsilon_1
+\left(2\,{D^2_{\scriptscriptstyle{\rm WKB*}}}
+2\,D_{\scriptscriptstyle{\rm WKB*}}
-\frac{71}{1083}\right)\,\epsilon_1^2
\nonumber
\right.
\\
&&
\left.
+\left(-{D^2_{\scriptscriptstyle{\rm WKB*}}}
-2\,D_{\scriptscriptstyle{\rm WKB*}}+\frac{\pi^2}{12}
+{b}_{\rm T}\,\frac{4}{57}-\frac{771}{361}\right)\,\epsilon_1\,\epsilon_2
\nonumber
\right.
\\
&&
\left.
+\left[-2\,\epsilon_1+2\left(2\,D_{\scriptscriptstyle{\rm WKB*}}+1\right)\,\epsilon_1^2
-2\,\left(D_{\scriptscriptstyle{\rm WKB*}}+1\right)\epsilon_1\,\epsilon_2\right]
\,\ln\left(\frac{k}{k_*}\right)\right.
\nonumber\\&&\left.
+\frac12\,\left(4\,\epsilon_1^2
-2\,\epsilon_1\,\epsilon_2\right)\,\ln^2\left(\frac{k}{k_*}\right)\right\}
\ .
\nonumber
\ee
We also obtain the spectral indices~(\ref{index_correct})
and their runnings~(\ref{runnings_correct}),
\be
&&
n_{\rm S,\scriptscriptstyle{\rm WKB*}}^{(2)}-1=
-2\,\epsilon_1-\epsilon_2-2\,\epsilon_1^2
-\left(2\,D_{\scriptscriptstyle{\rm WKB*}}+3\right)\,\epsilon_1\,\epsilon_2
-D_{\scriptscriptstyle{\rm WKB*}}\,\epsilon_2\,\epsilon_3
\nonumber\\&&
-2\,\epsilon_1\,\epsilon_2\ln\left(\frac{k}{k_*}\right)
-\epsilon_2\,\epsilon_3\ln\left(\frac{k}{k_*}\right)
\nonumber
\\
\label{n_2'order_next_WKB}
\\
&&
n_{\rm T,\scriptscriptstyle{\rm WKB*}}^{(2)}=
-2\,\epsilon_1-2\,\epsilon_1^2
-2\,\left(D_{\scriptscriptstyle{\rm WKB*}}+1\right)\,\epsilon_1\,\epsilon_2
-2\,\epsilon_1\,\epsilon_2\ln\left(\frac{k}{k_*}\right)
\nonumber
\ee
\be
\alpha_{\rm S,\scriptscriptstyle{\rm WKB*}}^{(2)}=
-2\,\epsilon_1\,\epsilon_2
-\epsilon_2\,\epsilon_3
\ ,\quad
\alpha_{\rm T,\scriptscriptstyle{\rm WKB*}}^{(2)}=
-2\,\epsilon_1\,\epsilon_2
\label{alpha_2'order_next_WKB}
\ ,
\ee
and the tensor-to-scalar ratio becomes
\be
R_{\scriptscriptstyle{\rm WKB*}}^{(2)}&\!\!=\!\!&
16\,\epsilon_1\left\{1+D_{\scriptscriptstyle{\rm WKB*}}\,\epsilon_2
+\left[D_{\scriptscriptstyle{\rm WKB*}}
+\left({b}_{\rm T}-{b}_{\rm S}\right)\,\frac{4}{57}
+\frac{71}{1083}\right]\,\epsilon_1\,\epsilon_2
\right.
\label{R_next-to-lead_WKB_2'order}
\\
&&\left.
+\left(\frac12\,{D^2_{\scriptscriptstyle{\rm WKB*}}}
-\frac{253}{1083}\right)\,\epsilon_2^2
+\left(\frac12\,{D^2_{\scriptscriptstyle{\rm WKB*}}}-\frac{\pi^2}{24}
-{d}_{\rm S}\,\frac{2}{57}+\frac{49}{722}\right)
\,\epsilon_2\,\epsilon_3
\right.
\nonumber
\\
&&\left.
+\left(\epsilon_2+\epsilon_1\,\epsilon_2
+D_{\scriptscriptstyle{\rm WKB*}}\,\epsilon_2^2
+D_{\scriptscriptstyle{\rm WKB*}}\,\epsilon_2\,\epsilon_3\right)\,
\ln\left(\frac{k}{k_*}\right)
+\frac12\,\left(\epsilon_2^2+\epsilon_2\,\epsilon_3\right)\,
\ln^2\left(\frac{k}{k_*}\right)\right\}
\ .
\nonumber
\ee
\end{subequations}
\par
Let us end this Section with a few remarks.
The undetermined coefficients $b_{\rm S,T}$ and $d_{\rm S}$
still appear in the PS (and their ratio $R$), but not in the spectral
indices and runnings, which are therefore uniquely specified by our
(heuristic) procedure.
In general, one may expect that the complete adiabatic corrections
to second order contain HFF calculated both in the super-horizon
limit and at the turning points (zeros) of the frequencies.
However, this is not relevant in the present context, since the possible
numerical coefficients that would multiply such terms can be eventually
re-absorbed in the definitions of ${b}_{\rm S,T}$ and ${d}_{\rm S}$.
This could only be an issue (in particular for the spectral indices)
if we considered third order terms (since it would then become important where
the second order terms are evaluated).
\subsection{MCE and second~slow-roll~order}
\label{lead_MCE_and_2SR}
We now consider the results~(\ref{spectra_S})-(\ref{corr_TP}) given by
the MCE to leading order (denoted by the subscript MCE) and evaluate them
to second~order in the HFF (labelled by the superscript $(2)$) for a general
inflationary scale factor.
A crucial point in our method is the computation of the function $\xi$
defined in Eq.~(\ref{new_EQ_int}) which can be found in detail in Section~III
of Ref.~\cite{WKB_SRprD}.
For the sake of brevity, we shall not reproduce that analysis here but
it is important to stress that, in contrast with the GFM and other slow-roll
approximations, it does not require {\em a priori\/} any expansion in the
HFF since Eq.~(34) of Ref.~\cite{WKB_SRprD} is exact and higher~order terms
are discarded {\em a fortiori\/}.
From that expression, upon neglecting terms of order higher than two in the HFF,
we obtain the power spectra
\begin{subequations}
\be
&&
\mathcal{P}_{\zeta,\scriptscriptstyle\scriptscriptstyle{\rm MCE}}^{(2)}
=
\frac{H^2}{\pi\epsilon_1m_{\rm Pl}^2}\!
\left\{1\!-\!2\left(C\!+\!1\right)\epsilon_1\!-\!C\,\epsilon_2
\!+\!\left(2C^2\!+\!2C\!+\!\frac{\pi^2}{2}\!-\!5\right)\epsilon_1^2
\!+\!\left(\frac12C^2\!+\!\frac{\pi^2}{8}\!-\!1\right)\epsilon_2^2
\right.
\nonumber
\\
&&
\,\,\,\,\,\,
+\!\!
\left.
\left(2\,C^2\!-\!2\,C\,D_{\scriptscriptstyle{\rm MCE}}
\!+\!D_{\scriptscriptstyle{\rm MCE}}^2\!-\!C\!-\!2\,C\,\ln(2)
\!+\!2\,D_{\scriptscriptstyle{\rm MCE}}\,\ln(2)
\!+\!\frac{7\pi^2}{12}\!-\!\frac{64}{9}\right)\epsilon_1\,\epsilon_2
\right.
\nonumber
\\
&&
\,\,\,\,\,\,
+\!\!
\left.
\left(-C\,D_{\scriptscriptstyle{\rm MCE}}
\!+\!\frac12\,D_{\scriptscriptstyle{\rm MCE}}^2\!-\!C\,\ln(2)
\!+\!D_{\scriptscriptstyle{\rm MCE}}\,\ln(2)\!+\!\frac{\pi^2}{24}
\!-\!\frac{1}{18}\right)\epsilon_2\,\epsilon_3
\right.
\nonumber
\\
&&
\,\,\,\,\,\,
+\!\!\left.
\left[-2\,\epsilon_1\!-\!\epsilon_2\!+\!2\left(2\,C\!+\!1\right)
\epsilon_1^2
\!+\!\left(4\,C\!-\!2\,D_{\scriptscriptstyle{\rm MCE}}\!-\!1\right)\epsilon_1\,\epsilon_2
\!+\!C\,\epsilon_2^2\!-\!D_{\scriptscriptstyle{\rm MCE}}\,\epsilon_2\,\epsilon_3\right]
\ln\left(\frac{k}{k_*}\right)
\right.
\nonumber
\\
%&&\left.
&&
\,\,\,\,\,\,
+\!\!
\left.
\frac12\left(4\,\epsilon_1^2\!
+\!2\,\epsilon_1\,\epsilon_2\!
+\!\epsilon_2^2
\!-\!\epsilon_2\,\epsilon_3\right)
\ln^2\left(\frac{k}{k_*}\right)\right\}
\label{PS_SlowRoll_0_2order}
\ee
\be
\mathcal{P}_{h,\scriptscriptstyle{\rm MCE}}^{(2)}
&\!=\!&
\frac{16H^2}{\pi m_{\rm Pl}^2}
\left\{1-2\left(C+1\right)\epsilon_1
+\left(2C^2+2C+\frac{\pi^2}{2}-5\right)\epsilon_1^2
\nonumber
\right.
\\
&&
%\,\,\,\,\,\,
+\!\!
\left.
\left(-2CD_{\scriptscriptstyle{\rm MCE}}+D_{\scriptscriptstyle{\rm MCE}}^2
-2C-2C\ln(2)+2D_{\scriptscriptstyle{\rm MCE}}\ln(2)
+\frac{\pi^2}{12}-\frac{19}{9}\right)\epsilon_1\epsilon_2
\nonumber
\right.
\\
&&
%\,\,\,\,\,\,
+\!\!
\left.
\left[-2\epsilon_1+2\left(2C+1\right)\epsilon_1^2
-2\left(D_{\scriptscriptstyle{\rm MCE}}+1\right)\epsilon_1\epsilon_2\right]\,
\ln\left(\frac{k}{k_*}\right)
\nonumber
\right.
\\
&&
%\,\,\,\,\,\,
+\!\!
\left.\frac12\left(4\epsilon_1^2
-2\epsilon_1\epsilon_2\right)\ln^2\left(\frac{k}{k_*}\right)
\right\}
\ ,
\label{PT_SlowRoll_0_2order}
\ee
\end{subequations}
where $D_{\scriptscriptstyle{\rm MCE}}\equiv\frac{1}{3}-\ln 3\approx -0.7652$
and $C\equiv\ln 2+\gamma_{\rm E}-2\approx -0.7296$, with $\gamma_{\rm E}$
the Euler-Mascheroni constant.
A clarification concerning the $g(x_0)$'s is in order.
Since the turning point does not coincide with the horizon crossing
where the spectra are evaluated~\cite{WKB_SRprD},
we have used the relation
\be
\epsilon_i(x_0)
\simeq
\epsilon_i-\epsilon_i\,\epsilon_{i+1}\,\ln\left(\frac32\right)
\ ,
\label{epsilon_n_TPtoHC}
\ee
in order to express the $g(x_0)$'s as functions of the crossing time
(HFF with no explicit argument are evaluated at the horizon crossing).
The spectral indices are given by
\begin{subequations}
\be
&&
\!\!\!\!\!\!\!\!\!\!\!\!\!\!\!\!\!\!
n_{\rm S,\scriptscriptstyle{\rm MCE}}^{(2)}-1
=
-2\,\epsilon_1-\epsilon_2-2\,\epsilon_1^2
-\left(2\,D_{\scriptscriptstyle{\rm MCE}}+3\right)\,\epsilon_1\,\epsilon_2
-D_{\scriptscriptstyle{\rm MCE}}\,\epsilon_2\,\epsilon_3
\label{n_2'order_S}
\\
&&
\!\!\!\!\!\!\!\!\!\!\!\!\!\!\!\!\!\!
n_{\rm T,\scriptscriptstyle{\rm MCE}}^{(2)}
=
-2\,\epsilon_1-2\,\epsilon_1^2
-2\,\left(D_{\scriptscriptstyle{\rm MCE}}+1\right)\,\epsilon_1\,\epsilon_2
\ ,
\label{n_2'order_T}
\ee
and their runnings by
\be
&&
\alpha_{\rm S,\scriptscriptstyle{\rm MCE}}^{(2)}=
-2\,\epsilon_1\,\epsilon_2
-\epsilon_2\,\epsilon_3
\label{alpha_2'order_S}
\\
&&
\alpha_{\rm T,\scriptscriptstyle{\rm MCE}}^{(2)}=
-2\,\epsilon_1\,\epsilon_2
\label{alpha_2'order_T}
\ .
\ee
\end{subequations}
The tensor-to-scalar ratio becomes
\be
\frac{R^{(2)}_{\scriptscriptstyle{\rm MCE}}}{16\,\epsilon_1}
&=&
1+C\,\epsilon_2
%\nonumber\\&&
+\left(C-\frac{\pi^2}{2}+5\right)
\epsilon_1\,\epsilon_2
+\left(\frac{1}{2}\,C^2-\frac{\pi^2}{8}+1\right)
\epsilon_2^2
\label{R_lead_MCE}
\\
&&
+\left(C\,D_{\scriptscriptstyle{\rm MCE}}
-\frac{1}{2}\,D_{\scriptscriptstyle{\rm MCE}}^2
+C\,\ln(2)
%\right.\nonumber\\&&\phantom{+\ \ }\left.
-D_{\scriptscriptstyle{\rm MCE}}\,\ln(2)
-\frac{\pi^2}{24}+\frac{1}{18}\right)
\epsilon_2\,\epsilon_3
\ .
\nonumber
\ee
The polynomial structure in the HFF of the results agrees with
that given by the Green Function Method (GFM)~\cite{SGong,LLMS} and the
WKB~approximation~\cite{martin_schwarz2,WKB1,WKB_let,WKB_SRprD} (the same polynomial
structure is also found for the spectral indices by means of the
uniform approximation~\cite{HHHJ}).
Let us also note other aspects of our results:
first of all, the factors $g(x_0)$ modify the standard
WKB leading order amplitudes~\cite{martin_schwarz2,WKB1} so as to
reproduce the standard first order slow-roll results;
secondly, we found that $C$ and $D_{\scriptscriptstyle{\rm MCE}}$
are ``mixed'' in the numerical factors in front of second order
terms (we recall that $D_{\scriptscriptstyle{\rm MCE}}$ differs from
$C$ by about $5\,\%$); further, $D_{\scriptscriptstyle{\rm 
MCE}}=D_{\scriptscriptstyle{\rm WKB}}$
of Refs.~\cite{martin_schwarz2,WKB1,WKB_let,WKB_SRprD}.
The runnings $\alpha_{\rm S}$ and $\alpha_{\rm T}$ are predicted to
be ${\mathcal O}(\epsilon^2)$~\cite{KT}, and in agreement with those
obtained by the GFM~\cite{SGong,LLMS}.
%
%
%\par
\subsection{Second~slow-roll~order MCE and WKB versus~GFM}
\label{mixed}
We shall now compare the slow-roll results obtained from the MCE
and other methods of approximation previously
employed~\cite{WKB_SRprD} with those obtained using the GFM in the
slow-roll expansion.%~\cite{WKB_lungo}.
We use the GFM just as a reference for the purpose of comparing the
other methods among each other and of illustrating deviations from pure
slow-roll results.
For a given inflationary ``observable'' $Y$ evaluated with the method
X, we denote the percentage difference with respect to its value given
by the GFM as  
\be
\Delta Y_{_{\rm X}}
\equiv
100\,\left|\frac{Y_{_{\rm X}}-Y_{_{\rm GFM}}}{Y_{_{\rm GFM}}}\right|\%
\ ,
\label{Y_X}
\ee
where ${\rm X}={\rm WKB}$ 
stands for the first WKB and second slow-roll
orders~\cite{WKB_let},
${\rm X}={\rm WKB*}$ for the second WKB and second slow-roll
orders~\cite{WKB_SRprD} and, of course ${\rm X}={\rm MCE}$ for the
result obtained in this paper. Note that for the case ${\rm X}={\rm 
WKB*}$ we shall set the three
undetermined parameters $b_{\rm S}=b_{\rm T}=d_{\rm S}=2$ in order
to minimize the difference with respect to the results of the GFM
(see Appendix~\ref{und_par} and Ref.~\cite{WKB_SRprD}).
\begin{figure}[!ht]
\includegraphics[width=0.41\textwidth]{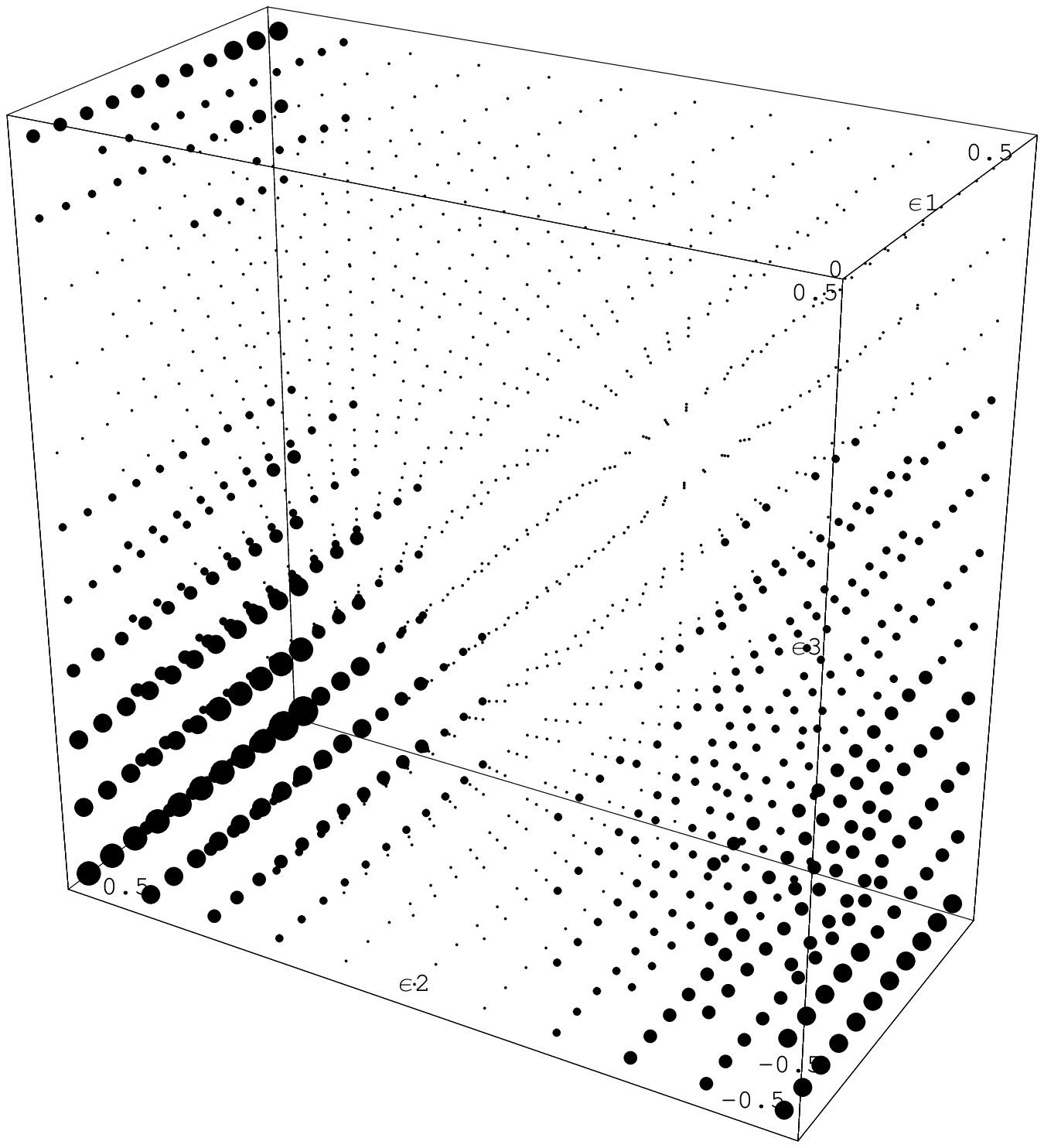}
%\hspace{1cm}
\includegraphics[width=0.41\textwidth]{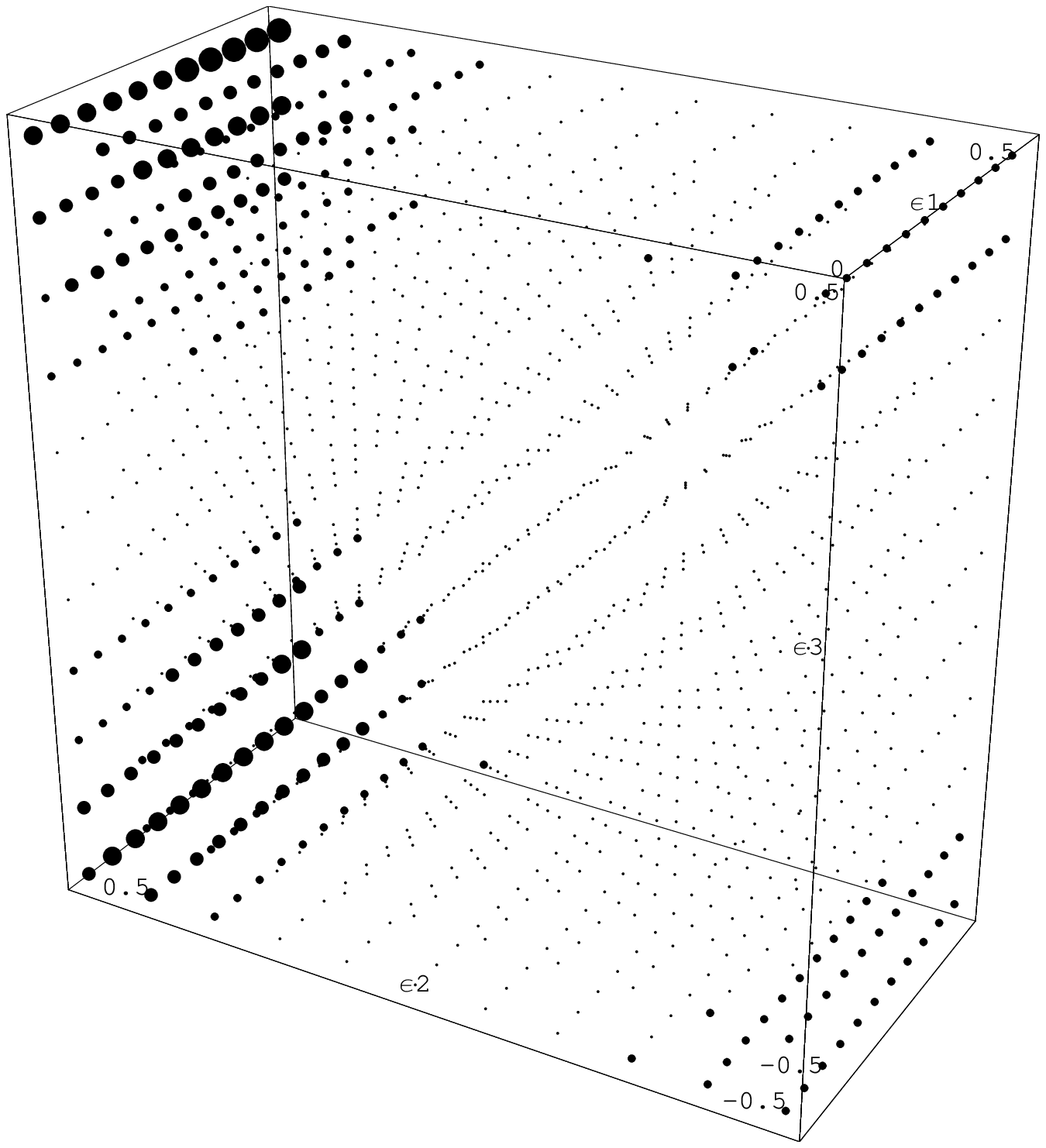}
\\
\null
\hspace{2cm}$\Delta R_{_{\rm WKB}}$
\hspace{6cm}$\Delta R_{_{\rm WKB*}}$
\\
\includegraphics[width=0.41\textwidth]{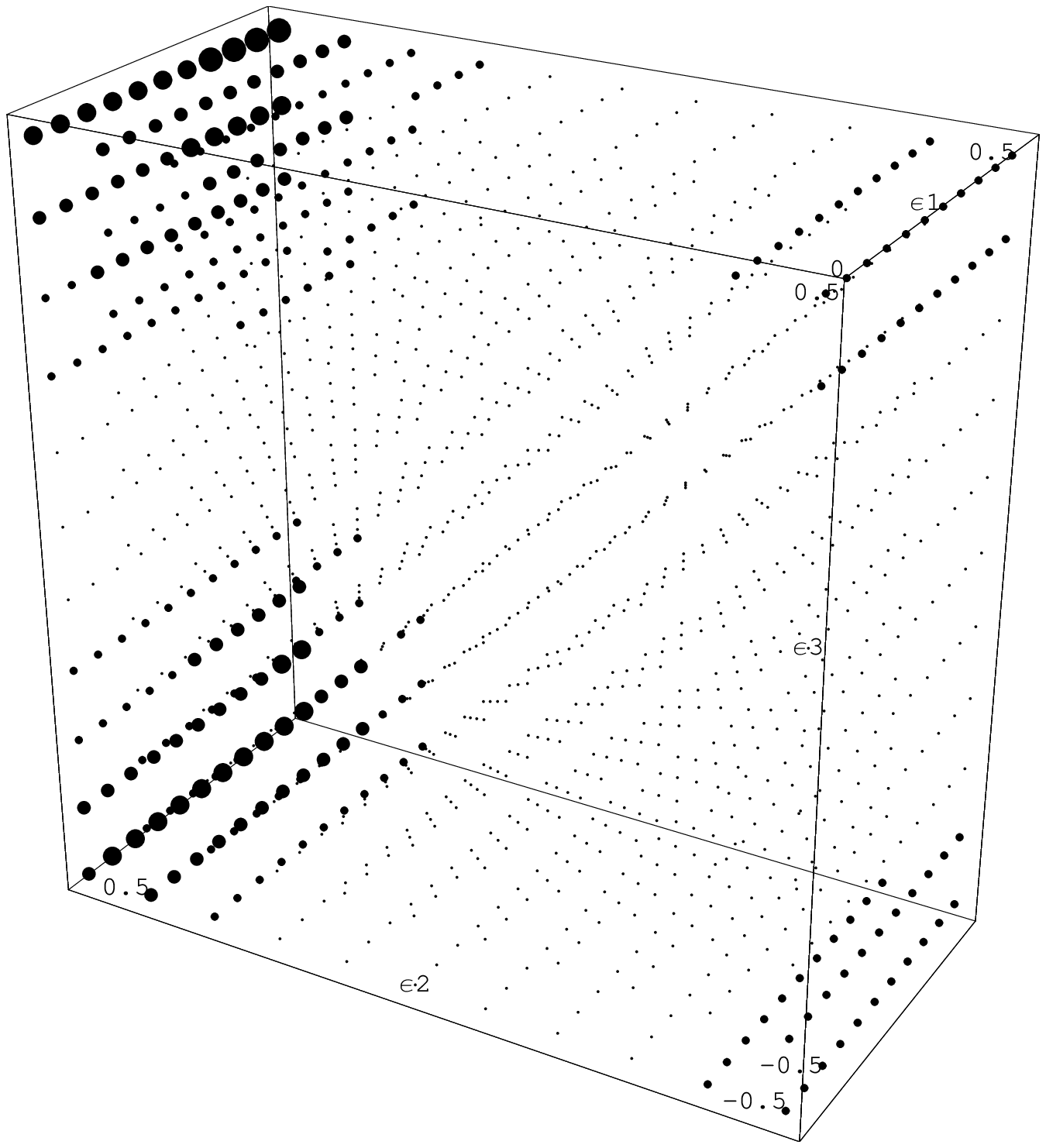}
\includegraphics[width=0.36\textwidth]{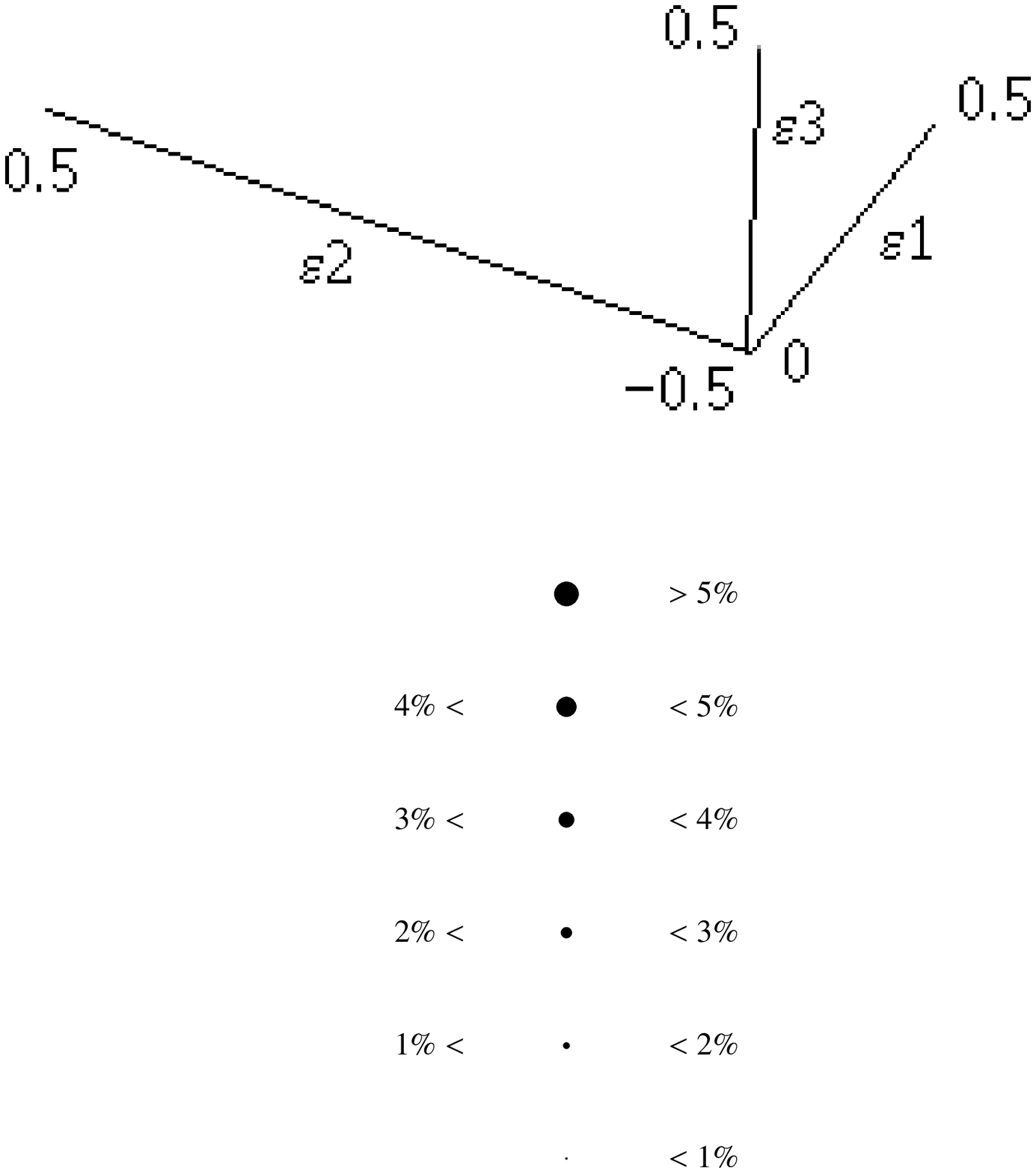}
\\
\null
\hspace{2cm}$\Delta R_{_{\rm MCE}}$
\caption{Percentage differences~(\ref{Y_X}) between scalar-to-tensor
ratios given by the GFM and those obtained from the WKB, ${\rm WKB*}$
and MCE.}
\label{compare_grap}
\end{figure}
%\par
In Fig.~\ref{compare_grap} we show, with dots of variable size,
the percentage differences~(\ref{Y_X}) for the scalar-to-tensor
ratios $R$ at the pivot scale $k=k_*$
for $0<\epsilon_1<0.5$, $|\epsilon_2|$ and $|\epsilon_3|<0.5$.
From the plots it appears that the level of accuracy of the MCE is
comparable to that of the ${\rm WKB*}$ and both are (almost everywhere)
more accurate than the WKB.
However, the MCE achieves such a precision at leading order and is thus
significantly more effective than the ${\rm WKB*}$. 
In Fig.~\ref{compare_ns} we show the difference in $n_{\rm S}$ and
the relative difference in $P_{\zeta} (k_*)$ between the MCE and GFM.
\begin{figure}[!ht]
\includegraphics[width=0.41\textwidth]{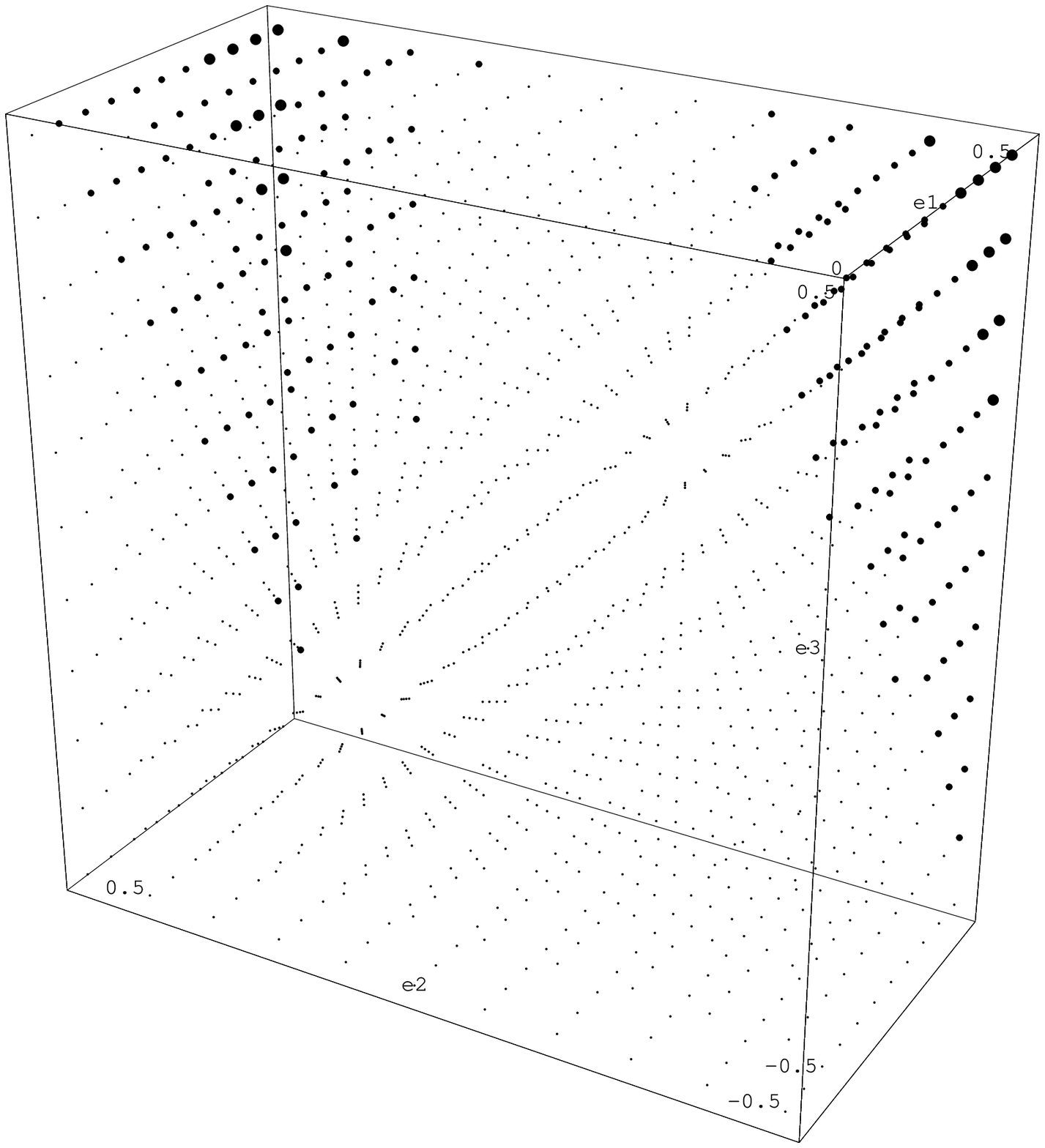}
\hspace{0.1cm}
\includegraphics[width=0.41\textwidth]{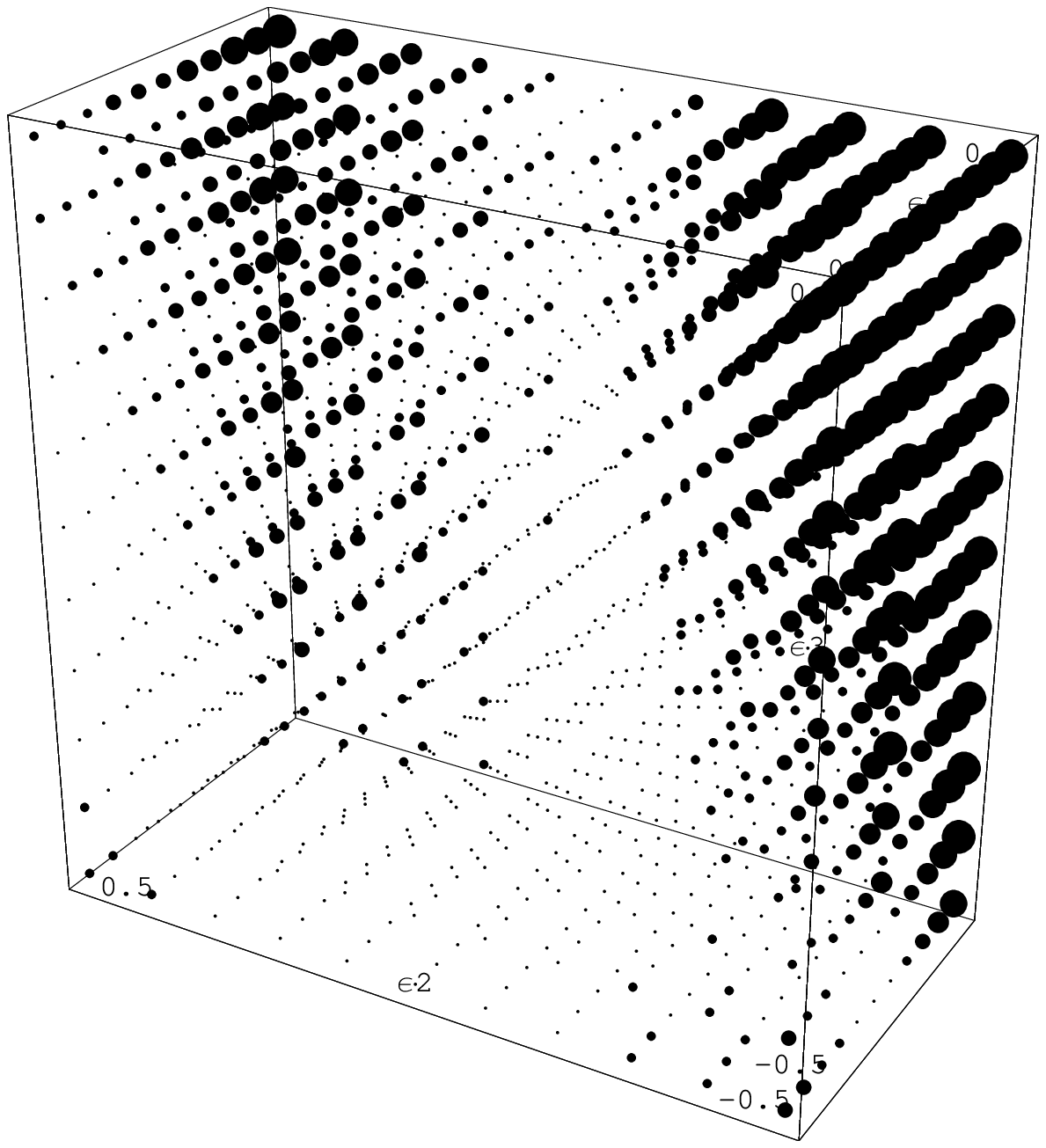}
\\
\null
\hspace{2cm} $|n_{\rm S\,\scriptscriptstyle{MCE}}-n_{\rm S\,\scriptscriptstyle{GFM}}|
\times 100$ 
\hspace{2cm} $\Delta P_{\zeta\,\scriptscriptstyle{\rm MCE}}(k_*)$
\caption{Left box: absolute difference between scalar spectral indeces $n_{\rm S}$
evaluated with the MCE and GFM (rescaled by a factor of $100$ for convenience).
Note that relevant differences of order $0.05$ (shown as $5$) occur at the
boundaries of the intervals considered for the $\epsilon_i$'s.
Right box: percentage difference~(\ref{Y_X}) for the amplitudes of scalar
perturbations $P_\zeta$ at the pivot scale $k=k_*$ between the MCE and GFM.
See Fig.~\ref{compare_grap} for the meaning of dot size.}
\label{compare_ns}
\end{figure}
\par
\begin{figure}[!ht]
\includegraphics[width=0.5\textwidth]{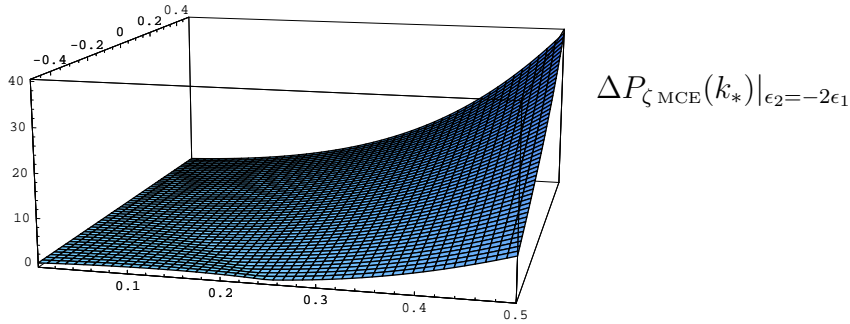}
\raisebox{3cm}{$\Delta P_{\zeta\,\scriptscriptstyle{\rm MCE}}(k_*)
|_{\epsilon_2=-2 \epsilon_1}$}
\caption{Percentage difference~(\ref{Y_X}) between the MCE and the GFM
for $P_\zeta(k_*)$ restricted on the hypersurface $\epsilon_2=-2\,\epsilon_1$,
which corresponds to a scale-invariant spectrum to first order in the slow-roll
expansion.
The graph is given for $0<\epsilon_1<0.5$ and $-0.5<\epsilon_3<0.5$.}
\label{figure_PS}
\end{figure}

In Fig.~\ref{figure_PS} we finally plot the relative difference in
$P_{\zeta} (k_*)$ for $\epsilon_2=-2\,\epsilon_1$ (the scale 
invariant case to first order in slow-roll parameters).
\subsection{GFM and slow-roll approximation}
Let us further compare our results with those obtained by the GFM.
As we stated before, the most general result of the present work
is given by the expressions for the power spectra~(\ref{spectra_S})
and~(\ref{spectra_T}) with the corrections~(\ref{corr_TP}).
In fact, the spectral indices and their runnings are the same as
found with the standard WKB leading order~\cite{martin_schwarz2,WKB_SRprD}.
On the other hand, the main results of the GFM are the expressions
for power spectra, spectral indices and runnings ``$\ldots$ in the
slow-roll expansion.'' that is, for small HFF (see,~e.g.,~Eqs.~(41)
and~(43) in Ref.~\cite{SGong}).
Here, and in some previous work~\cite{martin_schwarz2,WKB1}, we instead obtain
results which hold for a general inflationary context, independently
of the ``slow-roll conditions''~\footnote{Our approximation method
does not require small HFF.}
or ``slow-roll approximation''~\footnote{We expand instead according
to the scheme given in Ref.~\cite{WKB_SRprD}.}.
The slow-roll case is just a possible application of our general
expressions which can be evaluated for any model by simply specifying
the scale factor.
Our general, and non-local, expressions in fact take into account all
the ``history'' of the HFF during inflation.
For example, we note that the MCE at leading order reproduces the exact
result for power-law inflation (see Sec.~\ref{Power-law inflation}),
whereas the GFM reproduces those to the next-to-leading
order~\cite{SGong}.
\par
\begin{figure}[!ht]
\raisebox{3cm}{$\epsilon_i$}
\includegraphics[width=0.5\textwidth]{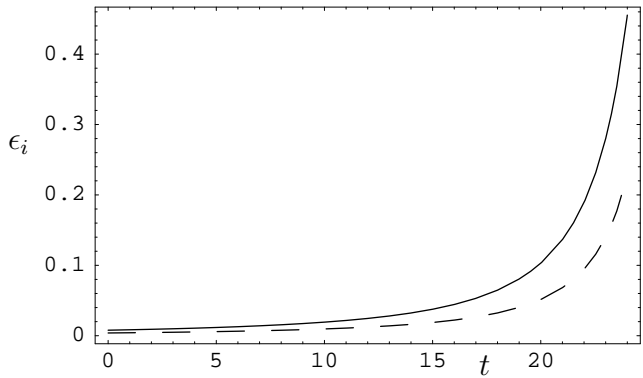}
\null
\hspace{-2.3cm}$t$
\caption{HFF in the slow-roll regime:
the dashed line is $\epsilon_1$ and the solid line is all the
$\epsilon_n$'s with $n\ge2$ (time in units of $1/m$).}
\label{eps1_epsn}
\end{figure}

Chaotic inflation~\cite{linde} is another example for which a Taylor
approximation of the HFF (such as the one required by the GFM~\cite{SGong}
or our Green's function perturbative expansion in Appendix~\ref{VENTURI_pert})
may lead to inaccurate results.
We consider a quadratic potential
\be
V(\phi)=\frac{1}{2}\,m^2\,\phi^2
\ ,
\label{chaos}
\ee
where $m$ is the mass of the inflaton $\phi$.
In a spatially flat Robertson-Walker background, the potential energy
dominates during the slow-rollover and the Friedman equation becomes
\be
H^2=\frac{4\,\pi}{3\,m_{\rm Pl}^2}\,\left[\dot\phi^2+m^2\,\phi^2\right]
\simeq\frac{4\,\pi\,m^2\,\phi^2}{3\,m_{\rm Pl}^2}
\ .
\label{hubble}
\ee
\begin{figure}[!ht]
\raisebox{3cm}{$H$}
\includegraphics[width=0.5\textwidth]{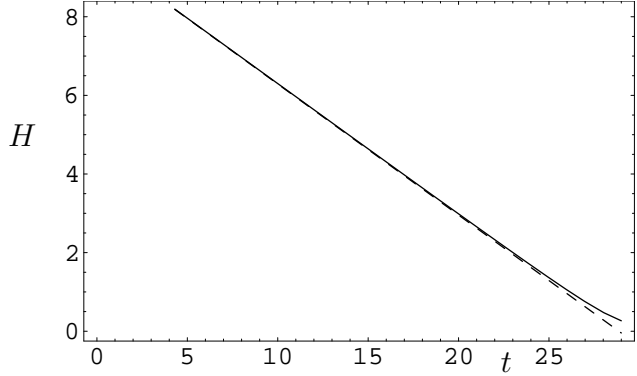}
\null
\hspace{-2.1cm}$t$
\caption{Numerical evolution of $H$ (solid line) and its
analytic approximation (dashed line).
The initial condition corresponds to $H_i\approx 8.2\,m$ (time
in units of $1/m$) when the inflaton starts in the slow-roll regime
with a value $4\,m_{\rm Pl}$.}
\label{fig:hubble}
\end{figure}
The Hubble parameter evolves as
\be
\dot H=-\frac{4\,\pi}{m_{\rm Pl}^2}\,\dot\phi^2
\ .
\label{hubbleder}
\ee
On using the equation of motion for the scalar field
\be
\ddot\phi+3\,H\,\dot\phi+m^2\,\phi=0
\ ,
\label{scalar_hom}
\ee
and neglecting the second derivative with respect to the cosmic time,
we have $3\,H\,\dot\phi\simeq-\,m^2\,\phi$.
Eq.~(\ref{hubbleder}) then yields
\be
\dot H\simeq-\frac{m^2}{3}\equiv\dot H_0
\ ,
\label{hdot}
\ee
which leads to a linearly decreasing Hubble parameter and, correspondingly,
to an evolution for the scale factor which is not exponentially linear in
time, i.e.
\begin{subequations}
\be
&&
H(t)\simeq H_0+\dot H_0\,t
\ ,
\label{evoH}
\\
&&
a(t)\simeq a(0)\,\exp\left(H_0\,t+\dot H_0\,\frac{t^2}{2}\right)
\equiv a(0)\,\exp\left[N(t)\right]
\ .
\label{evoA}
\ee
\end{subequations}
In Fig.~\ref{fig:hubble} the analytic approximation~(\ref{evoH}) is
shown to be very good on comparing with the exact (numerical) evolution.
We are interested in the HFF in the slow-roll regime.
From the definitions~(\ref{hor_flo_fun}) we then have
\begin{subequations}
\be
&&
\epsilon_1(t)=
-\frac{\dot H_0}{\left(H_0+\dot H_0\,t\right)^2}
\label{eps1_epsn_funct1}
\\
&&
\epsilon_n(t)=
-\frac{2\,\dot H_0}{\left(H_0+\dot H_0\,t\right)^2}
=2\,\epsilon_1(t)
\ ,
\quad
n\ge2
\ ,
\label{eps1_epsn_funct2}
\ee
\end{subequations}
which we plot in Fig.~\ref{eps1_epsn}.
\par
\begin{figure}[!ht]
\raisebox{3cm}{$\delta_1$}
\includegraphics[width=0.5\textwidth]{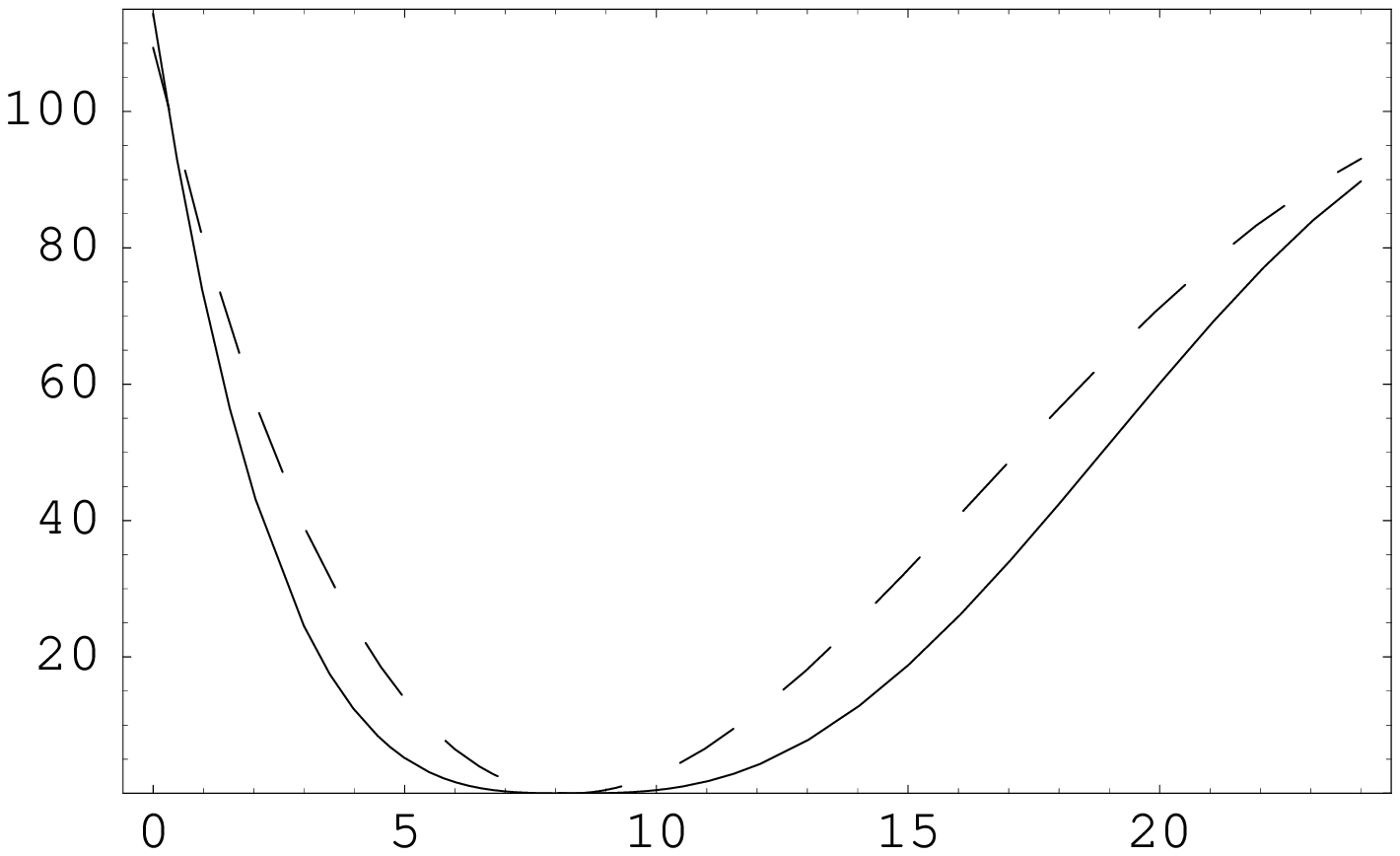}
\null
\hspace{-2.3cm}$t$
\caption{Percentage error for $\epsilon_1$ to first and second order in
$\Delta N$ (dashed and solid line respectively; time in units of $1/m$).}
\label{err1_err2}
\end{figure}

Let us take the end of inflation at $t_{\rm end}\simeq 24/m$
(so that the analytic approximation~(\ref{evoH}), (\ref{evoA}) is very
good till the end), corresponding to $N_{\rm end}=125$~e-folds.
We then consider the modes that leave the horizon 60~e-folds before
the end of inflation and take that time ($t=t_*\simeq 8/m$ or $N_*=125-60=65$)
as the starting point for a Taylor expansion to approximate the
HFF~\footnote{One can of course conceive diverse expansions, e.g.~using
other variables instead of $N$, however the conclusion would not change
as long as equivalent orders are compared.},
\be
\epsilon_i\left(\Delta N\right)\simeq
\epsilon_i+\epsilon_i\,\epsilon_{i+1}\,\Delta N
+\frac12\,\epsilon_i\,
\left(\epsilon_{i+1}^2+\epsilon_{i+1}\,\epsilon_{i+2}\right)\,
\Delta N^2
\label{epsn_expan}
\ ,
\ee
where the $\epsilon_i$'s in the r.h.s.~are all evaluated at the time
$t_*$ and $\Delta N=N-N_*$.
The percentage error with respect to the analytic
expressions~(\ref{eps1_epsn_funct1}) and~(\ref{eps1_epsn_funct2}) ,
\be
\delta_i\equiv100\,\left|\frac{\epsilon_i(t)
-\epsilon_i\left(\Delta N\right)}{\epsilon_i(t)}
\right|\%
\label{err_epsn_expan}
\ ,
\ee
with $\Delta N=N(t)-N_*$ is then plotted in Fig.~\ref{err1_err2}
for $i=1$.
It obviously becomes large rather quickly away from $t=t_*$ and
we can immediately conclude that, had we used a Taylor expansion to approximate
the HFF over the whole range of chaotic inflation ($-65<\Delta N<60$), we would
have obtained large errors from the regions both near the beginning and the end of
inflation.
\par
In general, we expect that the Taylor expansion of the HFF will lead to 
a poor determination of the numerical coefficients (depending on $C$ in
Eqs.~(\ref{PS_SlowRoll_0_2order}) and (\ref{PT_SlowRoll_0_2order})) in 
the second slow-roll order terms for those models in which the HFF 
vary significantly in time. 
In fact, we can calculate the integral of Eq.~(21) in Ref.~\cite{SGong}
with $\epsilon_1$ instead of the complete $g\left(\ln x\right)$ and $y_0(u)$
instead of $y(u)$.
The percentage difference between this integral calculated using the Taylor
expansion~(\ref{epsn_expan}) with respect to the same integral calculated with
$\epsilon_1$ in Eqs.~(\ref{eps1_epsn_funct1}) and~(\ref{eps1_epsn_funct2}) is
$92\,\%$ and $88\,\%$ to first and second orders in $\Delta N$, respectively.
The MCE method does not require such an expansion and does therefore not suffer
from this restriction.
\par
\begin{figure}[!ht]
\raisebox{3cm}{${\rm Log}(P_Q)$}
\includegraphics[width=0.5\textwidth]{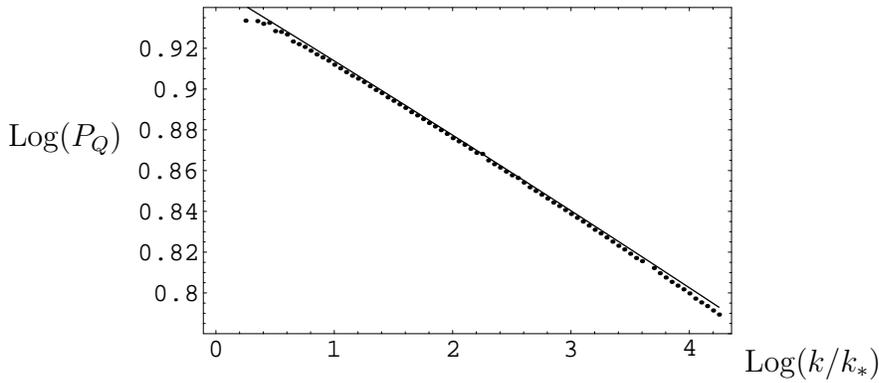}
\null
${\rm Log}(k/k_*)$
\caption{Spectrum of the Mukhanov variable $Q$ ($P_Q=k^3\,|Q_k|^2/(2\,\pi^2)$)
evaluated at the end of inflation for the chaotic model of Eq.~(\ref{chaos}).
The dots represent the numerical values and the solid line the analytic
fit based on Eqs.~(\ref{n_2'order_S}) and (\ref{alpha_2'order_S}) 
obtained by the second slow-roll order MCE approximation.
The first order and GFM analytic results are not shown 
since they are almost indistinguishable from the MCE result plotted.
$k_*$ crosses the Hubble radius at $\phi_*\simeq 3\,m_{\rm Pl}$
(the lines are normalized at $10^{2.25}\,k_*$).}
\label{fig:spe}
\end{figure}
We have also compared our analytic MCE results in Eqs.~(\ref{n_2'order_S})
and (\ref{alpha_2'order_S}) with the numerical evaluation of the spectrum
for scalar perturbations. 
In particular, we have evolved in cosmic time the modulus of the Mukhanov
variable $Q$ (recalling that $P_\zeta=k^3\,|Q_k|^2/(2\,\pi^2)$), 
which satisfies the associated non-linear Pinney equation~\cite{BFV}.
The initial conditions for the numerical evolution are fixed for 
wave-lengths well within the Hubble radius and correspond to the adiabatic 
vacuum, i.e.~$|Q_k(t_i)|=1/(a(t_i)\,\sqrt{k})$ and
$|\dot Q_k|(t_i) = - H(t_i)\,|Q_k(t_i)|$~\cite{FMVV_1}. 
The agreement of the analytic MCE approximation with the numerical results
is good, as shown in Fig.~\ref{fig:spe}. 
\subsection{Beyond slow-roll}
\begin{figure}[!hb]
\raisebox{3cm}{$\epsilon_i$}
\includegraphics[width=0.5\textwidth]{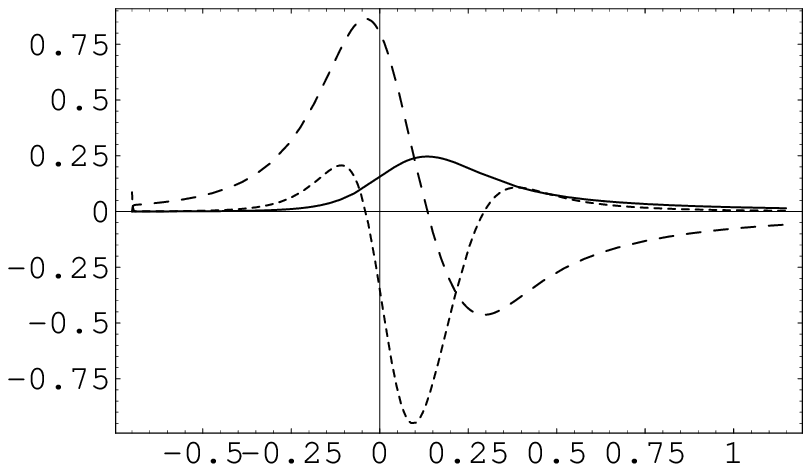}
\null
$\phi/m_{\rm Pl}$
\caption{Evolution of $\epsilon_i$ with the value of the inflaton
$\phi$ (in units of $m_{\rm Pl}$) in the arctan model:
$\epsilon_1$ (solid line), $\epsilon_2$ (long-dashed line) and
$\dot\epsilon_2/H=\epsilon_2\,\epsilon_3$ (short-dashed line).} 
\label{arctan_figure1}
\end{figure}

The slow-roll approximation is quite accurate for a wide class of 
potentials.
However, violations of the slow-roll approximation may occur 
during inflation, leading to interesting observational effects in the
power spectra of cosmological perturbations.
An archetypical model to study such violations is given by the potential
\be
V(\phi)=V_0\,\left[1
-\frac{2}{\pi}\,\arctan\left(N\,\frac{\phi}{m_{\rm Pl}}\right)\right]
\ ,
\label{arctan}
\ee
introduced with $N=5$ in Ref.~\cite{wang_mukh}.
\begin{figure}[!ht]
\raisebox{3cm}{${\rm Log}(P_Q)$}
\includegraphics[width=0.5\textwidth]{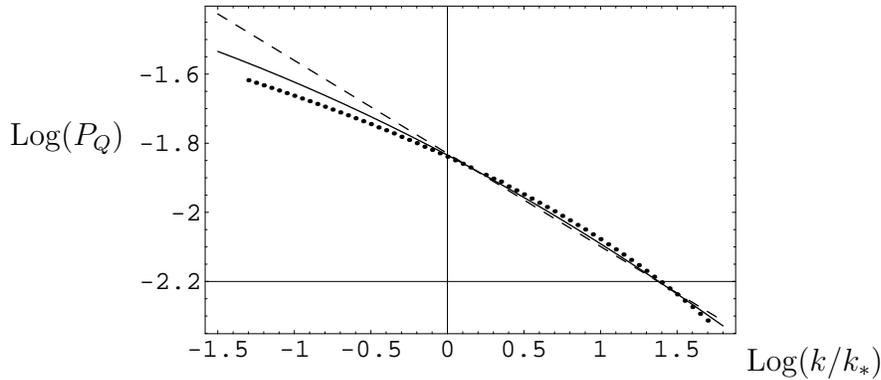}
\null
${\rm Log} (k/k_*)$
\caption{Spectrum of the Mukhanov variable $Q$ ($P_Q=k^3\,|Q_k|^2/(2\,\pi^2)$)
for the arctan model of Eq.~(\ref{arctan}).
The dots represent the numerical values, the solid line the analytic
results from Eqs.~(\ref{n_2'order_S}) and (\ref{alpha_2'order_S})
obtained by the second slow-roll order MCE approximation and
the dashed line those given by the first order slow-roll approximation.
$k_*$ crosses the Hubble radius at $\phi_*\simeq -0.3\,m_{\rm Pl}$
(the lines are normalized at $10^{3/20} k_*$) and the spectrum is
evaluated at $\simeq 55$~e-folds afterwards.} 
\label{arctan_figure2}
\end{figure}
In Fig.~\ref{arctan_figure2} we show that the MCE result expanded to
second slow-roll order provides a very good fit for the power spectrum
of scalar perturbations even in situations where the slow-roll parameters
are not very small (see Fig.~\ref{arctan_figure1}).
This example also shows that second slow-roll order results are much
better than first order ones (analogous results were also obtained with
the GFM in Ref.~\cite{LLMS}).
\nonumchapter{Conclusions}
\label{concl}
\markboth{Conclusions}{}
%
%
%WKB_PL
In this thesis, we have developed some approximations of the WKB type for the
purpose of estimating the spectra of cosmological perturbations during inflation.
In particular, in Section~\ref{imprWKBb} we found general formulae for the amplitudes,
spectral indices and $\alpha$-runnings of the fluctuations to next-to-leading order
both in the adiabatic expansion of Ref.~\cite{langer} and a new perturbative expansion
which makes use of the Green's function technique.
The most sophisticated Method of Comparison Equations was introduced in
Section~\ref{sMCEa} and then applied to cosmological perturbations
in Section~\ref{sMCEb}.
By construction (i.e.~by choosing a suitable comparison function), this
approach leads to the exact solutions for inflationary models with constant
horizon flow functions $\epsilon_i$'s (e.g.~power-law inflation).
The main result is that, on using this approach to leading order, we were
able to obtain the general expressions~(\ref{spectra_S})-(\ref{corr_TP})
for the inflationary power spectra which are more accurate than
those any other method in the literature can produce at the
corresponding order.
In fact, the MCE leads to the correct asymptotic behaviours (whereas
the standard slow-roll approximation fails~\cite{SL}) and solves the problems
in computing the amplitudes which were encountered with both the standard and
improved WKB method analised in Sections~\ref{stWKBb} and~\ref{imprWKBb}.
\par
In Section~\ref{Power-law inflation}, we have applied our methods to power-law inflation
in order to test it against exact results.
It is known that the spectral indices and their runnings are obtained exactly at
leading order in the WKB approximation, hence we focussed on the spectra
to next-to-leading order.
For the improved WKB method, we have found that the perturbative corrections
remain too small to yield any significant improvement, whereas the adiabatic
expansion to next-to-leading order reproduces the exact amplitudes with great
accuracy (see Ref.~\cite{HHHJM} for a similar result with the uniform approximation).
This result does not however mean that the perturbative expansion will not lead to
significant corrections in different inflationary scenarios.
One way of understanding the difference between the two
expansions in the power-law case may be the following.
It is known that the Born~approximation in quantum
mechanics (analogous to our perturbative method with
the Green's function) is good for high multipoles
(angular momenta) $\ell$ of the expansion in spherical
harmonics.
In our approach, the parameter $\beta$ plays the role
of the multipole index $\ell$ for the energy levels of
the hydrogen atom, as can be seen on comparing the
frequency~(\ref{omega2_PL}) with the expression
given in Ref.~\cite{langer}
(see also Ref.~\cite{martin_schwarz2}).
The approximation with the Green's function is therefore
expected to yield more significant corrections for large
values of $|\beta|$, which coincide with the regime of
fast-roll.
Indeed, the leading order becomes more and more accurate
for increasing $|\beta|$.
Since the interesting regime for inflation involves small
values of $\beta\sim -2$, it is instead the adiabatic
approximation which seems better since the horizon flow
functions evolve slowly and/or the states are
quasi-classical~\footnote{Cosmological perturbations
are amplified by inflation evolving from vacuum to highly
squeezed states, which resemble classical states in the
amplitude of fluctuations~\cite{polarski}.}.
This may be the reason whereby the adiabatic approximation
works better at the next-to-leading order.
However, in more general cases, either or both methods
may contribute significant corrections.
In this respect, let us remark that the adiabatic
expansion can be straightforwardly applied only for
a linear turning point~\footnote{The same limitation
seems to affect the method used in Ref.~\cite{HHHJM}.},
whereas the perturbative Green's function method is
not so restricted.
\par
In Section~\ref{Slow-Roll Inflation}, we have shown that the improved
WKB treatment of cosmological perturbations agrees with the standard
slow-roll approximation~\cite{SL} to within $0.1\,\%$,
finally resolving the issue of a $10\,\%$ error in the prediction of the amplitudes
to lowest order which was raised in Ref.~\cite{martin_schwarz2}.
The next issues are the inflationary predictions to second order
in the slow-roll parameters.
After the results on the running of the power spectra~\cite{KT},
second order results have been obtained using the Green's function
method with the massless solution in a de~Sitter space-time~\cite{SGong,LLMS}.
Here we have employed the leading WKB approximation to obtain the scalar
and tensor power spectra to second order in the slow-roll parameters.
The key technical point is to use Eq.~(\ref{xi_gen_exact_compl}),
which allows one to express the power spectra on large scales as a
function of the quantities evaluated at the Hubble crossing, but with no explicit
dependence on $\Delta N$.
As one of our main results, we find that the polynomial structure of
the power spectra in the $\epsilon_i$'s obtained from the WKB method
agrees with the one arising from the Green's function approach of
Refs.~\cite{SGong,LLMS}.
In Section~\ref{2+2} we employ the next-to-leading WKB approximation to second order in the
$\epsilon_i$'s and, by requiring that the power spectra do not explicitly
depend on $\Delta N$ (a property which instead was derived both in
Sections~\ref{1+2} and~\ref{2+1}), and using the expressions found
in Section~\ref{2+1}, we obtain unique expressions for the spectral
indices and runnings.
Our findings show that different ways of approximating cosmological
perturbations show up at the leading order in the amplitude, but in the
next-to-leading order in the derivatives of the power-spectra with respect
to wave number $k$.
The accuracy in the theoretical predictions on spectral indices to second
order in the slow-roll parameters, evaluated at $k=k_*$, is now striking:
the Green's function method yields a coefficient $C\simeq -0.7296$ which is
replaced by $D_{\rm WKB*}\simeq -0.7302$ in the improved WKB method, leading to a
precision of $1$ part on $1000$ in the predictions of the coefficients
of ${\cal O}(\epsilon_i^2)$ terms in the spectral indexes.
\par
In Section~\ref{2+2}, starting from the general results~(\ref{spectra_S})-(\ref{corr_TP}),
we have also computed the full analytic expressions for the inflationary power
spectra to second slow-roll order in Eqs.~(\ref{PS_SlowRoll_0_2order})
and~(\ref{PT_SlowRoll_0_2order}) and found that the dependence on the
horizon flow functions $\epsilon_i$'s is in agreement with that obtained
by different schemes of approximation, such as the GFM~\cite{SGong} and
the improved WKB approximation (see
Sections~\ref{Power-law inflation} and~\ref{Slow-Roll Inflation}).
Moreover, the results obtained with the MCE do not contain undetermined
coefficients, in contrast with the second slow-roll order results obtained
with the WKB$*$.
Let us conclude by remarking that, just like the WKB approach, the MCE does
not require any particular constraints on the functions $\epsilon_i$'s
and therefore has a wider range of applicability than any method which
assumes them to be small.
As an example, we have discussed in some detail the accuracy of the MCE for
the massive chaotic and arctan inflationary models.
We have shown that the MCE leads to accurate predictions even for a model
which violates the slow-roll approximation during inflation. 
\par
In Appendix~\ref{MCE_BH} we have also applied the Method of Comparison
Equations to a different problem, namely the scalar field dynamics on the
Schwarzschild background.
We have considered modes with energy lower than the peak of the
effective potential in the corresponding wave equation, so that there
are two zeros (turning points).
The comparison function has then been chosen of the Morse form
with an argument which we have estimated analytically as well.
This procedure was shown to be able to produced a fully analytical
approximation of the wave-functions with good accuracy over the
whole domain outside the Schwarzschild horizon.

\begin{appendix}
\chapter{First Order Fluctuations}
\label{appendice_metric_fluc}
We shall here summarise some technical aspects concerning the scalar, vector and
tensor fluctuations of the geometry.
As in the main text, we will assume a conformally flat (i.e., $\kappa =0$)
background metric of FRW type.
\par
First off, the fluctuations of the Christoffel connections can be written
to first order in the amplitude of the metric fluctuations as
\be
\delta\,\Gamma_{\alpha\beta}^{\mu}&=&
\frac12\,^{(0)}g^{\mu\nu}\,
\left(\partial_{\beta}\,\delta\,g_{\nu\alpha}
+\partial_{\alpha}\,\delta\,g_{\beta\nu}-\partial_{\nu}\,\delta\,g_{\alpha\beta}\right)
\nonumber\\
&&+\frac12\,\delta\,g^{\mu\nu}\,
\left(\partial_{\beta}\,^{(0)}g_{\nu\alpha}
+\partial_{\alpha}\,^{(0)}g_{\beta\nu}-\partial_{\nu}\,^{(0)}g_{\alpha\beta}\right)
\ ,
\label{Chris_pert}
\ee
where $^{(0)}$ denotes quantities evaluated for the background geometry
and the usual convention of summation over repeated indices is adopted.
Analogously, the first-order fluctuation of the Ricci tensor is given by
\be
\delta\,R_{\mu\nu}&=&
\partial_{\alpha}\,\delta\,\Gamma_{\mu\nu}^{\alpha}
-\partial_{\nu}\,\delta \Gamma_{\mu \beta}^{\beta}
\nonumber\\
&&+\delta\,\Gamma_{\mu\nu}^{\alpha}\,^{(0)}\Gamma_{\alpha\beta}^{\beta}
+^{(0)}\Gamma_{\mu\nu}^{\alpha}\,\delta\,\Gamma_{\alpha\beta}^{\beta}
-\delta\,\Gamma_{\alpha\mu}^{\beta}\,^{(0)}\Gamma_{\beta\nu}^{\alpha}
-^{(0)}\Gamma_{\alpha\mu}^{\beta}\,\delta\,\Gamma_{\beta\nu}^{\alpha}
\ ,
\label{Ricci_pert}
\ee
which leads to
\be
\delta\,R_{\mu}^{\nu}&=&
\delta\,g^{\nu\alpha}\,^{(0)}R_{\alpha\mu}
+^{(0)}g^{\nu\alpha}\,\delta R_{\alpha\mu}
\label{Ricci_pert_cov_contr}\\
\delta\,R&=&
\delta\,g^{\alpha\beta}\,^{(0)}R_{\alpha\beta}
+^{(0)}g^{\alpha\beta}\,\delta R_{\alpha\beta}
\ .
\label{R_pert}
\ee
Finally, the fluctuations of the Einstein tensor 
\be
\delta\,G_{\mu}^{\nu}=\delta\,R_{\mu}^{\nu}-\frac12\,\delta_{\mu}^{\nu}\,\delta\,R
\ ,
\label{Einstein_pert}
\ee
can be easily obtained from Eqs.~(\ref{Ricci_pert_cov_contr}) and (\ref{R_pert})
and the perturbation of the covariant conservation equation is
\be
\partial_{\mu}\,\delta T^{\mu\nu}
+^{(0)}\Gamma_{\mu\alpha}^{\mu}\,\delta\,T^{\alpha\nu}
+\delta\,\Gamma_{\mu\alpha}^{\mu}\,^{(0)}T^{\alpha\nu}
+^{(0)}\Gamma_{\alpha\beta}^{\nu}\,\delta\,T^{\alpha\beta}
+\delta\,\Gamma_{\alpha\beta}^{\nu}\,^{(0)}T^{\alpha\beta}=0
\ .
\label{cons_eq_pert}
\ee
\par
The Christoffel connections of the flat FRW background are explicitly given
by
\be
^{(0)}\Gamma_{00}^{0}={\cal H}
\ ,\quad^{(0)}
\Gamma_{i j}^{0}={\cal H}\,\delta_{i j}
\ ,\quad
^{(0)}\Gamma_{0i}^{j}={\cal H}\,\delta_{i}^{j}
\label{Chris_background}
\ee
while the components of the Ricci tensor and scalar are
\be
&&
^{(0)}R_{00}=-3\,{\cal H}'
\ ,\quad\quad\quad\quad\quad
^{(0)}R_{0}^{0}=-\frac{3}{a^2}\,{\cal H}'
\nonumber\\
&&
^{(0)}R_{ij}=\left({\cal H}'+2\,{\cal H}^2\right)\,\delta_{ij}
\ ,\quad
^{(0)}R_{i}^{j}=-\frac{1}{a^2}\left({\cal H}'+2\,{\cal H}^2\right)\,\delta_{i}^{j}
\nonumber\\
&&
^{(0)}R=-\frac{6}{a^2}\,\left({\cal H}^2+{\cal H}'\right)
\ .
\ee
\section{Scalar modes}
The scalar fluctuations of the geometry can be written to first-order as
\be
&&
\delta_{\rm s}\,g_{00}=2\,a^2\,\phi
\ ,\quad\quad\quad\quad\quad\quad\quad\,\,
\delta_{\rm s}\,g^{00}=-\frac{2}{a^2}\,\phi
\nonumber\\
&&
\delta_{\rm s}\,g_{ij}=2\,a^2\,\left(\psi\,\delta_{ij}
-\partial_{i}\,\partial_{j}\,E\right)
\ ,\quad
\delta_{\rm s}\,g^{ij}=
-\frac{2}{a^2}\,\left(\psi\,\delta^{ij}-\partial^{i}\,\partial^{j}\,E\right)
\nonumber\\
&&
\delta_{\rm s}\,g_{0i}=-a^2\,\partial_{i}\,B
\ ,\quad\quad\quad\quad\quad\quad\,\,
\delta_{\rm s}\,g^{0i}=-\frac{1}{a^2}\,\partial^{i}\,B
\ .
\label{scalar_fluc}
\ee
Inserting Eqs.~(\ref{scalar_fluc}) into Eq.~(\ref{Chris_pert}),
the first-order components of the (scalar) Christoffel
fluctuations can now be easily computed
\be
&&
\delta_{\rm s}\,\Gamma^{0}_{00}=
\phi'
\nonumber\\
&&
\delta_{\rm s}\,\Gamma_{i j}^{0}=
-\left[\psi'+2\,{\cal H}\left(\phi+\psi\right)\right]\,\delta_{ij}
+\partial_{i}\,\partial_{j}\,\left(E'+2\,{\cal H}\,E-B\right)
\nonumber\\
&&
\delta_{\rm s}\,\Gamma_{i0}^{0}=\delta_{\rm s}\,\Gamma_{0i}^{0}=
\partial_{i}\,\left(\phi+{\cal H}\,B\right)
\nonumber\\
&&
\delta_{\rm s}\,\Gamma^{i}_{00}=\partial^{i}\,\left(\phi+B'+{\cal H}\,B\right)
\nonumber\\
&&
\delta_{\rm s}\,\Gamma_{ij}^{k}=\partial^{k}\,\psi\,\delta_{ij}
-\partial_{i}\,\psi\,\delta^{k}_{j}-\partial_{j}\,\psi\,\delta^{k}_{i}
+\partial_{i}\,\partial_{j}\,\partial^{k}\,E-{\cal H}\,\partial^{k}\,B\,\delta_{ij}
\nonumber\\
&&
\delta_{\rm s}\,\Gamma_{0i}^{j}=-\psi'\,\delta_{i}^{j}+\partial_{i}\,\partial^{j}\,E'
\ .
\label{scalar_Chris}
\ee
The first-order fluctuations of the Ricci tensors are
\be
\delta_{\rm s}\,R_{00}&\!\!\!\!=\!\!\!\!&\nabla^2\,\left[\phi+\left(B-E'\right)'
+{\cal H}\,\left(B-E'\right)\right]
+3\left[\psi''+{\cal H}\left(\phi'+\psi'\right)\right]
\nonumber\\
\delta_{\rm s}\,R_{0 i}&\!\!\!\!=\!\!\!\!&\partial_{i}\,\left[\left({\cal H}'+2{\cal H}^2\right)B
+2\left(\psi'+{\cal H}\,\phi\right)\right]
\label{scalar_Ricci}\\
\delta_{\rm s}\,R_{ij}&\!\!\!\!=\!\!\!\!&-\delta_{ij}\,
\left[\psi''+2\left({\cal H}'+2\,{\cal H}^2\right)\,
\left(\psi+\phi\right)+{\cal H}\,\left(\phi'+5\psi'\right)-\nabla^2\,\psi
\right.
\nonumber
\\
&&
\phantom{-\delta_{ij}\ }
\left.
+
{\cal H}\,\nabla^2\,\left(B-E'\right)\right]
\nonumber
\\
&&+\partial_{i}\,\partial_{j}\,\left[\left(E'-B\right)'+2\left({\cal H}'
+2\,{\cal H}^2\right)\,E+2\,{\cal H}\,\left(E'-B\right)+\psi-\phi\right]
\ ,
\nonumber
\ee
while the first-order fluctuation of the Ricci scalar is
\be
\delta_{\rm s}\,R&=&
\frac{2}{a^2}\,\left\{3\,\psi''+6\,\left({\cal H}'+{\cal H}^2\right)\,\phi
+3\,{\cal H}\,\left(\phi'+3\,\psi'\right)\right.
\nonumber\\
&&\left.+\nabla^2\,\left[\phi-2\psi
+\left(B-E'\right)'+3\,{\cal H}\left(B-E'\right)\right]\right\}
\ .
\label{scalar_R}
\ee
Using Eqs.~(\ref{scalar_Ricci}) into Eqs.~(\ref{Ricci_pert_cov_contr}) and (\ref{R_pert})
the fluctuations of the Ricci tensor will be
\be
\delta_{\rm s}\,R_{0}^{0}\!\!\!\!&=&\!\!\!\!
\frac{1}{a^2}
\left\{\nabla^2\left[\phi+\left(B-E'\right)'
+{\cal H}\left(B-E'\right)\right]
+3\left[\psi''+{\cal H}\,\left(\phi'+\psi'\right)
+2\,{\cal H}'\phi\right]\right\}
\nonumber\\
\delta_{\rm s}\,R_{i}^{j}\!\!\!\!&=&\!\!\!\!
\frac{\delta_{i}^{j}}{a^2}\,\left[\psi''+2\,\left({\cal H}'+2\,{\cal H}^2\right)\,\phi
+{\cal H}\,\left(\phi'+5\,\psi'\right)-\nabla^2\,\psi
+{\cal H}\,\nabla^2\,\left(B-E'\right)\right]
\nonumber\\
&&
-\frac{1}{a^2}\,\partial_{i}\,\partial^{j}\,\left[\left(E'-B\right)'
+2\,{\cal H}\,\left(E'-B\right)+\psi-\phi\right]
\nonumber\\
\delta_{\rm s}\,R_{i}^{0}
\!\!\!\!&=&\!\!\!\!
\frac{2}{a^2}\,\partial_{i}\,\left[\psi'+{\cal H}\,\phi\right]
\nonumber\\
\delta_{\rm s}\,R_{0}^{i}
\!\!\!\!&=&\!\!\!\!
\frac{2}{a^2}\,\partial^{i}\,\left[-\left(\psi'+{\cal H}\,\phi\right)
+\left({\cal H}'-{\cal H}^2\right)\,B\right]
\ .
\label{scalar_Ricci_cov_contr}
\ee
Finally the fluctuations of the Einstein tensor are
\be
\delta_{\rm s}\,G_{0}^{0}&=&
\frac{2}{a^2}\,\left\{\nabla^2\,\psi
-{\cal H}\,\nabla^2\,\left(B-E'\right)-3\,{\cal H}\,
\left(\psi'+{\cal H}\,\phi\right)\right\}
\label{scalar_G_00}\\
\delta_{\rm s}\,G_{i}^{j}&=&
\frac{1}{a^2}\,\left\{-2\,\psi''
-2\,\left({\cal H}^2+2\,{\cal H}'\right)\,\phi-2\,{\cal H}\,\phi'-4\,{\cal H}\,\psi'
\right.
\nonumber\\
&&
-\left.\nabla^2\,\left[\phi-\psi+\left(B-E'\right)'
+2\,{\cal H}\left(B-E'\right)\right]\right\}\,\delta_{i}^{j}
\nonumber\\
&&
-\frac{1}{a^2}\,\partial_{i}\,\partial^{j}\,\left[\left(E'-B\right)'
+2\,{\cal H}\left(E'-B\right)+\psi-\phi\right]
\label{scalar_G_ij}\\
\delta_{\rm s}\,G_{i}^{0}&=&
\delta\,R_{i}^{0}
\ .
\label{scalar_G_0i}
\ee
\section{Vector modes}
The vector fluctuations of the geometry can be written as
\be
&&\!\!\!\!\!\!\!\!\!\!\!\!
\delta_{\rm v}\,g_{0i}=a^2\,S_{i}\ ,\quad\quad\quad\quad\quad\quad
\delta_{\rm v}\,g^{0i}=\frac{S^{i}}{a^2}
\nonumber\\
&&\!\!\!\!\!\!\!\!\!\!\!\!
\delta_{\rm v}\,g_{ij}=-a^2\,\left(\partial_{i}\,F_{j}+\partial_{j}\,F_{i}\right)
\ ,\quad
\delta_{\rm v}\,g^{ij}=\frac{1}{a^2}\,\left(\partial^{i}\,F^{j}+\partial^{j}\,F^{i}\right)
\label{vector_fluc}
\ee
and the corresponding fluctuations of the Chrristoffel connections
\be
\delta_{\rm v}\,\Gamma_{i0}^{0}&\!\!\!\!=\!\!\!\!&
-{\cal H}\,S_{i}
\nonumber\\
\delta_{\rm v}\,\Gamma_{ij}^{0}&\!\!\!\!=\!\!\!\!&
\frac{1}{2}\,\left(\partial_{i}\,S_{j}+\partial_{j},S_{i}\right)
+{\cal H}\,\left(\partial_{i}\,F_{j}+\partial_{j}\,F_{i}\right)
+\frac12\,\left(\partial_{i}\,F_{j}'+\partial_{j}\,F_{i}'\right)
\nonumber\\
\delta_{\rm v}\,\Gamma_{00}^{i}&\!\!\!\!=\!\!\!\!&
-{S^{i}}'-{\cal H}\,S^{i}
\nonumber\\
\delta_{\rm v}\,\Gamma_{ij}^{k}&\!\!\!\!=\!\!\!\!&
{\cal H}\,S^{k}\,\delta_{ij}
-\frac12\,\partial^{k}\,\left(\partial_{i}\,F_{j}+\partial_{j}\,F_{i}\right)
\nonumber\\
&\!\!\!\!\phantom{=}\!\!\!\!&
+\frac12\,\partial_{j}\,\left(\partial^{k}\,F_{i}+\partial_{i}\,F^{k}\right)
+\frac12\,\partial_{i}\,\left(\partial_{j}\,F^{k}+\partial^{k}\,F_{j}\right)
\nonumber\\
\delta_{\rm v}\,\Gamma_{i0}^{j}&\!\!\!\!=\!\!\!\!&
-\frac12\,\left(\partial_{i}\,S^{j}-\partial^{j}\,S_{i}\right)
+\frac12\,\left(\partial^{j}\,F_{i}'+\partial_{i}\,{F^{j}}'\right)
\label{vector_Chris}
\ee
where Eq.~(\ref{div_cond}) for $F_{i}$ and $S_{i}$ has been used.
The fluctuations of the Ricci tensor are
\be
\delta_{\rm v}\,R_{0 i}&\!\!\!\!=\!\!\!\!&
-\left[{\cal H}'+2\,{\cal H}^2\right]\,S_{i}
+\frac12\,\nabla^2\,S_{i}+\frac12\,\nabla^2\,F_{i}'
\nonumber\\
\delta_{\rm v}\,R_{i j}&\!\!\!\!=\!\!\!\!&
\frac12\,\left\{\left(\partial_{i}\,S_{j}+\partial_{j}\,S_{i}\right)'
+2\,{\cal H}\,\left(\partial_{i}\,S_{j}+\partial_{j}\,S_{i}\right)\right\}
+\frac12\,\left(\partial_{i}\,F_{j}''+\partial_{j}\,F_{i}''\right)
\nonumber\\
&\!\!\!\!\!\!\!\!&
+{\cal H}\,\left(\partial_{i}\,F_{j}'+\partial_{j}\,F_{i}'\right)
+\frac12\,\left(\partial_{i}\,F_{j}+\partial_{j}\,F_{i}\right)\,
\left(2\,{\cal H}'+4\,{\cal H}^2\right)
\ ,
\label{vector_Ricci}
\ee
\section{Tensor modes}
Let us now considering the tensor modes of the geometry,
that is the two polarisation of the graviton
\be
\delta_{\rm t}\,g_{i j}=-a^2\,h_{ij}\ ,\quad
\delta_{\rm t}\,g^{i j}=\frac{h^{ij}}{a^2}
\ .
\label{tensorial_fluc}
\ee
From Eq.~(\ref{Chris_pert}) the tensor contribution to the
fluctuations of the connections can be expressed as
\be
&&
\delta_{\rm t}\,\Gamma_{ij}^{0}=\frac12\,\left(h_{ij}'+2\,{\cal H}\,h_{i j}\right)
\nonumber\\
&&
\delta_{\rm t}\,\Gamma_{i0}^{j}=\frac12\,{h_{i}^{j}}'
\nonumber\\
&&
\delta_{\rm t}\,\Gamma_{ij}^{k}=\frac12\,\left[\partial_{j}\,h_{i}^{k}
+\partial_{i}\,h_{j}^{k}-\partial^{k}\,h_{ij}\right]
\ ,
\label{tensorial_Chris}
\ee
and the Ricci tensor becomes
\be
&&
\delta_{\rm t}\,R_{ij}=\frac12\,\left[h_{ij}''
+2\,{\cal H}\,h_{ij}+2\,\left({\cal H}'+2\,{\cal H}^2\right)\,h_{ij}-\nabla^2\,h_{ij}\right]
\label{tensorial_Ricci}\\
&&
\delta_{\rm t}\,R_{i}^{j}=-\frac{1}{2 a^2}\,\left[{h_{i}^{i}}''
+2\,{\cal H}\,{h_{i}^{j}}'-\nabla^2\,h_{i}^{j}\right]
\ .
\label{tensorial_Ricci_cov_contra}
\ee
\section{Energy-Momentum Tensor}
The background energy-momentum tensor of a perfect fluid is
\be
T_{\mu\nu}=\left(p+\rho\right)\,u_{\mu}\,u_{\nu}-p\,g_{\mu \nu}
\ .
\label{en-mom_tens_perf_fluid}
\ee
From the normalisation condition 
\be
g_{\mu\nu}\,u^{\mu}\,u^{\nu}=1
\ ,
\label{guu}
\ee 
one finds that $u_0=a$ and $\delta\,u^{0}=-\phi/a$.
Hence, the scalar perturbations become
\be
\delta_{\rm s}\,T_{0}^{0}=\delta\,\rho
\ ,\quad
\delta_{\rm s}\,T^{00}=\frac{1}{a^2}\,\left(\delta\,\rho-2\,\rho\,\phi\right)
\label{fluct_T_00}
\ee
and
\be
\delta_{\rm s}\,T_{i}^{j}&=&-\delta\,p\,\delta_{i}^{j}
\label{fluct_T_ij_cov_contr}\\
\delta_{\rm s}\,T^{ij}&=&\frac{1}{a^2}\,\left[\delta\,p\,\delta^{ij}
+2\,p\,\left(\psi\,\delta^{ij}-\partial^{i}\,\partial^{j}\,E\right)\right]
\label{fluct_T_ij}\\
\delta_{\rm s}\,T_{0}^{i}&=&\left(p+\rho\right)\,v^{i}
\label{fluct_T_0i_cov_contr}\\
\delta_{\rm s}\,T^{0i}&=&\frac{1}{a^2}\,\left[\left(p+\rho\right)\,v^{i}+\partial^{i}\,B\right]
\ .
\label{fluct_T_0i}
\ee
where we have defined $\delta\,u^{i}\equiv v^{i}/a$. The velocity field can be split
into divergenceless and  divergencefull parts
\be
v^{i}=\partial^{i}\,v+{\cal V}^{i}
\ ,\quad
\partial_{i}\,{\cal V}^{i}=0
\label{v_def}
\ee
For collisionless matter, the anisotropic stress must be considered
both in the Einstein equations and in the covariant conservation equations.
The anisotropic stress is then introduced as
\be
\delta_{s}\,T_{i}^{j}=-\delta\,p\,\delta_{i}^{j}+\Pi_{i}^{j}
\ ,
\label{anisotrop_stress_def}
\ee
where $\Pi_{i}^{j}\equiv T_{i}^{j}-\delta_{i}^{j}\,T_{k}^{k}/3$.
For the scalar fluctuations we adopt the notation
\be
\partial_{j}\,\partial^{i}\,\Pi_{i}^{j}\equiv-\left(p+\rho\right)\,\nabla^2\,\sigma
\ .
\label{anisotrop_stress_notation}
\ee
The perturbed velocity field can be defined in two ways,
\be
\delta_{\rm s}\,T_{0}^{i}=\left(p+\rho\right)\,u_{0}\,\delta\,u^{i}
\equiv\left(p+\rho\right)\,v^{i}
\equiv\left(p+\rho\right)\,\partial^{i}\,v
\label{v_def_1}
\ee
or
\be
\delta_{\rm s}\,T_{i}^{0}=\left(p+\rho\right)\,u^{0}\,\delta\,u_{i}
=\left(p+\rho\right)\,\overline{v}_{i}
\equiv\left(p+\rho\right)\,\partial_{i}\,\overline{v}
\ ,
\label{v_def_2}
\ee
where, from the normalization condition (\ref{guu}), $u_{0}= a$ and $u^{0} =1/a$.
The velocity fields $\overline{v}$ and $v$
defined in Eqs. (\ref{v_def_1}) and (\ref{v_def_2}) are not equivalent.
In fact $\delta u_{i}=a\,\overline{v}_{i}$ and $\delta u^{i}=v^{i}/a$.
Using
\be
\delta\,u^{i}&=&\delta\,\left(g^{i\alpha}\,u_{\alpha}\right)
\equiv\delta\,g^{i0}\,u_{0}+g^{ik}\,\delta\,u_{k}
\label{delta_u}\\
\delta\,g^{i0}&=&-\partial^{i}B/a^2
\ ,
\label{delta_g_i0}
\ee
and inserting the definitions of $\delta u^{i}$ and $\delta u_{k}$
in terms of $v^{i}$ and $\overline{v}_{k}$ we have
\be
v_{i}=-\overline{v}_{i}-\partial_{i}\,B
\ .
\label{v->overlinev}
\ee
The gauge transformations for $v_{i}$ and $\overline{v}_{i}$
are correspondingly different,
\be
&&
v_{i}\to\tilde{v}_{i}=v_{i}+\partial_{i}\,\epsilon'
\nonumber\\
&&
\overline{v}_{i}\to\tilde{\overline{v}}_{i}
=\overline{v}_{i}-\partial_{i}\,\epsilon_{0}
\ .
\label{v_overlinev_gauge_trensf}
\ee
In this thesis we use the velocity field as defined in Eq.~(\ref{v_def_1}).
\par
Let us now consider the energy-momentum tensor of
a scalar field $\varphi$ characterized by a potential $V(\varphi)$,
\be
T_{\mu\nu}=\partial_{\mu}\,\varphi\,\partial_{\nu}\,\varphi-g_{\mu\nu}\,
\left[\frac12\,g^{\alpha\beta}\,\partial_{\alpha}\,\varphi\,\partial_{\beta}\varphi
-V(\varphi)\right]
\ .
\label{en-mom_tensor_scalar_field}
\ee
Denoting with $\chi$ the first-order fluctuation of the scalar field $\varphi$,
we have
\be
\delta_{\rm s}\,T_{\mu\nu}&\!\!\!\!=\!\!\!\!&
\partial_{\mu}\,\chi\,\partial_{\nu}\,\varphi
+\partial_{\mu}\,\varphi\,\partial_{\nu}\,\chi
-\delta_{\rm s}\,g_{\mu\nu}\,
\left[\frac12\,g^{\alpha\beta}\,\partial_{\alpha}\,\varphi\,\partial_{\beta}\,\varphi-V\right]
\label{delta_T_scal_field}
\\
&\!\!\!\!\!\!\!\!&
-g_{\mu\nu}\,\left[\frac12\,\delta_{\rm s}\,g^{\alpha\beta}\,\partial_{\alpha}\,\varphi\,\partial_{\beta}\,\varphi
+g^{\alpha\beta}\,\partial_{\alpha}\,\chi\,\partial_{\beta}\,\varphi
-\frac{\partial\,V}{\partial\,\varphi}\,\chi\right]
\ ,
\nonumber
\ee
or, explicitly,
\be
\delta_{\rm s}\,T_{00}&=&
\chi'\,\varphi'+2\,a^2\,\phi\,V+a^2\,\frac{\partial\,V}{\partial\,\varphi}\,\chi
\nonumber\\
\delta_{\rm s}\,T_{0i}&=&
\varphi'\,\partial_{i}\,\chi
+a^2\,\partial_{i}\,B\,\left[\frac{{\varphi'}^2}{2\,a^2}-V\,\right]
\nonumber\\
\delta_{\rm s}\,T_{ij}&=&
\delta_{ij}\,\left[\varphi'\,\chi'
-\frac{\partial\,V}{\partial\,\varphi}\,\chi\,a^2
-\left(\phi+\psi\right)\,{\varphi'}^2+2\,a^2\,V\,\psi\right]
\nonumber\\
&&
+2\,a^2\left[\frac{{\varphi'}^2}{2\,a^2}-V\right]\,\partial_{i}\,\partial_{j}\,E
\ .
\label{delta_T_scal_field_explicit}
\ee
On using
\be
^{(0)}T_{00}=\frac{{\varphi'}^2}{2}+a^2\,V
\ ,\quad
^{(0)}T_{ij}=\left[\frac{{\varphi'}^2}{2}-a^2\,V\right]\,\delta_{ij}
\label{T_back_scal_field}
\ee
the perturbed components of the energy-momentum tensor can then
be written as
\be
\delta_{\rm s}\,T_{\mu}^{\nu}=
\delta_{\rm s}\,T_{\alpha\mu}\,^{(0)}g^{\alpha\nu}+^{(0)}T_{\alpha\mu}\,\delta_{\rm s}\,
g^{\alpha\nu}
\ ,
\label{delta_T_scal_field2}
\ee
that is
\be
\delta_{\rm s}\,T_{0}^{0}&=&
\frac{1}{a^2}\,\left(-\phi\,{\varphi'}^2+\frac{\partial\,V}{\partial\,\varphi}\,a^2\,\chi
+\chi'\,\varphi'\right)
\label{en-mom_expl_1}\\
\delta_{\rm s}\,T_{i}^{j}&=&
\frac{1}{a^2}\,\left(\phi\,{\varphi'}^2+\frac{\partial\,V}{\partial\,\varphi}\,a^2\,\chi
-\chi'\,\varphi'\right)\,\delta_{i}^{j}
\label{en-mom_expl_2}\\
\delta_{\rm s}\,T_{0}^{i}&=&
-\frac{1}{a^2}\,\varphi'\,\partial^{i}\,\chi
-\frac{{\varphi'}^2}{a^2}\,\partial^{i}\,B
\ .
\label{en-mom_expl_3}
\ee
The covariant conservation of the energy-momentum tensor implies the validity of
the Klein-Gordon equation which can be written as
\be
g^{\alpha\beta}\,\nabla_{\alpha}\,\nabla_{\beta}\,\varphi
+\frac{\partial\,V}{\partial\,\varphi}\,a^2=0
\ .
\label{Klein-Gordon_gen}
\ee
From Eq.~(\ref{Klein-Gordon_gen}) the perturbed Klein-Gordon equation is
\be
&&
\delta_{\rm s}\,g^{\alpha\beta}\,\left[\partial_{\alpha}\,\partial_{\beta}\,\varphi
-^{(0)}\Gamma_{\alpha\beta}^{\sigma}\,\partial_{\sigma}\,\varphi\right]
\nonumber\\
&&
+^{(0)}g^{\alpha\beta}\,\left[\partial_{\alpha}\,\partial_{\beta}\,\chi
-\delta_{\rm s}\,\Gamma^{\sigma}_{\alpha\beta}\,\partial_{\sigma}\,\varphi
-^{(0)}\Gamma_{\alpha\beta}^{\sigma}\,\partial_{\sigma}\,\chi\right]
+\frac{\partial^2 V}{\partial\varphi^2}=0
\ ,
\label{Klein-Gordon_pert}
\ee
or, more explicitly,
\be
&&
\chi''+2\,{\cal H}\,\chi'-\nabla^2\,\chi
+\frac{\partial^2\,V}{\partial\,\varphi^2}\,a^2\,\chi
\nonumber\\
&&
+2\,\phi\,\frac{\partial\,V}{\partial\,\varphi}\,a^2
-\varphi'\,\left(\phi'+3\,\psi'\right)+\varphi'\,\nabla^2\,\left(E'-B\right)=0
\ .
\label{Klein-Gordon_pert_expl}
\ee
\section{Gauge invariant scalar perturbations}
We show here the generalised evolution equations for scalar fluctuations
obtained without fixing a particular gauge.
From Eqs.~(\ref{scalar_G_00}), (\ref{fluct_T_00}), (\ref{scalar_G_0i})
and (\ref{fluct_T_0i_cov_contr}) the Hamiltonian and momentum constraints
become
\be
&&
\nabla^2\,\psi-{\cal H}\,\nabla^2\,\left(B-E'\right)-3\,{\cal H}\,
\left(\psi'+{\cal H}\,\phi\right)=4\,\pi\,G\,a^2\,\delta\,\rho
\label{00_without_gauge}\\
&&
\partial^{i}\,\left[\left(\psi'+{\cal H}\,\phi\right)
+\left({\cal H}^2-{\cal H}'\right)\,B\right]
=-4\,\pi\,G\,a^2\,\left(p+\rho\right)\,v^{i}
\ ,
\label{0i_without_gauge}
\ee
From Eqs.~(\ref{scalar_G_ij}) and (\ref{fluct_T_ij}) we obtain the spatial
components of the perturbed equations,
\be
4\,\pi\,G\,a^2\,\left[\delta\,p\,\delta_{i}^{j}-\Pi_{i}^{j}\right]
&\!\!\!\!=\!\!\!\!&
\left[\psi''+\left({\cal H}^2+2\,{\cal H}'\right)\,\phi
+{\cal H}\,\left(\phi'+2\,\psi'\right)\right]\,\delta_{i}^{j}
\label{ij_without_gauge}\\
&\!\!\!\!\!\!\!\!&
+\frac12\,\delta_{i}^{j}\,\nabla^2\,\left[\phi-\psi
+\left(B-E'\right)'+2\,{\cal H}\,\left(B-E'\right)\right]
\nonumber\\
&\!\!\!\!\!\!\!\!&
-\frac12\,\partial_{i}\,\partial^{j}\,\left[\phi-\psi
+\left(B-E'\right)'+2\,{\cal H}\,\left(B-E'\right)\right]
\nonumber
\ .
\ee
From Eq.~(\ref{ij_without_gauge}) we can separate the traceless part as
\be
4\,\pi\,G\,a^2\,\delta\,p&=&
\psi''+{\cal H}\,\left(\phi'+2\,\psi'\right)
+\left({\cal H}^2+2\,{\cal H}'\right)\,\phi
\label{ij_tracefull_and_traceless}\\
&&
+\frac13\,\nabla^2\,\left[\phi-\psi+\left(B-E'\right)'
+2\,{\cal H}\,\left(B-E'\right)\right]
\nonumber\\
12\,\pi\,G\,a^2\,\left(p+\rho\right)\,\sigma&=&
\nabla^2\,\left[\phi-\psi+\left(B-E'\right)'
+2\,{\cal H}\,\left(B-E'\right)\right]
\ .
\nonumber
\ee
The $(0)$ and $(i)$ components of Eq.~(\ref{cons_eq_pert}) can be written
\be
0&\!\!\!\!=\!\!\!\!&
\partial_{0}\,\delta_{\rm s}\,T^{00}+\partial_{j}\,\delta_{\rm s}\,T^{j0}
+\left(2\,\delta_{\rm s}\,\Gamma_{00}^{0}\,+\,\delta_{\rm s}\,\Gamma_{k0}^{k}\right)\,
^{(0)}T^{00}
\label{cons_eq_pert_0}\\
&\!\!\!\!\!\!\!\!&
+\left(2\,^{(0)}\Gamma_{00}^{0}+^{(0)}\Gamma_{k0}^{k}\right)\,\delta_{\rm s}\,T^{00}
+^{(0)}\Gamma_{ij}^{0}\,\delta_{\rm s}\,T^{ij}
+\delta_{\rm s}\,\Gamma_{ij}^{0}\,^{(0)}T^{ij}
\nonumber\\
0&\!\!\!\!=\!\!\!\!&
\partial_{0}\,\delta_{\rm s}\,T^{0j}+\partial_{k}\,\delta_{\rm s}\,T^{kj}
+\left(\delta_{\rm s}\,\Gamma_{0k}^{0}+\delta_{\rm s}\,\Gamma_{mk}^{m}\right)\,^{(0)}T^{kj}
\label{cons_eq_pert_i}\\
&\!\!\!\!\!\!\!\!&
+\left(^{(0)}\Gamma_{00}^{0}+^{(0)}\Gamma_{k0}^{k}\right)\,\delta_{\rm s}\,T^{0j}
+\delta_{\rm s}\,\Gamma_{00}^{j}\,^{(0)}T^{00}
+\delta_{\rm s}\,\Gamma_{km}^{j}\,^{(0)}T^{km}
+2\,^{(0)}\Gamma_{0k}^{j}\,\delta_{\rm s}\,T^{0k}
\ .
\nonumber
\ee
Inserting the perturbed connections~(\ref{scalar_Chris}) into Eqs.~(\ref{cons_eq_pert_0})
and (\ref{cons_eq_pert_i}) we have
\be
\!\!\!\!\!\!\!\!\!\!\!\!0&\!\!\!\!=\!\!\!\!&
\delta\,\rho'-3\,\psi'\,\left(p+\rho\right)+\left(p+\rho\right)\,\theta
+3\,{\cal H}\,\left(\delta\,\rho+\delta\,p\right)
+\left(p+\rho\right)\,\nabla^2\,E'
\label{cons_eq_pert_0_expl}\\
\!\!\!\!\!\!\!\!\!\!\!\!
0&\!\!\!\!=\!\!\!\!&\left(p+\rho\right)\,\theta'+\theta\,\left[p'+\rho'
+4\,{\cal H}\,\left(p+\rho\right)\right]+\left(p+\rho\right)\,\nabla^2\,B'
\nonumber\\
\!\!\!\!\!\!\!\!\!\!\!\!
&\!\!\!\!\!\!\!\!&
+\left[p'+{\cal H}\,\left(p+\rho\right)\right]\,\nabla^2\,B
+\nabla^2\,\delta\,p+\left(p+\rho\right)\,\nabla^2\,\phi
\ ,
\label{cons_eq_pert_i_expl}
\ee
where the divergence of the velocity field $\theta\equiv \partial_{i}\,v^{i}=\partial_{i}\,\partial^{i}\,v$
has been introduced.

\chapter{Slow-Roll {\em vs} Horizon Flow}
\label{appendice_SR-HFF}
In this Appendix, we compare the HFF we used in the text
with some other (equivalent) hierarchies.
\par
In Table~\ref{convers1}, we give the relations between the
slow-roll parameters defined by some other authors and the HFF
defined in Eq.~(\ref{hor_flo_fun}).
\begin{table}[ht]
\centerline{
\begin{tabular}{ c c }
\hline
\hline
\hspace{\stretch{1}} J.~Martin \textit{et al.} \cite{martin_sl_roll}$^\dag$
\hspace{\stretch{1}}
& \hspace{\stretch{1}} E.D.~Stewart \textit{et al.} \cite{SL,SGong}$^*$
\hspace{\stretch{1}}
\\
\hspace{\stretch{1}} S.~Habib \textit{et al.} \cite{H_MP}
\hspace{\stretch{1}}
& \hspace{\stretch{1}} S.~Habib \textit{et al.} \cite{HHHJM,HHHJ}
\hspace{\stretch{1}}
\\
\hline
   %\\
$\epsilon=\epsilon_1$
& $\epsilon=\epsilon_1$
\\ %\\
$\delta=\epsilon_1-\frac{1}{2}\,\epsilon_2$
& $\delta_{1}=\frac{1}{2}\,\epsilon_2-\epsilon_1$
\\ %\\
$\xi_2=\frac{1}{2}\,\epsilon_2\,\epsilon_3$
& $\delta_2=-\frac{5}{2}\,\epsilon_1\,\epsilon_2+2\,\epsilon_1^2
+\frac{1}{4}\,\epsilon_2^2+\frac{1}{2}\,\epsilon_2\,\epsilon_3$
\\ %\\
\hline
\hline
\end{tabular}
}
\caption{Relation between (some) slow-roll parameters and the HFF
used here and in Ref.~\cite{martin_schwarz2}.
$^*$In Ref.~\cite{SL} $\epsilon_1$, $\delta$, and
${\dddot{\phi}\,\delta}/H\,\ddot{\phi}$ are used instead of $\epsilon$,
$\delta_1$ and $\delta_2$ respectively.
$^\dag$In Ref.~\cite{martin_sl_roll} $\xi$ is used instead of
$\xi_2$.
\label{convers1}}
\end{table}
\par
We also compare the Hubble-slow-roll~(HSR) and the potential-slow-roll~(PSR)
parameters (see~Ref.~\cite{LPB1,LPB2}) with the HFF.
We recall that some authors redefine $\xi_V$ and $\xi_H$ as
$\xi_V^2$ and $\xi_H^2$ and write only $\epsilon$, $\eta$ and $\xi$,
which can be confusing.
The relations between the $\epsilon_i$'s and HSR parameters are
\be
&&
\epsilon_H=\epsilon_1
\ ,
\nonumber
\quad
\eta_H=\epsilon_1-\frac12\,\epsilon_2
\nonumber
\\
\label{HFFvsHSR}
\\
&&
\xi_H^2=\epsilon_1^2-\frac32\,\epsilon_1\,\epsilon_2
+\frac12\,\epsilon_2\,\epsilon_3
\nonumber
\ ,
\ee
and for the PSR parameters we have
\be
\epsilon_V
\!\!&\!\!\!\!\!\!=\!\!\!\!\!\!&\!\!
\epsilon_1\,\left[1+\frac{\epsilon_2}{2\,(3-\epsilon_1)}\right]^2
\sim
\epsilon_1
\nonumber
\\
\nonumber
\\
\eta_V
\!\!&\!\!\!\!\!\!=\!\!\!\!\!\!&\!\!
\frac{6\,\epsilon_1-\frac32\,\epsilon_2-2\,\epsilon_1^2
+\frac52\,\epsilon_1\,\epsilon_2
-\frac{1}{4}\,\epsilon_2^2
-\frac12\,\epsilon_2\,\epsilon_3}{3-\epsilon_1}
\sim
2\,\epsilon_1-\frac12\,\epsilon_2
\nonumber
\\
\nonumber
\\
\xi_V^2
\!\!&\!\!\!\!\!\!=\!\!\!\!\!\!&\!\!
\frac{\left(2\,\epsilon_1-6-\epsilon_2\right)
\left[8\,\epsilon_1^3-6\,\epsilon_1^2\,(4+3\,\epsilon_2)
+\epsilon_1\,\epsilon_2\,(18+6\,\epsilon_2+7\,\epsilon_3)
-\epsilon_2\,\epsilon_3\,(3+\epsilon_2+\epsilon_3+\epsilon_4)\right]}
{4\,\left(3-\,\epsilon_1\right)^2}
\nonumber
\\
\!\!&\!\!\!\!\!\!\sim\!\!\!\!\!\!&\!\!
4\,\epsilon_1^2
-3\,\epsilon_1\,\epsilon_2+\frac12\,\epsilon_2\,\epsilon_3
\ ,
\label{HFFvsPSR}
\ee
where we have shown both the exact formulas and approximate
relations to the order of interest.
\par
We complete this review with the first three exact connection
formulae between the PSR and HSR parameters~\cite{LPB1,LPB2}
\be
&
\strut\displaystyle\frac{\epsilon_V}{\epsilon_H}=
\frac{\left(3-\eta_H\right)^2}{\left(3-\epsilon_H\right)^2}
\ ,
\quad
\eta_V=
\strut\displaystyle\frac{3\left(\epsilon_H+\eta_H\right)-\eta_H^2-\xi_H^2}
{3-\epsilon_H}
&
\nonumber
\\
\label{PSRvsHSR}
\\
&
\xi_V^2=
\strut\displaystyle\frac{9\,\left(3\epsilon_H\eta_H+\xi_H^2-\epsilon_H\eta_H^2\right)
-3\left(4\eta_H\xi_H^2-\eta_H^2\xi_H^2+\sigma_H^3\right)+\eta_H\sigma_H^3}
{(3-\epsilon_H)^2}
\ ,
&
\nonumber
\ee
and so on with $\sigma_H$ the next HSR parameter.

\chapter{An application of MCE}
\label{app}
Here we use the Method of Comparison Equations to solve the
differential equation~(\ref{equation}) with the frequency
$\omega^2(x)=e^{x}-1+a\,\sin (x)$, which does not admit an exact solution.
In fact we shall solve in a semi-analytical way the equation
\begin{eqnarray}
\frac{d^2\chi(x)}{dx^2} = \frac{1}{\varepsilon}\left[e^{x}-1
+a\,\sin (x)\right]\chi(x)
\ .
\label{equation_app}
\end{eqnarray}
We shall give the semi-analytical results to leading and next-to-leading
order of our approximation, correspondingly to $0^{\rm th}$ and
$1^{\rm st}$-order in $\epsilon$. To build the semi-analytical solution we
need the first two terms of the asymptotic expansion~(\ref{asymp}), i.e.
\begin{eqnarray}
y(x)=y_0(x)+\epsilon\,y_1(x)
\ ,
\label{y_2terms}
\end{eqnarray}
and also the function
\begin{eqnarray}
\sigma(x)=\sigma_0+\int^x_0\frac{dx'}{y^2(x')}
\ .
\label{sigma-define}
\end{eqnarray}
The function $y(x)$ appears in Eq.~(\ref{form}) as a prefactor
(See Eq.~(\ref{y-define})), while $\sigma(x)$ emerges in the argument
of the analytical part of the solution~(\ref{form}), i.e. in $U\left[\sigma(x)\right]$.
The above quatities will be estimated numerically taking $\epsilon=1$ and
$a=1/10$~\footnote{We use a ``small'' value for $a$ in order to avoid other
turning points in the oscillating region.}. The initial conditions are
$u(-20)=1$ and $\dot{u}(-20)=0$, while the range of the plots is
$-20\le x\le 4$. We recall that the quantity $\sigma_0$ changes from
first to the second order of approximation, as we have shown
 in section~\ref{langer_TP} for the linear case. We shall use $\sigma_0=0$
for the leading order, while at the next-to-leading order we shall
recalculate its value utilising the previous order. Finally, the analytical
part will be a general solution of Eq.~(\ref{equation1}).
\par
In the following we shall consider two comparison functions; a linear one,
i.e. $\Theta^2(\sigma) = \sigma$, and a more complicated one,
i.e. $\Theta^2(\sigma) = {\rm e}^\sigma-1$
\subsection{Linear comparison function}
\begin{figure}[!ht]
\includegraphics[width=0.5\textwidth]{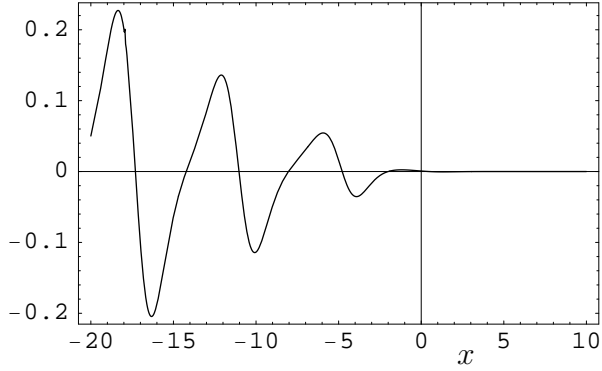}
\null
\hspace{-2.1cm}$x$
\caption{Function $y_1(x)$ for a linear comparison equation.}
\label{y1_fig}
\end{figure}

\begin{figure}[!ht]
\includegraphics[width=0.5\textwidth]{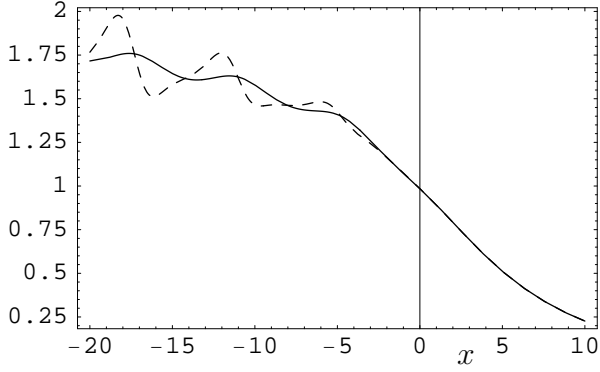}
\null
\hspace{-2.1cm}$x$
\caption{Function $y_0(x)$ (solid line) and  $y(x)=y_0(x)+\epsilon\,y_1(x)$
(dashed line) for a linear comparison equation.}
\label{y0_y01_fig}
\end{figure}

\begin{figure}[!ht]
\includegraphics[width=0.5\textwidth]{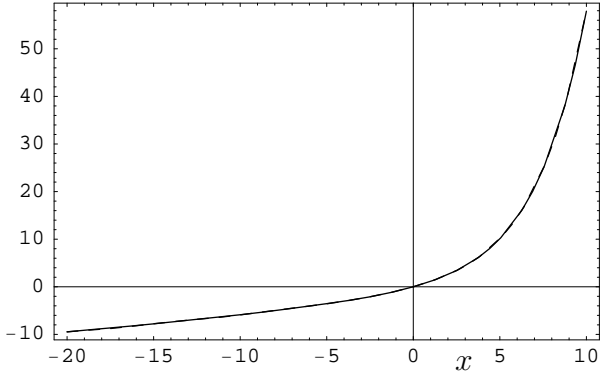}
\null
\hspace{-2.1cm}$x$
\caption{Function $\sigma(x)$, for a linear comparison equation, to
$0^{\rm th}$ and $1^{\rm st}$-order in $\epsilon$
(solid and dashed line respectively).
One can not distinguish the two curves.}
\label{sigma_fig}
\end{figure}

For the linear (i.e. Langer) case $y_0(x)$ and $y_1(x)$ are given by
Eqs.~(\ref{Langer02}) and (\ref{Langer1}), respectively.
In Fig.~\ref{y1_fig} we show the function $y_1(x)$, while in
Fig.~\ref{y0_y01_fig} we plot the function $y(x)$ at $0^{\rm th}$ and
$1^{\rm st}$-order in $\epsilon$.
The function $\sigma(x)$ at $0^{\rm th}$ and $1^{\rm st}$-order in $\epsilon$,
calculated from Eq.~(\ref{sigma-define}), is shown in Fig.~\ref{sigma_fig}.
\par
\begin{figure}[!ht]
\includegraphics[width=0.5\textwidth]{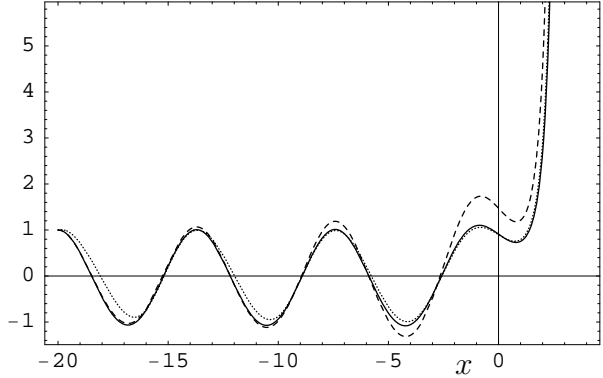}
\null
\hspace{-2.1cm}$x$
\caption{Solution $\chi(x)$ of Eq.~(\ref{equation_app}) for a linear comparison
equation; the solid line is the full numerical solution, while the dashed and
dotted ones are the $0^{\rm th}$ and $1^{\rm st}$-order in
$\epsilon$ respectively.}
\label{sol_fig}
\end{figure}

The analytical part of the solution follows from
\begin{eqnarray}
\frac{d^2U(\sigma)}{d\sigma^2} = \frac{1}{\varepsilon}\sigma U(\sigma)
\ ,
\label{equation1_app}
\end{eqnarray}
and is a linear combination of Airy functions Ai and Bi
\begin{eqnarray}
U\left[\sigma(x)\right] =A\cdot {\rm Ai}\left[\epsilon^{-1/3}\sigma(x)\right]
+B\cdot {\rm Bi}\left[\epsilon^{-1/3}\sigma(x)\right]
\ .
\label{analatic_sol_app}
\end{eqnarray}
Finally the complete solution becomes
\begin{eqnarray}
\chi(x)&=&y(x)\,\left\{
A\cdot {\rm Ai}\left[\epsilon^{-1/3}\sigma(x)\right]
+B\cdot {\rm Bi}\left[\epsilon^{-1/3}\sigma(x)\right]
\right\}
\ .
\label{analatic_sol_tot_app}
\end{eqnarray}
In Fig.~\ref{sol_fig} we exhibit the exact numerical solution and the first
and second order solutions in our approximation scheme.
\subsection{A more complicated comparison function}
\begin{figure}[!ht]
\includegraphics[width=0.5\textwidth]{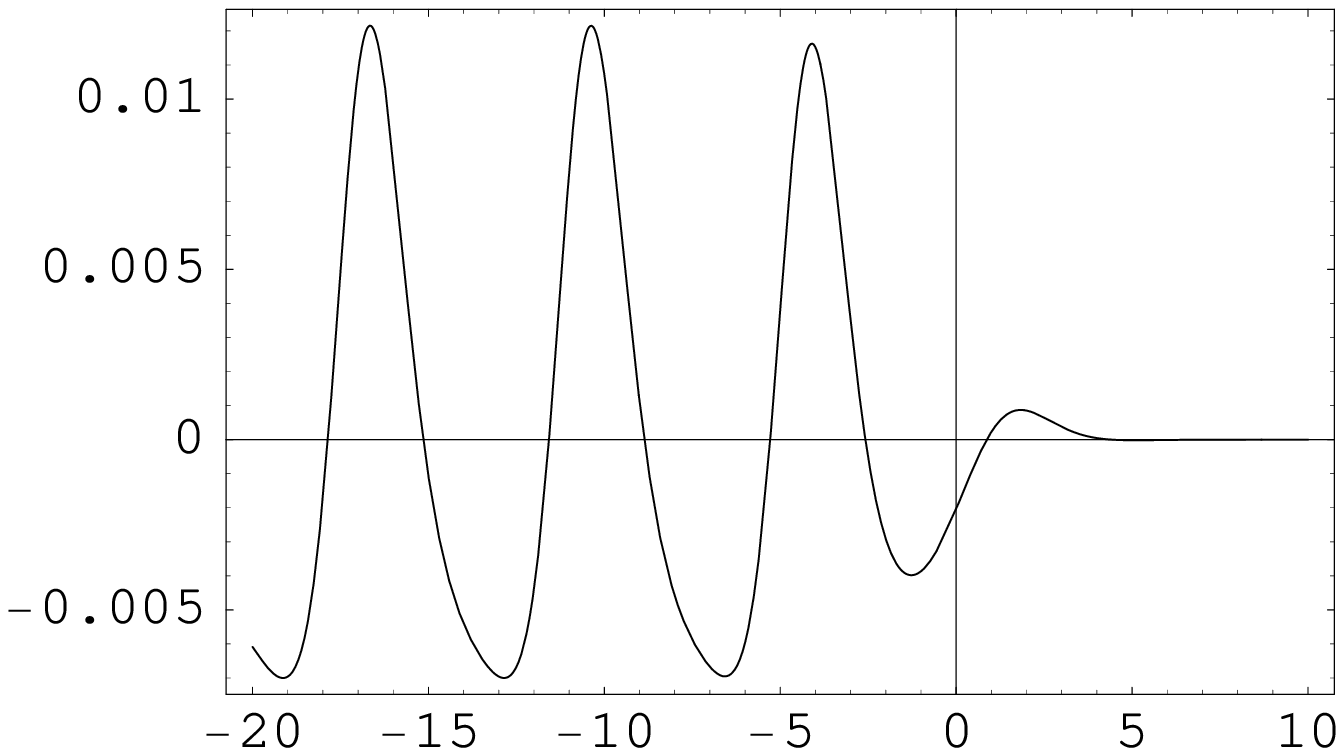}
\null
\hspace{-2.1cm}$x$
\caption{The function $y_1(x)$ for a more complicated comparison function.}
\label{y1_figE}
\end{figure}

\begin{figure}[!ht]
\includegraphics[width=0.5\textwidth]{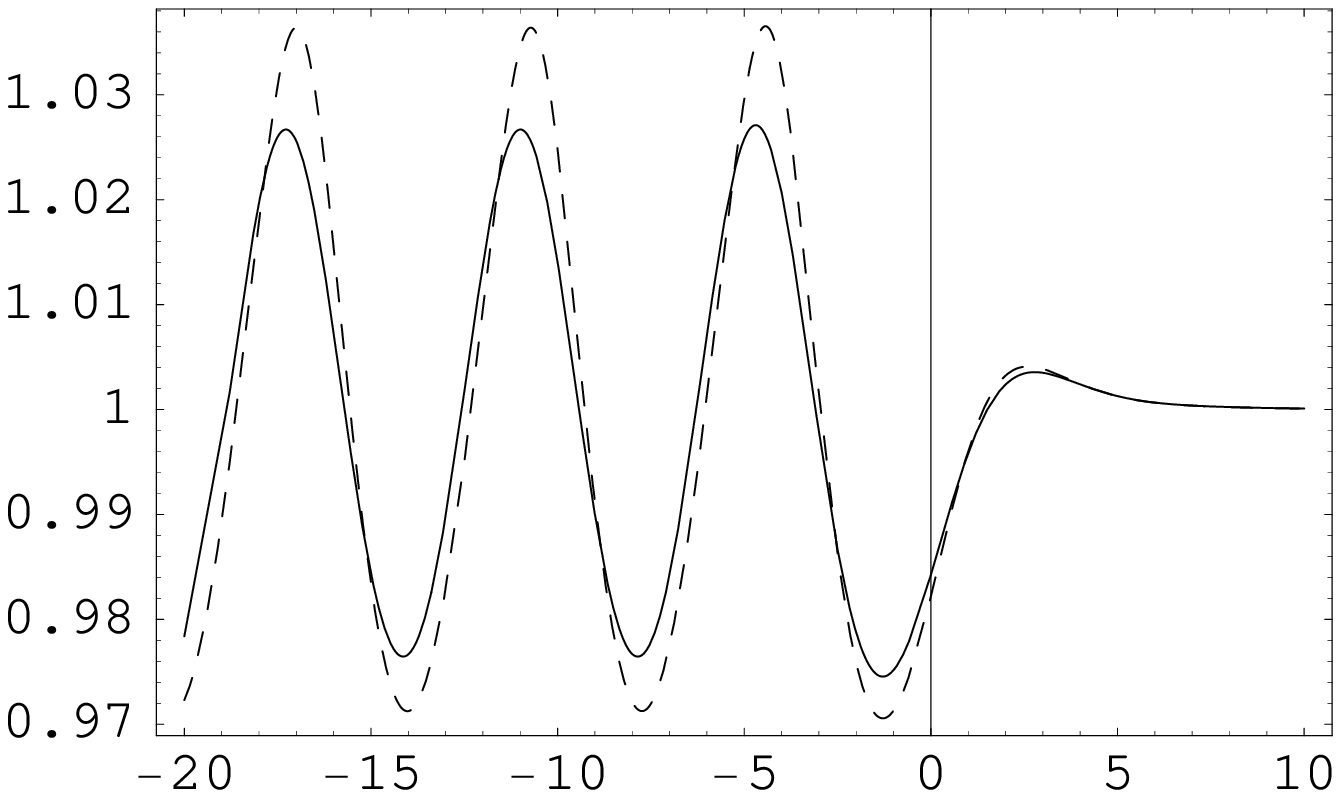}
\null
\hspace{-2.1cm}$x$
\caption{The functions $y_0(x)$ (solid line) and
$y(x)=y_0(x)+\epsilon\,y_1(x)$ (dashed line) for a more complicated
comparison function.}
\label{y0_y01_figE}
\end{figure}

\begin{figure}[!ht]
\includegraphics[width=0.5\textwidth]{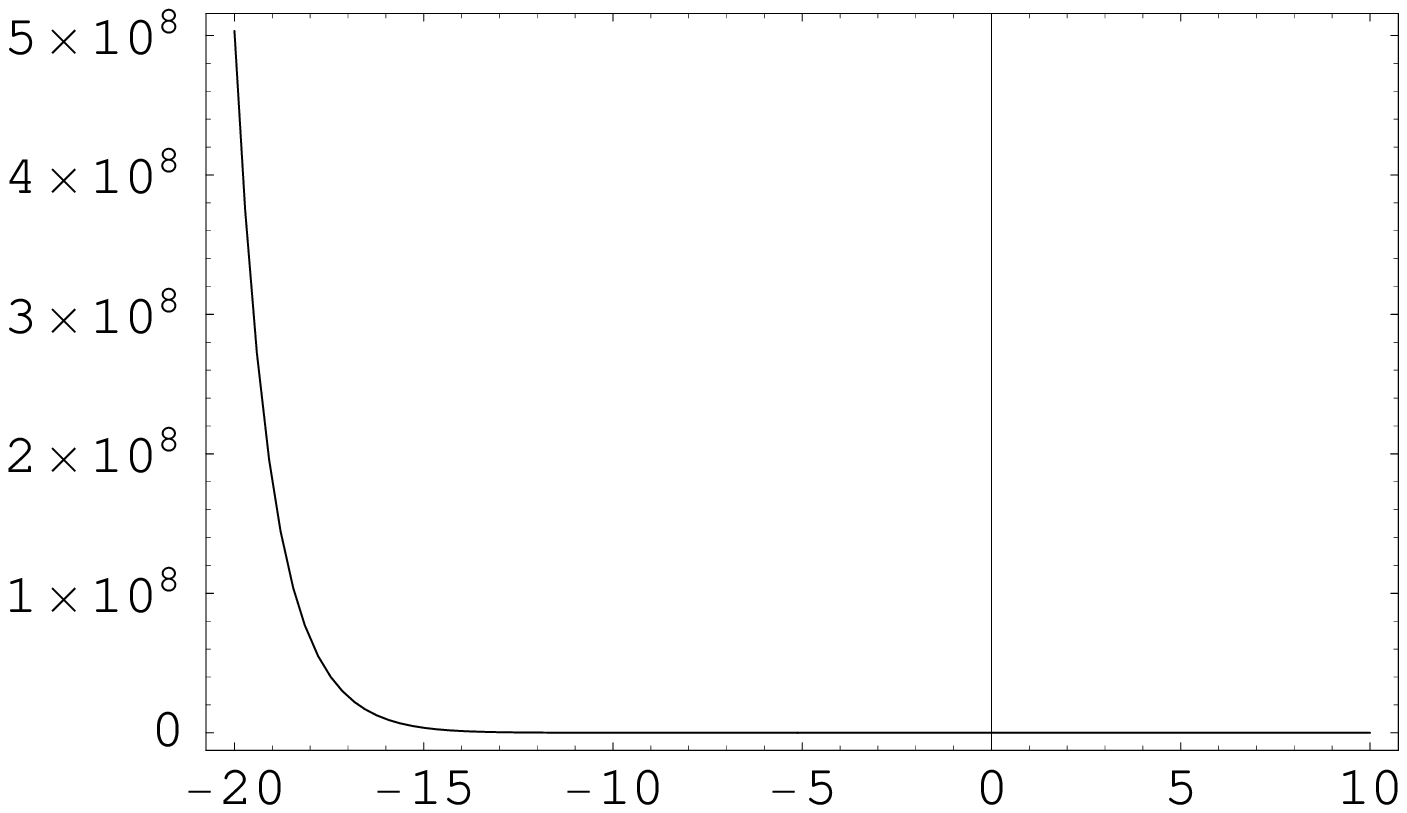}
\null
\hspace{-2.1cm}$x$
\caption{Function $G(x)$.} 
\label{G_fig}
\end{figure}

\begin{figure}[!ht]
\includegraphics[width=0.5\textwidth]{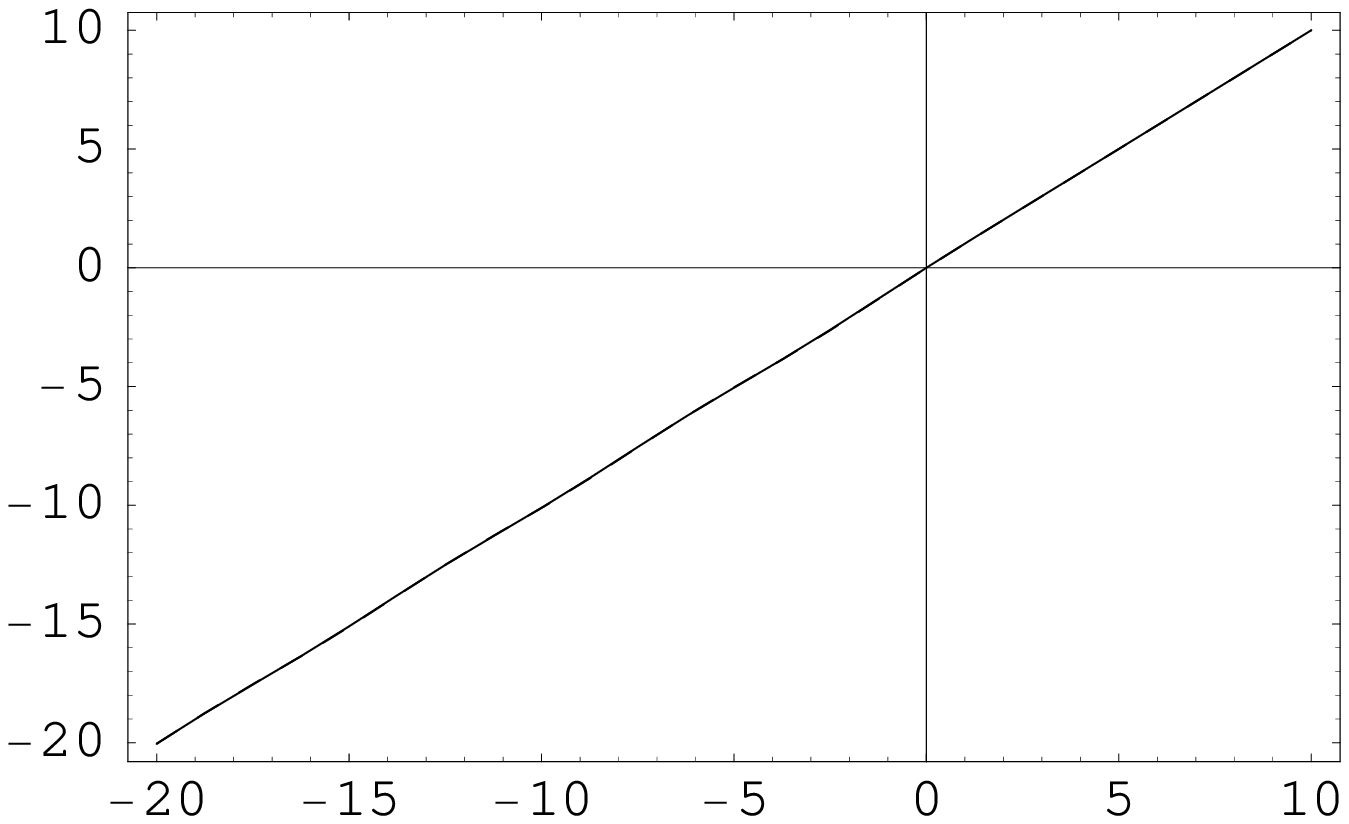}
\null
\hspace{-2.1cm}$x$
\caption{The function $\sigma(x)$,  for a more complicated comparison
function, to $0^{\rm th}$ and $1^{\rm st}$-orders in $\epsilon$, solid and
dashed lines respectively. As in the linear case one can not distinguish
the two curves.}
\label{sigma_figE}
\end{figure}

For the comparison function $\Theta^2(\sigma) = {\rm e}^\sigma-1$ the
quantities $y_0(x)$ is implicitly given by Eq.~(\ref{nonLanger01}), while
$y_1(x)$ is given by Eq.~(\ref{nonLangernsol}) on using
Eqs.~(\ref{nonLangerF1}) and (\ref{G-define}). In Fig.~\ref{y1_figE}
we show the function $y_1(x)$ while in Fig.~\ref{y0_y01_figE} we plot the
function $y(x)$ at $0^{\rm th}$ and $1^{\rm st}$-orders in $\epsilon$.
In Fig.~\ref{G_fig} we exhibit the function $G(x)$ from Eq.~(\ref{G-define}).
The function $\sigma(x)$ at $0^{\rm th}$ and $1^{\rm st}$-orders in
$\epsilon$, calculated from Eq.~(\ref{sigma-define}), is shown in
Fig.~\ref{sigma_figE}.
\par
\begin{figure}[!ht]
\includegraphics[width=0.5\textwidth]{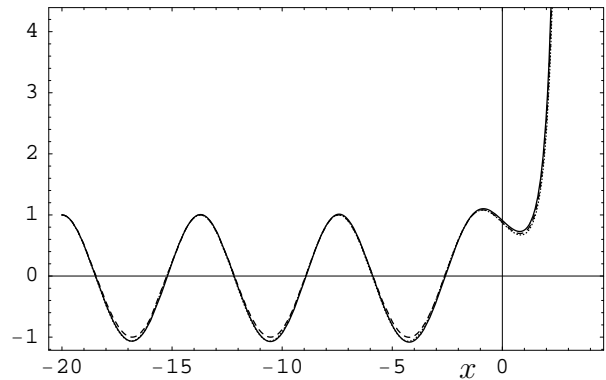}
\null
\hspace{-2.1cm}$x$
\caption{Solution $\chi(x)$ of Eq.~(\ref{equation_app}) for a more complicated
comparison function; the solid line is the full numerical solution, while the
dashed and dotted lines are respectively the $0^{\rm th}$ and
$1^{\rm st}$-orders in $\epsilon$.}
\label{sol_figE}
\end{figure}

The analytical part of the solution follows from
\begin{eqnarray}
\frac{d^2U(\sigma)}{d\sigma^2} = \frac{1}{\varepsilon}
\left({\rm e}^\sigma-1\right) U(\sigma)
\ ,
\label{equation1_appE}
\end{eqnarray}
and it is a linear combination of Bessel functions $I$
\begin{eqnarray}
U\left[\sigma(x)\right] =A\cdot I_{-\frac{2\,i}{\sqrt{\epsilon}}}
\left[\frac{2\,\sqrt{\sigma(x)}}{\sqrt{\epsilon}}\right]
+B\cdot I_{\frac{2\,i}{\sqrt{\epsilon}}}
\left[\frac{2\,\sqrt{\sigma(x)}}{\sqrt{\epsilon}}\right]
\ .
\label{analatic_sol_appE}
\end{eqnarray}
Finally the complete solution becomes
\begin{eqnarray}
\chi(x)&=&y(x)\,\left\{
A\cdot I_{-\frac{2\,i}{\sqrt{\epsilon}}}
\left[\frac{2\,\sqrt{\sigma(x)}}{\sqrt{\epsilon}}\right]
+B\cdot I_{\frac{2\,i}{\sqrt{\epsilon}}}
\left[\frac{2\,\sqrt{\sigma(x)}}{\sqrt{\epsilon}}\right]
\right\}
\ .
\label{analatic_sol_tot_appE}
\end{eqnarray}
In Fig.~\ref{sol_figE} we exhibit the exact numerical solution, the first and
second order solutions in our approximation scheme.
\par
In Fig.~\ref{sol_fig} we can note that the leading solution calculated with a
linear comparison function is close to the true solution in the oscillating
region, while the next-to-leading solution is better in the exponential
region.  It is quite natural because an iterative scheme is more stable in
regions where a function under consideration has a monotonic behaviour.
From Fig.~\ref{sol_fig} and Fig.~\ref{sol_figE} we can see that the more
complicated comparison function leads to better solutions than the linear
one. This feature is true because the more complicated comparison function
remains close to the original frequency in the whole range of interest, in
contrast with the linear case.
 \chapter{Other results in the literature}
\label{app_other_res}
We use Table~\ref{convers1} and Eqs.~(\ref{HFFvsHSR}),
to compare the results for spectral indices
- and for PS and runnings when available - in terms
of the HFF in Tables~\ref{LLMS_SG_tab} and~\ref{H_MP_tab_PRD71}.
\par
\begin{table*}[!ht]
\caption{Results obtained with the Green's function method~\cite{LLMS}
suggested in Ref.~\cite{SGong};
$C\equiv\ln 2+\gamma_E-2\simeq  -0.7296$ with $\gamma_E$
the Euler-Mascheroni constant.
We display the original results for spectral indices,
$\alpha$-runnings and tensor-to-scalar ratio evaluated at the pivot
scale $k_*$.}
\centerline{
\begin{tabular}{ l }
\hline
\hline
\centerline{
E.D.~Stewart \textit{et al.} \cite{SL},
E.D.~Stewart \textit{et al.} \cite{SGong},
}
\\
\centerline{
S.M.~Leach \textit{et al.} \cite{LLMS},
A.R.~Liddle \textit{et al.} \cite{LPB1,LPB2}.
}
\\
\hline
\\
$
\mathcal{P}_\zeta=
\frac{H^2}{\pi\,\epsilon_1\,m_{\rm Pl}^2}\,
\left\{
1
-2\left(C+1\right)\,
\epsilon_1
-C\,
\epsilon_2
+\left(2\,C^2+2\,C+\frac{\pi^2}{2}-5\right)\,
\epsilon_1^2
\right.$
\\
\\
$\phantom{aaaa}\left.
+\left(C^2-C+\frac{7\,\pi^2}{12}-7\right)\,
\epsilon_1\,\epsilon_2
+\left(\frac12\,C^2+\frac{\pi^2}{8}-1\right)\,
\epsilon_2^2
+\left(-\frac12\,C^2+\frac{\pi^2}{24}\right)\,
\epsilon_2\,\epsilon_3
\right.$
\\
\\
$\phantom{aaaa}\left.
+\left[-2\,\epsilon_1-\epsilon_2+2\left(2\,C+1\right)\,\epsilon_1^2
+\left(2\,C-1\right)\,\epsilon_1\,\epsilon_2
+C\,\epsilon_2^2
-C\,\epsilon_2\,\epsilon_3\right]\,
\ln\left(\frac{k}{k_*}\right)
\right.$
\\
\\
$\phantom{aaaa}\left.
+\frac12\,\left(4\,\epsilon_1^2
+2\,\epsilon_1\,\epsilon_2+\epsilon_2^2
-\epsilon_2\,\epsilon_3\right)
\,\ln^2\left(\frac{k}{k_*}\right)
\right\}$
\\
\\
$\mathcal{P}_h=
\frac{16\,H^2}{\pi\,m_{\rm Pl}^2}\,
\left\{
1-2\left(C+1\right)\,\epsilon_1
+\left(2\,C^2+2\,C+\frac{\pi^2}{2}-5\right)\,\epsilon_1^2
+\left(-C^2-2\,C+\frac{\pi^2}{12}-2\right)\,
\epsilon_1\,\epsilon_2
\nonumber
\right.$
\\
\\
$\phantom{aaaa}\left.
+\left[-2\,\epsilon_1+2\left(2\,C+1\right)\,\epsilon_1^2
-2\,\left(C+1\right)\epsilon_1\,\epsilon_2\right]
\,\ln\left(\frac{k}{k_*}\right)
+\frac12\,\left(4\,\epsilon_1^2
-2\,\epsilon_1\,\epsilon_2\right)\,\ln^2\left(\frac{k}{k_*}\right)
\right\}$
\\
\\
$n_{\rm S}-1=-2\,\epsilon_1-\epsilon_2-2\,\epsilon_1^2
-\left(2\,C+3\right)\,\epsilon_1\,\epsilon_2
-C\,\epsilon_2\,\epsilon_3$
\ ,\quad
$n_{\rm T}=-2\,\epsilon_1-2\,\epsilon_1^2-2\,\left(C+1\right)\,\epsilon_1\,\epsilon_2$
\\
\\
$\alpha_{\rm S}=-2\,\epsilon_1\,\epsilon_2-\epsilon_2\,\epsilon_3$
\ ,\quad
$\alpha_{\rm T}=-2\,\epsilon_1\,\epsilon_2$
\\
\\
$R=16\,\epsilon_1\left[1+C\,\epsilon_2
+\left(C-\frac{\pi^2}{2}+5\right)\,\epsilon_1\,\epsilon_2
+\left(\frac12\,C^2-\frac{\pi^2}{8}+1\right)\,\epsilon_2^2
+\left(\frac12\,C^2-\frac{\pi^2}{24}\right)\,\epsilon_2\,\epsilon_3
\right]$
\\
\\
\hline
\hline
\end{tabular}
}
\label{LLMS_SG_tab}
\end{table*}
In Table~\ref{LLMS_SG_tab} we summarize the results for scalar and tensor
perturbations given in Ref.~\cite{LLMS}, as calculated with the method
proposed in Ref.~\cite{SGong}.
We also refer to some other papers which give the same results,
although expressed in different hierarchies.
\par
Let us then compare with the results obtained by the uniform
approximation~\cite{H_MP,HHHJM,HHHJ}.
The integrand in Eq.~(9) of Ref.~\cite{H_MP} can be obtained
from our Eq.~(\ref{eq_wkb_expb}) by the change of variables
$x=\ln\left(-k\,\eta\right)$, $u={\rm e}^{-x/2}\,\mu$, which
of course differs from our Eq.~(\ref{transf}).
Therefore, the integrands in our Eq.~(\ref{xi_II}) and Eq.~(9) of
Ref.~\cite{H_MP} differ by ${\cal O}(\epsilon_i^4)$ terms, once
both are expressed in conformal time.
We also note that in Refs.~\cite{H_MP,HHHJM,HHHJ}, the authors just give
expressions for the spectral indices which depend on $k$ (i.e.~without
referring to a pivot scale $k=k_*$).
Further, it is not easy to see that they obtain the correct scaling in
$\ln\left(k/k_*\right)$, which yields the $\alpha$-runnings, and
the $\epsilon_i$'s are evaluated at the classical turning points for the
frequencies.
\par
\begin{table}[!ht]
\caption{Results obtained with the uniform approximation.
\label{H_MP_tab_PRD71}}
\centerline{
\begin{tabular}{ c }
\hline
\hline
S.~Habib \textit{et al.}~\cite{HHHJ}
\\
\hline
%\\
$n_{\rm S}-1=-2\,\epsilon_1-\epsilon_2-2\,\epsilon_1^2
-\left(\frac{20}{3}-2\,\pi\right)\,\epsilon_1\,\epsilon_2
-\left(\frac{11}{6}-\pi\right)\,\epsilon_2\,\epsilon_3$
\\
%\\
$n_{\rm T}=-2\,\epsilon_1-2\,\epsilon_1^2
-\left(\frac{14}{3}-\frac{3}{2}\pi\right)\,\epsilon_1\,\epsilon_2$
\\
%\\
\hline
\hline
\end{tabular}
}
\end{table}
In Tables~\ref{H_MP_tab_PRD71}, we
display the results for scalar and tensor perturbations
given in Ref.~\cite{HHHJ}.
As it was properly stated in Ref.~\cite{HHHJ}, the results for scalar
and tensor spectral indices given in Ref.~\cite{H_MP,HHHJM} were not
fully expanded to second order in the slow-roll parameters.

\chapter{Undetermined parameters in the WKB*}
\label{und_par}
\begin{figure}[!ht]
\includegraphics[width=0.5\textwidth]{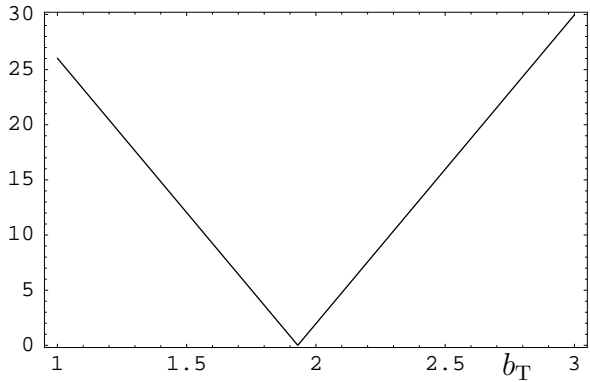}
\null
\hspace{-1.4cm}$b_{\rm T}$
\caption{Percentage difference between the numerical coefficients of
$\epsilon_1\,\epsilon_2$ in $P_h$ from ${\rm WKB*}$ and ${\rm GFM}$ as
a function of $b_{\rm T}$.
The difference vanishes for $b_{\rm T}\simeq 1.93$ ($\simeq 2$, the value
used in the main text).}
\label{param_plot}
\end{figure}

To clarify the impact of the undetermined parameters
$b_{\rm S}$, $b_{\rm T}$, and $d_{\rm S}=2$ on the
results to second order in the HFF, the percentage differences
between the numerical coefficients of some second order terms in the
power spectra are displayed in Table~\ref{perc_diff_tab_PS} and of some
of those in the scalar-to-tensor ratio in Table~\ref{perc_diff_tab_R}.
All the entries are for $b_{\rm S}=b_{\rm T}=d_{\rm S}=2$, as is used in
the main text, since other choices would give larger differences, as we
explicitly show in Fig.~\ref{param_plot} for the numerical coefficient
in front of $\epsilon_1\,\epsilon_2$ in $P_h$ (since the behaviour is
the same for all coefficients, we do not display other plots).
\begin{table}[!ht]
\begin{tabular}{|c|c|c|c|c|}
\hline
&$\epsilon_1^2$; $P_\zeta$, $P_h$
&$\epsilon_2^2$; $P_\zeta$
&$\epsilon_1\,\epsilon_2$; $P_\zeta (b_{\rm S}=2)$
&$\epsilon_1\,\epsilon_2$; $P_h (b_{\rm T}=2)$
\\
\hline
${\rm WKB}$
&$2.3\%$
&$3.0\%$
&$353.3\%$
&$37.2\%$
\\
\hline
${\rm WKB*}$
&$0.03\%$
&$0.06\%$
&$29.1\%$
&$2.0\%$
\\
\hline
\end{tabular}
\caption{Percentage differences between the numerical coefficients
of some second order terms in the power spectra given by ${\rm WKB}$
and ${\rm WKB*}$ with respect to those obtained by the GFM.}
\label{perc_diff_tab_PS}
\end{table}
\begin{table}[!ht]
\begin{tabular}{|c|c|c|c|}
\hline
&$\epsilon_2\,\epsilon_3$; $(d_{\rm S}=2)$
&$\epsilon_1\,\epsilon_2$; $(b_{\rm S}-b_{\rm T}=0)$
&$\epsilon_2^2$
\\
\hline
${\rm WKB}$
&$56.7\%$
&$1.5\%$
&$117.3\%$
\\
\hline
${\rm WKB*}$
&$1.3\%$
&$0.03\%$
&$1.5\%$
\\
\hline
\end{tabular}
\caption{Percentage differences between the numerical coefficients of
some second order terms in the scalar-to-tensor ratio given by ${\rm WKB}$
and ${\rm WKB*}$ with respect to those obtained from GFM.
Note that the numerical coefficient of $\epsilon_2\,\epsilon_3$ in $R$
is the same as that in $P_\zeta$.
Further, the coefficient of $\epsilon_1\,\epsilon_2$ does not depend
on any undetermined parameters if $b_{\rm S}=b_{\rm T}$.}
\label{perc_diff_tab_R}
\end{table}

\chapter{Perturbative approximations}
\label{VENTURI_pert}
In this Appendix we present a possible method of obtaining
the complete result to second order in the HFF.
We start by noting that any given function $f=f(x)=f(\epsilon,x)$ can
be written as the power series
\be
f(\epsilon,x)
=
\sum_{n=0}^\infty\,\frac{x^n}{n!}\,
\left(-\sum_i\,\frac{\hat\epsilon_i\,\hat\epsilon_{i+1}}{1-\hat\epsilon_1}\,
\frac{\partial}{\partial\hat\epsilon_i}\right)^n
f(\hat\epsilon_i,x)
\equiv
\sum_{n=0}^\infty\,\frac{x^n}{n!}\,f_n(x)
\ ,
\label{f_def}
\ee
where $\hat\epsilon_i=\epsilon_i(x_*)$ are the HFF evaluated at the
horizon crossing (i.e.~at $x=x_*$) and 
$f(\hat\epsilon_i,x)=f(\epsilon_i=\hat\epsilon_i,x)$.
On using Eqs.~(\ref{f_def}) to express both $\omega^2$ and
$\chi$ in Eq.~(\ref{new_eqa}), we obtain
\be
\chi_0(x)\,\sum_{n=1}^\infty\,\frac{x^n}{n!}\,\omega^2_n(x)
+\left[\frac{{\d}^2}{{\d}x^2}
+\sum_{m=0}^\infty\,\frac{x^m}{m!}\,\omega^2_m(x)
\right]\,
\sum_{n=1}^\infty\,\frac{x^n}{n!}\,\chi_n(x)
=0
\ ,
\label{exact_EQ_ter}
\ee
where we have used
\be
\left[\frac{{\d}^2}{{\d}x^2}+\omega^2_0(x)\right]
\,\chi_0(x)=0
\ ,
\label{0th_EQ}
\ee
with
$\omega^2_0(x)=\omega^2(\epsilon_i=\hat{\epsilon}_i;x)$
and $\chi_0(x)$ is a linear combination (i.e.~the MCE leading order solution)
of Eq.~(\ref{exact_SOL_bis}) for $\Theta^2(x)=\omega^2_0(x)$ and
$\sigma\to x$ in the argument of the Bessel functions.
One may solve Eq.~(\ref{exact_EQ_ter}) for higher orders by introducing
\be
\frac{x^n}{n!}\,\chi_n(x)\equiv g_n(x)\,\chi_0(x)
\ ,
\label{def_to_solve}
\ee
and using the standard formulae for second order differential equations with a
source term, so that
\be
g_n(x)=
-\int_{\infty}^x\,\frac{\d y}{\chi^2_0(y)}\,
\int_{\infty}^{y}\,\d z\,\chi^2_0(z)
\left[
\frac{z^n}{n!}\,\omega^2_n(z)
+\sum_{m=1}^{n-1}\,
g_m(z)\,\frac{z^{n-m}}{n-m}\,\omega^2_{n-m}(z)
\right]
\ .
\label{g_n_GEN}
\ee
The problem has therefore been reduced to computing $2\,n$ integrals.
\par
The above formalism can be modified in order to provide an
alternative way for calculating the power spectra,
spectral indices and runnings to second order in the HFF.
We start from Eq.~(\ref{eq_wkb_expb}) using the exact
relations
%
%\numparts
\be
\frac{z''_{\rm T}}{z_{\rm T}}
&=&
a^{2}\,H^{2}\,\left(2-\epsilon_1\right)
\nonumber
%\label{potT}
\\
\\
\frac{z''_{\rm S}}{z_{\rm S}}
&=&
a^{2}\,H^{2}\,\left(2-\epsilon_1+\frac{3}{2}\,\epsilon_2
-\frac{1}{2}\,\epsilon_{1}\,\epsilon_2
+\frac{1}{4}\,\epsilon_2^2
+\frac{1}{2}\,\epsilon_2\,\epsilon_3\right)
\ .
\nonumber
%\label{potS}
\ee
%\endnumparts
%
With a redefinitions of the wave-function and variable
\be
\psi={\rm e}^{-x/2}\,\mu
\ ,
\quad\quad
\tilde{x}=\ln\left(-k\,\eta\right)
\ee
we obtain
\be
\left[\frac{{\d}^2}{{\d}\tilde{x}^2}
+\tilde\omega^2(\tilde x)\right]\,\psi(\tilde x)
=0
\ ,
\ee
where
%\numparts
\be
\tilde\omega^2_{\rm T}&=&
{\rm e}^{2\,\tilde{x}}-\frac14
-\eta^2a^2H^2\left(2-\epsilon_1\right)
\nonumber
%\label{OUR_Green_EQ_Tpre}
\\
\\
\tilde\omega^2_{\rm S}&=&
{\rm e}^{2\,\tilde{x}}-\frac14
-\eta^2a^2H^2\left(2-\epsilon_1+\frac{3}{2}\,\epsilon_2-\frac{1}{2}\,\epsilon_{1}\,\epsilon_2
+\frac{1}{4}\,\epsilon_2^2+\frac{1}{2}\,\epsilon_2\,\epsilon_3\right)
\ .
\nonumber
%\label{OUR_Green_EQ_Spre}
\ee
%\endnumparts
%
On now using the approximate relation
\be
\eta\simeq-\frac{1}{a\,H}\,
\left(1+\epsilon_1+\epsilon_1^2+\epsilon_1\,\epsilon_2\right)
\ ,
\ee
the frequencies become
%
%\numparts
\be
\tilde\omega^2_{\rm T}&\simeq&
{\rm e}^{2\,\tilde{x}}-
\left(\frac94+3\,\epsilon_1+4\,\epsilon_1^2+4\,\epsilon_1\epsilon_2\right)
\nonumber
%\label{OUR_Green_EQ_T}
\\
\\
\tilde\omega^2_{\rm S}&\simeq&
{\rm e}^{2\,\tilde{x}}-
\left(\frac94+3\,\epsilon_1+\frac32\,\epsilon_2+4\,\epsilon_1^2
+\frac{13}{2}\,\epsilon_1\epsilon_2+\frac14\,\epsilon_2^2
+\frac12\,\epsilon_2\epsilon_3\right)
\ ,
\nonumber
%\label{OUR_Green_EQ_S}
\ee
%\endnumparts
%
where we only keep the second order terms in the (still
$\tilde{x}$-dependent) HFF.
We can now expand linearly each HFF around the horizon
crossing to obtain
%
%\numparts
\be
&&
\frac{{\d}^2\,\psi_{\rm T}}{{\d}\,\tilde{x}^2}
\!+\!\left[{\rm e}^{2\,\tilde{x}}\!-\!\left(\frac94
+3\,\hat{\epsilon}_1+4\,\hat{\epsilon}_1^2+4\,\hat{\epsilon}_1\hat{\epsilon}_2\right)\right]
\!
\psi_{\rm T}=-3\,\hat{\epsilon}_1\,\hat{\epsilon}_2\,\tilde{x}\,\psi_{\rm T}
\nonumber
%\label{OUR_Green_EQ_Tbis}
\\
\\
&&
\frac{{\d}^2\,\psi_{\rm S}}{{\d}\,\tilde{x}^2}
\!+\!\left[{\rm e}^{2\,\tilde{x}}\!-\!\left(\frac94
+3\,\hat{\epsilon}_1+\frac32\,\hat{\epsilon}_2+4\,\hat{\epsilon}_1^2
+\frac{13}{2}\,\hat{\epsilon}_1\hat{\epsilon}_2+\frac14\,\hat{\epsilon}_2^2
+\frac12\,\hat{\epsilon}_2\hat{\epsilon}_3\right)\right]\!
\psi_{\rm S}
\nonumber
\\
&&\quad\quad\quad\quad
=
-3\left(\hat{\epsilon}_1\,\hat{\epsilon}_2
+\frac12\,\hat{\epsilon}_2\,\hat{\epsilon}_3\right)\tilde{x}\,\psi_{\rm S}
\ ,
\nonumber
%\label{OUR_Green_EQ_Sbis}
\ee
%\endnumparts
%
which, beside the use of $\tilde x$ instead of $x$ and the appearance of
$\tilde\omega$ instead of $\omega$, is equivalent to the expansion in
Eq.~(\ref{f_def}) for $f=\tilde\omega^2$ and can be solved by means of a
standard perturbative approach (such as in Ref.~\cite{SGong}).
The method is close to GFM and gives the same results of Eq.~(41) in
Ref.~\cite{SGong} and Eq.~(29) in Ref.~\cite{LLMS} for the scalar and
tensor power spectra, respectively.
\par
Let us finally note that, for the GFM, the scalar spectral index and its running
are calculated by taking the derivative of Eq.~(41) in Ref.~\cite{SGong},
i.e.~the power spectra evaluated at $k=a\,H$, in terms of $\ln k$, while in our MCE
(but also in the WKB approximation~\cite{WKB_let,WKB_SRprD}) the 
dependences
in $\ln\left(k/k_*\right)$ are obtained explicitly (see, for example,
Eqs.~(\ref{PS_SlowRoll_0_2order}) and (\ref{PT_SlowRoll_0_2order})).

\chapter{MCE and Black Holes}
\label{MCE_BH}
In this Appendix we employ the Method of Comparison Equations to study the propagation
of a massless minimally coupled scalar field on the Schwarzschild background as an
example of a potential with two turning point in the wave equation.
In particular, we show that this method allows us to obtain explicit approximate
expressions for the radial modes with energy below the peak of the effective potential
which are fairly accurate over the whole region outside the horizon.
This case can be of particular interesting, for example,
for the problem of black hole evaporation.
\par
%
%
%
%
%\section{Introduction}
%
Wave equations on black hole backgrounds cannot be solved exactly
in terms of known simple functions (for a review, see, {\em e.g.}, Ref.~\cite{chandra}).
Although it is usually fairly easy to find asymptotic expressions near the (outer)
horizon or at spatial infinity, numerical analysis is what one eventually employs
to obtain a detailed profile of the wave-functions.
However, our understanding of the field dynamics on curved backgrounds
would still benefit of analytical approximations to the exact modes,
for example in the study of the (renormalized) energy-momentum tensor of quantum
fields~\cite{birrel} and, in particular, of the Hawking radiation~\cite{hawking1,hawking2},
or of quasi-normal modes~\cite{nollert}.
\par
By making use of the black hole symmetry, one can usually separate the wave-functions
and reduce the problem to the one task of solving a second order differential equation
for the radial part.
The latter takes the form of the one-dimensional Schr\"odinger equation for the
transmission through a potential barrier~\cite{chandra}
\be
\left[\frac{\d^2}{\d x^2}+Q_s(x)\right]\psi=0
\ ,
\label{gen}
\ee
and, for the general case of a Kerr-Newman black hole, is known as the Teukolsky
master equation for waves of spin-weight $s=0$, $1$ and $2$~\cite{teukolsky,chandra}.
\par
In the next Section, we shall briefly review the equation of motion
for massless scalar fields on a spherically symmetric black hole
background.
This includes the simplest case of linear perturbations of a
Schwarzschild black hole which will be used to show the
application of our method in all details (including the use of a convenient
radial coordinate and rescaling of the wave-function).
In Section~\ref{sMCE_BH}, we shall then apply the MCE and obtain
approximate expressions for scalar wave-modes with energy below the
peak of the potential.
This case is of particular interest, for example, for evaluating the grey-body
factors involved in the Hawking effect~\cite{hawking1,hawking2}.
It also presents a major complication with respect to the typical case of Cosmological Pertubations,
namely the presence of two zeros of the potential (``turning points'').
We shall use units with $c=G=1$.
The last three Sections will give some tecnical results which complete this Appendix.
\section{Scalar field on Schwarzschild background}
\label{sSBH}
Let us begin by recalling that the metric in a static and spherically
symmetric vacuum space-time can be written in general as
\be
\d s^2=h(r)\,\d t^2-\frac{\d r^2}{h(r)}-r^2\,\d\Omega^2
\ ,
\label{metric}
\ee
where $\d\Omega$ is the area element on the unit two-sphere.
The propagation of massless, minimally coupled scalar particles in this
background is governed by the Klein-Gordon equation
$\Phi^{;\mu}_{\ \ ;\mu}=0$.
Due to the symmetry of the background, the field $\Phi$ can be decomposed into
eigenmodes of normal frequency $\bar\omega$ and angular momentum
numbers $\ell$, $m$ as~\cite{chandra}
\be
\Phi(t,r,\Omega)={\rm e}^{-i\,\bar\omega\,t}\,Y^{m}_{\ell}(\Omega)\,R(r)
\ ,
\ee
where $Y_\ell^m$ are spherical harmonics.
The Klein-Gordon equation then separates and the dynamics
is described by the function $R$ which satisfies the radial equation
\be
\frac{\d}{\d r}\left[r^2\,h(r)\,\frac{\d R(r)}{\d r}\right]
+\left[\frac{\bar\omega^2\,r^2}{h(r)}-\ell\,(\ell+1)\right]R(r)=0
\ ,
\label{rad_eq}
\ee
where $\ell\,(\ell+1)$ is the separation constant.
No exact solution of the above equation is known even for as simple a background
as the four-dimensional Schwarzschild space-time with
\be
h(r)=1-\tilde{r}^{-1}
\ ,
\label{h_schw}
\ee
where we have introduced the dimensionless radial coordinate
$\tilde r=r/r_{\rm H}$ and $r_{\rm H}$ is the Schwarzschild radius.
\par
It is now convenient to introduce the ``tortoise-like''
coordinate~\footnote{A similar coordinate was already used
in Ref.~\cite{kanti_russ}.}
\begin{subequations}
\be
\d\,x=\frac{\d\,\tilde r}{\tilde r\,h(\tilde r)}
\ ,
\label{new_var}
\ee
and the new radial function
\be
R\left[\tilde r(x)\right]
={\rm exp}\left[-\frac12\,\int^{x}\,h(x')\,\d\,x'\right]\chi\left[\tilde r(x)\right]
\ ,
\label{new_funct}
\ee
\end{subequations}
so that the radial equation takes the Schr\"odinger form
\be
\frac{\d^2\,\chi(x)}{\d x^2}
+\left\{\tilde\omega^2\,\tilde r^2(x)-\ell\,(\ell+1)\,h[\tilde r(x)]
-\frac{1}{4}\,h^2[\tilde r(x)]-\frac{1}{2}\,\frac{\d h[\tilde r(x)]}{\d x}\right\}\chi(x)=0
\ .
\label{new_rad_eq}
\ee
where we also defined the dimensionless energy
$\tilde{\omega}\equiv\bar\omega\,r_{\rm H}$.
For the four-dimensional Schwarzschild case we have
\be
x(\tilde{r})=\ln (\tilde{r}-1)
\ ,
\label{hatr_r}
\ee
so that the new variable $x$ ranges from
$x(\tilde r_{\rm H}=1)=-\infty$ to $x(\tilde r=+\infty)=+\infty$
and $x$ and $\chi(x)$ are of Langer's form~\cite{langer}.
The Klein-Gordon equation finally becomes
\be
\left[
\frac{{\d}^2}{{\d}x^2}+\omega^2(x)
\right]
\,\chi(x)=0
\ ,
\label{exact_EQ_MCE}
\ee
where the ``frequency''
\be
\omega^2(x)
=
\tilde{\omega}^2\left(1+{\rm e}^{x}\right)^2
-\frac{\ell\,(\ell+1)\,{\rm e}^{x}}{1+{\rm e}^{x}}
-\frac{{\rm e}^{x}\left(2+{\rm e}^{x}\right)}{4\,\left(1+{\rm e}^{x}\right)^2}
\ ,
\label{freq_ex}
\ee
is not necessarily a positive quantity.
In fact, the above expression shows two qualitatively
different behaviors depending on the values of $\tilde{\omega}$
and $\ell$.
In particular, for a given angular momentum $\ell$, there exists
a critical value
\be
\tilde\omega_{\rm c}=\tilde\omega_{\rm c}(\ell)
\ ,
\label{om_c}
\ee
such that if the energy $\tilde{\omega}>\tilde\omega_{\rm c}$ there are
no turning points, otherwise there are two, say
$\omega(x_1)=\omega(x_2)=0$ with $x_1<x_2$.
The latter case is the one we want to analyze in detail
(see Sec.~\ref{app_TPs} for more details and
the solid line in Fig.~\ref{plot1} for an example).
\section{MCE solutions}
\label{sMCE_BH}
Referring to Section~\ref{sMCEa} for the basic formulae characterising the MCE,
we can find the approximate solution of Eq.~(\ref{exact_EQ_MCE}),
\be
\chi_{\rm MCE}(x)=\left(\frac{{\d}\sigma}{{\d}x}\right)^{-1/2}
U(\sigma)
\ ,
\label{chiMCE}
\ee
which is valid in the whole range of $x$ and including
the turning points.
\par
\begin{figure}[!ht]
\includegraphics[width=0.5\textwidth]{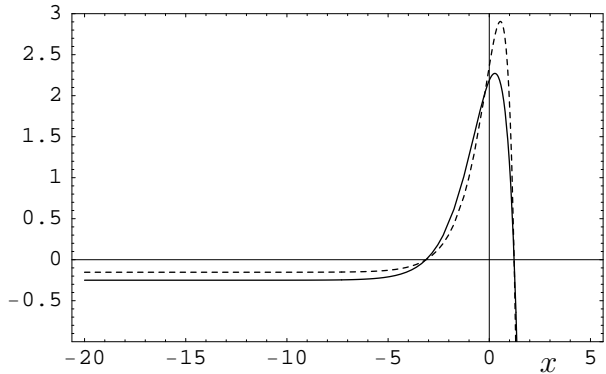}
\null
\hspace{-1cm}$x$
\caption{Comparison between the exact potential $-\omega^2(x)$ (solid line)
and the approximate Morse potential $-\Theta^2[\sigma(x)]$ obtained after
solving Eqs.~(\ref{A_B}) and~(\ref{third_cond}) (dotted line) for
$\tilde\omega=1/2$ and $\ell=2$.
The two turning points are $x_1\approx -3.13$ and $x_2\approx 1.23$.}
\label{plot1}
\end{figure}

We are now dealing with a problem of the form~(\ref{exact_EQ_MCE})
with two turning points, $x_1<x_2$.
In order to implement the MCE, we need a ``comparison frequency''
with the same behavior.
We shall use the Morse potential~\cite{RMNS}
\be
\Theta^2(\sigma)=
A\,{\rm e}^{2\,\sigma}
-B\,{\rm e}^{\sigma}
+D
\ .
\label{comp_freq}
\ee
The coefficients $A$ and $B$ are then fixed by imposing
that the turning points of the comparison function are
the same as those of the exact frequency, that is
\begin{subequations}
\be
\Theta(\sigma_1=x_1)=\Theta(\sigma_2=x_2)=0
\ ,
\label{A_B}
\ee
and the coefficient $D$ will then follow form the relation
\be
\xi\equiv
\int_{x_1}^{x_2}\sqrt{-\,\omega^2(y)}\,{\d}\,y
=
\int_{x_1}^{x_2}\sqrt{-\,\Theta^2(\rho)}\,{\d}\,\rho
\ .
\label{third_cond}
\ee
\end{subequations}
The system of non-linear equations~(\ref{A_B}) and (\ref{third_cond}) thus
yields
\begin{subequations}
\be
A&\!=\!&
\frac{4\,\xi^2}{\pi^2\left({w_1+w_2}-2\,\sqrt{w_1\,w_2}\right)^2}
\label{coeff_A}
\\
B&\!=\!&
\left(w_1+w_2\right)A
%\frac{4\,\xi^2\,(w_1+w_2)}{\pi^2 \left(w_1+w_2-2\,\sqrt{w_1\,w_2}\right)^2}
\label{coeff_AB}
\\
D&\!=\!&
w_1\,w_2\,A
%\frac{4\,\xi^2\,w_1\,w_2}{\pi^2 \left(w_1+w_2-2\,\sqrt{w_1\,w_2}\right)^2}
\ ,
\label{coeff_D}
\ee
\end{subequations}
where the exact expression for $\xi$ is given in Sec.~\ref{app_xi}.
A sample plot of both the exact frequency and the comparison function
thus obtained is given in Fig.~\ref{plot1}.
\par
The ``comparison solution'' is a linear combination of confluent hypergeometric
functions~\cite{abram} of the type ${}_1{\rm F}_1$ (for more details,
see Ref.~\cite{RMNS}),
\be
U(\sigma)&\!=\!&
C_+\,{\rm e}^{i\,\left(\sqrt{A}\,{\rm e}^{\sigma}+\sqrt{D}\,\sigma\right)}\,
{}_1{\rm F}_1
\left(\frac12+i\,\sqrt{D}+i\,\frac{B}{2\,\sqrt{A}},1+2\,i\,\sqrt{D},
-2\,i\,\sqrt{A}\,{\rm e}^\sigma\right)
\nonumber
\\
&&\!\!\!\!\!\!\!\!\!\!\!\!\!\!\!\!\!\!\!\!
+C_-\,{\rm e}^{i\,\left(\sqrt{A}\,{\rm e}^{\sigma}-\sqrt{D}\,\sigma\right)}\,
{}_1{\rm F}_1
\left(\frac12-i\,\sqrt{D}+i\,\frac{B}{2\,\sqrt{A}},1-2\,i\,\sqrt{D},
-2\,i\,\sqrt{A}\,{\rm e}^\sigma\right)
\ ,
\label{sol_morse}
\ee
and $\sigma(x)$ is given implicitly by Eq.~(\ref{new_EQ_int}) with
\be
&&\int_{x_1}^{\sigma}\sqrt{\Theta^2(\rho)}\,{\d}\,\rho
=
\sqrt{\Theta^2(\sigma)}
\label{sigma_of_x}\\
&&
-\ln\left[
\left(2\,A\,{\rm e}^\sigma-B
+2\,\sqrt{A\,\Theta^2(\sigma)}
\right)^{\frac{B}{2\,\sqrt{A}}}
\left(2\,D\,{\rm e}^{-\sigma}-B
+2\,{\rm e}^{-\sigma}\,\sqrt{\,D\,\Theta^2(\sigma)}
\right)^{\sqrt{D}}
\right]
\ .
\nonumber
\ee
\noindent
The above expression is rather complex.
Hence, we just consider the asymptotic relation between
$\sigma$ and $x$ to the left of the smaller  turning point
(that is, for $x\ll x_1$)
\begin{subequations}
\be
\sigma(x)
\simeq
\frac{\tilde\omega}{\sqrt{D}}\,x
\label{asy_-inf}
\ee
and to the right of the larger one ($x\gg x_2$)
\be
\sqrt{A}\,{\rm e}^{\sigma(x)}-\frac{B}{2\,\sqrt{A}}\,\sigma(x)
\simeq
\tilde\omega\left({\rm e}^{x} +x\right)
\ .
\label{asy_+inf}
\ee
\end{subequations}
\par
As an approximate expression for $\sigma(x)$
for $x<\kappa_1\,x_1$ (with $\kappa_1\gtrsim 1$)
we shall then use Eq.~(\ref{asy_-inf}) and the condition that
$\sigma(x_1)=x_1$.
Analogously, for $x>\kappa_2\,x_2$ (with $\kappa_2\gtrsim 1$)
we shall use the approximate solution of Eq.~(\ref{asy_+inf})
to next-to-leading order for large $x$.
Moreover, in the region between $\kappa_1\,x_1$ and $\kappa_2\,x_2$
we shall use an interpolating cubic function.
The reason for introducing two new parameters $\kappa_1$ and
$\kappa_2\gtrsim 1$ is that the asymptotic forms in
Eqs.~(\ref{asy_-inf}) and~(\ref{asy_+inf}) may differ significantly from the correct
$\sigma(x)$ around the turning points $x_1$ and $x_2$ and suitable values of
$\kappa_1$ and $\kappa_2$ usually improve the final result.
To summarize, we have
\be
\sigma\!=\!
\left\{
\begin{array}{ll}
\strut\displaystyle\frac{\tilde\omega}{\sqrt{D}}\,x
+x_1\left(1-\frac{\tilde\omega}{\sqrt{D}}\right)
\ ,
&
x_0<x<\kappa_1 x_1
\\
\\
C_0+C_1\,x+C_2\,x^2+C_3\,x^3
,
&
\kappa_1 x_1<x<\kappa_2 x_2
\\
\\
x+\ln\left(\strut\displaystyle\frac{\tilde\omega}{\sqrt{A}}\right)
\ ,
&
x>\kappa_2 x_2
\ ,
\end{array}
\right.
\label{sig_ap}
\ee
with $\kappa_1$ and $\kappa_2\gtrsim 1$ and
$x_0<\kappa_1\,x_1$ is the value of $x$ at which we wish to
impose the initial conditions for $\chi(x)$ and its derivative~\footnote{Of course,
if we are interested in setting the initial conditions at $x_0\gg x_2$ the expressions
must be adjusted correspondingly.}.
The explicit expressions for the coefficients $C_0$, $C_1$, $C_2$ and $C_3$
are rather cumbersome and will be given in Sec.~\ref{app_cub}.
\begin{figure*}[!ht]
\raisebox{3.5cm}{$\sigma$}
\includegraphics[width=0.45\textwidth]{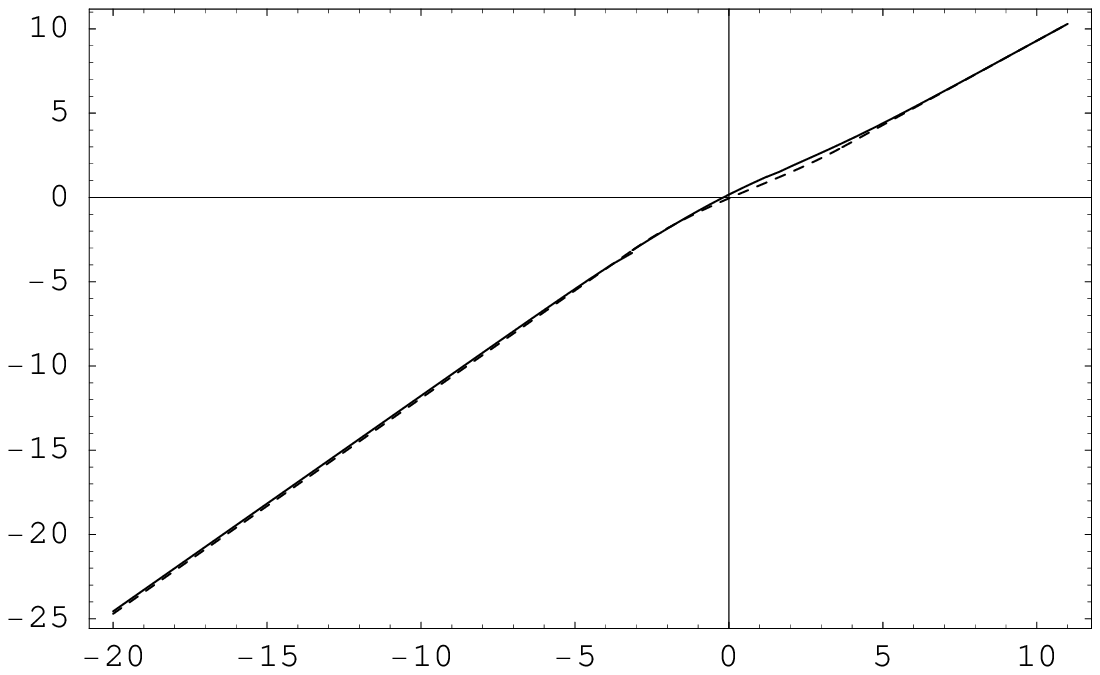}
\ \
\raisebox{3.5cm}{$\chi$}
\includegraphics[width=0.45\textwidth]{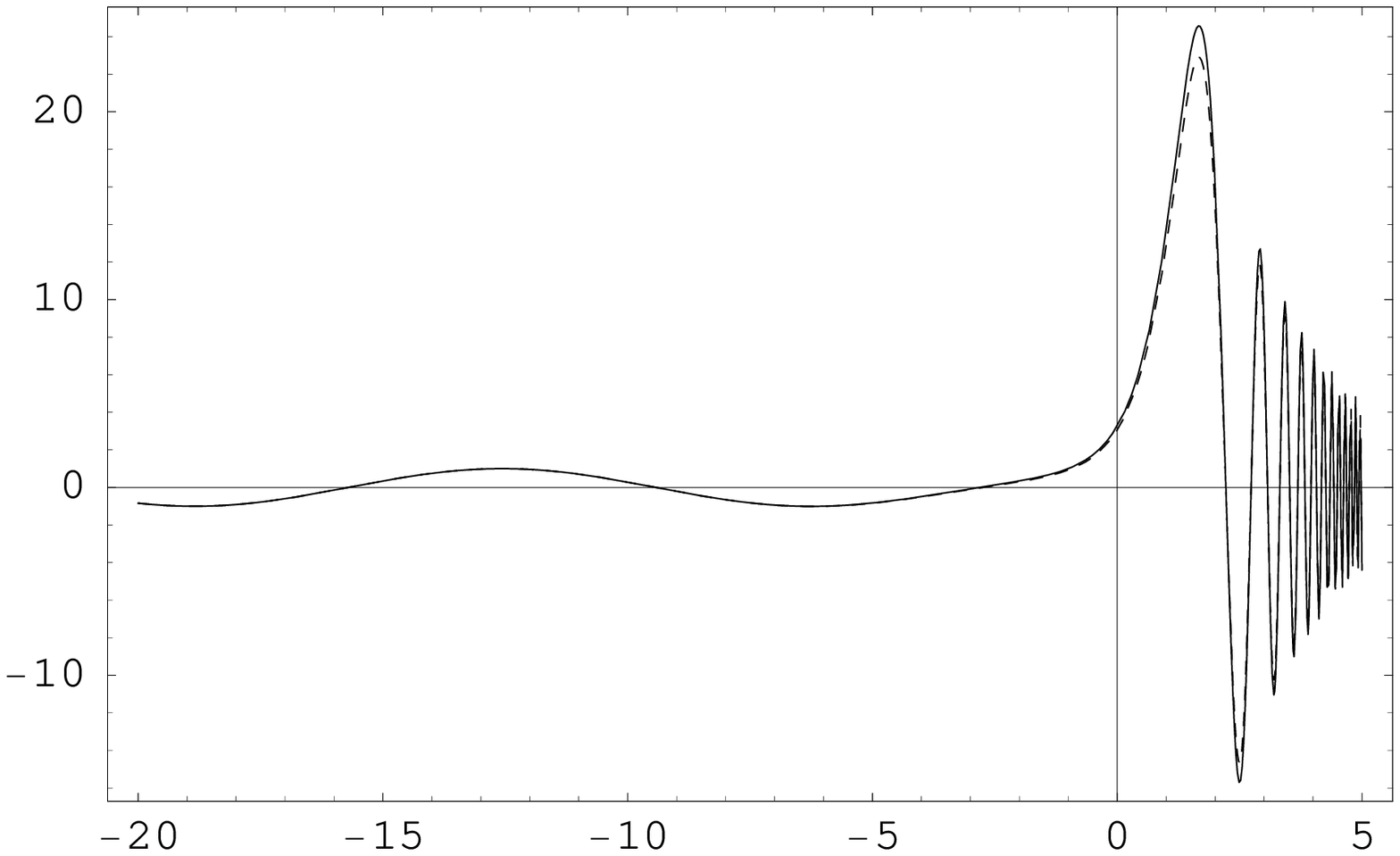}
\null
\hspace{-10.2cm}$x$\hspace{9cm}$x$
\null
\vspace{0.5cm}
\\
\includegraphics[width=0.32\textwidth]{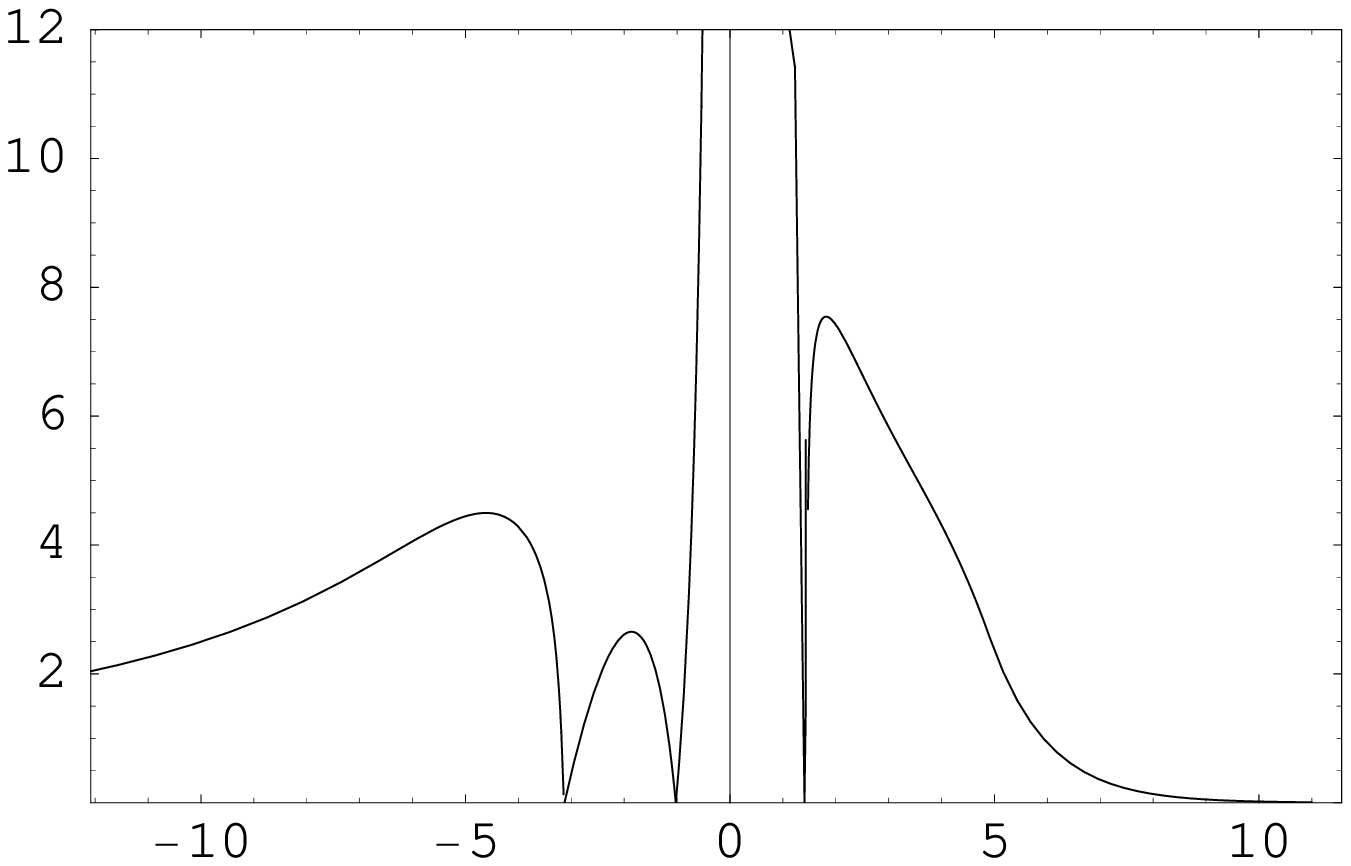}
\includegraphics[width=0.32\textwidth]{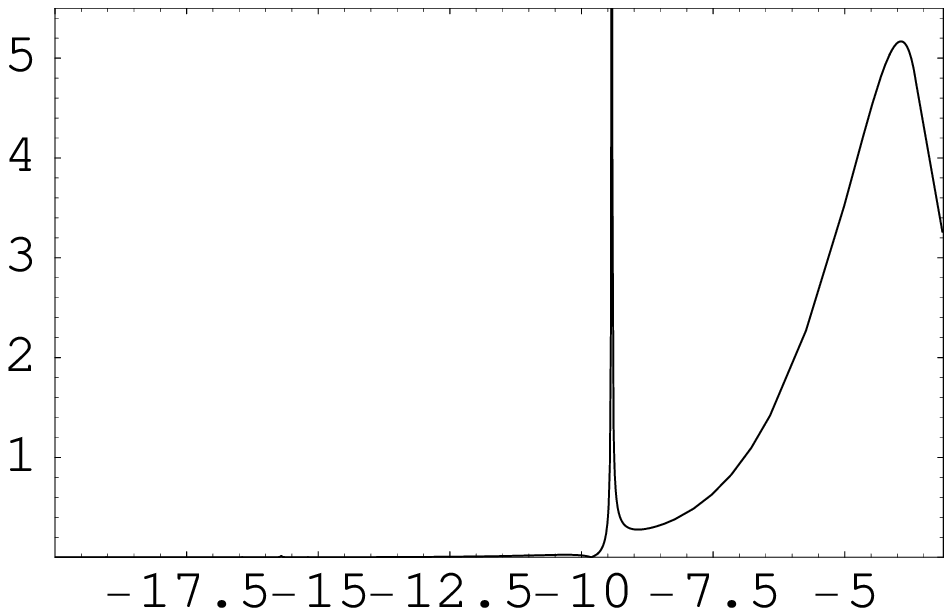}
\includegraphics[width=0.32\textwidth]{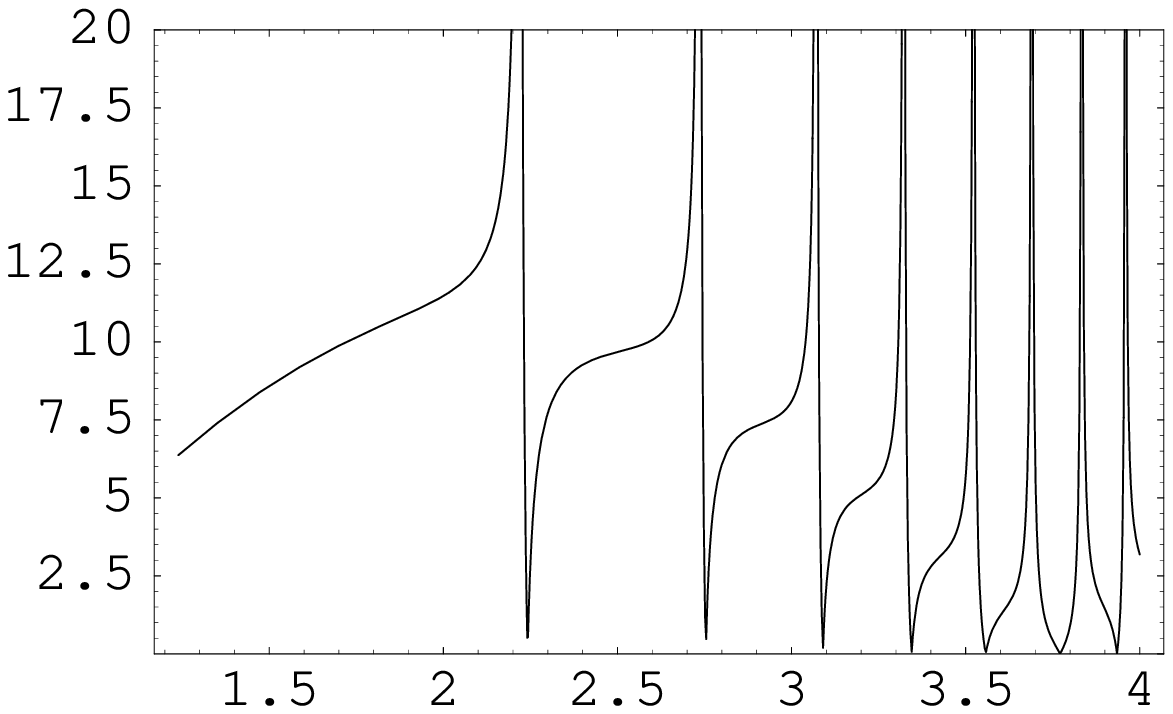}
\\
\null
\hspace{2.5cm}(a)\hspace{5.0cm}(b)\hspace{5cm}(c)
\caption{Upper left graph: The function $\sigma=\sigma(x)$
obtained by integrating Eq.~(\ref{new_EQ_int}) numerically (solid line)
and its approximate expression~(\ref{sig_ap}) with
$\kappa_1=1$ and $\kappa_2=3$  (dashed line).
Upper right graph:
Comparison between the exact solution (solid line)
and the approximate MCE solution~(\ref{chiMCE})
with $\sigma(x)$ evaluated numerically (dashed line) for
$C_+\approx 0.517-0.228\,i$ and $C_-\approx 0.517+0.228\,i$.
Panel~(a) shows the percentage error for the same approximate function
$\sigma$ given in the upper left graph.
Panel~(b) shows the percentage error for the approximate $\chi$ in the upper
right graph for $x<x_1$ and panel~(c) for $x_2<x$.
All plots are for $\tilde\omega=1/2$, $\ell=2$ as in Fig.~\ref{plot1}.}
\label{sigmaOFx}
\end{figure*}
\par
In the upper left panel of Fig.~\ref{sigmaOFx}, we plot $\sigma(x)$ for the same values of
$\tilde\omega$ and $\ell$ used in Fig.~\ref{plot1}.
The solid line represents the numerical solution of Eq.~(\ref{new_EQ_int})
and the dashed line the approximate expression~(\ref{sig_ap})
with $\kappa_1=1$ and $\kappa_2=3$.
It is clear that these values of $\kappa_1$ and $\kappa_2$ already lead to a very good
approximation for $\sigma$ around the turning points $x_1$ and $x_2$
and that larger values of $\kappa_1$ or $\kappa_2$
would not improve significantly the final result.
In the upper right panel of Fig.~\ref{sigmaOFx}, we compare an MCE approximate
wave-function with the exact solution (solid line) obtained numerically
for the same case.
The MCE solution shown by a dashed line is given by the
expression~(\ref{chiMCE}) with $U(\sigma)$ as in Eq.~(\ref{sol_morse}) and
$\sigma(x)$ determined by solving Eq.~(\ref{new_EQ_int}) numerically.
The coefficients $C_\pm$ are fixed by imposing that the MCE mode
and its derivative equal the corresponding values of the chosen numerical
solution at $x=-20$.
Since the two curves coincide almost everywhere, it is clear that solving
the approximate equation~(\ref{new_EQ_int})
for $\sigma(x)$ is, to all practical extents, equivalent to solving the original
equation~(\ref{exact_EQ_MCE}).
\par
In order to further clarify this point, in panel~(a) of Fig.~\ref{sigmaOFx}, we plot the
relative difference (in percent) between the approximate expression~(\ref{sig_ap})
and the exact numerical solution of Eq.~(\ref{new_EQ_int}).
We can see that the error is less than $10\%$ almost everywhere, except
near the zero of $\sigma(x)$.
In the same figure, we also show the percentage error for the approximate
solution shown in Fig.~\ref{sigmaOFx} on the left of $x_1$ and on the
right of $x_2$.
The latter error remains small, too, except around the zeros of $\chi$
due to the slight difference in phases.
\begin{figure*}[t]
\includegraphics[width=0.32\textwidth]{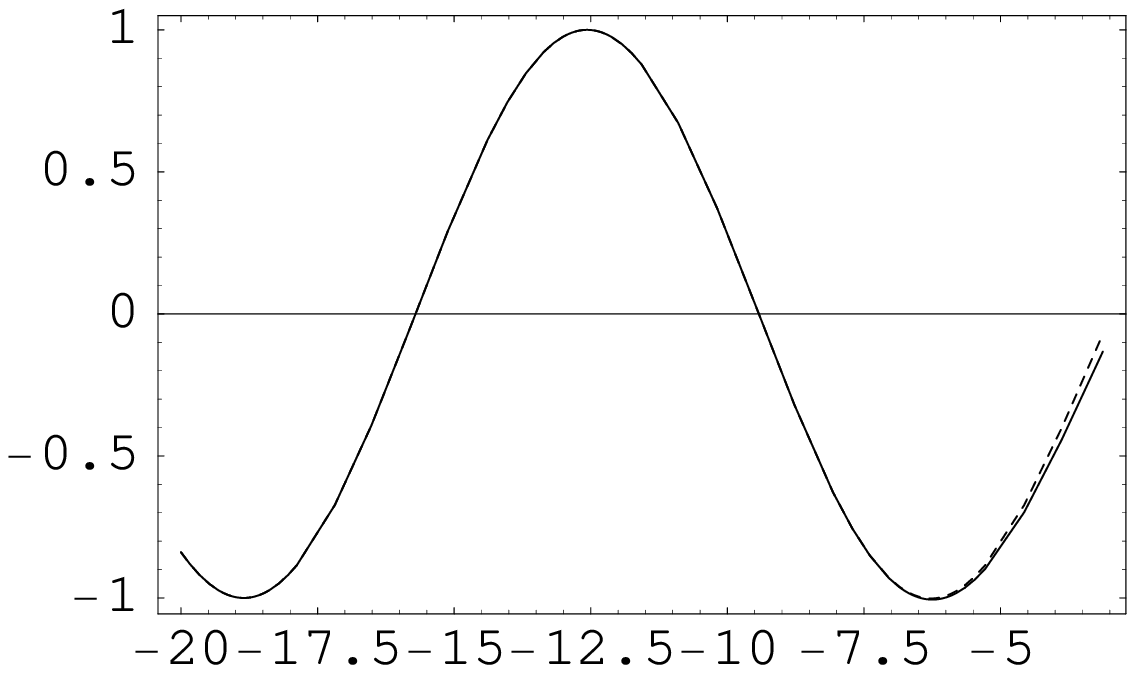}
\includegraphics[width=0.32\textwidth]{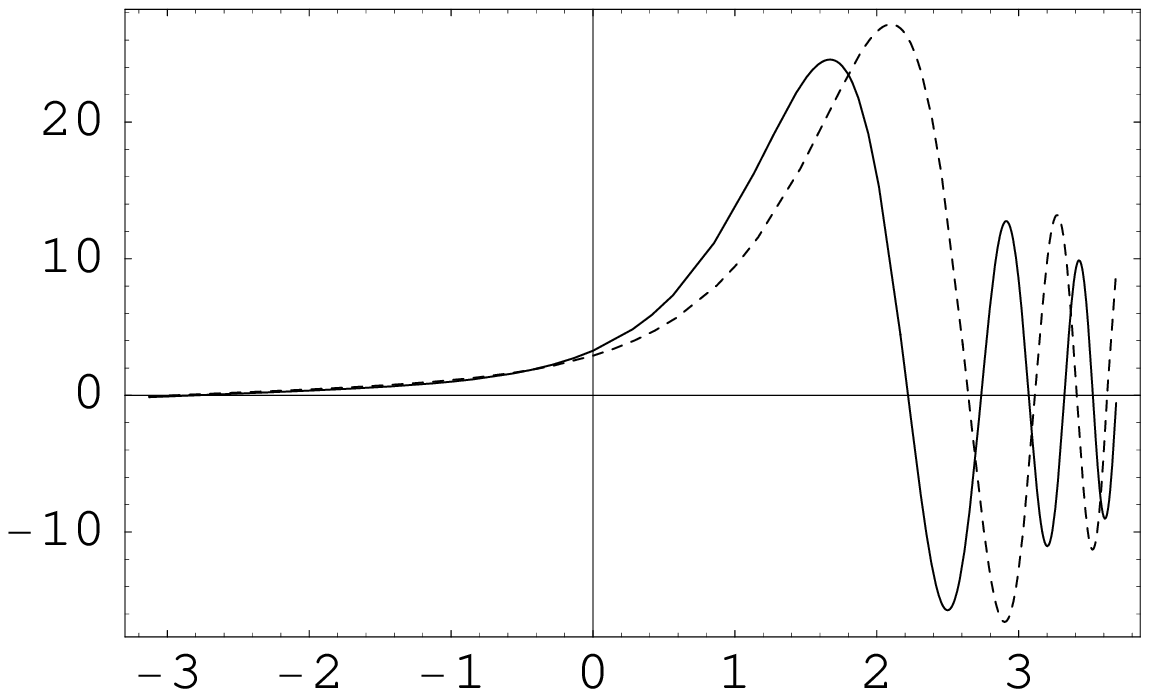}
\includegraphics[width=0.32\textwidth]{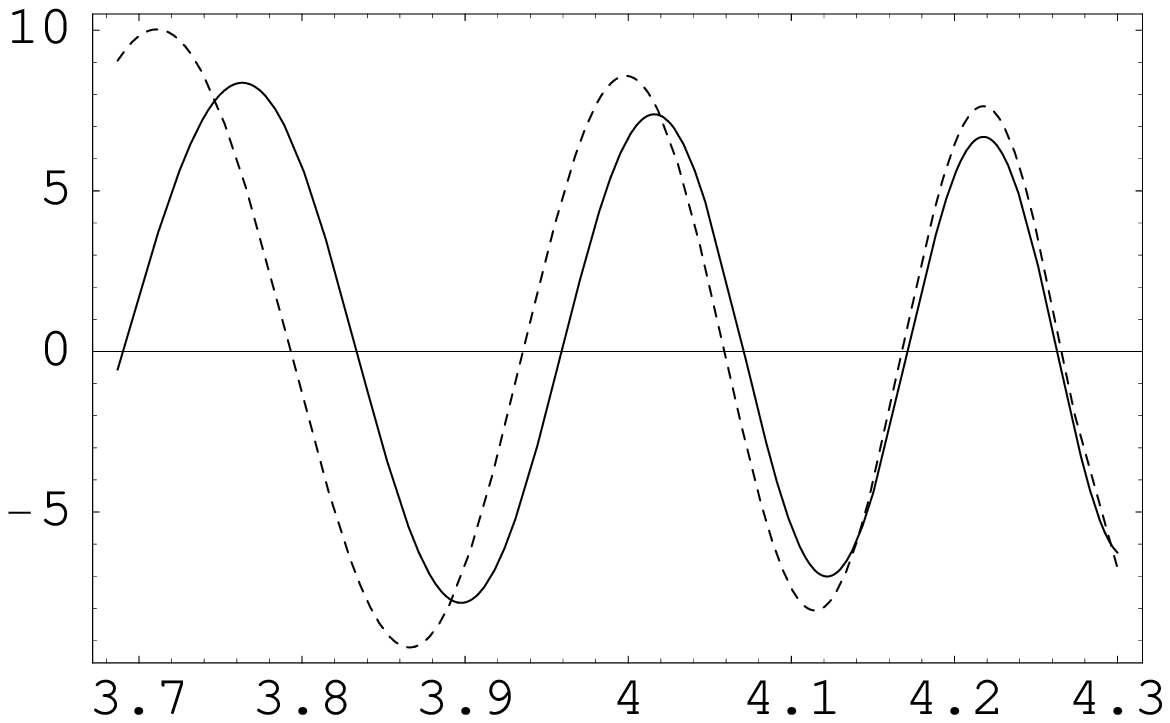}
\\
\null
\hspace{2.6cm}(a)\hspace{4.7cm}(b)\hspace{4.8cm}(c)
\caption{Comparison between the exact solution (solid line)
and the approximate MCE solution~(\ref{chiMCE})
with $\sigma(x)$ given in Eq.~(\ref{sig_ap}) with
$\kappa_1=1$ and $\kappa_2=3$
(dashed line) for $\tilde\omega=1/2$ and $\ell=2$ (same as in Figs.~\ref{plot1}
and \ref{sigmaOFx}).
%and $C_+\approx 0.517-0.228\,i$, $C_-\approx 0.517+0.228\,i$
Panel (a) shows the result for $x<\kappa_1\,x_1$, panel (b) for
$\kappa_1\,x_1<x<\kappa_2\,x_2$ and panel (c) for $x>\kappa_2\,x_2$.}
\label{final}
\end{figure*}
\par
Finally, in Fig.~\ref{final}, the dashed lines represent the same MCE solution
with $\sigma(x)$ given by its analytical approximation in Eq.~(\ref{sig_ap})
with $\kappa_1=1$ and $\kappa_2=3$.
This is the fully analytical expression obtained from the application
of the MCE and is still remarkably good, with a more significant error
around the larger turning point ($x_2\approx 1.23$).
From the three graphs in Fig.~\ref{final} and the upper left panel in Fig.~\ref{sigmaOFx},
it clearly appears that such an error develops in the
region between the two turning points, where the discrepancy between
the exact $\sigma(x)$ and the approximate expression~(\ref{sig_ap}) is
indeed larger.
In this respect, let us note that, had we imposed initial conditions
at $x\gg x_2$, the larger error would have occurred around the
smaller turning point ($x_1\approx-3.13$).
Moreover, when we used a linear (or quadratic) interpolating function
for $\kappa_1\,x_1<x<\kappa_2\,x_2$, the accuracy was (expectedly)
worse.
Should one need better accuracy, a higher order polynomial interpolation
between $\kappa_1\,x_1$ and $\kappa_2\,x_2$ could therefore be used instead. 
\par
Let us now make a few remarks.
Firstly, the application of the MCE is not as straightforward as
the standard WKB approximation.
In fact, the MCE usually requires solving some technical tasks specific to
the problem at hand in order to obtain the argument $\sigma=\sigma(x)$
of the known function $U=U(\sigma)$.
In the present case, with two turning points, all expressions become
more involved.
The best approximation for $\sigma(x)$ we found convenient to display
in Eq.~(\ref{sig_ap}) makes use of a cubic interpolation over
the region containing the two turning points and is given in terms of
two arbitrary parameters $\kappa_1$ and $\kappa_2$ which
can be fixed in order to minimize the error.
Of course, one could employ higher order interpolating functions
and further improve the result at the price of complexity if better
accuracy is needed. Our aim for the work presented in this Appendix was mainly to test the effectiveness
of the MCE to produce approximate wave modes on black hole backgrounds.
\par
In the next three technical Sections we give explicit expressions for the turning points, the calculation of the
quantity $\xi$ and some details of the cubic interpolation used to obtain $\sigma(x)$.
\subsection{Turning Points}
\label{app_TPs}
If we define $w={\rm e}^{x}$, we see that $\omega^2(x)=0$
for the exact frequency in Eq.~(\ref{freq_ex}) becomes
the quartic polynomial equation 
\begin{subequations}
\be
w^4 + 4\,w^3 + b\,w^2 + c\,w + 1 = 0
\ ,
\ee
with
\be
b
&\!=\!&
6 - \frac{1}{4\,{\tilde\omega }^2} - \frac{\ell\left(\ell+1\right)}{{\tilde\omega }^2}
\\
c
&\!=\!&
4 - \frac{1}{2\,{\tilde\omega }^2} - \frac{\ell\left(\ell+1\right)}{{\tilde\omega }^2}
\ .
\ee
\end{subequations}
Its four roots  can be written as
\begin{subequations}
\be
w_i=
-1
+\frac{\epsilon_i}{2}
\sqrt{
4-\frac{2}{3}\,b
+\frac{\gamma}{3}
}
-\frac{\theta_i}{2}
\sqrt{
8 - \frac{4}{3}\,b
+\epsilon_i\frac{4\,b -16 - 2\,c}{{\sqrt{4-\frac{2}{3}\,b
+\frac{\gamma}{3}
}}}
-\frac{\gamma}{3}
}
\ ,
\ee
where $i=1,\ldots,4$ and
\be
\!\!\!\!\!\!\!\!\!\!\!\!\!\!\!\!
\alpha
&\!=\!&
12 + b^2 - 12\,c
\\
\!\!\!\!\!\!\!\!\!\!\!\!\!\!\!\!
\beta
&\!=\!&
2\,b^3 - 36\,b\,c + 27\,c^2 + 432 - 72\,b
\\
\!\!\!\!\!\!\!\!\!\!\!\!\!\!\!\!
\gamma
&\!=\!&
\frac{2^{\frac{1}{3}}\,\alpha}{\left(\beta+\sqrt{\beta^2-4\,\alpha^3}\right)^{\frac{1}{3}}}
+\frac{\left(\beta+\sqrt{\beta^2-4\,\alpha^3}\right)^{\frac{1}{3}}}{2^{\frac{1}{3}}}
\ .
\ee
\end{subequations}
The coefficients $\epsilon_1=\epsilon_2=-\epsilon_3=-\epsilon_4=
\theta_1=-\theta_2=\theta_3=-\theta_4=1$, and the values of $\tilde\omega$ and $\ell$
determine the signs of the roots.
There can be at most two positive roots, in which case, say,
$w_3\le w_4\le 0\le w_1={\rm e}^{x_1}\le w_2={\rm e}^{x_2}$, and only
$x_1$ and $x_2$ can be valid turning points.
The critical value of $\tilde\omega$ in Eq.~(\ref{om_c}) is then determined by
the condition $0\le w_1=w_2$.
\subsection{Evaluation of $\xi$}
\label{app_xi}
In order to evaluate the integral in Eq.~(\ref{third_cond}), we start from the general
relation (again with $w={\rm e}^x$)
\be
&&\frac{1}{\tilde\omega}
\int_{x_1}^{\ln(w)}\sqrt{-\,\omega^2(y)}\,{\d}\,y
=
\sqrt{\frac{(w-w_3) (w-w_1) (w-w_2)}{w-w_4}}
+\frac{1}{\sqrt{(w_1-w_3)(w_4-w_2)}}
\nonumber
\\
&&\quad\quad
\times
\left\{
(w_3-w_1)(w_4-w_2)\,{\rm E}(z;W)
-(w_4-w_1)(w_4-w_2)\,{\rm F}(z;W)
\phantom{\frac{A}{B}}
\right.
\nonumber
\\
&&\quad\quad
\phantom{\times\ \ }
\left.
+(w_4-w_1)(w_3+w_4+w_1+w_2+2)\,
{\rm \Pi}\!\left[\frac{w_1-w_2}{w_4-w_2};z;W\right]
\right.
\nonumber
\\
&&\quad\quad
\phantom{\times\ \ }
\left.
-2\,(w_4-w_1)(w_3+1)(w_2+1)\,{\rm \Pi}\!
\left[\frac{(w_4+1)(w_1-w_2)}{(w_1+1)(w_4-w_2)};z;W\right]
\right.
\nonumber
\\
&&\quad\quad
\phantom{\times\ \ }
\left.
+2\,(w_4-w_1)w_3\,w_2\,{\rm \Pi}\!
\left[\frac{w_4\,(w_1-w_2)}{w_1\,(w_4-w_2)};z;W\right]
\right\}
\ ,
\label{xi_gen}
\ee
where E, F and ${\rm \Pi}$ are elliptic functions~\cite{abram},
\be
z={\rm ArcSin}\!\left[\sqrt{\frac{(w-w_1)(w_4-w_2)}{(w-w_4)(w_1-w_2)}}\right]
\ ,
\ee
\be
W=\frac{(w_3-w_4)(w_1-w_2)}{(w_3-w_1)(w_4-w_2)}
\ ,
\ee
and $w_i={\rm e}^{x_i}$ are the roots of $\omega^2(x)$
as given in the previous Section.
The above expression evaluated at $w=w_2$ yields
\be
&&
\xi=
\frac{\tilde\omega}{\sqrt{(w_1-w_3)(w_4-w_2)}}
\left\{(w_3-w_1)(w_4-w_3)\,{\rm E}(W)
+(w_4-w_1)\,(w_2-w_4)\,{\rm K}(W)
\right.
\nonumber
\\
&&
\left.
+(w_4-w_1)\,(w_3+w_4+w_1+w_2+2)\,{\rm \Pi}\!
\left[\frac{w_1-w_2}{w_4-w_2};W\right]
\right.
\nonumber
\\
&&
\left.
-2 (w_3+1)(w_2+1)(w_4-w_1)\,{\rm \Pi}\!
\left[\frac{(w_4+1)(w_1-w_2)}{(w_1+1)(w_4-w_2)};\!W\!\right]
\right.
\nonumber
\\
&&
\phantom{\times\ \ }
\left.
+2\,w_3\, w_2\,(w_4-w_1)\,{\rm \Pi}\!
\left[\frac{w_4\,(w_1-w_2)}{w_1\,(w_4-w_2)};W\right]
\right\}
\ ,
\label{xi}
\ee
in which we used E$(\pi/2;q)=$E$(q)$,
F$(\pi/2;q)=$K$(q)$ and
$\Pi(p;\pi/2;q)=\Pi(p,q)$~\cite{abram}.
\subsection{Cubic interpolation}
\label{app_cub}
In Section~\ref{sMCE_BH}, we use the analytic approximation for $\sigma(x)$
in Eq.~(\ref{sig_ap}) which involves the cubic
interpolation
\be
\sigma=C_0+C_1\,x+C_2\,x^2+C_3\,x^3
\ ,
\ee
for $\kappa_1\,x_1<x<\kappa_2\,x_2$.
The above coefficients $C_i$, with $i=0,\ldots,3$, can be easily expressed
in terms of the parameters
$A$, $B$ and $D$
for the Morse potential~(\ref{comp_freq}) as
\be
\!\!\!\!\!\!\!\!\!\!\!
&&
C_0=
{\kappa_1}\,{x_1}\,
\frac{\left(2\,{\sqrt{D}-1} \right) \,{{\kappa_2}}^2\,{{x_2}}^2\,
\left( {\kappa_2}\,{x_2} - {\kappa_1}\,{x_1} \right)
-2\,{\sqrt{D}}\,\ln \left(2\,{\sqrt{A}}\right)
{\kappa_1}\,{x_1}
\left(3\,{\kappa_2}\,{x_2}- {\kappa_1}\,{x_1} \right)}
{2\,{\sqrt{D}}\,{\left( {\kappa_2}\,{x_2} - {\kappa_1}\,{x_1} \right) }^3}
\nonumber
\\
\!\!\!\!\!\!\!\!\!\!\!
&&
C_1=
\frac{\left( {\kappa_2}\,{x_2} - {\kappa_1}\,{x_1} \right)
\left[
{\kappa_2}\,{x_2}\,\left( 2\,{\kappa_1}\,{x_1} + {\kappa_2}\,{x_2} \right)
-2\,{\sqrt{D}}\,{\kappa_1}\,{x_1}\,
\left( 4\,{\kappa_2}\,{x_2} -{\kappa_1}\,{x_1} \right)
\right] }
{2\,{\sqrt{D}}\,{\left( {\kappa_2}\,{x_2} - {\kappa_1}\,{x_1} \right) }^3}
\nonumber
\\
\!\!\!\!\!\!\!\!\!\!\!
&&\quad\quad\quad
+\frac{12\,{\sqrt{D}}\,\ln \left(2\,{\sqrt{A}}\right)\,{\kappa_1}\,{\kappa_2}\,{x_1}\,{x_2} 
}
{2\,{\sqrt{D}}\,{\left( {\kappa_2}\,{x_2} - {\kappa_1}\,{x_1} \right) }^3}
\\
\!\!\!\!\!\!\!\!\!\!\!
&&
C_2=
\frac{6\,{\sqrt{D}}\,\ln \left(2\,{\sqrt{A}}\right)
\left( {\kappa_1}\,{x_1} + {\kappa_2}\,{x_2} \right)  +
\left(2\,{\sqrt{D}}-1 \right)
\left({{\kappa_1}}^2\,{{x_1}}^2 + {\kappa_1}\,{\kappa_2}\,{x_1}\,{x_2}
-2\,{{\kappa_2}}^2\,{{x_2}}^2 \right) }{2\,{\sqrt{D}}
{\left( {\kappa_1}\,{x_1} - {\kappa_2}\,{x_2} \right) }^3}
\nonumber
\\
\!\!\!\!\!\!\!\!\!\!\!
&&
C_3=
\frac{4\,{\sqrt{D}}\,\ln \left(2\,{\sqrt{A}}\right)
-\left(2\,{\sqrt{D}}-1\right) \,\left( {\kappa_2}\,{x_2} - {\kappa_1}\,{x_1} \right) }
{2\,{\sqrt{D}}\,{\left( {\kappa_2}\,{x_2} - {\kappa_1}\,{x_1} \right) }^3}
\nonumber
\ .
\ee
Finally, one can express everything in terms of the original parameters
$\tilde\omega$ and $\ell$ by making use of Eqs.~(\ref{coeff_A})-(\ref{coeff_D}).

\end{appendix}

\bibliographystyle{unsrt}
\bibliography{biblio}

\begin{thebibliography}{10}

\bibitem{guth}
A.~H. Guth.
\newblock {\em Phys. Rev. D}, 23:347, 1981.

\bibitem{weinberg}
S.~Weinberg.
\newblock {\em Gravitation and Cosmology}.
\newblock John Wiley \& Sons, New York, US, 1972.

\bibitem{KTbook}
E.~W. Kolb and M.~S. Turner.
\newblock {\em The Early Universe}.
\newblock Addison-Wesley, US, 1990.

\bibitem{KS}
H.~Kodama and M.~Sasaki.
\newblock {\em Prog. Theor. Phys. Suppl.}, 78:1, 1984.

\bibitem{mukh_thory}
\mbox{V. F. Mukhanov, H. A. Feldman and R. H. Brandenberger}.
\newblock {\em Phys. Rep.}, 215:203, 1992.

\bibitem{sachs}
R.~K. Sachs and A.~M. Wolfe.
\newblock {\em Astrophys. J.}, 147:73, 1967.

\bibitem{silk1a}
J.~Silk.
\newblock {\em Nature}, 215:1155, 1967.

\bibitem{silk1b}
J.~Silk.
\newblock {\em Astrophys. J.}, 151:459, 1968.

\bibitem{PeeblesYu}
P.~J.~E. Peebles and J.~T. Yu.
\newblock {\em Astrophys. J.}, 162:815, 1970.

\bibitem{doro}
\mbox{A. G. Doroshkevich, Ya. B. Zeldovich, and R. A. Sunyaev}.
\newblock {\em Sov. Astron.}, 22:523, 1978.

\bibitem{wilson}
M.~L. Wilson and J.~Silk.
\newblock {\em Astrophys. J.}, 243:14, 1981.

\bibitem{sakharov}
A.~D. Sakharov.
\newblock {\em Sov. Phys. JETP}, 22:241, 1965.

\bibitem{cobe}
C.~L.~Bennet {\it et al.}
\newblock {\em Astrophys. J. Lett.}, 464:L1, 1996.

\bibitem{cob2}
M.~Tegmark.
\newblock {\em Astrophys. J. Lett.}, 464:L35, 1996.

\bibitem{boom1}
P.~de~Bernardis~{\it et al.}
\newblock {\em Astrophys. J.}, 564:559, 2002.

\bibitem{boom2}
C.~B.~Netterfield {\it et al.}
\newblock {\em Astrophys. J.}, 571:604, 2002.

\bibitem{maxima}
A.~T.~Lee {\it et al.}
\newblock {\em Astrophys. J.}, 561:L1, 2001.

\bibitem{dasi}
N.~W.~Halverson {\it et al.}
\newblock {\em Astrophys. J.}, 568:38, 2002.

\bibitem{archeops}
A.~Beno\^it {\it et al.}
\newblock {\em Astron. Astrophys.}, 399:L19, 2003.

\bibitem{DICK}
C.~Dickinson {\it et al.}
\newblock {\em Mon. Not. Roy. Astron. Soc.}, 353:732, 2004.

\bibitem{mason}
B.~S.~Mason {\it et al.}
\newblock {\em Astrophys. J.}, 591:540, 2003.

\bibitem{pearson}
T.~J.~Pearson {\it et al.}
\newblock {\em Astrophys. J.}, 591:556, 2003.

\bibitem{kuo}
C.~L.~Kuo {\it et al.}
\newblock {\em Astrophys. J.}, 600:32, 2004.

\bibitem{WMAP02}
L.~Page {\it et al.}
\newblock {\em Astrophys. J. Suppl.}, 148:223, 2003.

\bibitem{WMAP01}
C.~L.~Bennett {\it et al.}
\newblock {\em Astrophys. J. Suppl.}, 148:1, 2003.

\bibitem{LSS01}
M.~Tegmark {\it et al.}
\newblock {\em Astrophys. J.}, 606:702, 2004.

\bibitem{CL01}
N.~Bachall {\it et al.}
\newblock {\em Astrophys. J.}, 585:182, 2003.

\bibitem{BBN01}
B.~Fields and S.~Sarkar.
\newblock {\em Phys. Lett. B}, 592:202, 2004.

\bibitem{bardeen}
J.~M. Bardeen.
\newblock {\em Phys. Rev. D}, 22:1882, 1980.

\bibitem{mukh1}
V.~F. Mukhanov.
\newblock {\em Sov. Phys. JETP Lett.}, 41:493, 1985.

\bibitem{mukh2}
V.~F. Mukhanov.
\newblock {\em Sov. Phys. JETP Lett.}, 67:1297, 1988.

\bibitem{bender}
C.~M. Bender and S.~A. Orszag.
\newblock {\em Advanced mathematical methods for scientists and engineers}.
\newblock McGraw-Hill Book Company, 1978.

\bibitem{birrel}
N.~D. Birrel and P.~C.~W. Davies.
\newblock {\em Quantum Fields in Curved Space}.
\newblock Cambridge University Press, Cambridge, England, 1982.

\bibitem{jeffreys}
H.~Jeffreys.
\newblock {\em Proc. Lon. Math. Soc. (2)}, 23:428, 1923.

\bibitem{langer34}
R.~E. Langer.
\newblock {\em Bull. Am. Math. Soc.}, 40:545, 1934.

\bibitem{hunt}
P.~Hunt and S.~Sarkar.
\newblock {\em Phys. Rev. D}, 70:103518, 2004.

\bibitem{langer}
R.~E. Langer.
\newblock {\em Phys. Rev.}, 51:669, 1937.

\bibitem{schiff}
L.~I. Schiff.
\newblock {\em Quantum Mechanics}.
\newblock McGraw-Hill Book Company, 1952.

\bibitem{abram}
M.~Abramowitz and I.~Stegun.
\newblock {\em Handbook of mathematical functions with formulas, graphs, and
  mathematical table}.
\newblock Dover Publishing, 1965.

\bibitem{langer49}
R.~E. Langer.
\newblock {\em Trans. Am. Math. Soc.}, 67:461, 1949.

\bibitem{MG}
S.~C. Miller and R.~H. Good.
\newblock {\em Phys. Rev.}, 91:174, 1953.

\bibitem{dingle}
R.~B. Dingle.
\newblock {\em Appl. Sci. Res. B}, 5:345, 1956.

\bibitem{berry}
M.~V. Berry and K.~E. Mount.
\newblock {\em Rep. Prog. Phys.}, 35:315, 1972.

\bibitem{hecht}
C.~E. Hecht and J.~E. Mayer.
\newblock {\em Phys. Rev.}, 106:1156, 1957.

\bibitem{mori}
H.~Moriguchi.
\newblock {\em J. Phys. Soc. Japan}, 14:1771, 1959.

\bibitem{pechukas}
P.~Pechukas.
\newblock {\em J. Chem. Phys.}, 54:3864, 1971.

\bibitem{KLV}
\mbox{A. Kamenshchik, M. Luzzi and G. Venturi}.
\newblock Remarks on the method of comparison equations (generalized wkb
  method) and the generalized ermakov-pinney equation.
\newblock {\em arXiv:math-ph/0506017}, 2005.

\bibitem{Ermakov}
V.~P. Ermakov.
\newblock {\em Univ. Izv. Kiev, series III}, 9:1, 1880.

\bibitem{Pinney}
E.~Pinney.
\newblock {\em Proc. Amer. Math. Soc.}, 1:681, 1950.

\bibitem{Grad}
I.~S. Gradstein and I.~M. Ryzhik.
\newblock {\em Tables of integrals, series and products}.
\newblock Academic Press, New York, 1980.

\bibitem{Lewis}
H.~R. Lewis.
\newblock {\em J. Math. Phys.}, 9:1976, 1968.

\bibitem{Milne}
W.~E. Milne.
\newblock {\em Phys. Rev.}, 35:863, 1930.

\bibitem{Espinosa}
P.~Espinosa.
\newblock {\em Ermakov-Lewis dynamic invariants with some applications}.
\newblock MS Thesis, Instituto de F\'{i}sica - Universidad de Guanajuato,
  Le\`{o}n, 2000.

\bibitem{Fano}
U.~Fano and A.~R.~P. Rau.
\newblock {\em Atomic Collisions and Spectra}.
\newblock Academic Press, Orlando, 1986.

\bibitem{Gray}
C.~J. Eliezer and A.~Gray.
\newblock {\em SIAM J. Appl. Math.}, 30:46, 1976.

\bibitem{ioffe}
M.~V. Ioffe and H.~J. Korsch.
\newblock {\em Phys. Lett. A}, 311:200, 2003.

\bibitem{Ray}
J.~R. Ray and J.~L. Reid.
\newblock {\em Phys. Lett. A}, 71:317, 1979.

\bibitem{martin_schwarz2}
J.~Martin and D.~J. Schwarz.
\newblock {\em Phys. Rev. D}, 67:083512, 2003.

\bibitem{WKB1}
\mbox{R. Casadio, F. Finelli, M. Luzzi and G. Venturi}.
\newblock {\em Phys. Rev. D}, 71:043517, 2005.

\bibitem{MCE}
\mbox{R. Casadio, F. Finelli, A. Kamenshchik, M. Luzzi and G. Venturi}.
\newblock {\em JCAP}, 04:011, 2006.

\bibitem{wang_mukh}
\mbox{L. Wang, V. F. Mukhanov and P. J. Steinhardt}.
\newblock {\em Phys. Lett. B}, 414:18, 1997.

\bibitem{HHHJM}
\mbox{S. Habib, A. Heinen, K. Heitmann, G. Jungman and C. Molina-Par\'{\i}s}.
\newblock {\em Phys. Rev. D}, 70:083507, 2004.

\bibitem{olver}
F.~W.~J. Olver.
\newblock {\em Asymptotics and special functions}.
\newblock AKP Classics, Wellesley, MA, 1997.

\bibitem{lythstewart}
D.~H. Lyth and E.~D. Stewart.
\newblock {\em Phys. Lett. B}, 274:168, 1992.

\bibitem{abbott}
L.~F. Abbott and M.~B. Wise.
\newblock {\em Nucl. Phys. B}, 244:541, 1984.

\bibitem{martin_PL}
J.~Martin and D.~J. Schwarz.
\newblock {\em Phys. Rev. D}, 57:3302, 1998.

\bibitem{SL}
E.~D. Stewart and D.~H. Lyth.
\newblock {\em Phys. Lett. B}, 302:171, 1993.

\bibitem{LLMS}
\mbox{S. M. Leach, A. R. Liddle, J. Martin and D. J. Schwarz}.
\newblock {\em Phys. Rev. D}, 66:023515, 2002.

\bibitem{martin_sl_roll}
J.~Martin and D.~J. Schwarz.
\newblock {\em Phys. Rev. D}, 62:103520, 2000.

\bibitem{WKB_let}
\mbox{R. Casadio, F. Finelli, M. Luzzi and G. Venturi}.
\newblock {\em Phys. Lett. B}, 625:1, 2005.

\bibitem{WKB_SRprD}
\mbox{R. Casadio, F. Finelli, M. Luzzi and G. Venturi}.
\newblock {\em Phys. Rev. D}, 72:103516, 2005.

\bibitem{lewin}
L.~Lewin.
\newblock {\em Dilogarithms and associated functions}.
\newblock Macdonald, 1958.

\bibitem{H_MP}
\mbox{S. Habib, K. Heitmann, G. Jungman and C. Molina-Par\'{\i}s}.
\newblock {\em Phys. Rev. Lett.}, 89:281301, 2002.

\bibitem{HHHJ}
\mbox{S. Habib, A. Heinen, K. Heitmann and G. Jungman}.
\newblock {\em Phys. Rev. D}, 71:043518, 2005.

\bibitem{SGong}
E.~D. Stewart and J.~O. Gong.
\newblock {\em Phys. Lett. B}, 510:1, 2001.

\bibitem{KT}
A.~Kosowsky and M.~S. Turner.
\newblock {\em Phys. Rev. D}, 52:1739, 1995.

\bibitem{linde}
A.~D. Linde.
\newblock {\em Phys. Lett. B}, 129:177, 1983.

\bibitem{BFV}
\mbox{C. Bertoni, F. Finelli and G. Venturi}.
\newblock {\em Phys. Lett. A}, 237:331, 1998.

\bibitem{FMVV_1}
\mbox{F. Finelli, G. Marozzi, G. P. Vacca, and G.Venturi}.
\newblock {\em Phys. Rev. D}, 65:103521, 2002.

\bibitem{polarski}
D.~Polarski and A.~A. Starobinsky.
\newblock {\em Class. Quant. Grav.}, 13:377, 1996.

\bibitem{LPB1}
\mbox{A. R. Liddle, P. Parsons and J. D. Barrow}.
\newblock {\em Phys. Rev. D}, 50:7222, 1994.

\bibitem{LPB2}
\mbox{J. E. Lidsey, A. R. Liddle, E. W. Kolb, E. J. Copeland, T. Barreiro and
  M. Abney}.
\newblock {\em Rev. Mod. Phys.}, 69:373, 1997.

\bibitem{chandra}
S.~Chandrasekhar.
\newblock {\em The mathematical theory of black holes}.
\newblock Oxford University Press, Oxford, 1992.

\bibitem{hawking1}
S.~W. Hawking.
\newblock {\em Nature}, 248:30, 1974.

\bibitem{hawking2}
S.~W. Hawking.
\newblock {\em Comm. Math. Phys.}, 43:199, 1975.

\bibitem{nollert}
H.~P. Nollert.
\newblock {\em Class. Quantum Grav.}, 16:R159, 1999.

\bibitem{teukolsky}
S.~A. Teukolsky.
\newblock {\em Astrophys. J.}, 185:635, 1973.

\bibitem{kanti_russ}
P.~Kanti and J.~March-Russell.
\newblock {\em Phys. Rev. D}, 66:024023, 2002.

\bibitem{RMNS}
\mbox{G. Rawitscher, C. Merow, M. Nguyen and I. Simbotin}.
\newblock {\em Am. J. Phys.}, 70:935, 2002.

\end{thebibliography}

\end{document}